%% file: main.tex
\documentclass[11pt,twoside,a4paper]{article}

\usepackage[utf8]{inputenc}
\usepackage[english]{babel}
\usepackage{hyperref}
\usepackage{graphicx}
\usepackage{wrapfig}
\usepackage{braket}

\usepackage{tcolorbox}

\usepackage{xcolor}
\hypersetup{
    colorlinks,
    linkcolor={blue!100!black},
    citecolor={blue!50!black},
    urlcolor={blue!80!black}
}

\usepackage{amsmath}

\numberwithin{equation}{section}
\numberwithin{figure}{section}
\numberwithin{table}{section}

\newcommand*\VF[1]{\mathbf{#1}}

\usepackage{helvet}

\usepackage{parskip}
\usepackage{titlesec}
\titlespacing{\section}{0pt}{\parskip}{\parskip}
\titlespacing{\subsection}{0pt}{\parskip}{\parskip}
\titlespacing{\subsubsection}{0pt}{\parskip}{\parskip}

\usepackage[margin=2cm]{geometry}

\usepackage{titlesec}
\titleformat{\section}
  {\normalfont\fontsize{11}{15}\bfseries}{\thesection}{1em}{}
  \titleformat{\subsection}
  {\normalfont\fontsize{11}{15}\bfseries}{\thesection}{1em}{}

\usepackage{etoolbox}
\makeatletter
\patchcmd{\@maketitle}{\LARGE}{\fontsize{14}{0}\selectfont}{}{}
\makeatother

\title{\textbf{Notes on Images and Communications}}
\author{Denis Martynov}
\date{}

\usepackage{setspace}

\begin{document}


\include{intro.tex}
\newpage
\include{acknowledgements.tex}
\newpage
\tableofcontents
\newpage
\include{week1.tex}
\newpage
\include{week2.tex}
\newpage
\include{week3.tex}
\newpage
\include{week4.tex}
\newpage
\include{week5.tex}
\newpage
\include{week6.tex}
\newpage
\include{week7.tex}
\newpage
\include{week8.tex}
\newpage
\include{week9.tex}
\newpage
\include{week10.tex}
\newpage
\include{week11.tex}
\newpage
\include{exercises}
\newpage
\bibliographystyle{unsrt}
\bibliography{bibliography.bib}

\end{document}

%% file: intro.tex
\begin{center} 
  {\fontsize{14pt}{4pt} \textbf{Notes on Images and Communication}}
  
  Denis Martynov

  Institute for Gravitational Wave Astronomy, School of Physics and Astronomy, University of Birmingham, Birmingham B15 2TT, United Kingdom
\end{center}

\vspace{5mm}

This is an active module taught to Bachelor’s and Master’s students at the University of Birmingham since 2021, covering selected topics in applied optics with an emphasis on imaging, lasers, and classical and quantum communication. The module covers both the theoretical foundations and experimental aspects of these topics, and explores a range of instrumentation examples, including optical and radio telescopes, adaptive optics, laser cutting systems, optical tweezers, laser interferometers, optical atomic clocks, optical coatings, coaxial cables and optical fibres, frequency combs, and quantum key distribution technologies.

We discuss five major sections
\begin{itemize}
    \item \textbf{Imaging} is devoted to the fundamentals of imaging across different wavelengths, with an emphasis on resolution and aberrations. We examine the operation of photographic film and optical sensors, including CCD and CMOS platforms. We also discuss how images are digitised, compressed, and processed, as well as convolution and deconvolution algorithms.

    \item \textbf{Applications of lasers}  discusses the spatial and temporal coherence of light, the propagation of laser radiation in optical systems, and its applications, including material processing, scattering phenomena, and optical trapping. We also examine lasing media, including atomic dopants in crystals, semiconductors, and gas mixtures.
    
    \item \textbf{Interference and precision measurements} is devoted to applications of laser beams that exploit their phase information. We discuss the quantum-limited resolution of laser interferometers and their applications in both fundamental physics and industrial settings, including gravitational-wave detectors, time keeping, gyroscopes, and coatings.
        
    \item \textbf{Classical communication}  considers communication via coaxial cables, low-orbit satellites, and optical fibres. We discuss the advantages and limitations of these techniques, including wavelength multiplexing using frequency combs, as well as modal, chromatic, and polarisation dispersion of optical pulses, together with absorption effects.

    \item \textbf{Quantum communication} discusses the limitations of classical communication, which relies on computational complexity for security. We then examine techniques for quantum key distribution and the generation of entangled photons using nonlinear crystals. We also discuss the practical challenges of quantum communication links, including exponential loss in transmission channels, and explore emerging approaches toward the quantum internet.
\end{itemize}

%% file: acknowledgements.tex
\textbf{Acknowledgements:} D.M.\ would like to thank the many students at Birmingham whose questions and feedback have helped shape and improve this module over the years. In particular, the \textit{Quantum Communication} section emerged directly from topics requested in the module questionnaire. D.M.\ gratefully acknowledges Prof.\ Kai Bongs and Dr.\ Conor Mow-Lowry, who taught the module prior to 2021 and established a substantial part of its agenda. D.M.\ also acknowledges the support of the Institute for Gravitational Wave Astronomy at the University of Birmingham.

%% file: week1.tex
\section{Imaging: Geometrical optics, ABCD matrices, diffraction limit}
\label{Lecture1}

We define imaging as a linear mapping between the intensity distribution in the source plane and the corresponding distribution on the recording surface:
\begin{equation}
\label{eq:w1_image_def}
I_{\rm image}(x,y) = \alpha\, I_{\rm source}\!\left(\frac{x}{m}, \frac{y}{m} \right),
\end{equation}
where $m$ is the magnification and $\alpha$ is a proportionality constant. In the ideal case described by Eq.~\ref{eq:w1_image_def}, the recorded intensity distribution reproduces the source distribution. In practice, however, both technical and fundamental limitations reduce the image quality: even a point source produces a finite spot in the image plane. This behaviour is characterised by the point spread function, which limits the resolution of the system. The aim of this lecture is to examine the principles of image formation and to analyse the effects of such imperfections in imaging systems.

In this lecture, we discuss
\begin{itemize}
\item history and principles of image formation techniques,
\item how optical systems transform rays with ABCD matrices,
\item defocusing and resolution of the optical system,
\item diffraction-limited resolution of the imaging system,
\item depth of focus of imaging systems.
\end{itemize}

\subsection{Historical outlook on image formation}

\begin{figure}[t!]
\centering
\includegraphics[height=3.35cm]{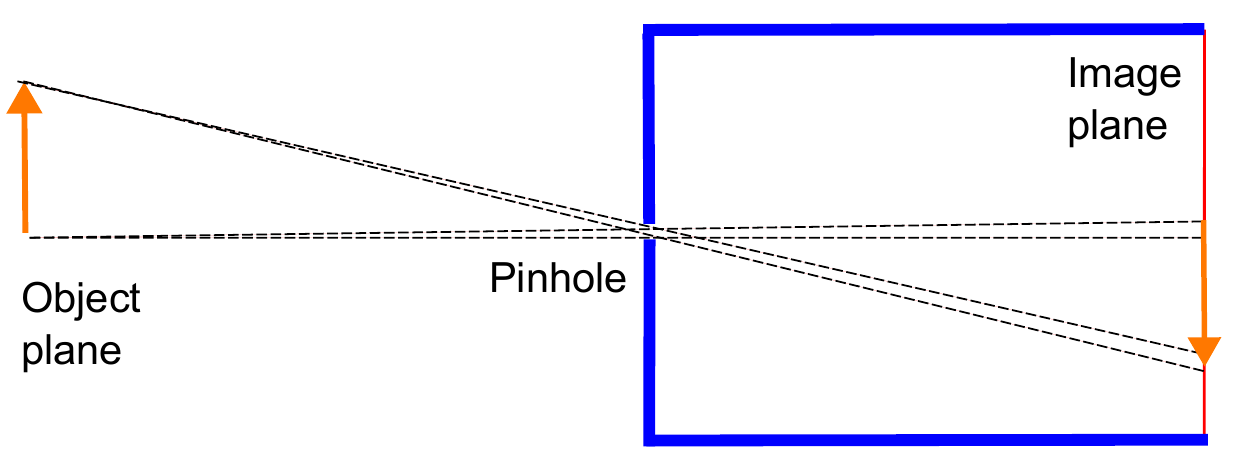} 
\hspace{1mm}
\includegraphics[height=3.35cm]{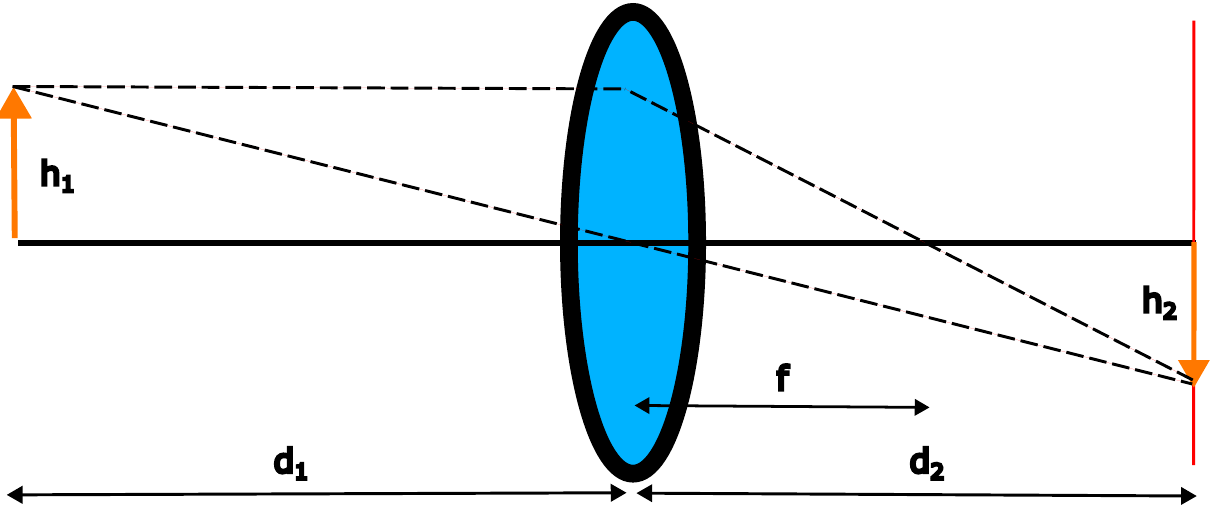} 
\caption{Ray traces in imaging with a Camera Obscura (left) and a lens (right).}
\label{fig:w1_cam_obs_lens}
\end{figure}

One of the first imaging devices was the Camera Obscura~\cite{gernsheim1955history}, which consisted of a dark room with a pinhole in one of the walls. No lenses were used in the first iterations of the system. Instead, light rays were constrained by the aperture as shown in Fig.~\ref{fig:w1_cam_obs_lens}\,(left). The camera achieved an intensity map from Eq.~\ref{eq:w1_image_def} with a precision given by the pinhole diameter. The Camera Obscura proved that light rays travel in straight lines (to a good approximation) and provided useful measurements of the Sun, eclipses of the Moon, and bright scenery on Earth. However, the images were dark because pinholes allowed a small amount of light into the room.

Lenses started to replace pinholes in Camera Obscuras in the 16th century. Though lenses were utilised in eye sight correction in the 13th century, it took another hundreds of years to embed lenses in Cameras Obscure. The lenses have improved the sharpness and brightness of the images because more light was accumulated by the lens aperture than by a pinhole. The Camera Obscura became widely used by artists for accurate drawings. The cameras also served as a scientific tool for studying light, optics, and astronomical phenomena, such as solar eclipses. The separation between the lens and the source, $d_1$, and the lens and the image, $d_2$, as shown in Fig.~\ref{fig:w1_cam_obs_lens}\,(right) are related via the lens equation
\begin{equation}\label{eq:w1_lens_eq}
    \frac{1}{d_1} + \frac{1}{d_2} = \frac{1}{f},
\end{equation}
where $f$ is the focal length of the lens. We can find the magnification factor using the equation
\begin{equation}\label{eq:w1_lens_mag}
    m = \frac{h_2}{h_1} = -\frac{d_2}{d_1} = \frac{f}{f-d_1},
\end{equation}
where the minus sign implies that the image is inverted relative to the original object if $d_1 > f$. In the geometric optics approach, the magnification can be arbitrarily large for convex lenses if $d_1 \approx f$. For concave lenses, however, the magnification factor is less than 1 because $f<0$. For $f=-d_1$, the magnification factor is 1/2.

Lenses can create ideal images, given by Eq.~\ref{eq:w1_image_def}, in the geometrical optics approach when rays travel in straight lines. However, the wave nature of light leads to point spreading and aberrations. Between the 17th and 19th centuries, the wave nature of light became evident through both theory and experiment~\cite{bornwolf1999principles}. In the 17th century, Huygens proposed that light propagates as a wave, offering an alternative to the geometrical approach. Newton, in contrast, emphasised a corpuscular theory, though his studies of dispersion hinted at wavelength-dependent effects. By the early 19th century, Young’s double-slit experiment demonstrated interference patterns, and the prediction and experimental confirmation of the bright (Poisson - Arago) spot in the centre of a shadow from a circular disk provided evidence of diffraction, a phenomenon in which light bends around obstacles.

Advances in the chemistry of salts led to the first recorded photographs~\cite{newhall1982history}. In the late 18th century, Schulze discovered that silver salts darken upon light exposure. Early 19th-century pioneers developed practical methods: Niépce produced the first durable image using bitumen of Judea coated on plates, and Daguerre exploited silver halides on polished metal plates to reduce exposure times and increase image clarity. Throughout the late 19th century, improvements in light-sensitive silver halide emulsions, glass plate coatings, and fixing chemicals enabled shorter exposures and reproducible images. By the early 20th century, these advances culminated in gelatin-silver photographic films, which became the standard for photography.

Recording the full intensity map from Eq.~\ref{eq:w1_image_def} requires an infinite number of light-sensitive elements in the image plane that is not practical. A typical size of the silver halide crystals is $\approx 1$\,um and defines the "pixel" size in the image plane. Digital photography emerged in the late 20th century with a similar pixel size. Early breakthroughs came with the invention of the Charge-Coupled Device (CCD) at Bell Labs, which converts incoming photons into electrical charge and transfers it across the chip for readout with low noise~\cite{howes1997charge}. Later developments introduced Complementary Metal-Oxide-Semiconductor sensors~\cite{theuwissen2007cmos}, which integrate amplification and readout circuitry at each pixel and allow faster operation and lower power consumption compared to CCD cameras.

Quantum imaging developed in the late 20th and early 21st centuries as advances in quantum optics enabled the use of entangled photons to form images with capabilities beyond classical limits. Early experiments demonstrated Ghost imaging~\cite{PhysRevA.52.R3429}, where an image can be reconstructed using correlations between photons even when the detector does not directly view the object, and Quantum illumination~\cite{PhysRevLett.101.253601}, which enhances detection sensitivity in noisy environments. These techniques rely on sources like spontaneous parametric down-conversion in non-linear crystals to generate correlated photon pairs and exploit entanglement to improve signal-to-noise ratios and resolution.

\subsection{Ray-transfer matrices}

In this section, we introduce ray-transfer (ABCD) matrices to describe the propagation of optical rays through complex systems of lenses. Remarkably, the same formalism can also be used to analyse the propagation of Gaussian laser beams through optical systems, as will be discussed in  Lecture~\ref{Lecture5}.

\begin{figure}[h]
\centering
\includegraphics[height=3.3cm]{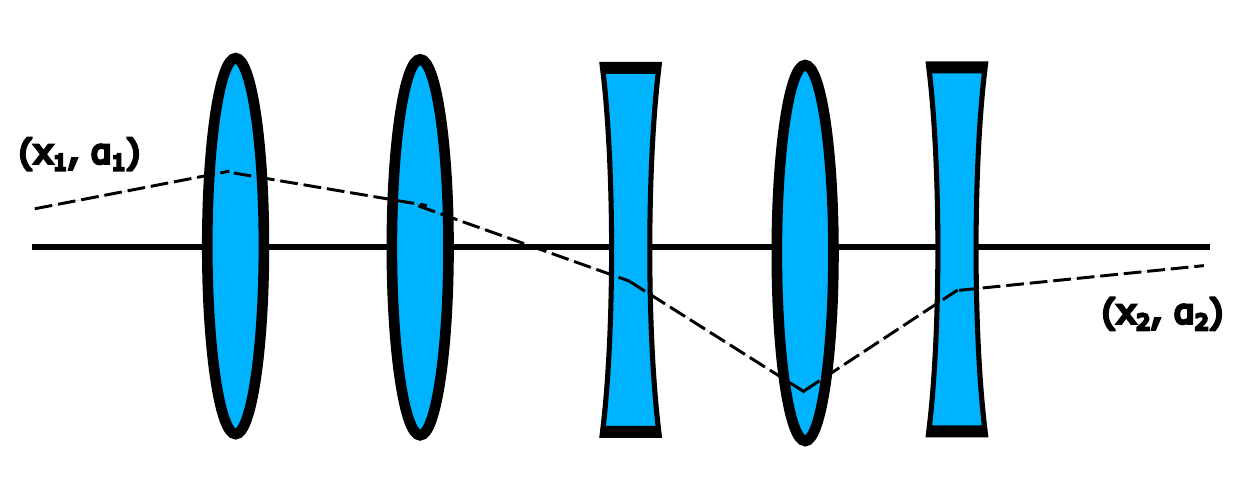} 
\hspace{3mm}
\includegraphics[height=3.3cm]{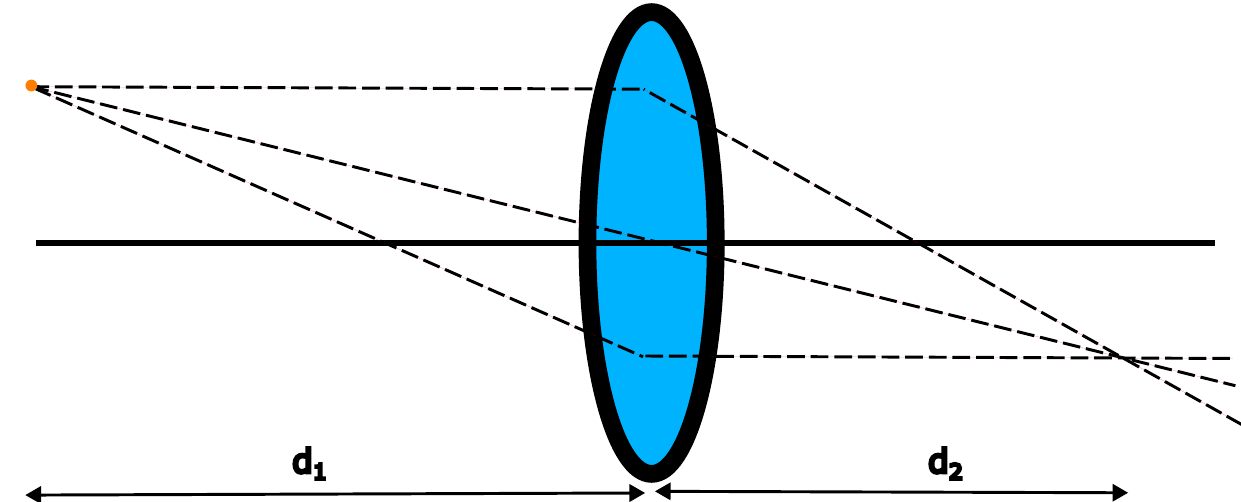} 
\caption{A ray travelling through a complex optical system (left) and a convergence of all rays in one dot in the image plane (right).}
\label{fig:w1_ABCD}
\end{figure}

Let's consider a ray of light, which starts from point $x_1$ and propagates at an angle $\alpha_1 \ll 1$ relative to the horizontal axis. We would like to compute the parameters $x_2$ and $\alpha_2$ after the ray travels a distance $d$. In free space, the angle does not change $\alpha_1 = \alpha_2$, and the distance to the horizontal axis becomes $x_2 = x_1 + \alpha_1 d$. We can write these two equations in matrix form
\begin{equation}
    \begin{pmatrix}
        x_2\\
        \alpha_2
    \end{pmatrix}
    =
    \begin{pmatrix}
        1 & d\\
        0 & 1
    \end{pmatrix}
    \begin{pmatrix}
        x_1\\
        \alpha_1
    \end{pmatrix}.
\end{equation}
Similarly, a thin lens does not change the position of the ray relative to the horizontal axis $x_1 = x_2$ but bends the ray's angle by $-x_1/f$. In the matrix form, we achieve the equation
\begin{equation}
    \begin{pmatrix}
        x_2\\
        \alpha_2
    \end{pmatrix}
    =
    \begin{pmatrix}
        1 & 0\\
        -\frac{1}{f} & 1
    \end{pmatrix}
    \begin{pmatrix}
        x_1\\
        \alpha_1
    \end{pmatrix}.
\end{equation}

The ABCD formalism allows us to express ray propagation in matrix form and compute the overall $2 \times 2$ transfer matrix of an optical system. For example, consider a ray that starts at position $x_1$ with an initial angle $\alpha_1$. After propagating a distance $d_1$, the ray passes through a sequence of lenses with focal lengths $f_1$, $f_2$, etc., separated by distances $d_2$, $d_3$, and so on, as shown in Fig.~\ref{fig:w1_ABCD}\,(left). We can determine the final position and angle of the ray by multiplying the appropriate ABCD matrices to obtain the overall transfer matrix $M$, as given by
\begin{equation}
    \begin{pmatrix}
        x_2\\
        \alpha_2
    \end{pmatrix}
    =
    \begin{pmatrix}
        1 & d_{N+1}\\
        0 & 1
    \end{pmatrix}
    \begin{pmatrix}
        1 & 0\\
        -\frac{1}{f_N} & 1
    \end{pmatrix}
    \begin{pmatrix}
        1 & d_N\\
        0 & 1
    \end{pmatrix}
    ...
    \begin{pmatrix}
        1 & 0\\
        -\frac{1}{f_1} & 1
    \end{pmatrix}
    \begin{pmatrix}
        1 & d_1\\
        0 & 1
    \end{pmatrix}
    \begin{pmatrix}
        x_1\\
        \alpha_1
    \end{pmatrix}
    = M
    \begin{pmatrix}
        x_1\\
        \alpha_1
    \end{pmatrix}.
\end{equation}

The ray-transfer formalism can also determine the image location $d_{N+1}$ and the corresponding magnification $x_2 / x_1$ after propagation through a complex optical system. Since all rays originating from a point source must converge to a single point in the image plane, the image position should be independent of the initial ray angle. This condition implies that the image distance $d_{N+1}$ can be found from the equation $M_{12}(d_{N+1}) = 0$, and the magnification is given by $m = M_{11}$.

This formalism can be applied, for example, to calculate the magnification of microscopes, which are used to enlarge small objects such as cells, bacteria, or viruses to make them visible to the human eye. As an illustration, we consider the lens system shown in Fig.~\ref{fig:w1_ABCD} (right) and compute the corresponding total ABCD matrix, given by
\begin{equation}
    \begin{pmatrix}
        x_2\\
        \alpha_2
    \end{pmatrix}
    =
     \begin{pmatrix}
        1 & d_2\\
        0 & 1
    \end{pmatrix}
    \begin{pmatrix}
        1 & 0\\
        -\frac{1}{f} & 1
    \end{pmatrix}
    \begin{pmatrix}
        1 & d_1\\
        0 & 1
    \end{pmatrix}
    \begin{pmatrix}
        x_1\\
        \alpha_1
    \end{pmatrix}
    =
    \begin{pmatrix}
        1-\frac{d_2}{f} & d_1+d_2 - \frac{d_1 d_2}{f}\\
        -\frac{1}{f} & 1-\frac{d_1}{f}
    \end{pmatrix}
    \begin{pmatrix}
        x_1\\
        \alpha_1
    \end{pmatrix}
    =
     M
    \begin{pmatrix}
        x_1\\
        \alpha_1
    \end{pmatrix},
\end{equation}
which shows that the condition $M_{12}(d_2) = 0$ reproduces the lens equation, consistent with Eq.~\ref{eq:w1_lens_eq}, and $M_{11}$ corresponds to the magnification factor, in agreement with Eq.~\ref{eq:w1_lens_mag}.

\subsection{Image defocusing}

\begin{figure}[t]
\centering
\includegraphics[height=3.3cm]{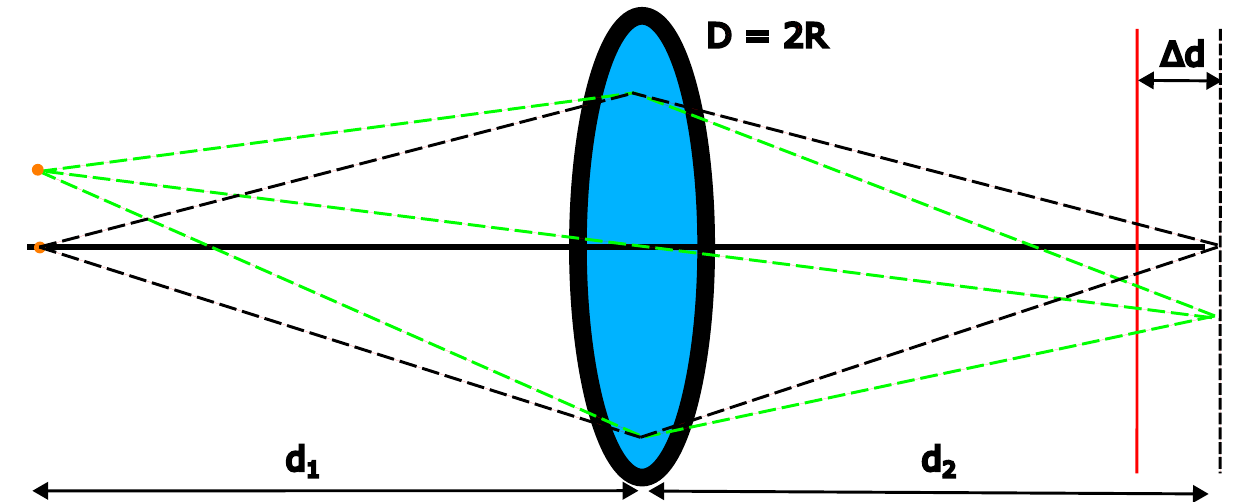} 
\hspace{3mm}
\includegraphics[height=3.3cm]{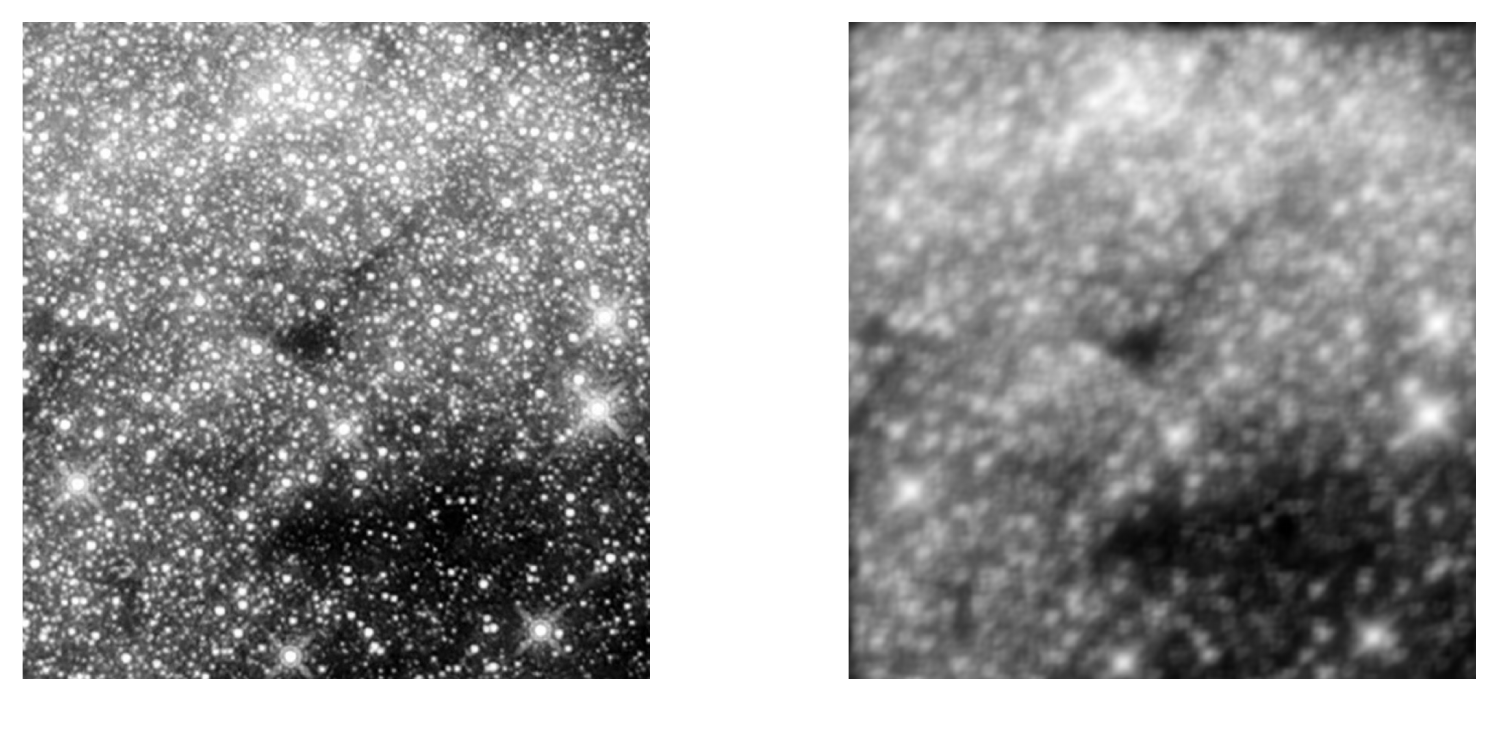} 
\caption{Defocusing (left), example of a sharp image of stars (centre), and a blurred image of stars caused by defocusing (right).}
\label{fig:w1_defocus}
\end{figure}

We can achieve infinite resolution in the geometrical optics approach if the lens equation for $d_1$, $d_2$, and $f$ is satisfied with infinite precision. In this case, we can distinguish two dots in the source plane separated by infinitely small distances $(\Delta x, \Delta y)$ because the images are also dots of zero size separated by $(m\Delta x, m\Delta y)$. In this section, we discuss the first practical limitation of the imaging resolution, which is present even in the geometrical optics model: defocusing. We define defocusing or defocus as an optical condition where an image is formed away from the detector plane and appears blurred, as shown in Fig.~\ref{fig:w1_defocus}. The radius of the spot is related to the displacement of the recording surface from the image plane, $\Delta d$, and the radius of the imaging lens, $R$, according to the equation
\begin{equation}
r = \frac{\Delta d}{d_2}R.
\end{equation}

\begin{figure}[b]
\centering
\includegraphics[height=3.3cm]{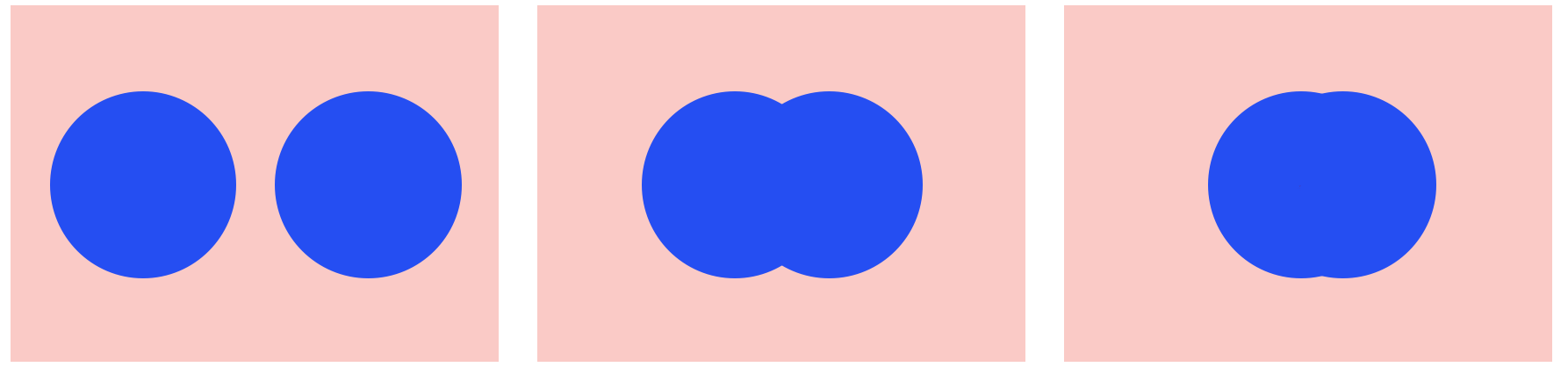} 
\caption{Examples of blurred images on the sensing surface, illustrating the definition of resolvable features in the image: resolvable (left), marginally resolvable (centre), and unresolvable (right).}
\label{fig:w1_res_defocus}
\end{figure}

We call two points in the source plane distinguishable by our imaging system if their spots on the recording surface are separated by a distance $l \geq r$, as shown in Fig.~\ref{fig:w1_res_defocus}. Otherwise, we will call the two points indistinguishable by the imaging system because they produce one, slightly extended, spot. The criterion is a simplification because the ultimate resolution is determined by the noise present in the system, such as photodetector's thermal noise or photon shot noise, which we will discuss in Lecture~\ref{Lecture3}. Blurriness acts as a low-pass filter and suppresses high-frequency features in the image plane that may still be observed if the noise level is low. However, the criterion is a good benchmark for imaging systems, and we will follow this definition in this module.
In the small angle approximation, the angular resolution of the imaging system, $\theta$, is then given by the equation
\begin{equation}\label{eq:l1_ang_res}
\theta = \frac{L}{d_1} = \frac{l}{d_2} = \frac{r}{d_2},
\end{equation}
where $L$ is the separation between two points in the source plane that we can still resolve with our imaging system.

As an example, let's consider an eye. The cornea and adjustable lens act as one strong lens and create an image on the retina, which consists of photosensitive cells, as we will discuss in Lecture~\ref{Lecture3}. A typical separation between the lens and the retina is $d_2 = 2.5$\,cm and is determined by the structure of the eye. Therefore, an eye needs to change the focal length of its lens during refocusing from close to distant objects. An average human eye can focus on objects as close as $d_1 = 6$\,cm and, therefore, achieve $f = 1.8$\,cm. Looking far ($d_1 \gg f$), the same eye needs to tune its focal length to $f = d_2 = 2.5$\,cm. If an eye cannot adjust its focal length in the range, then defocusing occurs either when a person looks too far or too close. Eq.~\ref{eq:l1_ang_res} also explains why people who do not see well tend to squint: reducing $R$ improves the angular resolution of their eyes. However, the amount of light that hits their sensing elements also reduces, and, therefore, the technique works only for relatively bright objects. Also, a smaller $R$ will limit the angular resolution caused by diffraction, as we discuss below.

\subsection{Diffraction-limited resolution and depth of focus}

\begin{figure}[t]
\centering
\includegraphics[height=3.3cm]{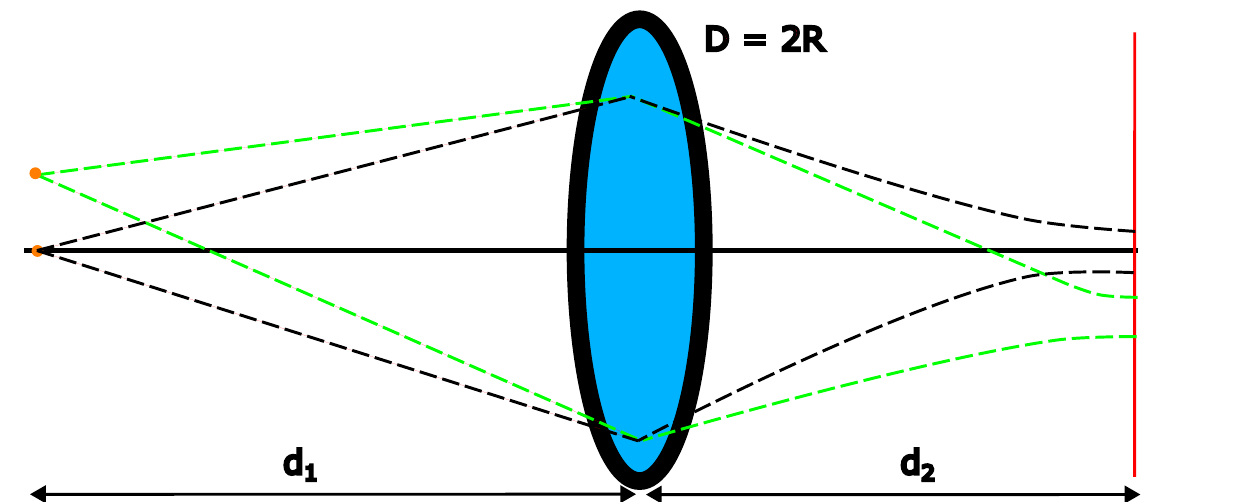} 
\hspace{3mm}
\includegraphics[height=3.3cm]{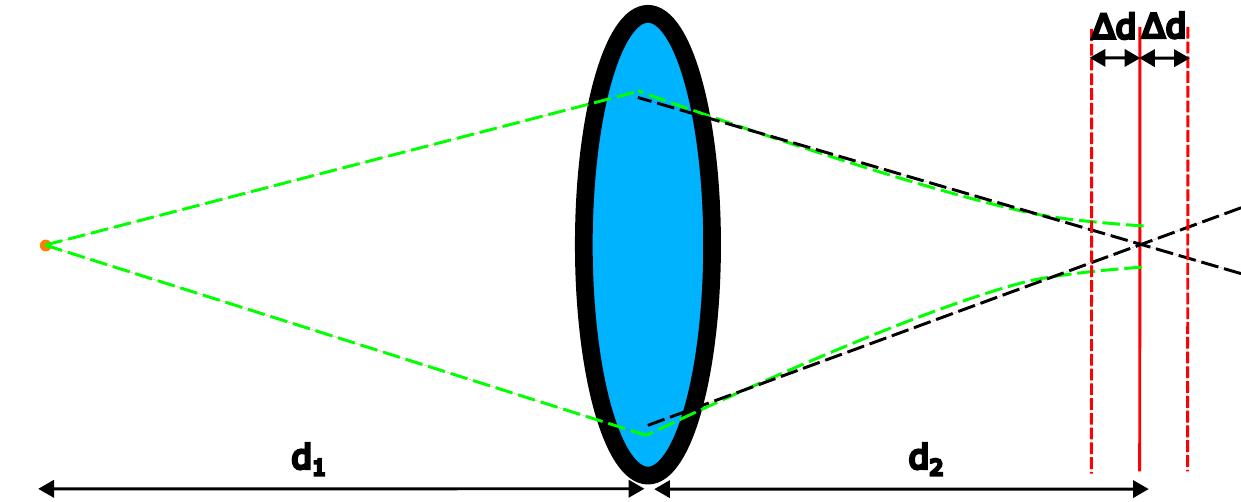} 
\caption{Diffraction on a lens (left) and depth of focus (right).}
\label{fig:w1_diffraction_depth}
\end{figure}

The wave nature of light makes the light rays diverge on their way from the lens to the sensing element (see Fig.~\ref{fig:w1_diffraction_depth}), and the images of point sources have a finite radius $r$ even without defocusing. If the source plane is far away from the camera, $d_1 \gg f$, then the image is formed at the focal plane of the lens, and the spot in the imaging plane, produced by a point source, has a radius $r$ that is given by the equation
\begin{equation}\label{eq:l2_r}
    r = 1.22 \frac{\lambda}{D}f,
\end{equation}
where a factor of 1.22 will be derived in Lecture~\ref{Lecture2}, $\lambda$ is the wavelength of light, and $D=2R$ is the diameter of the lens. High-quality imaging systems achieve $D / f \sim 1$ and, therefore, $r \sim \lambda$.

Similar to defocusing, finite spot sizes in the imaging plane limit the angular resolution of the imaging system according to the equation
\begin{equation}\label{eq:diff_ang_res}
    \frac{L}{d_1} = \frac{l}{d_2} = \frac{r}{d_2} = 1.22\frac{\lambda}{D}.
\end{equation}

We refer to the quantity given by Eq.~\ref{eq:diff_ang_res} as the diffraction-limited angular resolution of the imaging system, which sets a fundamental benchmark for its performance. Diffraction also sets the precision of focusing. We define the depth of focus, $\Delta d$, as the axial displacement between the image plane and the detector plane over which the angular resolution is not significantly degraded by defocus relative to the diffraction limit. For distant objects, where $d_2 \approx f$, the depth of focus is obtained from
\begin{equation}
\frac{\Delta d \, D}{2f} \leq 1.22\frac{\lambda}{D}f \hspace{1cm}\Rightarrow\hspace{1cm} \Delta d \leq 2.44 \lambda \left( \frac{f}{D} \right)^2,
\end{equation}
which shows that for imaging systems with $f \sim D$, the position of the sensing element must be controlled with a precision on the order of the wavelength, $\lambda$. This presents a significant challenge for many imaging systems, including extreme ultraviolet photolithography~\cite{kazazis2023euvl}, which requires positioning precision on the order of 10\,nm.

%% file: week2.tex
\section{Imaging: Fourier transform of an aperture and aberrations}
\label{Lecture2}

In Lecture~\ref{Lecture1}, we discussed imaging within the framework of geometrical optics. However, the wave nature of light imposes fundamental limits on imaging resolution, because light diffracts as it propagates through free space. In this lecture, we derive the propagation of electromagnetic waves through an aperture and establish the diffraction-limited resolution introduced in Lecture~\ref{Lecture1}. We discuss Fraunhofer diffraction, a powerful result showing that the electric field in the far field, or equivalently in the focal plane of a lens, is given by the Fourier transform of the aperture function. We then consider imaging aberrations and methods for their correction, including deconvolution techniques in post-processing and adaptive optics for real-time wavefront control.

In this lecture, we discuss
\begin{itemize}
\item Maxwell's wave equation and a plane wave solution,
\item Huygens principle, Fresnel and Fraunhofer diffraction,
\item properties of the Fourier transform,
\item aberrations and point spread function,
\item deconvolution algorithms,
\item adaptive optics.
\end{itemize}

\subsection{Wave equation and a plate wave solution}

We review the wave equation of light starting from the four Maxwell equations~\cite{griffiths2017introduction}: Gauss's law for the electric field, Gauss's law for the magnetic field, Faraday's law of induction, and Ampère's circuital law.

Gauss's law for electricity relates the distribution of electric charge to the resulting electric field. It considers the charge enclosed by an imaginary closed surface $S$ and states that
\begin{equation}\label{eq:l2_maxwell_1}
    \oint \limits_S \VF{E} \cdot \VF{dS} = \frac{1}{\epsilon_0} \int \limits_V \rho \, dV,
\end{equation}
where $\VF{E}$ is the electric field on the boundary, $\epsilon_0$ is the vacuum permittivity, $V$ is the volume enclosed by the surface $S$, and $\rho$ is the charge density. Equation~(\ref{eq:l2_maxwell_1}) shows that the flux of the electric field through a closed surface is proportional to the total enclosed charge. Gauss's law can be derived from Coulomb's law and vice versa.

Gauss's law for magnetism is based on the empirical observation that no magnetic monopoles have been found. As a result, the net magnetic flux through any closed surface is zero:
\begin{equation}
    \oint \limits_S \VF{B} \cdot \VF{dS} = 0.
\end{equation}
This reflects the fact that magnetic field lines form closed loops: they emerge from the north pole of a magnet and return to its south pole. If a magnet is cut in two, each piece forms a new dipole with its own north and south poles.

Faraday's law of induction describes how a time-varying magnetic flux generates an electric field:
\begin{equation}
     \oint \limits_l \VF{E} \cdot \VF{dl} = -\frac{d}{dt}\int \limits_S \VF{B} \cdot \VF{dS},
\end{equation}
where $S$ is any surface bounded by the closed loop $l$. This law is the operating principle behind transformers, inductors, and electrical motors.

Ampère's circuital law relates the circulation of the magnetic field around a closed loop to the electric current passing through the enclosed surface:
\begin{equation}
     \oint \limits_l \VF{B} \cdot \VF{dl} = \mu_0 \int \limits_S \left(\VF{J} + \epsilon_0\frac{d\VF{E}}{dt} \right) \cdot \VF{dS},
\end{equation}
where $\mu_0$ is the vacuum permeability and $\VF{J}$ is the current density. Historically, the magnetic effect of electric current was first observed by Hans Christian Ørsted in 1820, when he noticed that a compass needle near a current-carrying wire deflected perpendicular to the wire during a lecture demonstration~\cite{orsted1820electromagnetic}.

We can write Maxwell's equations in the differential form by using Gauss divergence and Kelvin-Stokes theorems. In a vacuum, there are no charges or electric currents: $\rho = 0$, $J=0$ and Maxwell equations have the forms:
\begin{equation}\label{eq:l2_Maxwell_diff}
\nabla \cdot \VF{E} = 0 \hspace{1cm} \nabla \cdot \VF{B} = 0 \hspace{1cm} \nabla \times \VF{E} = -\frac{\partial \VF{B}}{\partial t} \hspace{1cm} \nabla \times \VF{B} = \frac{1}{c^2}\frac{\partial \VF{E}}{\partial t}
\end{equation}

We can eliminate $\VF{B}$ fields from the equations by taking the curl of Faraday's law of induction and using Ampere's law to get the following equation
\begin{equation}
    \nabla \times \nabla \times \VF{E} = -\frac{\partial}{\partial t} (\nabla \times B) = -\frac{1}{c^2}\frac{\partial^2 \VF{E}}{\partial t^2}.
\end{equation}
Applying the "curl of curl" rule: $\nabla \times \nabla \times \VF{E} = \nabla (\nabla \cdot \VF{E}) - \nabla^2 \VF{E}$ and Gauss's law in vacuum, we get the wave equation for the electric field in vacuum
\begin{equation}
    \frac{1}{c^2}\frac{\partial^2 \VF{E}}{\partial t^2} - \nabla^2 \VF{E} = 0.
\end{equation}

Similarly, we take the curl of the fourth Maxwell equation
\begin{equation}
\nabla \times \nabla \times \VF{B} = \frac{1}{c^2}\frac{\partial}{\partial t}(\nabla \times E)
\end{equation}
and use the curl-of-curl rule on the left side and Faraday's law of induction on the right side. We then get the equation
\begin{equation}
\nabla (\nabla \cdot \VF{B}) - \nabla^2 \VF{B} = -\frac{1}{c^2}\frac{\partial^2 \VF{B}}{\partial t^2}.
\end{equation}
Since the first term on the left is zero due to the Gauss law for the magnetic field, we get the wave equation for the magnetic field
\begin{equation}
\frac{1}{c^2}\frac{\partial^2 \VF{B}}{\partial t^2} - \nabla^2 \VF{B} = 0.
\end{equation}

In this module, we consider solutions to the wave equation in the form of (i) plane waves, as discussed below, (ii) Gaussian beams, as discussed in Lecture~\ref{Lecture5}, and (iii) guided modes, as discussed in Lecture~\ref{Lecture10}. The plane-wave solution has the form
\begin{equation}
    \VF{E}(\VF{r}, t) = \VF{E}_0 e^{i (\omega t - \VF{k} \cdot \VF{r})} + \text{c.c.},
\end{equation}
where $\VF{k}$ is the wave vector, $\omega$ is the angular frequency, and $\text{c.c.}$ denotes the complex conjugate of the preceding term.

\subsection{Huygens Fresnel principle, Fresnel and Fraunhofer diffraction}

Since light has a wave nature, we can now compute its diffraction on a lens' aperture. Consider a plane wave propagating along the Z-axis. At plane $z=0$, we have a thin wall with a circular hole constrained by $x'^2 + y'^2 \leq R^2$, where $x'$ and $y'$ are coordinates in the X, Y plane at $z=0$, and $R = D/2$ is the radius of the hole, and $D$ is its diameter. We find the profile of the electric field on a far screen at $z \rightarrow \infty$ by applying the Huygens-Fresnel principle. The principle states that every point on a wavefront is the source of spherical wavelets, and the secondary wavelets emanating from different points mutually interfere. The sum of these spherical wavelets forms the wavefront as given by the equation

\begin{equation}\label{w2:eq_HF}
    E(x,y,z) \sim \int_S E(x', y') e^{-ikl}dS,
\end{equation}
where $l$ is the distance between points $(x', y')$ on the aperture and $(x, y)$ on the screen, $k=2\pi/\lambda$ is the wavenumber, and we integrate over the aperture area $S$. The proportionality sign in Eq.~\ref{w2:eq_HF} includes the decay of the electric field proportional to the distance travelled by the wave.

\begin{figure}[h]
\centering
\includegraphics[height=4.2cm]{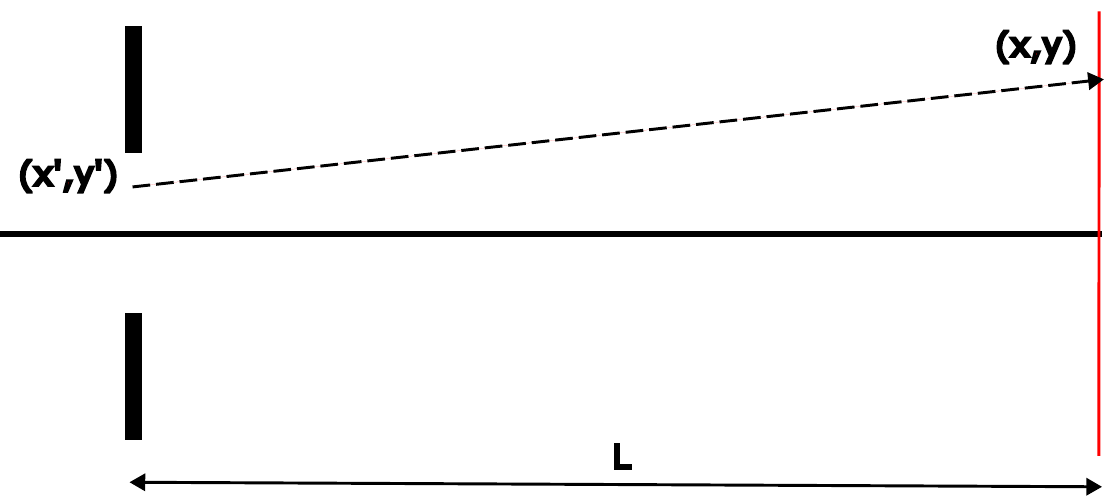} 
\hspace{3mm}
\includegraphics[height=4.2cm]{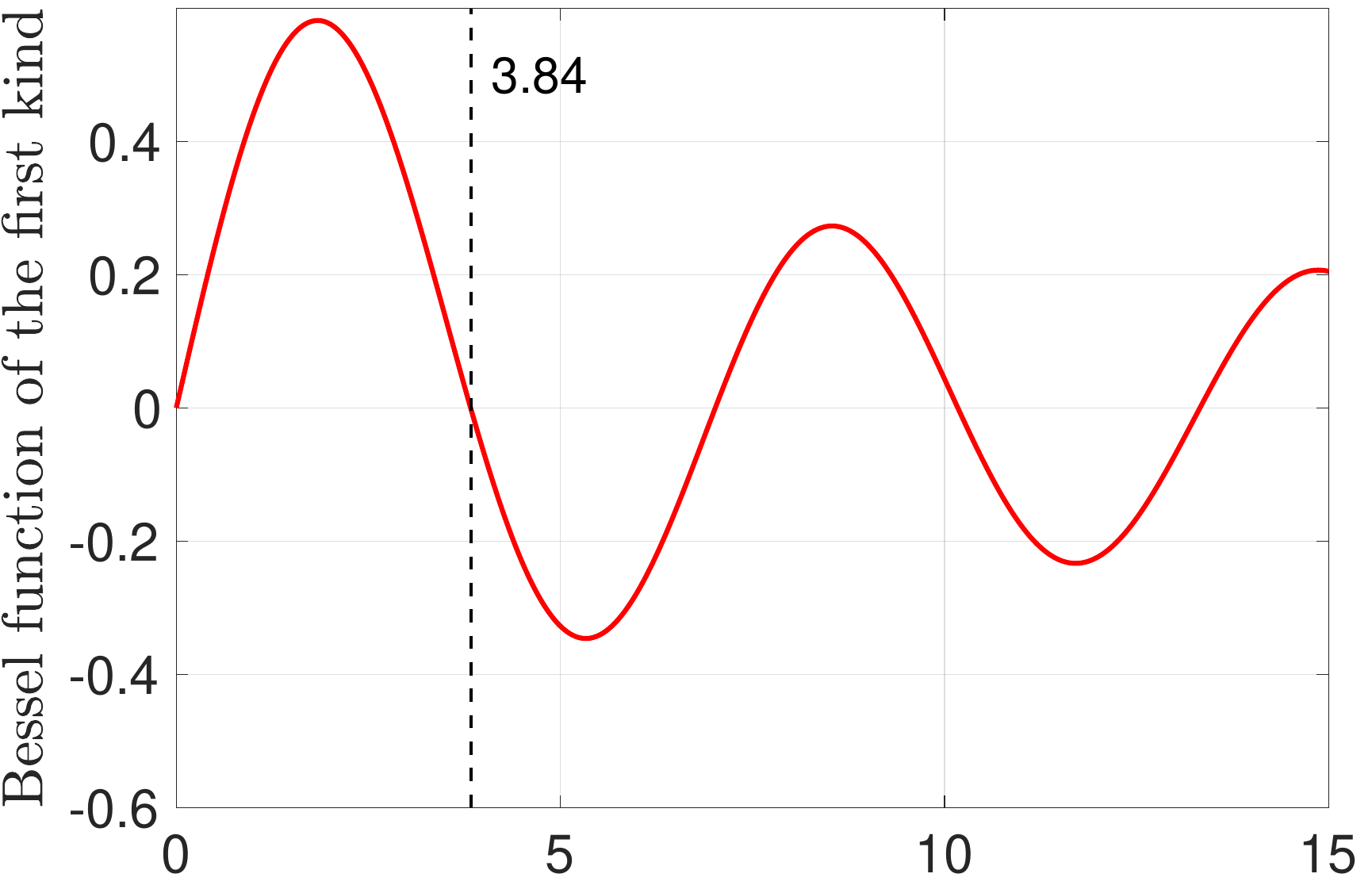} 
\caption{Diffraction on an aperture (left) and Bessel function of the first kind (right).}
\label{fig:w2_FR_Bessel}
\end{figure}

Since the screen is far from the wall, we apply approximations $z \gg x,y \gg x',y'$ and simplify the previous equation to
\begin{equation}
    l = \sqrt{z^2 + (x -x')^2 + (y - y')^2} \approx z \left(1 + \frac{(x-x')^2}{2z^2} + \frac{(y-y')^2}{2z^2} \right)
\end{equation}

\begin{equation}\label{w2:eq_FT_aperture}
    E(x,y,z) \sim \int_S E(x', y') \exp\left(i\frac{k}{z}(xx' + yy')\right)dx' dy',
\end{equation}
which is a 2D Fourier transform of the electric field in the aperture plane.

The result in Eq.~\ref{w2:eq_FT_aperture} is valid for an arbitrary aperture shape and transmissivity and can be applied, for example, to a grating with a phase or amplitude modulation. In this lecture, we solve Eq.~\ref{w2:eq_FT_aperture} for a circular aperture with full transmission, because the result applies to a circular lens, which we considered in Lecture~\ref{Lecture1}. Since $E(x', y')$ is a constant, the equation below gives the electric field on the far screen
\begin{equation}\label{w2:eq_FR_polar}
    E(x,y,z) \sim \int_S \exp\left(i\frac{k}{z}(xx' + yy')\right)dx' dy' = \int_0^R r'dr' \int_0^{2\pi} \exp\left(i\frac{krr'}{z}(\cos(\phi'-\phi))\right) d\phi',
\end{equation}
where we introduced polar coordinates on the aperture and on the screen: $x' = r\cos{\phi'}, y' = r\sin{\phi'}, x = r\cos{\phi}, y = r\sin{\phi}$. We also introduce an angle $\theta = \frac{r}{z}$. In the small-angle approximation assumed in this lecture ($z \gg x,y$), angle $\theta$ is the direction in which the light propagates after passing through the aperture.

We first integrate Eq.~\ref{w2:eq_FR_polar} over the azimuthal angle $\phi'$, and then over the radial coordinate $r'$. The angular dependence appears through a factor of the form $\exp\!\left(ik r' \theta \cos(\phi' - \phi)\right)$, which integrates to the zeroth-order Bessel function of the first kind~\cite{watson1922bessel}:
\begin{equation}\label{w2:eq_FR_sol}
    E(\theta) \sim \int_0^R r' dr'\, J_0(k\theta r')
    = \int_0^{k\theta R} J_0(p)\, \frac{p\,dp}{k^2\theta^2}
    = \frac{1}{k^2\theta^2} \int_0^{k\theta R} d\!\left[p J_1(p)\right]
    = \frac{R\, J_1(kR\theta)}{k\theta},
\end{equation}
where we introduced the substitution $p = k r'\theta$. In the derivation, we used the standard recurrence relation for Bessel functions,
\begin{equation}
\frac{d}{dx}\!\left[x^{n+1}J_{n+1}(x)\right] = x^{n+1}J_n(x).
\end{equation}

Eq.~\ref{w2:eq_FR_sol} gives the dependence of the electric field on the screen as a function of the angle $\theta$. The Bessel function of the first order is plotted in Fig.~\ref{fig:w2_FR_Bessel}\,(right). The Bessel function $J_1(kR\theta)$ is zero at $\theta=0$. However, the denominator is also zero, and we can show that the electric field is maximised for $\theta=0$ if we consider the expansion of the Bessel function around zero. The second zero of $E(kR\theta)$ occurs at $3.84$. Therefore, we find that minimum angle $\theta_0$ for which $E(\theta_0)=0$ is given by the equation
\begin{equation}\label{eq:l2_theta}
    \theta_0 = \frac{3.84}{\pi}\frac{\lambda}{D} \approx 1.22 \frac{\lambda}{D}.
\end{equation}

We now need to make the final step: add a circular lens. So far, we only introduced its aperture and found that its effect is a diverging beam as given by Eq.~(\ref{eq:l2_theta}). Since the wave is now diverging, the lens will not focus it in one single spot in the focal plane. Since the rays travelling through the centre of the lens are not bent, we can apply a geometrical argument to find the radius of the spot at the focal plane of the lens according to the equation
\begin{equation}\label{eq:w2_FG_lens_sol}
    r = \theta_0 f = 1.22 \frac{\lambda}{D}f.
\end{equation}

We can prove Eq.~\ref{eq:w2_FG_lens_sol} more accurately using the wave optics formalism. Since the lens introduces a phase shift to the incident wave given by the equation
\begin{equation}
\delta_{\mathrm{lens}}(r) = -\frac{k}{2f}\,r'^2 = -\frac{k}{2f}\,(x'^2 + y'^2),
\end{equation}
we can substitute the plane wave from Eq.~\ref{w2:eq_FT_aperture} with the wave with a curved wavefront and get the equation
\begin{equation}\label{eq:w2_FH_lens_wave}
    E(x,y,z) \sim \int_S \exp\left(-i\frac{k}{2}\,\left(\frac{1}{z} - \frac{1}{f}\right)\,(x'^2 + y'^2) \right) \exp(i\frac{k}{z}(xx' + yy'))dx' dy'
\end{equation}
and we achieve the Fourier transform of the aperture in the plane $z=f$. The square of the field gives the intensity on the sensing element and is known as the Airy pattern.

\subsection{Fourier transform and its properties}

As discussed in the previous section, the Fourier transform can describe optical propagation through apertures and lenses. Its efficient numerical implementation via the Fast Fourier Transform (FFT) enables rapid processing of electric fields and images, with applications ranging from wavefront reconstruction and adaptive optics in astronomy to real-time image filtering~\cite{goodman2005introduction}.

In this section, we define the direct and inverse Fourier transforms in one and two dimensions according to the equations
\begin{equation}\label{eq::w2_FT_def}
\begin{split}
    &\tilde{f}(k) = \int_{-\infty}^{+\infty} f(x)\, e^{-ikx}\, dx, \hspace{1cm}
    f(x) = \frac{1}{2\pi} \int_{-\infty}^{+\infty} \tilde{f}(k)\, e^{ikx}\, dk, \\
    &\tilde{f}(k_x, k_y) = \iint_{-\infty}^{+\infty} f(x,y)\, e^{-i(k_x x + k_y y)}\, dx\, dy, \\
    & f(x,y) = \frac{1}{(2\pi)^2} \iint_{-\infty}^{+\infty} \tilde{f}(k_x, k_y)\, e^{i(k_x x + k_y y)}\, dk_x\, dk_y,
\end{split}
\end{equation}
where $k$ is the spatial frequency, $(x,y)$ are spatial coordinates, and $\tilde{f}(k)$ is the Fourier transform of $f(x)$, $k_x$ and $k_y$ are the spatial frequency components in the $x$ and $y$ directions.

Though the definitions given by Eqs.~\ref{eq::w2_FT_def} have a convenient mathematical form, and we will use them when studying the properties of the Fourier transform, signals are always recorded over a finite interval $X$ rather than an infinite domain in real systems. For example, $X$ may represent the spatial extent of an image or the temporal duration of a recorded signal. To avoid divergences and the artificial accumulation of power with increasing observation range, it is therefore useful to define a finite-range Fourier transform,
\begin{equation}
\tilde{f}(k) = \frac{1}{\sqrt{X}} \int_{0}^{X} f(x)\, e^{-ikx}\, dx.
\end{equation}
This normalisation ensures that $\tilde{f}(k)$ remains finite in the limit of large observation ranges and provides a direct connection to experimentally measurable quantities. For example, while the Fourier transform of idealised white noise diverges at all frequencies in the infinite-domain definition, the finite-range Fourier transform remains finite and well-defined. In particular, $|\tilde{f}(k)|^2$ corresponds (up to finite-bandwidth effects and normalisation conventions) to the power spectral density, while $|\tilde{f}(k)|$ is proportional to the amplitude spectral density. This figure of merit is often preferred in experimental analysis because it provides a direct measure of fluctuations per unit bandwidth.

\begin{figure}[t]
\centering
\includegraphics[height=4.8cm]{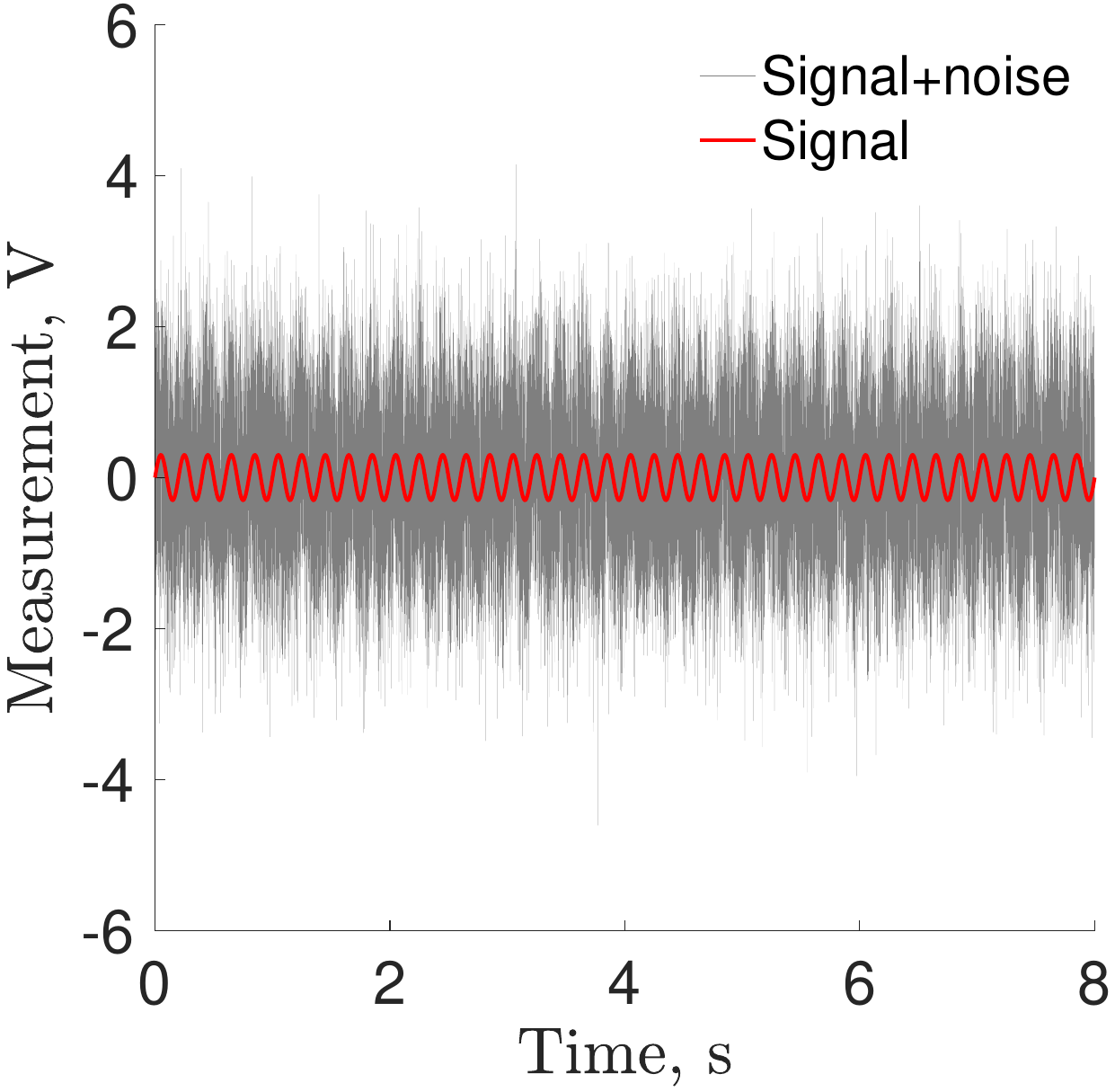} 
\hspace{2mm}
\includegraphics[height=4.8cm]{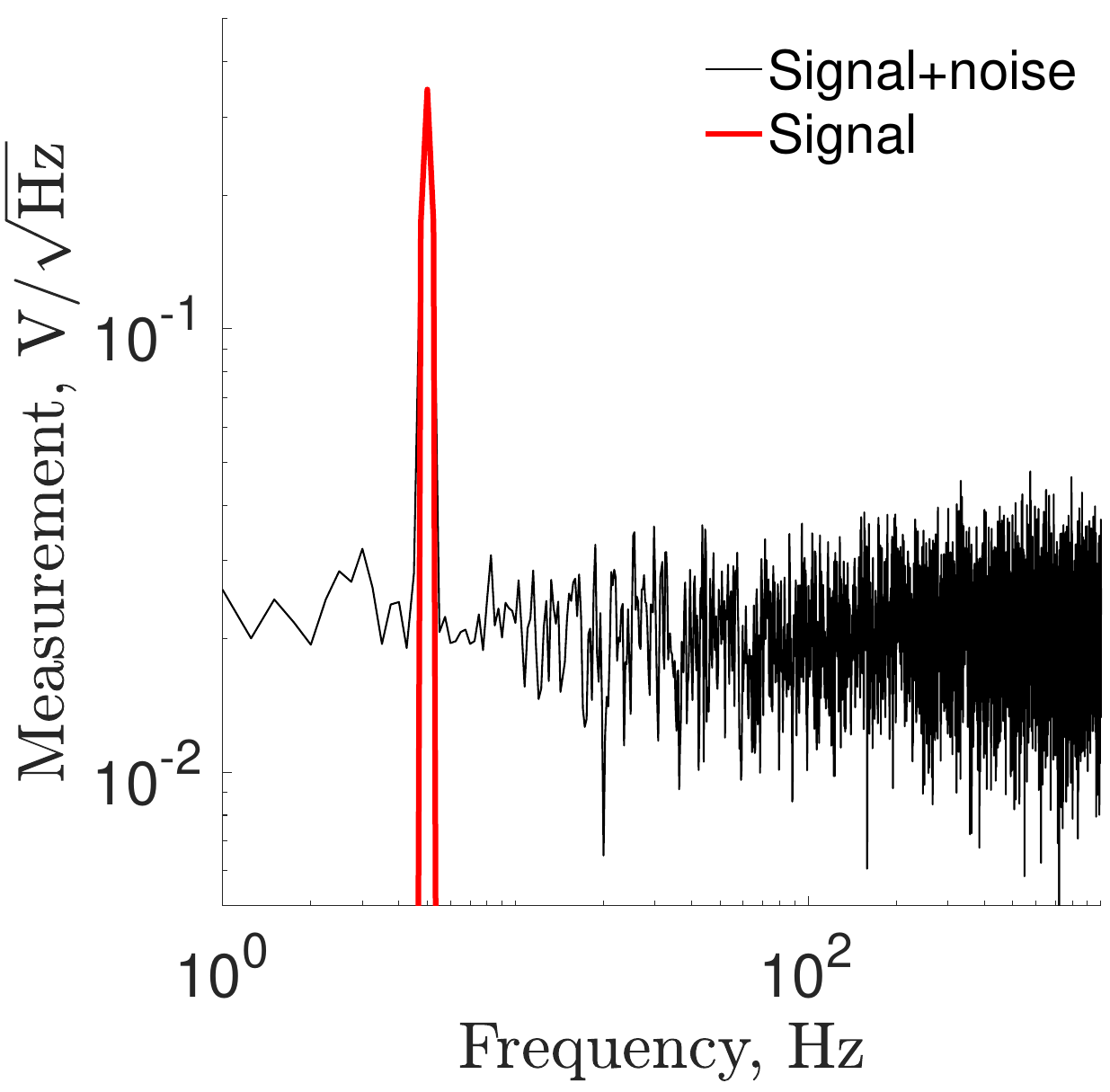} 
\hspace{2mm}
\includegraphics[height=4.8cm]{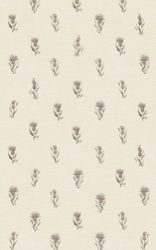}
\hspace{2mm}
\includegraphics[height=4.8cm]{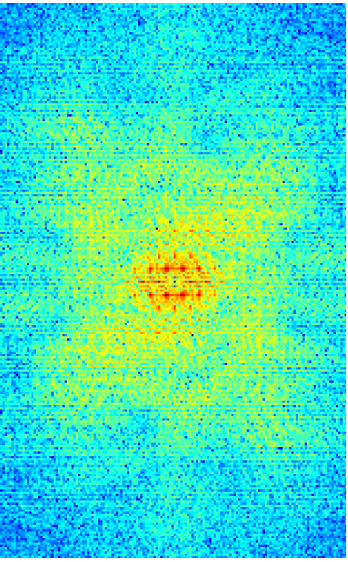}
\caption{Examples of signals in the spatial (or time) and frequency domains. (i) A one-dimensional noisy signal with a small embedded signal; the noise RMS is larger than the signal amplitude, making the signal difficult to identify in the time domain. (ii) Amplitude spectral density representation of the same signal in the frequency domain: the noise power is distributed across all frequencies, while the signal power is localised at 5\,Hz and becomes clearly visible. (iii) A photograph of a wallpaper representing a two-dimensional spatial intensity distribution. (iv) Amplitude of the Fourier transform of the image, where peaks correspond to spatial frequencies of the periodic structures in the scene.}
\label{fig:w2_FT_ex}
\end{figure}

Although the spatial and frequency domains contain the same information, the Fourier transform makes many physical problems easier to understand. In the frequency domain, complex operations like convolution become simple multiplication and can take advantage of the Fast Fourier Transform algorithm. The features of the image, such as periodicity and filtering behavior, that are hard to spot in the spatial domain, become visible in the frequency domain. Examples of the signal representation in the spatial and frequency domains are shown in Fig.~\ref{fig:w2_FT_ex}.

There are a few important properties of the Fourier transform:
\begin{itemize}
\item Linearity:
\begin{equation}
   {\rm FT}(af(x) + bg(x)) = a\tilde{f}(k) + b\tilde{g}(k),
\end{equation}

\item Scaling:
\begin{equation}
   {\rm FT}(f(\lambda x)) = \frac{1}{|\lambda|}\tilde{f}\left(\frac{k}{\lambda} \right),
\end{equation}

\item Parity:
\begin{equation}
   {\rm FT}(f(-x)) = \tilde{f}(-k),
\end{equation}
which is identical to the special case of the scaling property with $\lambda=-1$,

\item Shift:
\begin{equation}
{\rm FT}(f(x-a)) = \int \limits_{-\infty}^{+\infty} f(x-a) e^{-ikx} dx = \int \limits_{-\infty}^{+\infty} f(y) e^{-iky} e^{-ika} dy = e^{-ika} \tilde{f}(k),
\end{equation}
where we introduced the substitution $y=x-a$.

\item Derivative:
\begin{equation}
   {\rm FT}(f'(x)) = ik\tilde{f}(k),
\end{equation}
which can be verified by differentiating the inverse Fourier transform:
\begin{equation}
f'(x) = \left(\frac{1}{2\pi} \int \limits_{-\infty}^{+\infty} \tilde{f}(k) e^{ikx} dk \right)' = \frac{1}{2\pi} \int \limits_{-\infty}^{+\infty} (\tilde{f}(k) ik) e^{ikx} dk.
\end{equation}
\end{itemize}

\section*{Aberrations, point spread function, and the convolution}

The Fourier transform can also analyse an important class of image imperfections. In this module, we define aberrations as systematic deviations of an optical system’s wavefront from the ideal wavefront. Such aberrations lead to blur and distortion. They arise because real lenses and mirrors do not perfectly conform to the assumptions of geometrical optics and lead to point spreading.

As we discussed in Lecture~\ref{Lecture1}, the point spread function (PSF) shows how a point source is spread into a finite pattern in the imaging plane due to defocusing, diffraction, and aberrations. It represents the limit of image sharpness because every object can be viewed as a collection of points whose images are blurred according to the point spread function. The shape and size of the point spread function determine the system’s resolution and contrast, with an ideal diffraction-limited system producing an Airy pattern. We can compute the point spread function from the Fourier transform of the field at the lens' aperture according to the equation
\begin{equation}
{\rm PSF} = \left|\rm{FT}(E(x', y')) \right|^2.
\end{equation}
As an example, we consider the point spread function due to defocussing, which is given by the equation
\begin{equation}
    \rm{PSF} = \left|\rm{FT} \left(\exp(-i\frac{\pi}{\lambda}\,\left(\frac{1}{z} - \frac{1}{f}\right)\,(x'^2 + y'^2) \right)\right|^2,
\end{equation}
which is a consequence of Eq.~\ref{eq:w2_FH_lens_wave}, and, in the case of $z \neq f$, we get blurring due to defocusing in addition to diffraction. As discussed in Lecture~\ref{Lecture1}, we need to tune the distance $z$ between the lens and the recording surface with a precision of the depth of focus to reduce the point spreading due to defocusing below the point spreading that is caused by diffraction.

If the point spread function is shift-invariant, then the aberrated image, $I_2$, can be computed according to the equation
\begin{equation}\label{eq:w2_I_PSF}
I_2(x,y) = \int I_1(x_0, y_0) {\rm PSF}(x-x_0, y-y_0) dx_0 dy_0 = I_1 * \rm{PSF},
\end{equation}
where $I_1$ is the ideal image without aberrations and "$*$" implies the convolution of two functions. The equation implies that the measured intensity $I_2$ at any point on the sensing surface $(x,y)$ is the sum of point spreading from all points $(x_0, y_0)$ on the surface. A shift-invariant point spread function means that the imaging system responds to a point source in the same way regardless of where that point is located in the field of view, and the blur pattern does not change with coordinate in the image plane.

The convolution operation, shown in Eq.~\ref{eq:w2_I_PSF}, is widely utilised in imaging because it describes how an object is blurred by the system’s point spread function, and enables modelling and correction of effects such as optical defocus in microscopy, motion blur in photography, and atmospheric distortion in astronomy. In this module, we define convolution as a mathematical operation on two functions ($f$ and $g$) that produces a third function ($f * g$) that expresses how the shape of one is modified by the other, and is given by the equation
\begin{equation}
    (f * g)(x) = \int \limits_{-\infty}^{\infty} f(u) g(x-u) du = \int \limits_{-\infty}^{\infty} f(x-u) g(u) du.
\end{equation}

Convolution has several interesting properties:
\begin{itemize}
\item Commutativity:
\begin{equation}
f * g = \int \limits_{-\infty}^{\infty} f(u) g(x-u) du = \int \limits_{-\infty}^{\infty} f(x-y) g(y) dy = g * f,
\end{equation}
\item Distributivity:
\begin{equation}
\begin{split}
f * (g + h) &=  \int \limits_{-\infty}^{\infty} f(u) (g(x-u) + h(x-u) du \\
&= \int \limits_{-\infty}^{\infty} f(u) g(x-u) du + \int \limits_{-\infty}^{\infty} f(u) h(x-u) du = f * g + f * h,
\end{split}
\end{equation}
\item Differentiation:
\begin{equation}
(f * g)' = \left(\int \limits_{-\infty}^{\infty} f(u) g(x-u) du \right)' = \int \limits_{-\infty}^{\infty} f(u) g'(x-u) du = f * g',
\end{equation}
\item Integration:
\begin{equation}
\begin{split}
\int \limits_{-\infty}^{\infty} (f*g)dx &= \int \limits_{-\infty}^{\infty} \int \limits_{-\infty}^{\infty} f(u) g(x-u) du dx \\
&= \int \limits_{-\infty}^{\infty} f(u) du \int \limits_{-\infty}^{\infty} g(x-u) dx = \int \limits_{-\infty}^{\infty} f(u) du \int \limits_{-\infty}^{\infty} g(y) dy,
\end{split}
\end{equation}
\item Convolution theorem:
\begin{equation}\label{eq:w2_conv_theorem}
    {\rm FT}(f * g) = \tilde{f} \tilde{g},
\end{equation}
which is widely applied in imaging because it turns the image formation process, given by Eq.~\ref{eq:w2_I_PSF}, into a simple multiplication in the frequency domain. The process also simplifies the analysis of how different spatial frequencies are transmitted by the imaging system and reveals its filtering behavior. For example, Eq.~\ref{eq:w2_I_PSF} simplifies to
\begin{equation}
\tilde{I_2} = \tilde{I}_1 \tilde{p},
\end{equation}
where $\tilde{p}$ is the Fourier transform of the point spread function.
\end{itemize}

Modelling a specific imaging aberration can be implemented in the following steps: (i) compute the Fourier transform of the field at the aperture, (ii) obtain the point spread function by taking the squared magnitude of the Fourier-transformed field, (iii) compute the Fourier transforms of both the original image and the point spread function, and (iv) obtain the final image by taking the inverse Fourier transform of their product.

In this module, we consider five examples of aberrations:
\begin{itemize}

\item \textbf{Spherical aberrations}. Most lenses are spherical because they are easier to manufacture: a spherical surface can be produced by grinding and polishing processes that naturally create uniform curvature, making them practical for mass production. However, spherical surfaces are not quadratic and do not form perfect lenses. The wavefront phase error added by a spherical lens is given by the equation
\begin{equation}
\delta(r) = \frac{2 \pi}{\lambda} A (r^4 - r^2),
\end{equation}
where $A$ is a coefficient that sets the strength (magnitude) of the spherical aberration. The point spread function is shown in Fig.~\ref{fig:w2_PSF}\,(i). Optical engineers reduce aberrations by using aspherical lenses or combining multiple lenses, reducing the size of the aperture at every step.

\item \textbf{Coma}. The aberration appears for off-axis objects even for parabolic mirrors. Light from an off-axis point source is imaged as an asymmetric, comet-shaped blur rather than a sharp point, with a bright head and a trailing tail. The phase error is given by the equation
\begin{equation}
\delta(r, \phi) = \frac{2 \pi}{\lambda} B (r^3 - r) \cos(\phi),
\end{equation}
$B$ is a coefficient that sets the strength of the coma aberration, and $\phi$ is the angular coordinate around the optical axis of the lens. The point spread function is shown in Fig.~\ref{fig:w2_PSF}\,(ii). Coma occurs because rays passing through different parts of a lens are focused to different positions depending on their angle relative to the optical axis, leading to a position-dependent distortion of the point spread function. Coma becomes more pronounced toward the edges of the field of view and is especially noticeable in systems with large apertures, stretching point-like sources.

\item \textbf{Astigmatism}. The aberration occurs when the imaging system has different effective focal lengths in two perpendicular directions, often called the tangential and sagittal planes. As a result, rays in one plane come to focus at a different distance than rays in the orthogonal plane, and the wave front phase error is given by the equation
\begin{equation}
\delta(r, \phi) = \frac{2 \pi}{\lambda} C r^2 \cos(2\phi),
\end{equation}
$C$ is a coefficient that sets the strength of the astigmatism aberration. The point spread function is shown in Fig.~\ref{fig:w2_PSF}\,(iii).

\begin{figure}[t]
\centering
\includegraphics[height=3.7cm]{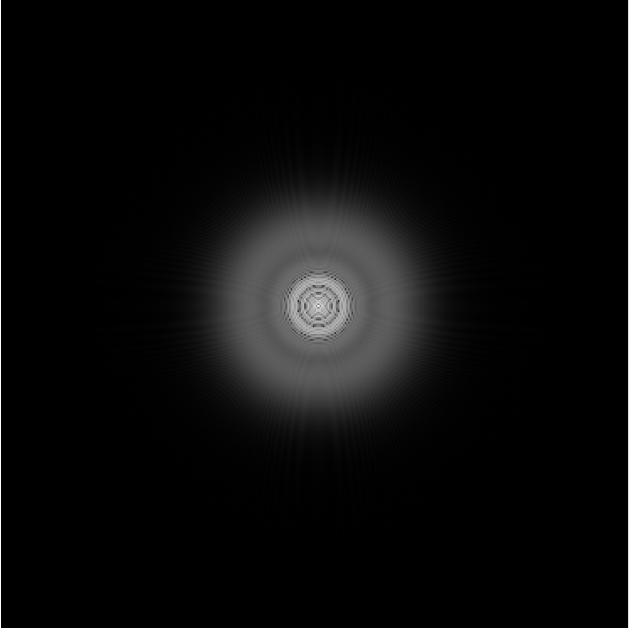} 
\hspace{0mm}
\includegraphics[height=3.7cm]{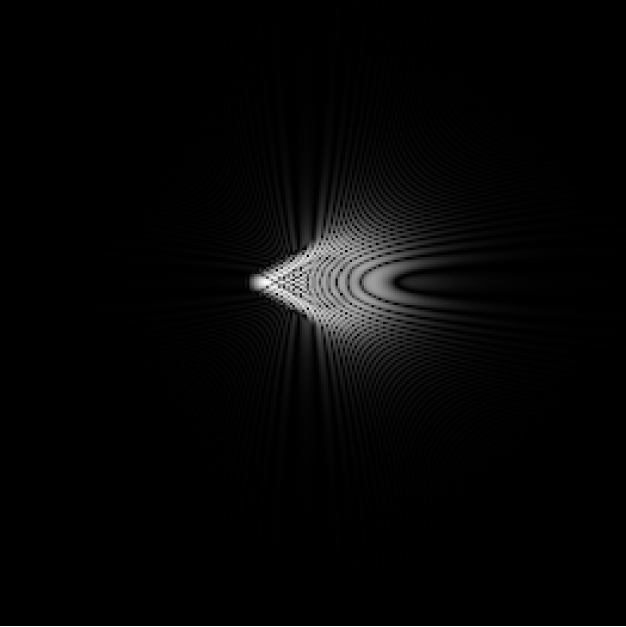} 
\hspace{0mm}
\includegraphics[height=3.7cm]{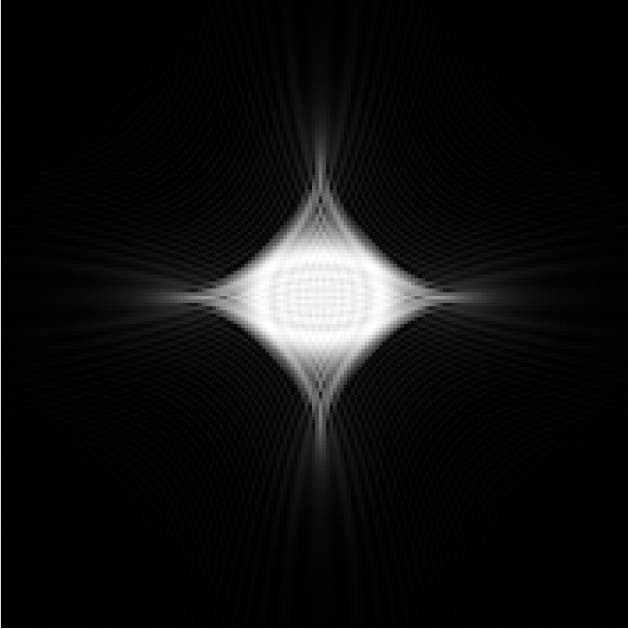}
\hspace{0mm}
\includegraphics[height=3.7cm]{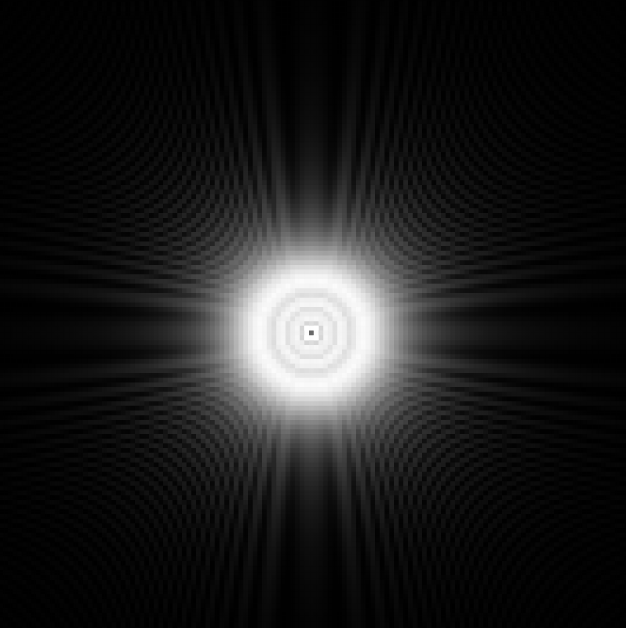}
\caption{Examples of Point spread functions from (i) spherical aberrations, (ii) coma, (iii) astigmatism, and (iv) field curvature.}
\label{fig:w2_PSF}
\end{figure}

\item \textbf{Field curvature}. The aberration occurs when the imaging plane of the system forms a curved surface. It causes portions of the field to appear out of focus when the sensor is flat. The wavefront error depends quadratically on the radius and is given by the equation
\begin{equation}
\delta(r) = \frac{2 \pi}{\lambda} D r^2,
\end{equation}
$D$ is a coefficient that sets the strength of the field curvature. The point spread function is shown in Fig.~\ref{fig:w2_PSF}\,(iv). Though the phase error has a quadratic dependence of $r$, which is similar to defocusing, defocus is a global misplacement of the sensor, and the field curvature is an intrinsic lens property that causes different parts of the field to focus at different distances.

\item \textbf{Distortion}. This is an optical aberration in which the magnification varies across the field of view, causing straight lines in the object to appear curved without introducing blur. Unlike other aberrations, distortion does not affect the point spread function but instead modifies the geometric mapping between object points and image points in a nonlinear way.
\end{itemize}

\section*{Deconvolution algorithms}

Since imperfections in imaging systems introduce optical aberrations, a natural question is whether the measured image can be corrected. If the lens profile, including its imperfections, is known, one can attempt to decouple these effects from the measured image, $I_2$, and recover the unaberrated image, $I_1$. In the frequency domain, this direct deconvolution procedure is given by
\begin{equation}
\tilde{I_2} = \tilde{I_1} \tilde{p} \hspace{5mm}=>\hspace{5mm} \tilde{I_1} = \frac{\tilde{I_2}}{\tilde{p}},
\end{equation}
where, similar to above, $\tilde{p}$ is the Fourier transform of the point spread function.

In a real system, directly dividing the image spectrum by the Fourier transform of the point spread function is generally not optimal, since this operation strongly amplifies high-frequency noise. In the presence of noise, $n$, the direct deconvolution can be written as
\begin{equation}\label{eq:w2_deconv_direct}
\tilde{I}_2 = \tilde{I}_1 \tilde{p} + \tilde{n} 
\hspace{5mm}\Rightarrow\hspace{5mm}
\tilde{I}_1 = \frac{\tilde{I}_2 - \tilde{n}}{\tilde{p}},
\end{equation}
where $\tilde{n}$ is the Fourier transform of the noise. This noise typically arises from photon shot noise or thermal fluctuations in the photodetector electronics, as will be discussed in Lecture~\ref{Lecture3}, and is characterised by an approximately flat amplitude spectral density. Since the point spread function acts as a low-pass filter, its Fourier transform $\tilde{p}$ suppresses high spatial frequencies; consequently, dividing by $\tilde{p}$ leads to strong amplification of high-frequency noise components in $\tilde{I}_1$.

However, it is still possible to improve an aberrated image if the point spread function is known. A well-known example of successful deconvolution is the correction of images from the Hubble Space Telescope~\cite{nasa_hubble_mirror_flaw}. The telescope initially suffered from spherical aberration caused by a small manufacturing error in its primary mirror, where the outer edge was too flat by approximately $2.2$\,um. This curvature error led to significant blurring. In this module, we will consider two deconvolution algorithms that address such imaging imperfections.

\textbf{Richardson–Lucy deconvolution}. Our goal is to run a processing algorithm to deconvolve the measured image $I_2(x,y)$ with the point spread function $p(x,y)$ and recover the original image $I_1(x,y)$ in the presence of noise. In practice, $p$ can be derived from the measurements of the lens profile, as has been done with the Hubble flawed primary mirror.

\begin{figure}[t]
\centering
\includegraphics[height=5.3cm]{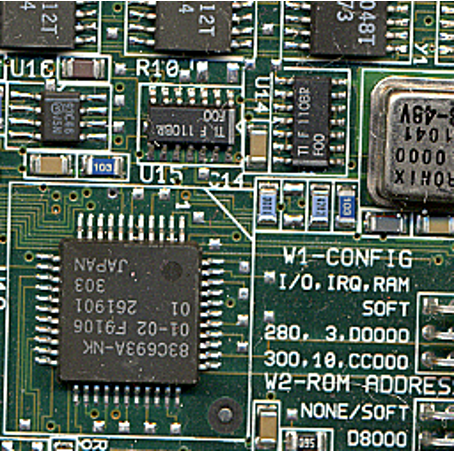} 
\hspace{2mm}
\includegraphics[height=5.3cm]{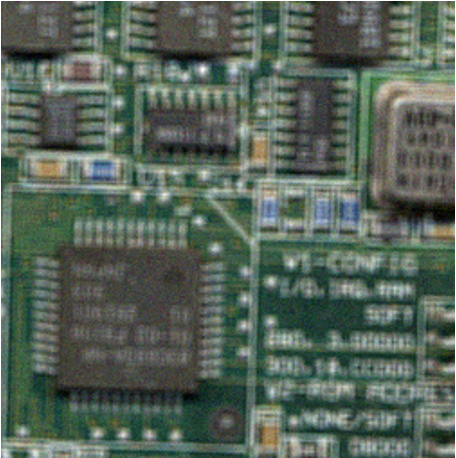} 
\hspace{2mm}
\includegraphics[height=5.3cm]{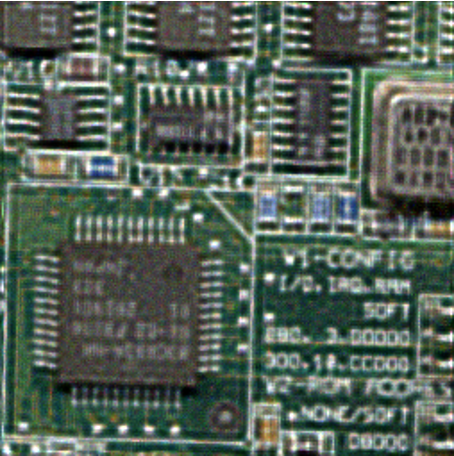}
\caption{Application of a Richardson–Lucy deconvolution to a blurred image in the presence of noise. The left panel shows the ideal image, the central panel shows the measured image, and the right panel shows the results of the deconvolution.}
\label{fig:w2_RL}
\end{figure}

In the algorithm, we perform an iterative procedure~\cite{Richardson:72, Lucy:1974yx} to compute the estimation of the true image $\hat{I_1}^{(n)}$ for iteration number $n$. The deconvolution algorithm iteratively estimates the true image by (i) convolving the current guess with the point spread function to simulate the measured image, (ii) dividing the actual measured image by this simulation, (iii) convolving that ratio with the flipped point spread function, $p^*(-x,-y)$, and (iv) multiplying the result with the current estimate to produce the next iteration according to the equation
\begin{equation}
    \hat{I_1}^{(n+1)} =  \hat{I_1}^{(n)} \left[\frac{I_2}{\hat{I_1}^{(n)} * p} * p^*(-x,-y) \right].
\end{equation}

The application of an algorithm to an aberrated image in the presence of noise is shown in Fig.~\ref{fig:w2_RL}. The Richardson–Lucy algorithm is derived from the maximum-likelihood estimation for Poisson noise and is well-suited for photon-limited imaging. The algorithm enforces positivity of the image estimate.

\textbf{Wiener deconvolution.} The algorithm assumes Gaussian noise and requires knowledge or an estimate of the signal-to-noise power ratio~\cite{Wiener_49}. Wiener deconvolution is performed in the frequency domain and seeks the optimal frequency-dependent gain $G$ used to estimate the original image according to
\begin{equation}
    \hat{\tilde{I}}_1 = G\,\tilde{I}_2.
\end{equation}

The gain $G$ is chosen to minimise the mean-square error
\begin{equation}
    e = E\!\left|\hat{\tilde{I}}_1 - \tilde{I}_1\right|^2
    = E\!\left|G\tilde{I}_2 - \tilde{I}_1\right|^2
    = E\!\left|G(\tilde{p}\tilde{I}_1 + \tilde{n}) - \tilde{I}_1\right|^2
    = E\!\left|(G\tilde{p}-1)\tilde{I}_1 + G\tilde{n}\right|^2,
\end{equation}
where $E[\cdot]$ denotes an ensemble expectation value. In the context of a single image, this expectation value should be interpreted as an average over a statistical ensemble of possible images and noise realisations consistent with the same imaging conditions.

Since the noise is assumed to be statistically independent of the signal, the error can be written as
\begin{equation}
\begin{split}
    e &= (G\tilde{p}-1)(G^*\tilde{p}^*-1)\,E\!\left[|\tilde{I}_1|^2\right]
    + GG^*\,E\!\left[|\tilde{n}|^2\right] \\
    &= (G\tilde{p}-1)(G^*\tilde{p}^*-1)\,S_{II}
    + GG^*\,S_{nn},
\end{split}
\end{equation}
where $S_{II} = E[|\tilde{I}_1|^2]$ and $S_{nn} = E[|\tilde{n}|^2]$ are the power spectral densities of the signal and noise, respectively. Cross terms such as $E[\tilde{I}_1 \tilde{n}^*]$ vanish because the signal and noise are uncorrelated.

As discussed above, the power spectral density describes how the variance (or power) of the image is distributed across spatial frequencies, independently of phase. It quantifies how strongly each spatial frequency component contributes, on average, to the image intensity fluctuations.

We find $G$ and $G^*$ by minimising the error $e$ relative to these gains. The derivative is given by the equation
\begin{equation}
\frac{de}{dG^*} = \tilde{p}^* (G\tilde{p}-1) S_{II} + G S_{nn} = 0
\end{equation}

and the solution to the optimisation problem in the frequency domain is given by the equation
\begin{equation}
    G = \frac{\tilde{p}^* S_{II}}{\tilde{p}\tilde{p}^* S_{II} + S_{nn}}.
\end{equation}

\begin{figure}[t]
\centering
\includegraphics[height=5.3cm]{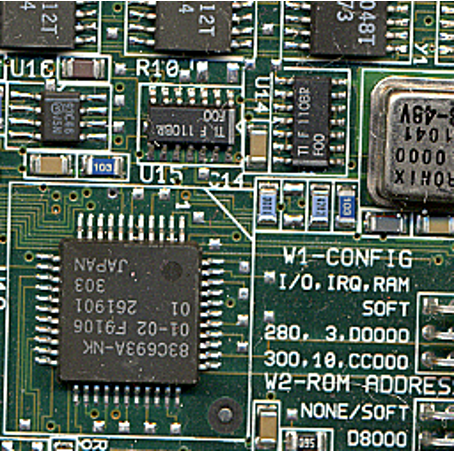} 
\hspace{2mm}
\includegraphics[height=5.3cm]{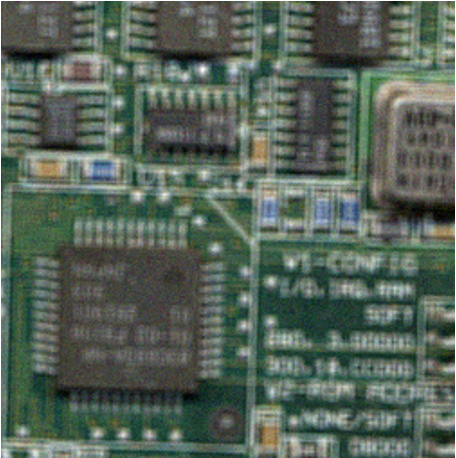} 
\hspace{2mm}
\includegraphics[height=5.3cm]{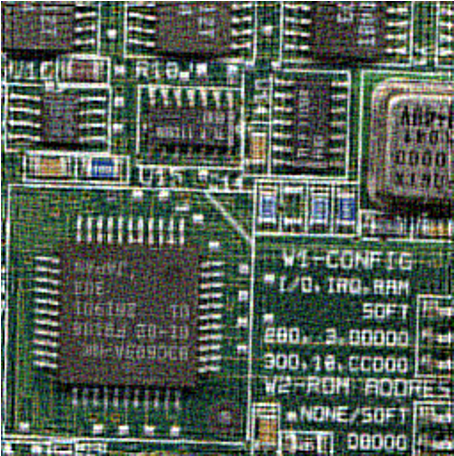}
\caption{Application of the Wiener deconvolution algorithm to a blurred image in the presence of noise. The left panel shows the ideal image, the central panel shows the measured image, and the right panel shows the results of the deconvolution.}
\label{fig:w2_Wiener}
\end{figure}

The Wiener gain $G$ weights the contribution of signal and noise to the measured image. For the frequencies with a small noise contribution, $S_{nn} \ll S_{II}$, we get $G = 1/\tilde{p}$. This gain corresponds to the direct deconvolution given by Eq.~\ref{eq:w2_deconv_direct}. For the frequencies with a large noise, when $S_{nn} \gg S_{II}$, the optimal gain is $G \rightarrow 0$, and we disregard these frequencies in the image estimation. Fig.~\ref{fig:w2_Wiener} shows an example of the application of the Wiener deconvolution to a noisy and blurry image.

Since the true image is unknown, the image power spectral density and the noise power spectral density can be estimated from the measured data with statistical approaches. For example, $S_{II}$ can be approximated from multiple realisations of similar images, from regions assumed to be signal-dominated, or by modeling the object’s expected spatial correlations. The noise power spectral density, $S_{nn}$, can be estimated from background regions, calibration frames, or high-frequency components where the signal is negligible.

\subsection*{Adaptive optics}

Deconvolution algorithms can help recover the original image from the measured one and reduce the effects of aberrations. However, it is best to avoid aberrations in the first place to reduce the effect of noise. This can be done with adaptive optics. For example, after the discovery of the spherical aberration in Hubble’s primary mirror, engineers corrected the problem not by replacing the primary telescope mirror but by installing adaptive optics between the primary mirror and the cameras~\cite{Feinberg:93}. These optics introduced compensating wavefront errors opposite to those caused by the flawed mirror. This correction restored the telescope’s diffraction-limited performance and allowed Hubble to achieve sharp, high-resolution imaging.

Another example explored in this module is the correction of atmospheric aberrations. Consider a ground-based telescope imaging distant stars. As light propagates through the atmosphere, fluctuations in density, humidity, and temperature induce spatial and temporal variations in the refractive index. These variations introduce random lens-like distortions along the propagation path. The goal of adaptive optics is to correct these aberrations in real time~\cite{beckers1993adaptive}.

\begin{figure}[t]
\centering
\includegraphics[height=7cm]{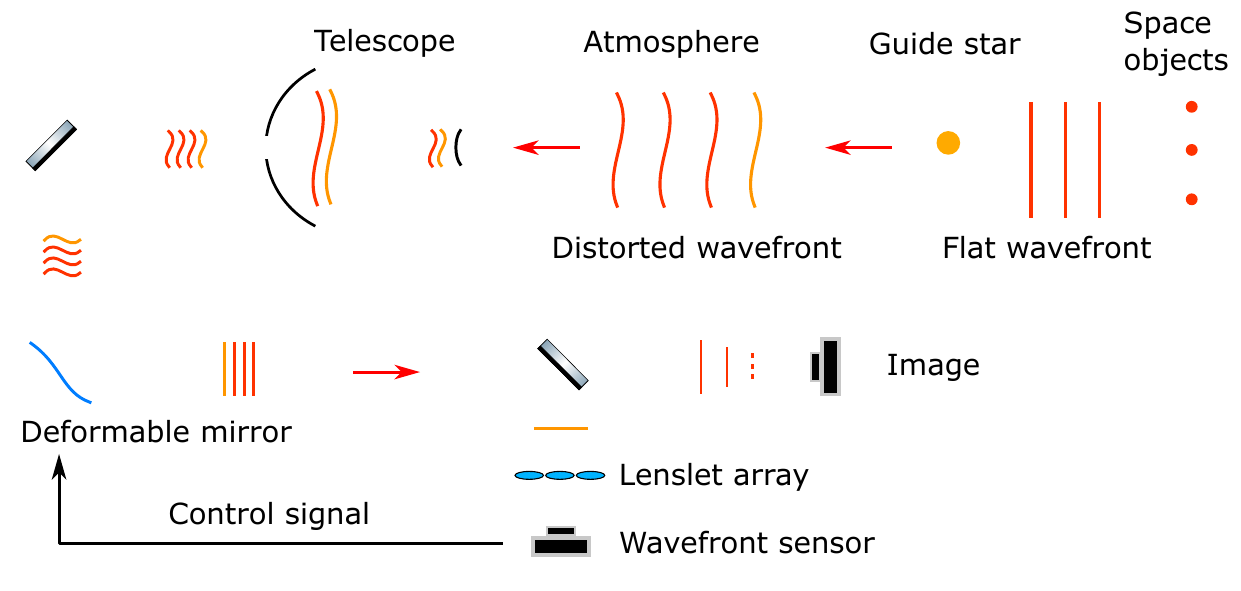} 
\caption{Simplified diagram of real-time correction of atmospheric aberrations using a wavefront sensor and a deformable mirror. The correction is applied before the measurement of scientific targets and prior to the coupling of shot noise into the final image. This approach provides a key advantage over deconvolution-based methods, which are applied after noise has already been introduced into the system.}
\label{fig:w2_adaptive}
\end{figure}

This is achieved using a reference source, a wavefront sensor, and a corrective element such as a deformable mirror. The incoming wavefront is first measured using a wavefront sensor, from which the phase distortions are reconstructed and subsequently compensated by adjusting the shape of a deformable mirror, as shown in Fig.~\ref{fig:w2_adaptive}. Wavefront sensors typically consist of microlens arrays (e.g.\ Shack--Hartmann sensors) that convert local wavefront slopes into measurable displacements of focal spots~\cite{hardy1998adaptive}. Deformable mirrors implement dynamic phase correction by mechanically adjusting a reflective surface using, for example, piezoelectric actuators.

The reference source can be either a natural guide star or an artificial star. Natural stars are often too faint to provide optimal correction. Artificial guide stars are created by projecting a laser beam into the upper atmosphere, where it excites sodium atoms and produces a bright fluorescence signal. Since light from astronomical objects propagates through nearly the same turbulent atmosphere as the return light from the excited sodium layer, correcting the wavefront of the reference source enables correction across the telescope's field of view.

%% file: week3.tex
\section{Imaging: Sensing technologies and pixel-limited resolution}
\label{Lecture3}

Up to this point, we treated the sensing elements as if they were continuous, capable of detecting light at any exact position $(x,y)$ on the sensing surface. However, real sensing elements have finite resolution and cannot distinguish between light hitting $(x,y)$ or $(x+\Delta x, y+\Delta y)$ when $\Delta x, \Delta y$ are smaller than a particular distance. The distance is determined by the properties of the sensing element itself, which we explore in this lecture. We examine how this finite resolution limits the overall resolution of imaging systems, compare it with the diffraction-limited resolution, and discuss the physics of photodetection in semiconductors. We also focus on image processing techniques, including compression and convolution.

In this lecture, we discuss
\begin{itemize}
\item pixel-limited resolution of imaging systems,
\item physics of the sensing technologies,
\item image compression and the Nyquist-Shannon theorem,
\item image processing and convolution kernels.
\end{itemize}

\subsection*{Pixel-limited resolution}

All sensing surfaces consist of a finite number of sensing elements. For example, the human eye contains $\sim 10^8$ photoreceptor cells, with a density of $\sim 1.5 \times 10^5$\,mm$^{-2}$ in the most sensitive region, the fovea~\cite{rodieck1998first}. The number of pixels in modern CCD cameras is $\sim 10^7$, with pixel densities comparable to those found in the eye, depending on the sensor design.

\begin{figure}[t]
\centering
\includegraphics[height=3.9cm]{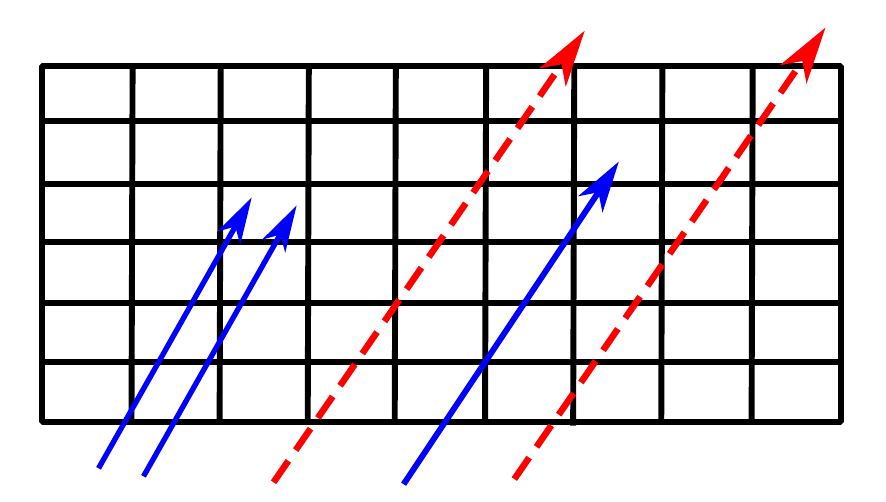} 
\hspace{2mm}
\includegraphics[height=3.9cm]{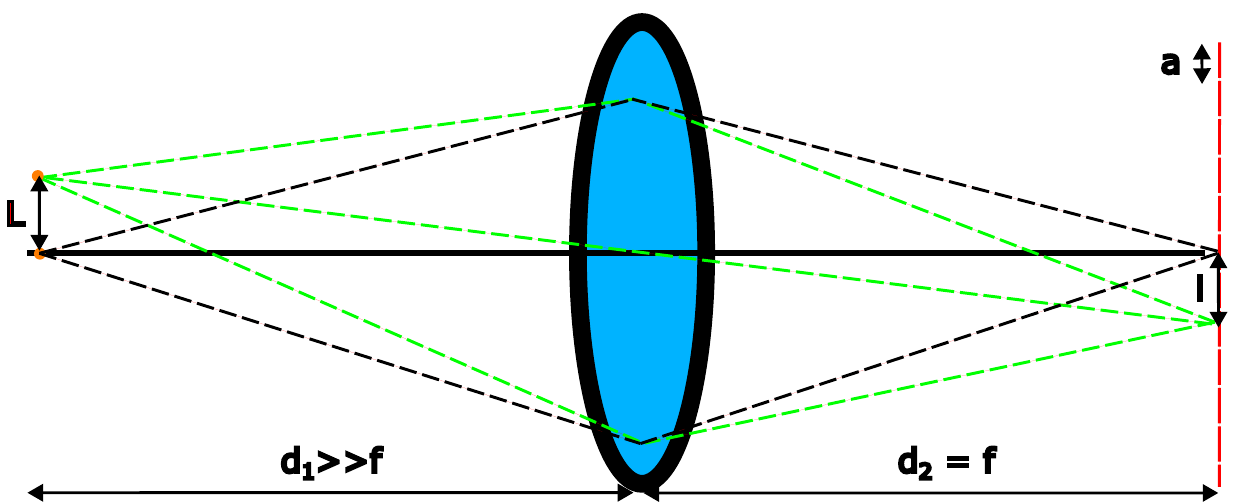} 
\caption{Example of pixel arrangement on a sensing surface (left). Two blue rays that hit the same pixel are indistinguishable by the sensing element. The surface absorbs light of a particular wavelength and may transmit other wavelengths, such as dashed red rays. An illustration of the pixel-limited resolution (right). An imaging system can distinguish two point sources in the source plane if their images are formed on different pixels in the sensing surface.}
\label{fig:w3_pixels_res}
\end{figure}

Each cell or pixel converts absorbed photons into an electric signal: charge or current. The "precise position" of the photon hitting the pixel does not influence the total electric signal, as shown in Fig.~\ref{fig:w3_pixels_res}\,(left), and the signal is proportional to the total number of photons that hit the pixel during the exposure time. The precise position of a ray is a term from geometrical optics and is used loosely here because of the finite spatial extent of light–matter interactions. For example, the characteristic cross-section for an atom–light interaction is on the order of $\frac{3}{2\pi}\lambda^2$, which sets a fundamental limit on how precisely the absorption event can be spatially defined.

In this module, we define the pixel-limited resolution as follows: we will say that we can resolve two point sources with our imaging system if their images are formed in different pixels, as shown in Fig.~\ref{fig:w3_pixels_res}\,(right). The pixel-limited resolution for distant sources, when $d_1 \gg f$, is given by the equation
\begin{equation}\label{eq:w3_pixel_res_def}
    \frac{L}{d_1} = \frac{l}{f} = \frac{a}{f},
\end{equation}
where $L$ and $l$ are separations between the point sources and their images, as usual, and $a$ is the pixel size.

The definition of the pixel-limited resolution given by Eq.~\ref{eq:w3_pixel_res_def} is a useful approximation, but imaging systems can surpass it. A key limitation of the definition is that it treats pixels as hard boundaries for resolution, whereas in reality, the image of a point source is spread over multiple pixels by the system’s point spread function, as discussed in Lecture~\ref{Lecture2}. As a result, two nearby sources can still be distinguished even if their light falls on the same pixel, if their combined intensity pattern across neighboring pixels can be reliably separated.  Modern techniques exploit this: by fitting the measured intensity distribution to a model point spread function, it is possible to estimate source positions with subpixel precision, even orders of magnitude smaller than the pixel size, as in high-precision astrometry for exoplanet detection~\cite{Gai_2022, Lizzana_pixel}. However, such improvements depend on high signal-to-noise ratios, system calibration, and on a particular application. Therefore, we will utilise our definition of the pixel-limited resolution as a robust approximation.

A good imaging system balances the pixel-limited and diffraction-limited resolution for a particular wavelength. As an example of such a well-balanced imaging system, we consider a human eye. From the density of photosensitive cells, we get an effective pixel size of $a=2.5$\,um and the pixel-limited resolution is $\approx 10^{-4}$\,rad. The radius of the focused beam due to diffraction is $\approx 4.2$\,um for the iris diameter of 4\,mm and is comparable to the pixel size, $a$. A human eye can also resolve colours by utilising four types of cells: S, M, L, and rods. S cells or blue cones are mostly sensitive to blue light with a central wavelength of $420$\,um. M cells or green cones have a central wavelength of $540$\,um. L cells or red cones are mostly sensitive to red light with a central wavelength of $580$\,um. Rods have a broader sensitivity than S, M, and L cells with a central wavelength of $500$\,um. A human brain can distinguish different colours by computing the amount of activated cells of a particular type.

\subsection{Photodetection technologies}

Recording light has historically faced several key challenges related to sensitivity and practicality. Early materials such as plant extracts, natural dyes, and bitumen exhibited only weak light sensitivity, requiring long exposure times. Silver salts, such as silver chloride, bromide, and iodide, provided a much stronger photochemical response. However, achieving image stability was equally critical: once an image was formed, it needed to remain unchanged after exposure, which required fixing processes to prevent further photo reactions. In addition, the process needed to support reproducibility, such as creating multiple copies of photographs via negatives.

Photographic plates were the earliest practical imaging detectors, using light-sensitive chemical emulsions to record images with high spatial resolution, and they played a central role in early astronomy and spectroscopy. However, they are relatively insensitive, non-linear, and require chemical development, so they have largely been replaced by electronic detectors. Charge-coupled devices (CCDs) marked a major advance by converting incoming photons into stored charge that is read out with low noise and good linearity. Complementary metal–oxide–semiconductor (CMOS) sensors, while initially noisier than CCD, have improved and now dominate most imaging applications due to their low power consumption, fast readout, and ability to integrate processing electronics on-chip. CMOS sensors are widely used in consumer cameras, machine vision, and increasingly in scientific instruments. Today, CCDs remain important in niche areas requiring the highest image quality.

\subsection{Photographic plates}

We start with photographic plates because, despite being an older technology, they embody fundamental processes of light detection that are closely related to those in modern semiconductor devices. After hitting the emulsion, photons trigger electronic excitations and lead to chemical changes in silver halide grains. The process forms a latent image that is invisible immediately after exposure. The plate is then developed in a chemical bath, which reduces the exposed silver ions to metallic silver. The process creates a visible image and prevents the grains from further reaction to light. 

Photographic plates can also serve as negatives, and multiple positive copies can be made by their projection. Photographic negatives appear matte because the metallic silver forms as tiny, irregular grains scattered throughout the emulsion, rather than as a smooth surface in silver-coated mirrors. These microscopic grains absorb and scatter light during the projection process and help reconstruct the positive image: more photons during the exposure lead to dark spots on the negative, and then these dark silver spots block light during the projection and lead to bright spots on the positive image.

Silver halides are semiconductors with a crystalline lattice structure and a bandgap of $2-3.5$\,eV. Bandgaps are energy ranges in a solid where no electron states can exist. The gaps separate allowed continuous energy bands. 
In this section, we consider examples of why the bandgaps are formed in crystals. We first start with a free electron in a 1D box of length $L$. The time-independent Schr\"odinger equation and its solution are given by the equations below
\begin{equation}\label{eq:w3_free_electron}
\left( -\frac{\hbar^2 \nabla^2}{2 m_e} \right) \psi(x) = E \psi(x) \hspace{5mm}=>\hspace{5mm}
 \psi(x) = \frac{1}{\sqrt{L}}e^{ikx}, \hspace{5mm} E = \frac{\hbar^2 k^2}{2 m_e}
\end{equation}
where $k$ is the electron's wave number, $1/\sqrt{L}$ is the normalisation factor of the wave function $\psi(x)$, $E$ is the energy of the free electron, and $m_e$ is the electron mass. Eq.~\ref{eq:w3_free_electron} shows that all energy states are possible for a particular wave number, $k$, as shown in Fig.~\ref{fig:w3_el_states}\,(ii), and no bandgaps exist in this case.

We now consider an electron in an isolated atom. The energy states can be found from the time-independent Schr\"odinger equation
\begin{equation}\label{eq:w3_bound_electron}
\left( -\frac{\hbar^2 \nabla^2}{2 m_e} + V(x) \right) \psi(x) = E \psi(x),
\end{equation}
where $V(x)$ is the potential field on the atom, including its nucleus and other electrons. The energy states are discrete in the bound case, as shown in Fig.~\ref{fig:w3_el_states}\,(i).

\subsubsection{Energy splitting of interacting oscillators}

\begin{figure}[t]
\centering
\includegraphics[height=3.3cm]{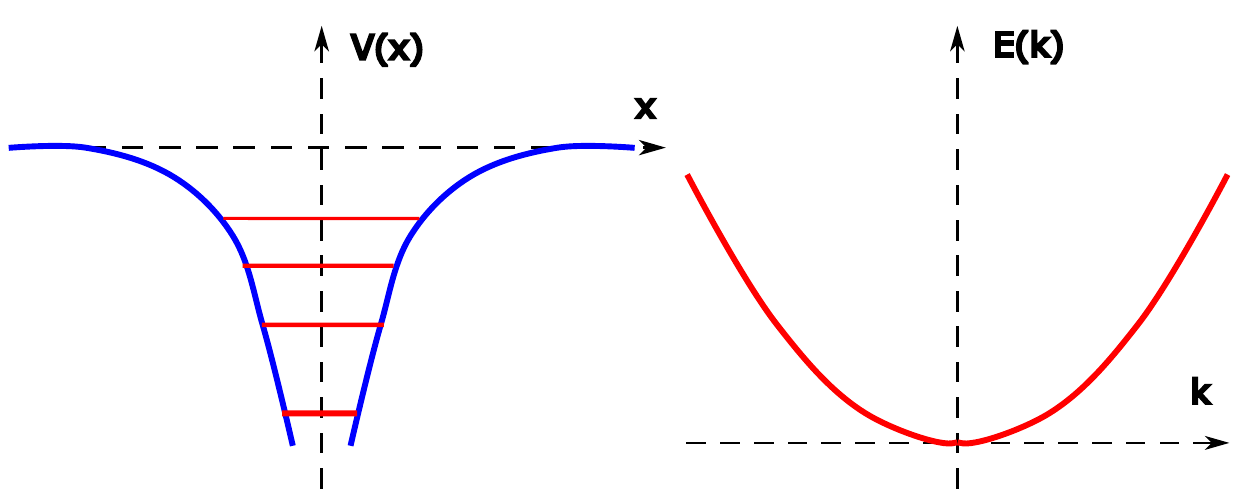} 
\hspace{1mm}
\includegraphics[height=3.3cm]{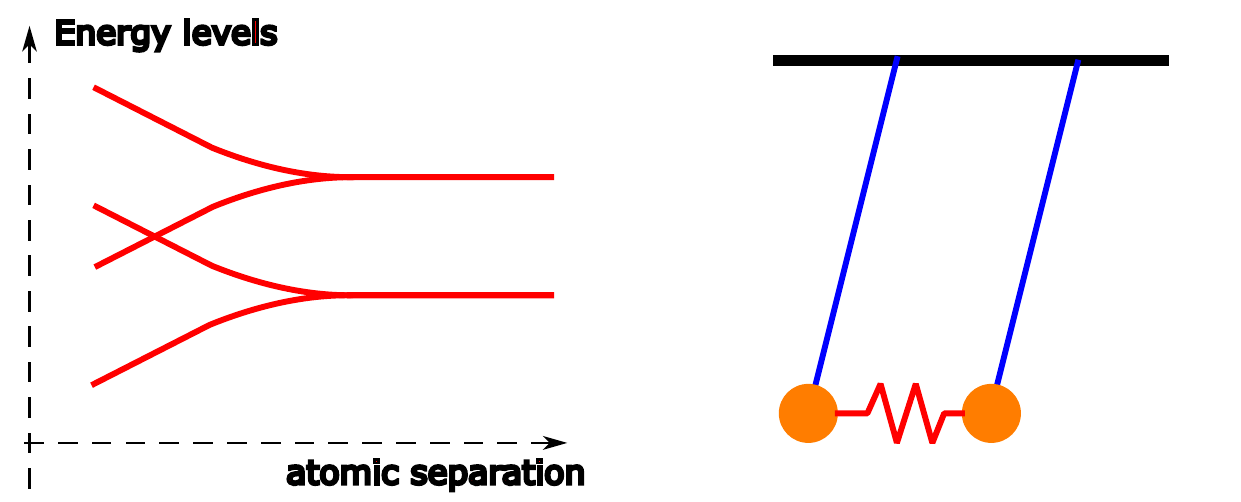} 
\caption{(i) Energy states of a bound electron in an atom with a potential $V(x)$. (ii) Energy states of a free electron. (iii) Splitting of the atomic energy levels in interacting atoms. (iv) Splitting of the energy levels of coupled oscillators.}
\label{fig:w3_el_states}
\end{figure}

The discrete energy states of electrons start to split when atoms start to interact with each other, as shown in Fig.~\ref{fig:w3_el_states}\,(iii). This is analogous to a coupled pendulum system, shown in Fig.~\ref{fig:w3_el_states}(iv). If we have two identical oscillators, then they have the same eigen frequencies. However, if the oscillators interact with each other, for example, via a spring, then the energy degeneracy breaks. We can explore this by considering a Lagrangian of the coupled system, which is given by the equation
\begin{equation}
\mathcal{L} = \frac{1}{2}m\dot{x_1}^2 + \frac{1}{2}m\dot{x_2}^2 - \left(\frac{1}{2L}mgx_1^2 + \frac{1}{2L}mgx_2^2 + \frac{1}{2}\kappa(x_1 - x_2)^2  \right),
\end{equation}
where $m$ is the mass of each suspended mass, $g$ is the free-fall acceleration, $x_1$ and $x_2$ are coordinates of each oscillator, $L$ is the pendulum length, and $\kappa$ is the spring stiffness. We find the dynamics of the coupled system by the Euler-Lagrange equations
\begin{equation}\label{eq:w3_coupled_x12}
\begin{split}
& \ddot{x_1} + \frac{g}{L}x_1 + \frac{\kappa}{m}(x_1 - x_2) = 0 \\
& \ddot{x_2} + \frac{g}{L}x_2 + \frac{\kappa}{m}(x_2 - x_1) = 0.
\end{split}
\end{equation}

We can solve the coupled equations from Eq.~\ref{eq:w3_coupled_x12} by introducing new variables $y_1 = x_1 + x_2$, which is the common motion of the pendulums, and $y_2 = x_1 - x_2$, which is the differential motion of the pendulums. The solutions to the equations are given by the equations
\begin{equation}\label{eq:w3_coupled_y12}
\begin{split}
& \ddot{y_1} + \frac{g}{L}y_1 = 0 \hspace{18mm} => \hspace{5mm} \omega^2_{\rm comm} = \frac{g}{L} \\
& \ddot{y_2} + \frac{g}{L}y_2 + \frac{2\kappa}{m}y_2 = 0 \hspace{5mm} => \hspace{5mm} \omega^2_{\rm diff} = \frac{g}{L} + \frac{2\kappa}{m},
\end{split}
\end{equation}
where $\omega_{\rm comm}$ and $\omega_{\rm diff}$ are the eigen frequencies of the common mode and differential motion of the oscillators. Eq.~\ref{eq:w3_coupled_y12} shows that $\omega_{\rm comm} \neq \omega_{\rm diff}$ for $\kappa \neq 0$.

\subsubsection{Electron energy levels in a crystal}

Similar to the interacting oscillators, when many atoms come together in a solid, their discrete electron energy levels split into a large number of closely spaced levels that effectively form continuous energy bands. For example, the number of atoms in a silver halide grain is $10^7 - 10^{10}$. The periodic arrangement of atoms in the crystal lattice creates a repeating potential that leads to constructive and destructive interference of electron wavefunctions, producing allowed and forbidden energy regions. For an electron in a periodic potential $V(x+q) = V(x)$, the Bloch theorem~\cite{kittel2004introduction} states that the electron wavefunction is given by the equation
\begin{equation}
\psi_k(x) = e^{ikx} u_k(x),
\end{equation}
where $u_k(x+q) = u_k(x)$ is a periodic function that may be different for every electron wave number. The theorem implies that the probability density for finding an electron at a given position is periodic in space, i.e.\ $|\psi_k(x)|^2 = |u_k(x)|^2$, reflecting the underlying periodicity of the crystal lattice. The electron wavefunctions are delocalised over the entire lattice rather than tied to a single atom, and the electrons extend across the whole crystal as Bloch waves. However, in real crystals, impurities, thermal vibrations, or disorder can partially localise electrons, and the full delocalisation that we consider below is an idealisation valid for a perfectly periodic crystal.

\begin{figure}[t]
\centering
\includegraphics[height=3.3cm]{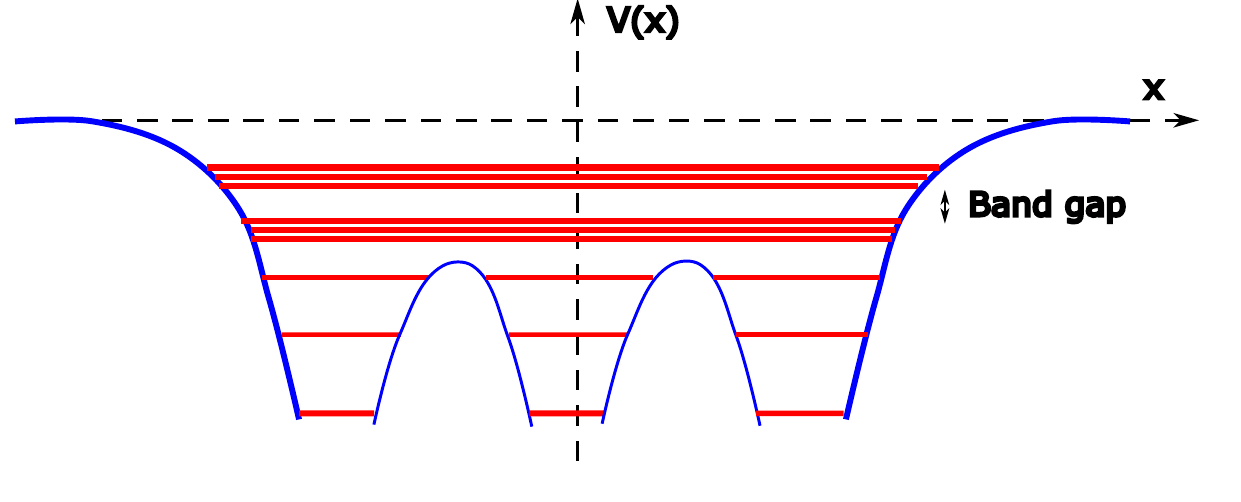} 
\hspace{1mm}
\includegraphics[height=3.3cm]{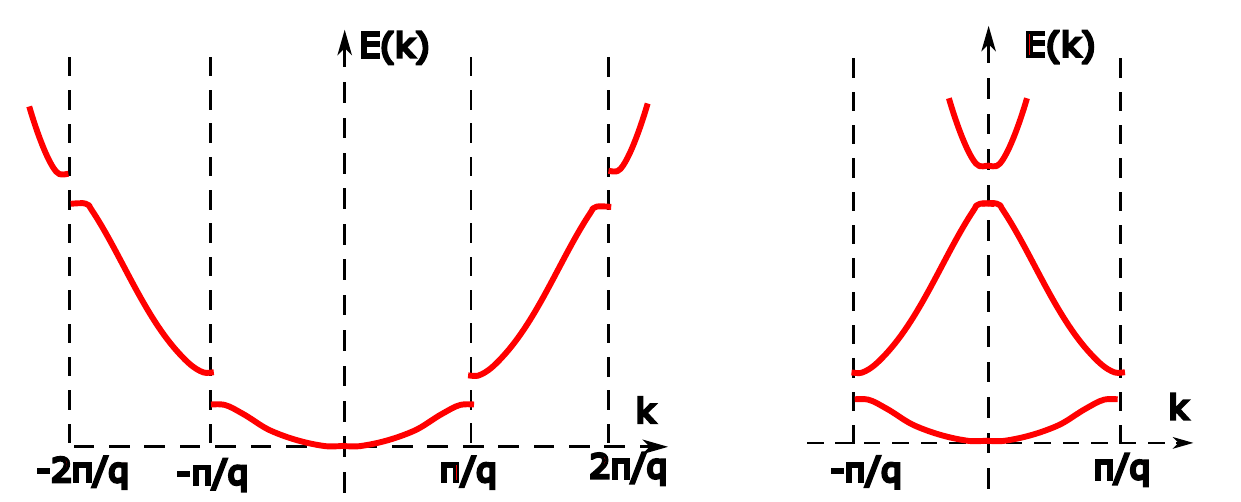} 
\caption{Splitting of electron energy levels in a crystal (left). Sketch of the solutions of the Schr\"odinger equation for electron energy levels (right).}
\label{fig:w3_bandgaps}
\end{figure}

Since the potential $V(x)$ is periodic, it can be expanded as a Fourier series,
\begin{equation}
V(x) = \sum_{n=-\infty}^{\infty} \tilde{V}(nG) e^{inGx}, \hspace{1cm}
G = \frac{2\pi}{q},
\end{equation}
where $G$ is a reciprocal lattice vector and $q$ is the lattice period. As an example, consider the case where $\tilde{V}(nG)=0$ except for $n=\pm 1$, yielding a sinusoidal potential
\begin{equation}
V(x) = V_0 \cos(Gx).
\end{equation}
In this case, the time-independent Schr\"odinger equation reduces to the Mathieu equation, which has analytical solutions.
The resulting band structure exhibits energy gaps at $k = \pm G/2, \pm 3G/2, \dots$, as shown in Fig.~\ref{fig:w3_bandgaps}\,(centre).

In the weak-potential limit, the first bandgap is proportional to $|V(G)|$, while higher-order gaps arise from higher-order coupling processes. More generally, for a periodic potential, the $j$-th bandgap is determined by the magnitude of the corresponding Fourier component $|\tilde{V}(jG)|$.

In crystals, the energy $E(k)$ can be wrapped into the first Brillouin zone, as shown in Fig.~\ref{fig:w3_bandgaps}\,(right), because the crystal’s periodicity makes wavevectors $k$ and $k + nG$ physically equivalent. Adding a reciprocal lattice vector $G = 2\pi/q$ does not change the Bloch wavefunction except for a phase factor, and the energies repeat in k-space. Unlike a free electron in a box, where larger $k$ directly corresponds to higher momentum and energy, in a periodic lattice, the electron’s crystal momentum is only defined modulo $G$, and higher k values outside the first zone can be folded back into it without changing physical observables.

Electrons occupy energy bands according to the Pauli exclusion principle, filling lower-energy states first. The valence band is the highest band that is fully occupied at zero temperature. The conduction band is the next higher band that is empty or partially filled and provides states for electron conduction. In an insulator, the bandgap between valence and conduction bands is large ($\sim 10$\,eV), and thermal energy at room temperature of $\approx 25$\,mK is insufficient to excite electrons across the gap, preventing conductivity. In a semiconductor, the bandgap is smaller ($\sim 1-2$ eV), and a modest fraction of electrons can thermally excite into the conduction band at room temperature, enabling limited conductivity. The conductivity can be controlled by temperature or doping of the semiconductor. The number of electrons per unit volume, $n$, in a semiconductor is given by the equation~\cite{shur1990physics}
\begin{equation}
n = \int_{E_c}^\infty D(E) f(E) dE,
\end{equation}
where energy $E_c$ corresponds to the bottom of the conducting band, $D(E)$ is the density of states available in the conducting band, and will be discussed in more detail in Lecture~\ref{Lecture6} when we will consider laser transitions in semiconductors. Function $f(E)$ is the Fermi-Dirac occupation probability given by the equation
\begin{equation}
f(E) = \frac{1}{\exp(-\frac{E-\mu}{k_B T})+1},
\end{equation}
where $\mu$ is the chemical potential of the crystal. The potential represents the energy cost required to add or remove an electron from the system, and at absolute zero, it coincides with the Fermi energy, the highest occupied energy level.

\subsubsection{Photodetection process}

Thermal electrons in silver halides do not degrade photodetection significantly because, at typical operating temperatures, the thermal energy $k_B T$ is too small to promote electrons across the relatively large bandgap in the absence of light. Thermally generated excitations tend to recombine rapidly and do not produce sustained charge separation in silver halide crystals.

In contrast, photon absorption creates electron--hole pairs if the photon energy exceeds the bandgap as given by the equation
\begin{equation}
\frac{hc}{\lambda} \geq E_g \hspace{5mm} => \hspace{5mm} \lambda \leq \frac{1.24}{E_g}\,{\rm um},
\end{equation}
where $E_g$ should be in units of eV in the last equation.
For AgBr with a bandgap of about 2.6\,eV, this corresponds to $\lambda \leq 480$\,nm and photons with a longer wavelength cannot trigger photodetection.

After photodetection in a silver halide crystal, the excited electron migrates through the lattice and becomes trapped at a sensitivity speck, where the time-independent Schr\"odinger equation is
\begin{equation}
\left( -\frac{\hbar^2 \nabla^2}{2 m_e} + V(x) + V_{\rm speck}(x) \right) \psi(x) = E \psi(x),
\end{equation}
where $V_{\rm speck}(x)$ is the potential near the speck. The electron reduces nearby silver ions and forms a small cluster of neutral silver atoms. In quantum terms, the speck collapses the electron’s wave function and performs a quantum measurement. Each newly trapped electron adds another silver atom to the growing cluster, and the presence of the cluster itself enhances further growth.

\subsection{CCD cameras}

Charge-coupled device (CCD) cameras represent a major advance from photographic plates by converting light directly into electronic signals. The technology was recognised with the Nobel Prize in Physics 2009, awarded to W. S. Boyle and G. E. Smith. Originally invented as memory devices, CCDs store information as charge in tiny potential wells (pixels), where the presence of charge represents a “1” and its absence a “0”. In imaging applications, incident photons generate electron-hole pairs, and the electrons are collected in these wells and then transferred across the chip for readout. 

\begin{wrapfigure}{r}{0.4\textwidth}
    \centering
    \includegraphics[width=0.38\textwidth]{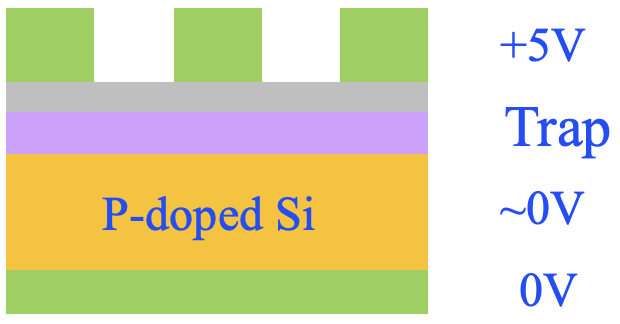} 
    \caption{A scheme of a CCD pixel. The dielectric layer is shown in gray, and the depletion region is shown in pink.}
    \label{fig:w3_MOS}
\end{wrapfigure}

Each CCD pixel has a metal–oxide–semiconductor (MOS) structure: a P-doped semiconductor substrate forms the bulk, above which lies a thin oxide layer (typically SiO$_2$) that electrically insulates the surface from a patterned metal electrode held at a positive potential, as shown in Fig.~\ref{fig:w3_MOS}. P-type semiconductors, such as boron-doped silicon, are created by doping an intrinsic semiconductor with an electron acceptor element. In P-type semiconductors, holes are the majority carriers and electrons are the minority carriers. The opposite is true in N-type semiconductors, such as phosphorus-doped silicon.

The positive voltage repels holes under the electrode in the P-type semiconductor and creates a depletion region: a zone free of mobile charge carriers. The zone acts as a potential well for photoelectrons. When light generates electron–hole pairs in the semiconductor, the electrons are attracted into the depletion layer under the positively biased gate and form the pixel’s signal. We find the thickness, $W$, of the depletion layer by solving the Poisson equation 
\begin{equation}
\frac{d^2\phi}{dx^2} = -\frac{\rho}{\epsilon} = \frac{eN_A}{\epsilon},
\end{equation}
where $\phi(x)$ is the electrostatic potential, $\phi(0) = V_0$, $\phi(W) = 0$,  $V_0$ is the positive potential applied to the metal gate of the pixel, $\rho = -eN_A$ is the charge density, $N_A$ is the density of the acceptor impurities and is $\sim 10^{14}-10^{16}$\,cm$^{-3}$ for CCD cameras, and $\epsilon$ is the permiability of the semiconductor. The solution to the Poisson equation in the range $(0, W)$ is given by the electrostatic potential
\begin{equation}
\phi(x) = \frac{eN_A}{2\epsilon}(x-W)^2\hspace{5mm} => \hspace{5mm} W = \sqrt{\frac{2 \epsilon V_0}{e N_A}} \sim 1\,{\rm um}.
\end{equation}

The photoelectrons are created in the depletion region and are accelerated towards the positive gate by the electrostatic potential. However, the electrons cannot reach the positive terminal because of the dielectric layer. A photon can also produce electron-hole pairs outside of the depletion region in the P-type substrate, but the photoelectrons recombine with holes because there is no electric field outside of the depletion region in the first-order approximation. 

After the exposure time is complete, the electrons stored in each CCD pixel’s depletion region are read out by shifting them through the array. The transfer is done by sequentially applying clocked voltages to the gate electrodes, which move the charge packets from pixel to pixel along the vertical and then horizontal shift registers toward a readout amplifier. At the output node, each packet is transferred onto a capacitor where its charge induces a voltage proportional to the number of electrons.

\subsection{CMOS cameras}

Complementary metal–oxide–semiconductor (CMOS) image sensors were developed as an evolution of CCD technology. The CMOS architecture integrates photodetectors and readout electronics on the same chip. Unlike CCDs, which transfer charge across the entire array to a single output node, CMOS sensors allow each pixel to have its own amplifier and readout circuitry, enabling random access and faster readout speeds.

In CMOS, “complementary” refers to the use of both N-type and P-type metal–oxide–semiconductor field-effect transistors in a single circuit. By combining N-channel and P-channel transistors in a complementary way, the circuit can efficiently switch between logic states with a low static power consumption, because at any moment, only one type of transistor is on while the other is off. The same fabrication techniques are utilised for almost all contemporary electronics~\cite{rabaey2003digital}, including microprocessors and memory chips.

\begin{wrapfigure}{r}{0.4\textwidth}
    \centering
    \vspace{-4mm}
    \includegraphics[width=0.38\textwidth]{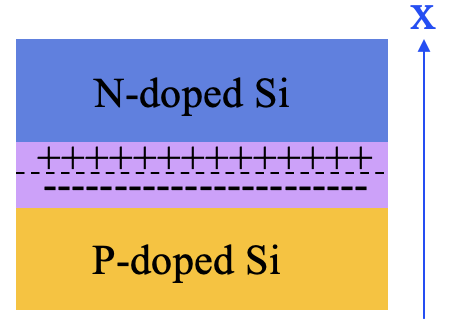} 
    \caption{A scheme of a PN-junction. The depletion region is shown in pink.}
    \label{fig:w3_CMOS}
\end{wrapfigure}

The key component of each CMOS pixel is a photodetector: a PN-junction. If P- and N-doped semiconductors are in contact, then they form a depletion layer as shown in Fig.~\ref{fig:w3_CMOS}. The layer forms in a PN junction because when P-type and N-type semiconductors are joined, electrons from the N-side diffuse into the P-side and recombine with holes, and holes from the P-side diffuse into the N-side and recombine with electrons. This diffusion leaves behind fixed ionised donor and acceptor atoms and creates a region devoid of mobile charge carriers.

We find the width of the depletion layer by solving the Poisson equation in the P-doped and N-doped regions
\begin{equation}\label{eq:w3_cmos_poisson}
\begin{split}
& \frac{d^2\phi_P}{dx^2} = \frac{eN_A}{\epsilon} \hspace{1mm}=>\hspace{1mm} E_P = \frac{d\phi_P}{dx} = \frac{eN_A}{\epsilon} (x - W_P)\hspace{1mm}=>\hspace{1mm} \phi_P(x) = \frac{eN_A}{2\epsilon}(x-W_P)^2 \\
&\frac{d^2\phi_N}{dx^2} = -\frac{eN_D}{\epsilon} \hspace{1mm}=>\hspace{1mm} E_N = \frac{d\phi_N}{dx} = -\frac{eN_D}{\epsilon} (x + W_P) \hspace{1mm}=>\hspace{1mm} \phi_N(x) = -\frac{eN_D}{2\epsilon}(x+W_N)^2 + V_{bi},
\end{split}
\end{equation}
where $W_P$ and $W_N$ are the widths of the depletion layer in the P-doped and N-doped semiconductors, $N_A$ and $N_D$ are the concentrations of the acceptors and donors,  $E_P$ and $E_N$ is the electric field in the junction, and $V_{bi}$ is the integration constant and is the equilibrium electrostatic potential difference across the PN junction’s depletion region that balances carrier diffusion and prevents net current flow.

The charge redistributes inside the PN-junction, but it stays electrically neutral: $W_P N_A = W_N N_D$ and, therefore, $E_N(0) = E_P(0)$, and the electrostatic potential must be continuous: $\phi_P(0) = \phi_N(0)$. Note that we have introduced the constant of integration $V_{bi}$ in Eq.~\ref{eq:w3_cmos_poisson} to satisfy this condition. We find that the depletion layer thickness is given by the equation
\begin{equation}
\begin{split}
&W_P = \sqrt{\frac{2 \epsilon}{e}\frac{N_D}{N_A(N_A + N_D)} V_{bi}}, \hspace{1cm} W_N = \sqrt{\frac{2 \epsilon}{e}\frac{N_A}{N_D(N_A + N_D)} V_{bi}}, \\
&W = W_P + W_N = \sqrt{\frac{2 \epsilon}{e}\frac{N_A + N_D}{N_A N_D} V_{bi}}.
\end{split}
\end{equation}

As in the CCD case, photodetection occurs in the depletion layer. Incident photons with sufficient energy generate electron–hole pairs in the depletion region, where the built-in electric fields $E_N$ and $E_P$ separate them: electrons are driven toward the N-side and holes toward the P-side, preventing recombination and creating a photocurrent. Carriers generated just outside the depletion region can still contribute if they diffuse into it before recombining. This separation of charge produces a measurable current (or voltage under open-circuit conditions) proportional to the light intensity.

The built-in potential of a PN junction is found from equilibrium carrier statistics by requiring that the Fermi level is constant across the junction and is given by the equation~\cite{sze2012semiconductor}
\begin{equation}
V_{bi} = \frac{k_B T}{e}\ln\frac{N_A N_D}{n_i^2},
\end{equation}
where $n_i$ is the intrinsic carrier concentration and represents the number of thermally generated charge carriers in a pure semiconductor at equilibrium. For silicon at room temperature, $n_1 \approx 1.5\times 10^{10}$\,cm$^{-3}$ and assuming typical doping levels of $N_A = N_D = 10^{16}$\,cm$^{-3}$, we get $V_{bi} \approx 0.7$\,V and the thickness of the depletion layer of $W \approx 0.4$\,um. The maximum value of the electric field in the PN-junction is at $x=0$ and equals $E_P(0) = E_N(0) \approx 3$\,MV/m. This is a large electric field which is comparable to the breakdown field of air at standard conditions.

\subsection*{Digital images}

Once an image is captured with a CCD or CMOS camera, it is typically stored digitally and can be further processed for tasks such as edge detection, sharpening, or smoothing. During the exposure, each pixel collects light, which is converted into an electric charge or current proportional to the local light intensity. After the exposure, this charge is measured and recorded, providing a quantitative representation of the image. Mathematically, we can represent the image as a 2D array or matrix of size $N_x \times N_y$, where $N_x$ and $N_y$ are the number of pixels along the X and Y axes. Each matrix element is usually an integer from 0 to 255, corresponding to an 8-bit storage format: 0 represents no detected light, while 255 represents the maximum detectable light, i.e., the saturation of the pixel. Saturation occurs when the number of incoming photons exceeds the number of electron-hole pairs the pixel’s depletion region can generate, and no additional light can be recorded.

Photons of different wavelengths excite electrons in the depletion regions of PN junctions. For example, silicon photodiodes are sensitive to light in the 300–1100\,nm range because the silicon bandgap is $1.1$\,eV. The range extends beyond the visible spectrum and is broader than what the human eye can see. However, unlike our eyes, which have different types of photoreceptor cells to distinguish colors, silicon diodes alone cannot differentiate colors. To record color images, mosaic filters are placed over the pixels, allowing only blue, green, or red light to pass through each pixel. After exposure, interpolation algorithms reconstruct the full-color information from these filtered measurements.

Color images are represented using multiple matrices, one for each primary color. In visible-light imaging, three matrices correspond to the blue, green, and red channels, analogous to the S, M, and L cones in the human eye. When displayed on a screen, each pixel emits the appropriate intensity of blue, green, and red light to stimulate the corresponding cones, reproducing the perceived colors. In specialised applications such as optical telescopes, additional matrices may be stored for each measured wavelength, allowing detailed spectral analysis of astrophysical objects.

\subsection*{Image compression}

Typical images from commercial cameras range from 5 to 10\,MB in size. To reduce storage requirements, images can be compressed, either losslessly or lossily. Lossless compression algorithms, such as entropy encoding, preserve all original information while storing it more efficiently~\cite{gonzalez2018digital}. For example, if a row contains 10 pixels with no light, the corresponding matrix elements are all zeros: 0000000000. Instead of storing 10 bytes, a lossless algorithm can encode this as “10 zeros in sequence,” requiring only 2 bytes.

\begin{figure}[t]
\centering
\includegraphics[height=3.7cm]{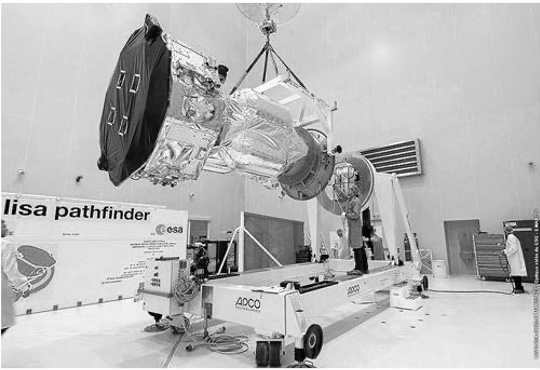} 
\hspace{1mm}
\includegraphics[height=3.7cm]{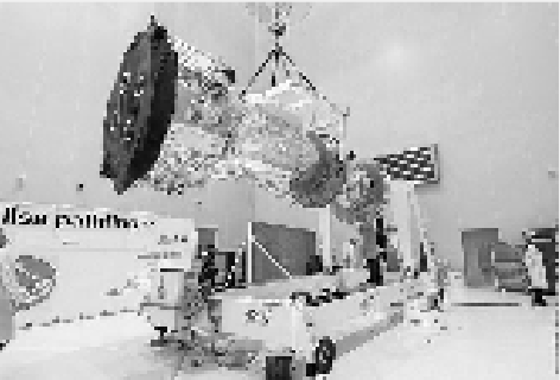} 
\hspace{1mm}
\includegraphics[height=3.7cm]{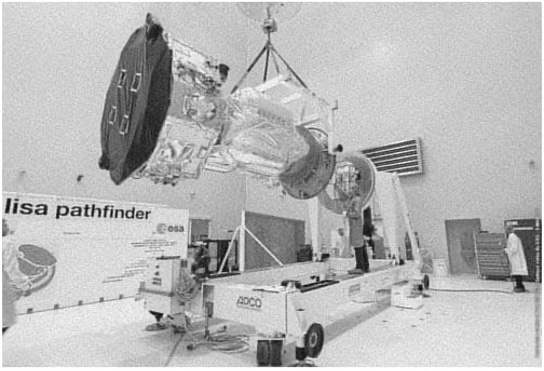} 
\caption{(Uncompressed image of LISA Pathfinder (left). Incorrect compression of the image, where high-frequency components alias into the compressed image (centre). Correct compression of the image (right).}
\label{fig:w3_compression}
\end{figure}

In this section, we focus on lossy compression, which further reduces image size at the expense of resolution. The two main steps are: (i) filtering the image to remove high spatial frequencies, and (ii) downsampling. Step (i) is crucial to avoid aliasing, which occurs when high-frequency signals are incorrectly mapped to lower frequencies during downsampling, degrading image quality. An example of incorrect and correct compression is shown in Fig.~\ref{fig:w3_compression}.

Mathematically, this process relies on the Nyquist-Shannon sampling theorem~\cite{gonzalez2008digital}, which states that a signal containing no frequencies higher than $F_{\rm max}$ can be completely determined by its samples spaced $1/(2F_{\rm max})$ apart. For instance, a square image with $1024 \times 1024$ pixels contains $1,048,576$ pixels. To reduce its size by a factor of 16, the sampling frequency along each axis must be reduced by a factor of 4. First, high-frequency components are removed with a spatial-domain filter or by transforming the image into the frequency domain and suppressing Fourier components outside the range $(-f_s/8, f_s / 8)$. Next, the image is downsampled by keeping every 4th pixel along both axes.

Practical compression algorithms often split the image into blocks of $8 \times 8$ or $16 \times 16$ pixels, transform each block into the frequency domain, suppress high frequencies, and then convert back to the spatial domain. Block-based processing reduces computational load compared to transforming the entire image at once while still enabling effective compression~\cite{wallace1992jpeg}.

\subsection*{Convolution kernels}

In image processing, convolution is used to achieve effects such as blurring, sharpening, edge detection, and more. Convolution involves applying a kernel, a small matrix of weights, to an image represented as a matrix~\cite{gonzalez2018digital}. For each pixel, we align the kernel with the surrounding pixels of the same size, flip the kernel both horizontally and vertically, perform an element-wise multiplication between the flipped kernel and the image patch, and then sum the results to obtain the new pixel value. This flipping distinguishes true convolution from cross-correlation, which applies the kernel directly without flipping. For example, a simple matrix given by the equation
\begin{equation}
    M_I = 
    \begin{pmatrix}
    0 & 0 & 0 \\
    0 & 1 & 0 \\
    0 & 0 & 0
    \end{pmatrix},
\end{equation}
is the identity convolution kernel, where the original pixel value is multiplied by 1, and all its neighbours are multiplied by zero. A smoothing kernel can have different representations, such as the one given by the equation
\begin{equation}
    M_s = \frac{1}{9}
    \begin{pmatrix}
    1 & 1 & 1 \\
    1 & 1 & 1 \\
    1 & 1 & 1
    \end{pmatrix},
\end{equation}
and takes an average of the central pixel with its neighbours. Edge detection subtracts the values of neighbouring pixels from a particular pixel to highlight edges in the image. The corresponding kernel matrix is given by the equation
\begin{equation}
    M_E = 
    \begin{pmatrix}
    -1 & -1 & -1 \\
    -1 & 8 & -1 \\
    -1 & -1 & -1
    \end{pmatrix}.
\end{equation}

\begin{figure}[t]
\centering
\includegraphics[height=3.7cm]{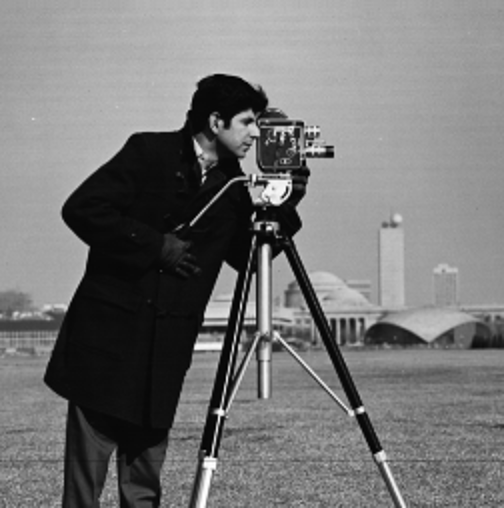} 
\hspace{1mm}
\includegraphics[height=3.7cm]{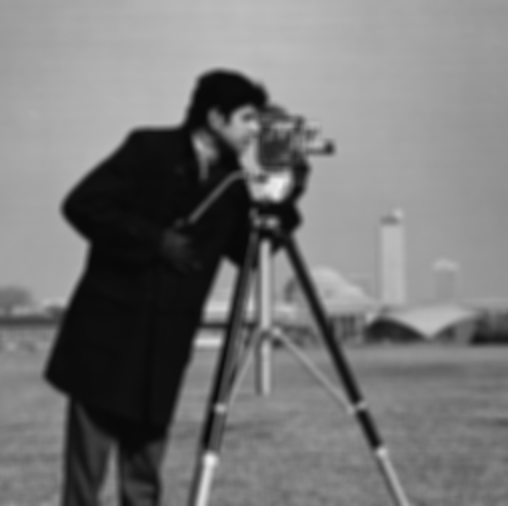} 
\hspace{1mm}
\includegraphics[height=3.7cm]{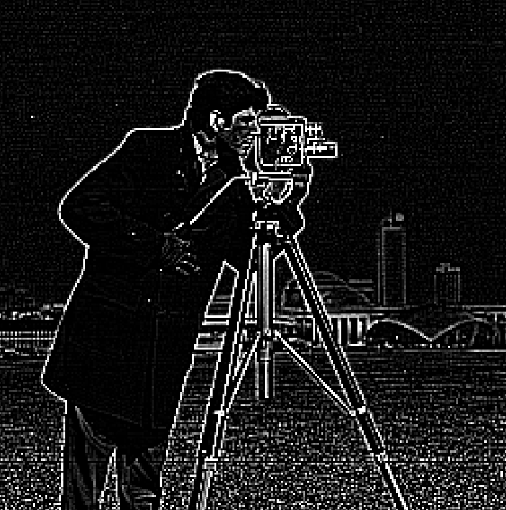} 
\hspace{1mm}
\includegraphics[height=3.7cm]{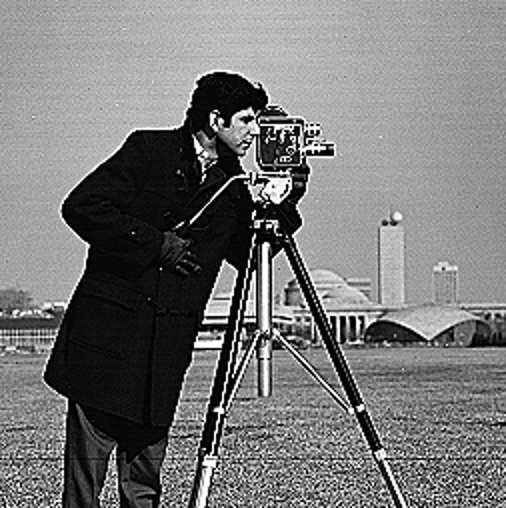} 
\caption{(i) Original image. (ii) Smoothed image. (iii) Edges of the objects. (iv) Sharpened image.}
\label{fig:w3_kernels}
\end{figure}

An image can also be sharpened by applying a kernel $M_E + M_I$, which computes the edges of the image and then adds them to the original image. Examples of the application of the convolution kernels are shown in Fig.~\ref{fig:w3_kernels}. Convolution kernels can be one-dimensional and, for example, add motion effects, such as the kernel given by the equation
\begin{equation}
    M_m = \frac{1}{7}
    \begin{pmatrix}
    1 & 1 & 1 & 1 & 1 & 1 & 1 
    \end{pmatrix}.
\end{equation}

\subsection{Noise in imaging}

Noise is a fundamental factor that limits the quality of images in all imaging systems. Noise can obscure fine details, reduce contrast, and limit the dynamic range of an image. Quantifying noise is needed for designing imaging systems, choosing exposure settings, and applying post-processing techniques. Noise arises from various sources, including photon shot noise, which reflects the discrete nature of light, and thermal noise in electronic circuits.

\subsubsection{Thermal noise in PN-junctions}

Thermal noise in a PN junction arises from the random thermal motion of electrons and holes in the semiconductor. Even when no external voltage is applied, carriers constantly move and scatter, producing fluctuating currents across the junction. The rate of thermal generation of electron–hole pairs in a semiconductor depends on the intrinsic carrier concentration, $n_i$, and the carrier lifetime, $\tau_g$. The rate is given by the equation~\cite{sze2012semiconductor}
\begin{equation}
\frac{dN}{dt} = \frac{n_i W S}{\tau_g},
\end{equation}
where $S$ is the area of the PN-junction and $WS$ is the volume of the PN-junction. The electron-hole pair production is random and follows the Poisson distribution. Therefore, the deviation of the thermally produced number of electrons and holes is given by the equation
\begin{equation}
\langle N \rangle = \frac{n_i W S \tau}{\tau_g} \hspace{5mm}=>\hspace{5mm} \sigma_N = \sqrt{\langle N \rangle},
\end{equation}
where $\tau$ is the exposure time, $\langle N \rangle$ is the average number of electron-hole pairs, and $\sigma_N$ is the variance of the thermal noise.
For a high-quality PN-junction, $\tau_g \approx 1$\,ms and the electron-hole production rate in one pixel with $S=(3\,{\rm um})^2$ is $dN/dt \approx 40$\,s$^{-1}$ at room temperature. Since the exposure time is typically smaller than 1\,sec in normal light, the thermal noise level is typically low. However, in low-light or long-exposure applications, the thermal generation becomes noticeable, and cooling the sensor is required to reduce the thermal noise.

\subsubsection{Shot noise of light}

Shot noise in imaging arises from the quantum nature of light and the fact that photons arrive at a sensor in a random, Poisson-distributed manner. Even under constant illumination, the number of photons detected by each pixel fluctuates around the average, producing an inherent noise that cannot be eliminated. The average number of photons, $\langle N_\nu \rangle$, is related to the classical power, $P_{\rm cl}$ by the equation
\begin{equation}
\langle N_\nu \rangle h\nu = P_{\rm cl} \tau,
\end{equation}
where $\nu$ is the frequency of light, and $\tau$ is the exposure time. Similar to the thermal noise, the variance is given by the equation
\begin{equation}
\sigma_N = \sqrt{\langle N_\nu \rangle},
\end{equation}
and is particularly significant in low-light conditions, where the number of detected photons is small. Unlike thermal or readout noise, which can often be mitigated, for example, by cooling the sensor or improving the electronics, shot noise represents a fundamental limit associated with the discrete nature of light. For thermal light sources, this noise cannot be reduced below the Poissonian level. Although quantum optics techniques, which we will consider in Lecture~\ref{Lecture11}, can suppress shot noise below this limit, such approaches require non-classical light and do not apply to conventional imaging of thermal objects.

%% file: week4.tex
\section{Imaging: Non-visible wavelengths}
\label{Lecture4}

Each color corresponds to a specific wavelength, for example, 400\,nm light appears blue, 530\,nm green, and 650\,nm red. However, the human eye can only perceive a tiny fraction of the electromagnetic spectrum. Fortunately, detectors can observe wavelengths outside the visible range. For instance, X-ray observatories image the universe in the 0.12–12\,nm range, and infrared cameras can capture images around 10\,um. Just as in visible-light imaging, we can assign colors to different wavelengths to create a visual representation. Unlike the visible spectrum, these assignments can be inverted. For example, shorter, more energetic wavelengths might be displayed as red and longer wavelengths as blue. Such choices are largely conventional, based on human perception, with red evoking warmth (like fire) and blue evoking cold (like ice).

In this lecture, we discuss
\begin{itemize}
\item gamma-ray imaging,
\item X-ray imaging,
\item infrared imaging,
\item terahertz imaging,
\item radio imaging,
\item synthetic radar aperture imaging.
\end{itemize}

\subsection{Gamma-ray imaging}

Gamma-ray imaging detects high-energy photons with energies above 100\,keV (wavelengths shorter than $\sim 10$\,pm), produced by some of the most energetic processes in the universe~\cite{longair2011high}. Key astrophysical sources include supermassive black holes at galactic centers, where infalling matter emits intense gamma radiation; massive star collapses, such as supernovae and gamma-ray bursts, which release a large amount of energy in short timescales; and solar flares, where accelerated particles generate high-energy photons. Unlike visible or X-ray light, there are currently no practical focusing optics for gamma rays, and telescopes rely on indirect detection methods with an angular resolution of  $\approx 0.1$\,degrees.

\begin{wrapfigure}{r}{0.33\textwidth}
    \centering
    \includegraphics[width=0.31\textwidth]{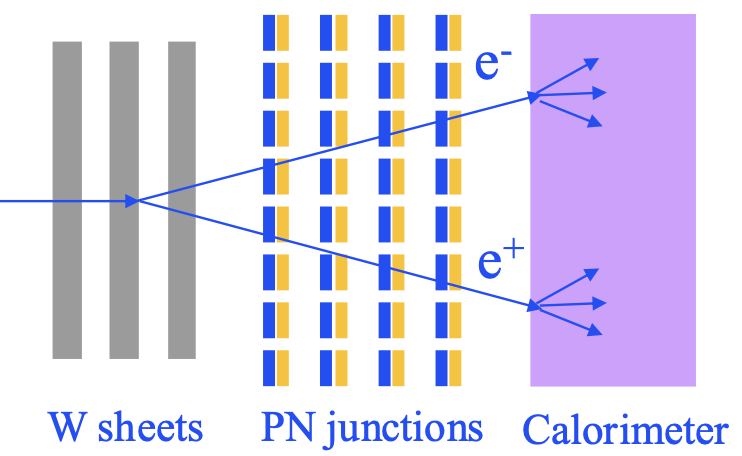} 
    \caption{A sketch of a gamma-ray detector with tungsten sheets, PN junctions, and a calorimeter.}
    \label{fig:w4_gamma}
\end{wrapfigure}
Notable missions include Fermi Gamma-ray Space Telescope~\cite{Atwood_2009}, INTEGRAL~\cite{KNODLSEDER2004189} (INTErnational Gamma-Ray Astrophysics Laboratory), and Swift~\cite{gehrels2004swift}, which mapped gamma-ray sources and studied phenomena such as gamma-ray bursts, pulsars, and active galactic nuclei.

One may detect gamma rays with tungsten (W) sheets, which produce energetic electrons and positrons following interactions with incident gamma rays. Tungsten is commonly employed in gamma-ray detectors because high-energy photons interact more strongly with high-$Z$ (high atomic number) materials. Tungsten has a large atomic number ($Z = 74$) and a high density ($\rho \approx 19.3\,\mathrm{g/cm^3}$), and therefore provides a high probability for interactions. At high gamma-ray energies (above a few MeV) electron–positron pair production dominates:
\begin{equation}
\gamma + \text{nucleus} \rightarrow e^- + e^+ + \text{nucleus},
\end{equation}
because the reaction requires at least $1.022$\,MeV of energy for the electron-positron production. Inelastic Compton scattering dominates at lower gamma-ray energies:
\begin{equation}
\gamma + e^- \rightarrow e^- + \gamma.
\end{equation}

The charged particles are then detected with silicon PN-junction detectors, which reconstruct the photon’s incident direction and energy. When the charged particles pass through a PN junction, they create electron–hole pairs in the depletion region, similar to photons. The electron-hole pairs are quickly separated by the built-in electric field and produce a measurable electrical signal. Another method employs calorimeters, which absorb the entire gamma-ray energy in a dense material, producing a measurable temperature rise or scintillation signal proportional to the photon energy. By combining tungsten sheets, PN-junctions, and calorimeters, as shown in Fig.~\ref{fig:w4_gamma}, gamma-ray telescopes can determine both the energy and trajectory of high-energy photons.

\subsection{X-rays}

X-rays are emitted by some of the hottest objects in the universe. According to Wien’s law, objects with temperatures around $10^6$\,K emit radiation peaking at wavelengths of a few nanometers, in the X-ray regime. For example, X-ray observatories study black holes, where infalling matter in the accretion disk is heated to very high temperatures and emits intense X-rays. By analysing this radiation, astronomers can infer the presence, mass, and dynamics of black holes. Another important class of sources is active galactic nuclei, including quasars, which are powered by accretion onto supermassive black holes at the centers of galaxies and produce strong X-ray emission due to gravitational heating. Major X-ray missions such as the Chandra X-ray Observatory~\cite{slane2025chandra}, XMM-Newton~\cite{santos-lleo2009first}, and NuSTAR~\cite{harrison2013nuclear} provided high-resolution imaging and spectroscopy of the high-energy universe.

\begin{wrapfigure}{r}{0.33\textwidth}
    \centering
    \includegraphics[width=0.31\textwidth]{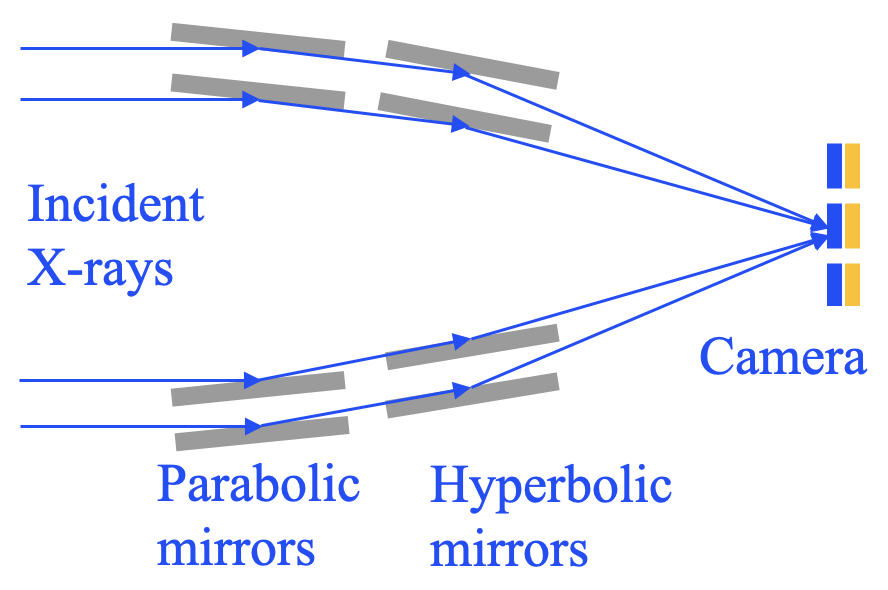} 
    \caption{A sketch of an X-ray detector with parabolic and hyperbolic mirrors.}
    \label{fig:w4_xray}
\end{wrapfigure}
Focusing optics in X-ray observatories rely on grazing-incidence reflection, because X-rays penetrate normal mirror surfaces instead of reflecting from them. In designs such as Wolter type I telescopes~\cite{wolter1952glancing}, as shown in Fig.~\ref{fig:w4_xray}, light hits nested mirror shells at shallow angles and are directed toward a common focal point. To reduce aberrations considered in Lecture~\ref{Lecture2}, the mirrors are shaped as sections of paraboloids and hyperboloids, with different radii of curvature in the tangential and sagittal directions. The tangential curvature controls the focal length along the optical axis, and the sagittal curvature minimises astigmatism. By stacking multiple mirror shells, the collecting area is increased.

X-ray detection with silicon pixels relies on CCD cameras with thick-depletion layers of $\sim 100$\,um that absorb incoming X-ray photons via the photoelectric effect. When an X-ray photon is absorbed, it ejects a high-energy photoelectron, which collides with other electrons in the valence band, creating thousands of electron–hole pairs per photon~\cite{knoll2010radiation}. Each pair requires about 3.6\,eV, more than the silicon bandgap of 1.1\,eV, because a fraction of the electron energy is lost to phonon excitations. The total number of produced electrons is given by the equation
\begin{equation}
N_e = \frac{N_\nu}{3.6\,{\rm eV}}.
\end{equation}
By counting these electrons, the detector can measure not only the presence of X-rays but also the photon energy, enabling energy-resolved (colour) imaging.

X-ray observatories can, in principle, achieve significantly finer resolution than telescopes operating in the visible or infrared bands. For example, a 1\,m-diameter X-ray telescope observing at $\lambda = 1$\,nm has a diffraction-limited resolution of $\sim 1$\,nrad, which is about 200 times finer than that of the Hubble Space Telescope. However, achieving this limit in practice is challenging because X-ray telescopes rely on grazing-incidence optics, which do not form filled-aperture imaging systems. As a result, the practical diffraction-limited resolution of X-ray telescopes is typically of order $\sim 1$\,urad.
The smallest pixel sizes in modern CCD and CMOS detectors are on the order of a few micrometres. For the Chandra X-ray Observatory, the pixel size is $a \approx 6$\,um, and a pixel-limited angular resolution of approximately $600$\,nrad is achieved using a long focal length of $f = 10$\,m, which is comparable to the diffraction-limited performance of the telescope.

\subsection{Infrared radiation}

Infrared imaging captures radiation with wavelengths longer than visible light, emitted by objects cooler than the Sun. For example, the central wavelength of radiation from the Earth or the human body is around 10\,um, corresponding to a temperature of $\approx 300$\,K, while fires at $\approx 1200$\,K emit primarily near $2.5$\,um according to the Wien's law~\cite{rybicki2004radiative}. Infrared telescopes are required for observing celestial objects that emit or absorb infrared radiation, including cool stars, interstellar dust clouds, and the atmospheres of exoplanets. On Earth, infrared sensors are used for environmental monitoring and climate studies.

Silicon-based cameras are common in visible-light imaging but are largely insensitive to infrared light due to silicon’s relatively large bandgap of 1.1\,eV and photons with wavelengths longer than $\approx 1.1$\,um transmit through the sensor without generating electron–hole pairs. Therefore, infrared imaging relies on narrow-bandgap semiconductors such as InAs or PbSe. The James Webb Space Telescope~\cite{gardner2006james}, for example, observes distant galaxies and detects their redshifted light from 0.6\,um to 28\,um with mercury-cadmium-telluride detectors, where cadmium telluride has a bandgap of 1.5\,eV and mercury telluride is a semimetal with a zero bandgap. It is feasible to combine the two in one crystalline alloy semiconductor~\cite{rauscher2007detectors} via molecular beam epitaxy because their lattice constants are similar (within 0.3\%). The bandgap of the detector can be tuned to $\approx 0.04$\,eV, enabling observations up to $\approx 30$\,um.

\subsection{Terahertz}

We now consider longer-wavelength radiation in the terahertz (THz) band, around $0.1$\,mm. This regime occupies a challenging intermediate range between infrared and radio-frequency imaging. The THz frequency is too high for conventional electronic circuits to process directly, while its photon energy is too low for efficient detection with standard silicon or infrared imaging technologies. Traditional detection methods include semiconductor devices such as Schottky diodes~\cite{mattauch1957mass}, which rely on a metal--semiconductor junction, and bolometers~\cite{richards1994bolometers}, which measure temperature changes induced by absorbed radiation. More recent sensor technologies include transition-edge sensors~\cite{irwin2005tes}, which exploit superconductors operated near their critical transition, and kinetic inductance detectors~\cite{day2003kekid}, which measure changes in superconducting inductance. Another important class of detectors is based on nonlinear systems such as superconductor--insulator--superconductor junctions~\cite{tinkham2004superconductivity}, which require a local oscillator to heterodyne THz signals into measurable electrical currents.

THz radiation can be generated via optical rectification, where an ultrafast laser pulse is shone onto a nonlinear medium~\cite{tonouchi2007cutting}, such as Au/Ti layers or nonlinear crystals with large second-order susceptibility. We will consider nonlinear crystals in Lecture~\ref{Lecture11}.
THz radiation is non-ionising and is safe for medical diagnostics, particularly for detecting skin cancer and studying biological tissues. THz scanners can penetrate clothing and certain materials, making them ideal for security screening at airports to detect concealed objects, and for non-destructive quality control in industrial settings. In astrophysics, THz waves probe the high-frequency tail of the cosmic microwave background, interstellar dust, and molecular clouds, providing information about the composition and structure of the universe. THz frequencies also offer the potential for ultra-high-speed wireless communication.

\subsection{Radio frequencies}

Radio imaging at wavelengths $\lambda \geq 1$\,mm faces two inherent limitations: (i) individual radio telescopes typically act as single-pixel detectors, and (ii) their angular resolution is low because the wavelength is larger than in the optical band. Interferometric arrays overcome these limitations because radio telescopes measure both the amplitude and phase of incoming waves. By combining signals from antennas separated by thousands of kilometers, astronomers can synthesise a virtual aperture as large as the array’s maximum baseline, achieving angular resolutions comparable to or exceeding those of optical telescopes. This technique, known as aperture synthesis, enables detailed imaging of celestial sources at radio frequencies.

\begin{wrapfigure}{r}{0.42\textwidth}
    \centering
    \vspace{-9mm}
    \includegraphics[width=0.4\textwidth]{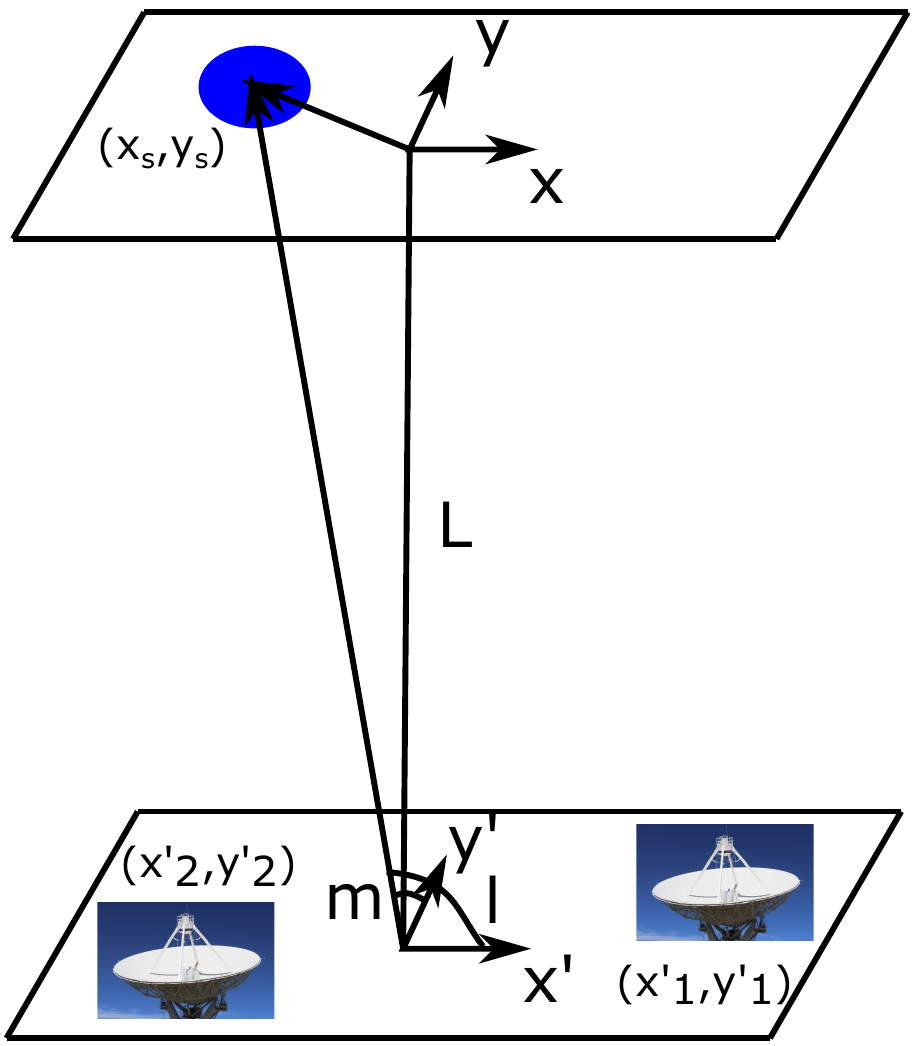}
    \caption{Scheme of a source in the $xy$-plane and telescopes in the $x'y'$-plane.}
    \label{fig:w4_syn_ap}
\end{wrapfigure}

Radio imaging plays a key role in astronomy. The historic imaging of the M87 supermassive black hole was achieved through a global network of radio telescopes operating together as a virtual Earth-sized telescope~\cite{akiyama2019m87shadow}. Observing at a wavelength of $1.3$\,mm, the network achieved high-resolution imaging of the black hole’s environment. The resulting image revealed the silhouette of the event horizon. This observation confirmed theoretical predictions about these objects at the centers of galaxies.

The Earth-size aperture in radio imaging can be synthesised utilising the van Cittert–Zernike theorem~\cite{goodman2005introduction}. Under certain assumptions discussed below, the theorem states that the mutual coherence function ($V$) and the sky intensity distribution ($I$) are Fourier pairs as given by the equation
\begin{equation}
V(x'_1 - x'_2, y'_1 - y'_2) = \frac{1}{T} \int \limits_0^T E(x'_1, y'_1, t) E^*(x'_2, y'_2, t) dt \sim \tilde{I},
\end{equation}
where $T$ is the integration time.

We prove the theorem by considering an incoherent source in the $xy$-plane and detectors in the $x'y'$-plane on Earth, as shown in Fig.~\ref{fig:w4_syn_ap}. The electric field emitted by a source of area $S$ in the $xy$-plane is measured at two locations on Earth, $(x'_1, y'_1)$ and $(x'_2, y'_2)$. For a given point source in the $xy$-plane, we express the electric field in terms of its inverse Fourier transform,
\begin{equation}
E(x,y,t) = \frac{1}{2\pi}\int\limits_{\omega_1}^{\omega_2} \tilde{E}(x,y,\omega)\, e^{i\omega t}\, d\omega,
\end{equation}
which decomposes the random field $E(x,y,t)$ into a superposition of monochromatic components within the frequency band of interest $(\omega_1,\omega_2)$, as discussed below. In this representation, the randomness of the field $E(x,y,t)$ is encoded in the frequency-dependent complex amplitude $\tilde{E}(x,y,\omega)$, in particular through its random phase for each $\omega$.

According to the Huygens-Fresnel principle discussed in Lecture~\ref{Lecture2}, the telescopes observe a delayed version of the electric field at the source by $r_1 / c$ and $r_2/c$ and scaled by the distances $r_1$ and $r_2$ from the source to the telescopes
\begin{equation}
\begin{split}
    E(x'_1, y'_1, t) &= \alpha \int \limits_S \int\limits_{\omega_1}^{\omega_2} \tilde{E} (x,y,\omega ) \frac{e^{i \omega (t- r_1 / c)}}{r_1} d\omega dx dy \\
    E(x'_2, y'_2, t) &= \alpha  \int \limits_S \int\limits_{\omega_1}^{\omega_2}  \tilde{E} (x,y,\omega) \frac{e^{i \omega (t- r_1 / c)}}{r_2} d\omega dx dy,
\end{split}
\end{equation}
where $\alpha$ is the normalisation constant, $\tilde{E} (x,y,\omega )$ is Fourier transform of the electric field at point $(x,y)$, and $\omega_1$ and $\omega_2$ are detection limits. The source may have a broad spectrum of frequencies. However, the antenna can apply bandpass filters to its detected signals and separate them into narrow frequency bands, such as $(\omega_1 : \omega_2)$.

The distances $r_1$ and $r_2$ between the point source at $(x,y)$ and the telescopes are given by the equations
\begin{equation}
r_1^2 = L^2 + (x - x'_1)^2 + (y - y'_1)^2 \hspace{1cm}
r_2^2 = L^2 + (x - x'_2)^2 + (y - y'_2)^2,
\end{equation}
where $L$ is the astronomical distance between xy- and x'y'-planes and satisfies the condition $L \gg x,y \gg x', y'$ because the distance to the source is much larger than the size of the source, and the size of the source is much larger than Earth.


We compute the mutual coherence function $V$ by making several assumptions, including incoherence of the source, and a narrow frequency bandwidth of filtered electric fields. The mutual coherence function averages the product of electric fields over the exposure time $T\gg 2\pi/\omega_1$, and oscillating terms average out according to the equation
\begin{equation}
\begin{split}
V &= \frac{\alpha^2}{T} \int \limits_0^T  \int \limits_S \int\limits_S \int\limits_{\omega_1}^{\omega_2} \int\limits_{\omega_1}^{\omega_2} \tilde{E} (x_1,y_1,\omega_a) \tilde{E}^* (x_2,y_2,\omega_b) \frac{e^{i \omega_a (t- r_1 / c)}}{r_1} \frac{e^{-i \omega_b (t- r_2 / c)}}{r_2} d\omega_a d\omega_b dx_1 dy_1 dx_2 dy_2 dt \\
&= \frac{\alpha^2}{T} \int \limits_S \int\limits_S  \int\limits_{\omega_1}^{\omega_2} \tilde{E} (x_1,y_1,\omega) \tilde{E}^* (x_2,y_2,\omega) \frac{e^{i \omega (r_2 / c- r_1 / c)}}{r_1 r_2} d\omega dx_1 dy_1 dx_2 dy_2.
\end{split}
\end{equation}

We then utilise the assumption that the source is incoherent, which is well justified for most astrophysical sources. This assumption implies that the phases of the electric fields emitted from different coherence patches of characteristic size $S_0 \sim \lambda^2 \ll S$ are uncorrelated, and their cross terms average to zero when integrating over the source area. We then get the equation
\begin{equation}
V \approx \frac{\alpha^2 S_0}{T} \int\limits_S  \int\limits_{\omega_1}^{\omega_2} \tilde{E} (x,y,\omega) \tilde{E}^* (x,y,\omega) \frac{e^{i \omega (r_2 / c- r_1 / c)}}{r_1 r_2} d\omega dx dy.
\end{equation}

The final approximation we make is that the frequency band $\Delta \omega = \omega_2 - \omega_1$ is so narrow that we can treat it as a single frequency in the integral. Since the intensity of the source is related to the electric field according to the equation 
\begin{equation}
I(x,y) \sim \int\limits_{\omega_1}^{\omega_2} \tilde{E} (x,y,\omega) \tilde{E}^* (x,y,\omega) d\omega,
\end{equation}
the equation for $V$ then simplifies to 
\begin{equation}
V = \beta \int\limits_S I(x,y) \frac{e^{i \omega (r_2 / c- r_1 / c)}}{r_1 r_2} dx dy,
\end{equation}
where $\beta$ is the product of all previous constants, which we introduce for simplicity.

We then simplify $r_1$ and $r_2$ using the Taylor expansion and neglect second-order terms, such as $(x' / L)^2$ and $(y' / L)^2$  for each telescope. We get the equations
\begin{equation}
r_1 \approx L_s \left(1 - \frac{x x'_1}{L_s} - \frac{y y'_1}{L_s} \right) \hspace{1cm} r_2 \approx L_s \left(1 - \frac{x x'_2}{L_s} - \frac{y y'_2}{L_s}\right),
\end{equation}
where $L_s^2 = L^2 + x_s^2 + y_s^2$ and $(x_s, y_s)$ are coordinates of the centre of the source in the xy-plane. We then notice that $l=x/L_s$ and $m = y/L_s$ are directional cosines from Earth to the source. The equation for $V$ then simplifies to
\begin{equation}
\begin{split}
V &= \beta \int\limits_S I(x,y) e^{i \frac{\omega}{c} (l(x'_1 - x'_2) + m(y'_1 - y'_2))} \frac{dx}{L_s} \frac{dy}{L_s} \\
&= \beta \int\limits_{\Delta l, \Delta m} I(l,m) e^{i \frac{\omega}{c} (l(x'_1 - x'_2) + m(y'_1 - y'_2))} dl dm,
\end{split}
\end{equation}
where $\Delta l$ and $\Delta m$ are dimensions of the source as seen from Earth in units of directional cosines.

We make an important observation here. The mutual coherence function depends only on the difference in coordinates of the two telescopes, $V = V(x'_1 - x'_2, y'_1 - y'_2)$, rather than on all four coordinates $x'_1, x'_2, y'_1, y'_2$. Therefore, we can introduce new dimensionless variables
\begin{equation}
u = \frac{2\pi}{\lambda}(x'_1 - x'_2) \hspace{1cm} v = \frac{2\pi}{\lambda}(y'_1 - y'_2)
\end{equation}
and come to the final equation for the mutual coherence function
\begin{equation}
V(u,v) = \beta \int\limits_{\Delta l, \Delta m} I(l,m) e^{i (lu + mv)} dl dm,
\end{equation}
which shows that the mutual coherence function and the source intensity are Fourier conjugate pairs. 

A practical radio interferometric imaging procedure consists of measuring the electric field at multiple telescopes and filtering the signal into narrow frequency bands to preserve phase information. Pairs of telescopes are then correlated to compute the visibility function 
$V(u,v)$, with each baseline sampling a point in the spatial frequency plane. As the Earth rotates, additional samples of $V(u,v)$ are obtained. The sky brightness distribution can then be reconstructed by taking the inverse Fourier transform of the measured visibilities. However, since the sampling is incomplete and is not a continuous function of $(u,v)$, the resulting image should be further processed using deconvolution techniques to recover the intensity distribution in the object plane~\cite{thompson2017interferometry}.

\subsection{Synthetic aperture radar}

Creating a large synthetic aperture, as discussed in the previous section, is achievable for radio waves, because typical mechanical perturbations such as ground vibrations from ocean waves are $\sim 1$\,um and are therefore negligible compared to the wavelength. In addition, atmospheric aberrations, discussed in Lecture~\ref{Lecture2}, are far less problematic at radio wavelengths than in the visible band because phase distortions are typically small compared to long radio wavelengths.

Synthetic aperture techniques can also be implemented with a single moving antenna: by scanning the sensor around the target and combining the recorded signals coherently, one effectively synthesises a large aperture. This principle underlies synthetic aperture radar (SAR), which is commonly deployed on satellites~\cite{skolnik2008radar}. As the satellite follows a well-characterised orbit, it observes the Earth's surface from multiple positions, thereby constructing a large effective aperture that enables high-resolution radar imaging of terrain.  Earth-orbiting satellites typically operate SAR systems with $\lambda\approx 3$\,cm. The pulses can penetrate clouds, rain, and other atmospheric obstacles, allowing reliable observation under virtually all weather conditions and at night because SAR utilises active imaging.

Synthetic aperture radar was ideally suited to study Venus at $\lambda=12.6$\,cm, because the planet has a dense atmosphere composed primarily of carbon dioxide. The Magellan mission~\cite{pettengill1991magellan} employed SAR to map approximately 98\% of Venus’ surface. The Magellan mission achieved a synthetic aperture radar resolution of about $100$\,m per pixel for most of Venus’s surface. In addition, altimetry data provided vertical resolution of roughly $50$\,m. The instrument revealed detailed geological features such as volcanoes, highland regions, and large impact craters.

%% file: week5.tex
\section{Applications of lasers: Coherence of light, Gaussian beams}
\label{Lecture5}

In the previous lectures, we primarily focused on thermal light, such as that emitted by stars, bulbs, or torches. Thermal light contains a broad range of wavelengths, propagates in all directions, and is generally incoherent. In this lecture, we turn our attention to laser light, which is nearly monochromatic, directional, and coherent. Monochromatic means that the light consists of a single wavelength, or equivalently, a single frequency. Directional indicates that the light propagates predominantly in one direction, like the beam from a laser pointer. Spatial coherence describes the correlation of the light wave at different points in space, while temporal coherence describes the correlation of the wave at different moments in time. In practice, laser light is never perfectly monochromatic or fully coherent, but it can often be approximated as such.

In this lecture, we discuss
\begin{itemize}
\item spatial and temporal coherence of light,
\item filtering of light,
\item Gaussian beams,
\item laser cutting and welding,
\item scattering of light,
\item optical tweezers.
\end{itemize}

\subsection{Coherence of light}

Coherence characterises the predictability of a light wave over space and time. Light emitted by thermal sources, where photons are produced randomly in time, exhibits low coherence and a broad spectrum, meaning the phase of the wave varies unpredictably and many wavelengths are present. In contrast, light whose waves maintain a well-defined phase relationship over both time and space possesses high temporal and spatial coherence~\cite{mandel1995optical}. Coherence can be quantified by the complex degree of the first-order coherence, defined by the equation
\begin{equation}\label{eq:w5_coh_def}
g^{(1)}(\VF{r_1}, t_1, \VF{r_2}, t_2) = \frac{<E^*(\VF{r_1}, t_1) E(\VF{r_2}, t_2)>}{\sqrt{<|E(\VF{r_1}, t)|^2><|E(\VF{r_2}, t_2))|^2>}},
\end{equation}
where $|g^{(1)}| \approx 1$ corresponds to strong coherence, while $|g^{(1)}| \approx 0$ indicates weak or random phase relationships.

\begin{figure}[t]
\centering
\includegraphics[height=4cm]{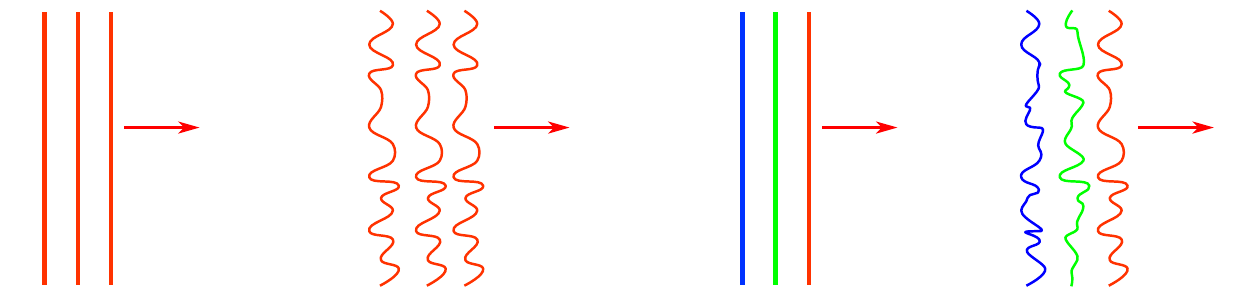} 
\caption{Wavefronts illustrating different coherence properties: (i) spatially and temporally coherent light, (ii) temporally coherent but spatially incoherent light, (iii) spatially coherent but temporally incoherent light, and (iv) fully incoherent light.}
\label{fig:w5_coh}
\end{figure}

The coherence of light can vary in space and time. Light that is coherent both spatially and temporally, such as an ideal plane wave of a single frequency, maintains a well-defined phase across its wavefront and over time. Light that is spatially but not temporally coherent, such as emission from distant stars, has a predictable wavefront across space but contains multiple frequencies, so its phase varies unpredictably over time. Conversely, light that is temporally but not spatially coherent, such as laser light reflected from a rough surface, retains phase correlation over time but is scrambled across the wavefront. Finally, incoherent light, typical of thermal sources near the emitter, has random phase variations both in space and time and exhibits a broad spectrum. Examples of coherent and incoherent wavefronts are shown in Fig~\ref{fig:w5_coh}.

\subsection{Filtering thermal light}

Thermal light with peak intensity in the visible band is emitted by hot objects with surface temperatures between roughly $4000$\,K and $8000$\,K. For example, the Sun’s surface, at a temperature of $\approx 5778$\,K, produces a spectrum that peaks at a wavelength of $\lambda=550$\,nm, with a comparable spectral bandwidth $\Delta \lambda \sim \lambda$, which covers much of the visible range.

While thermal light is incoherent, it can be partially converted into more coherent light using filtering techniques. Spatial filtering, for example, with a pinhole of the diameter equal to the light's spatial coherence length, passes the coherent portion of the wavefront and blocks the rest of the light. Spectral filtering, such as with a diffraction grating, removes most frequency components, narrows the spectrum, and increases temporal coherence. Both approaches improve coherence but come at the cost of reduced optical power, because much of the original light is blocked or discarded in the process, as we discuss below.

\begin{wrapfigure}{r}{0.42\textwidth}
    \centering
    \includegraphics[width=0.4\textwidth]{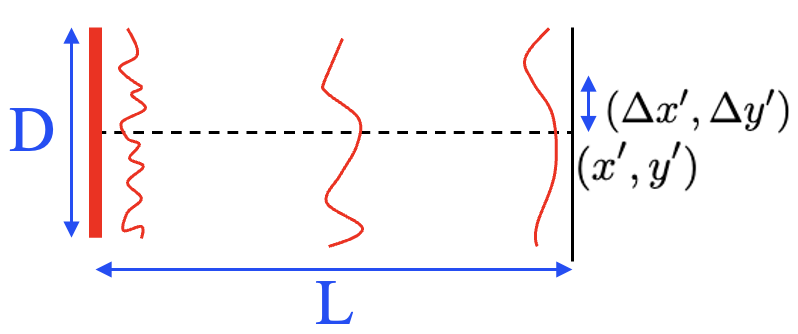} 
    \caption{A sketch of a wavefront from an incoherent source. The spatial coherence improves as the waves travel away from the source.}
    \label{fig:w5_spatial_coh}
\end{wrapfigure}

We first consider spatial filtering of sunlight. A thermal source is spatially incoherent because light is produced by uncorrelated sources. However, as light travels away from the point source, its spatial coherence improves. The spatial coherence length of sunlight can be estimated by considering two points on a wavefront near Earth separated by $(\Delta x', \Delta y')$, as shown in Fig.~\ref{fig:w5_spatial_coh}. The coordinate system in the source plane is $(x, y)$ and in the observer plane is $(x', y')$. The distances satisfy the relationships $L \gg D \gg \delta x', \Delta y'$, which is a valid approximation for the Sun. The sunlight is temporarily incoherent, but since our goal is to find the spatial coherence length, we consider only one wavelength, $\lambda$, from the thermal spectrum. According to the Huygens-Fresnel principle, the electric fields in the two points shown in Fig.~\ref{fig:w5_spatial_coh} are given by the equations
\begin{equation}
\begin{split}
& E(x', y') = \alpha \int_S E(x, y) \exp(-ikl_1) dx dy \\
& E(x'+\Delta x', y'+\Delta y') = \alpha \int_S E(x, y) \exp(-ikl_2) dx dy,
\end{split}
\end{equation}
where $\alpha$ is the proportionality coefficient that takes into account the $1/L$ decay of the field, $S$ is the area of the source, $k=2\pi/\lambda$ is the wave number. The distances $l_1$ and $l_2$ travelled by the rays are given by the equations
\begin{equation}
\begin{split}
l_1 &= \sqrt{L^2 + (x - x')^2 + (y - y')^2} \approx L\left( 1 + \frac{x^2 + y^2}{2L^2} - \frac{xx' + yy'}{L^2} \right) \\
l_2 &= \sqrt{L^2 + (x - x' - \Delta x')^2 + (y - y' - \Delta y')^2} \approx L\left( 1 + \frac{x^2 + y^2}{2L^2} - \frac{xx' + yy'}{L^2} - \frac{x\Delta x' + y \Delta y'}{L^2} \right) \\
& = l_1 - \frac{x\Delta x' + y \Delta y'}{L},
\end{split}
\end{equation}
which show that the difference between the electric fields at points $(x', y')$ and $(x'+\Delta x', y'+\Delta y')$ is caused by the phase term $\exp(ik (x\Delta x' + y \Delta y')/L)$. The term decreases with larger distance $L$ or smaller source size, $D$. The further we travel away from the thermal source, the smaller the over phase shift between the two points separated by $(\Delta x', \Delta y')$. As a simplified estimation of the spatial coherence length in the observer plate, we consider two maximally separated points on the source, and request that the total phase shift between two points in the observer plane separated by $l_{\rm coh}$ is smaller than $2\pi$ to achieve the phase predictability between the two points. Then we get the equation
\begin{equation}\label{eq:w5_l_coh_simple}
k\frac{D l_{\rm coh}}{L} \approx 2\pi \hspace{5mm}=>\hspace{5mm} l_{\rm coh} \approx \frac{\lambda}{\theta_s},
\end{equation}
where $\theta_s = D/L$ is the angular size of the source.

We now prove the result given by Eq.~\ref{eq:w5_l_coh_simple} more rigorously and find the degree of coherence given by Eq.~\ref{eq:w5_coh_def}. The cross correlation term between the electric fields at points $(\Delta x', \Delta y')$ and $(x'+\Delta x', y'+\Delta y')$ is given by the equation
\begin{equation}
\begin{split}
E^*(x', y')E(x'+\Delta x', y'+\Delta y') &= \int_S \int_S E^*(x_1, y_1) E(x_2, y_2) \exp\left(-ik(l_1 - l_2)\right) dx_1 dy_1 dx_2 dy_2 \\
&\sim \int_S E^*(x, y) E(x, y) \exp\left(-ik\frac{x\Delta x' + y \Delta y'}{L}\right) dx dy \\
&\sim \int_S I(x, y) \exp\left(-ik\frac{x\Delta x' + y \Delta y'}{L}\right) dx dy,
\end{split}
\end{equation}
where we have utilised similar assumptions and mathematical operations relevant for the incoherence sources as we discussed in Lecture~\ref{Lecture4} while proving the Van Cittert–Zernike theorem. Similar to the diffraction on the aperture, considered in Lecture~\ref{Lecture2}, we achieved a 2D Fourier transform of the source aperture. However, in this case, we implement the Fourier transform of the intensity function. Since the intensity of the thermal source is uniform $I(x,y) = I_0$ and since the denominator of Eq.~\ref{eq:w5_coh_def} is proportional to the intensity of the field in the observer plane with the same proportionality coefficient as its numerator, we find that the first-order coherence is given by the equation
\begin{equation}
g^{(1)}(\Delta x', \Delta y') = \frac{1}{S}\int_S \exp\left(-ik\frac{x\Delta x' + y \Delta y'}{L}\right) dx dy,
\end{equation}
where $1/S$ normalises the coherence function because $g^{(1)}(0,0) = 1$. We solve the equation in polar coordinates for the circular aperture using Bessel functions, similar to the Fraunhofer diffraction considered in Lecture~\ref{Lecture2}, and get the equation
\begin{equation}\label{eq:w5_g_sol}
g^{(1)}(r') = \frac{2J_1(kRr'/L)}{kRr'/L} = \frac{2J_1(\pi \theta_s r'/\lambda)}{\pi \theta_s r'/\lambda},
\end{equation}
where $R=D/2$ and $r'$ is the separation between the two points in the observation plane. The first zero of the Bessel function $J_1(\pi \theta_s r'/\lambda)$ is achieved at 3.84, and the spatial coherence length is given by the equation
\begin{equation}
g^{(1)}(l_{\rm coh}) = 0 \hspace{5mm}=>\hspace{5mm} \frac{\pi \theta_s l_{\rm coh}}{\lambda} = 3.84 \hspace{5mm}=>\hspace{5mm} l_{\rm coh} = \frac{3.84}{\pi}\frac{\lambda}{\theta_s} \approx \frac{\lambda}{\theta_s}.
\end{equation}

If a wavefront is passed through a pinhole whose area is on the order of the coherence area ($\sim l_{\rm coh}^2$), the transmitted light becomes spatially filtered and significantly more coherent. This is because the pinhole selects a region over which the phase variations are small and removes contributions from other parts of the wavefront that would introduce random phase differences. As a result, the emerging light has a well-defined phase across the aperture, behaving much more like a spatially coherent beam. While different wavelengths remain present, each spectral component becomes more spatially coherent, and their superposition produces a beam with improved spatial coherence.

Temporal filtering can be achieved using a blazed diffraction grating, which separates different wavelengths of light by diffracting them at different angles, as given by the equation
\begin{equation}
\sin(\theta_d) = \frac{\lambda}{d},
\end{equation}
where $\theta_d$ is the diffraction angle and $d$ is the grating period. When broadband light passes through the grating, each wavelength is sent in a distinct direction, $\theta_d$. By placing a slit or aperture at a specific angle, one can select a narrow range of wavelengths, effectively reducing the spectral bandwidth of the light.

The coherence of light is often desirable, as we will see in this lecture. However, it is not always beneficial in imaging~\cite{goodman2007speckle}. Coherent light can produce speckle patterns, a grainy noise caused by interference of light. The noise is highly sensitive to phase differences on the object and leads to unwanted fringes or ghost images. To mitigate these effects, the spatial coherence of the illumination must be reduced below the scale of the smallest features being imaged. This can be achieved by using an extended light source, introducing a diffuser, or moving or rotating the source during imaging, all of which help suppress unwanted interference while preserving the overall illumination.

\subsection{Gaussian beams}

In this section, we consider the propagation of coherent laser beams, for which the Gaussian beam provides a fundamental model in optics. A key motivation for this model arises from diffraction in the Fraunhofer limit: at a far-field screen, the electric field profile is given by the Fourier transform of the beam at its initial plane. The Gaussian function is special because it is an eigenfunction of the Fourier transform, meaning that a Gaussian beam retains its functional form upon propagation, spreading in a predictable way while preserving its Gaussian shape.

We solve Maxwell's wave equation
\begin{equation}
\frac{\partial^2 \VF{E}}{c^2 \partial t^2} - \left(\frac{\partial^2 \VF{E}}{ \partial x^2} +\frac{\partial^2 \VF{E}}{ \partial y^2} + \frac{\partial^2 \VF{E}}{ \partial z^2} \right) = 0
\end{equation}
with an assumption that the beam is directional. For simplicity, we consider an electromagnetic wave with electric field fluctuations along the X-axis. In mathematical terms, this implies that we can search for the solutions in the form
\begin{equation}
    \VF{E}(x,y,z,t) = \VF{e_{x}} E(x,y,z) \exp(i\omega t - ikz),
\end{equation}
where $\VF{e_x}$ is the unit vector along the X-axis, $\omega$ is the angular frequency of light, $k$ is the wave vector, and $E(x,y,z)$ is a slow function of $z$ and satisfies the inequality 
\begin{equation}\label{eq:l5_slow}
    \frac{\partial E}{\partial z} \ll kE,
\end{equation}
which states that the profile of the electric field changes slowly at distances $\sim \lambda$ along the Z-axis. The equation implies that the beam is directional, and the slow changes of the beam profile are captured by the term $E(x,y,z)$. We can then simplify the wave equation to the equation
\begin{equation}
    -\frac{\omega^2}{c^2} E \exp(i\omega t - ikz) - \left(\frac{\partial^2 E}{ \partial x^2} +\frac{\partial^2 E}{ \partial y^2} -k^2 E -2ik \frac{\partial E}{ \partial z} \right)\exp(i\omega t - ikz)  = 0
\end{equation}
and since $k = \omega/c$, the equation further simplifies to the paraxial wave equation
\begin{equation}
    \frac{\partial^2 E}{\partial x^2} + \frac{\partial^2 E}{\partial y^2} = 2ik\frac{\partial E}{\partial z},
\end{equation}

\begin{wrapfigure}{r}{0.35\textwidth}
    \centering
    \vspace{-7mm}
    \includegraphics[width=0.33\textwidth]{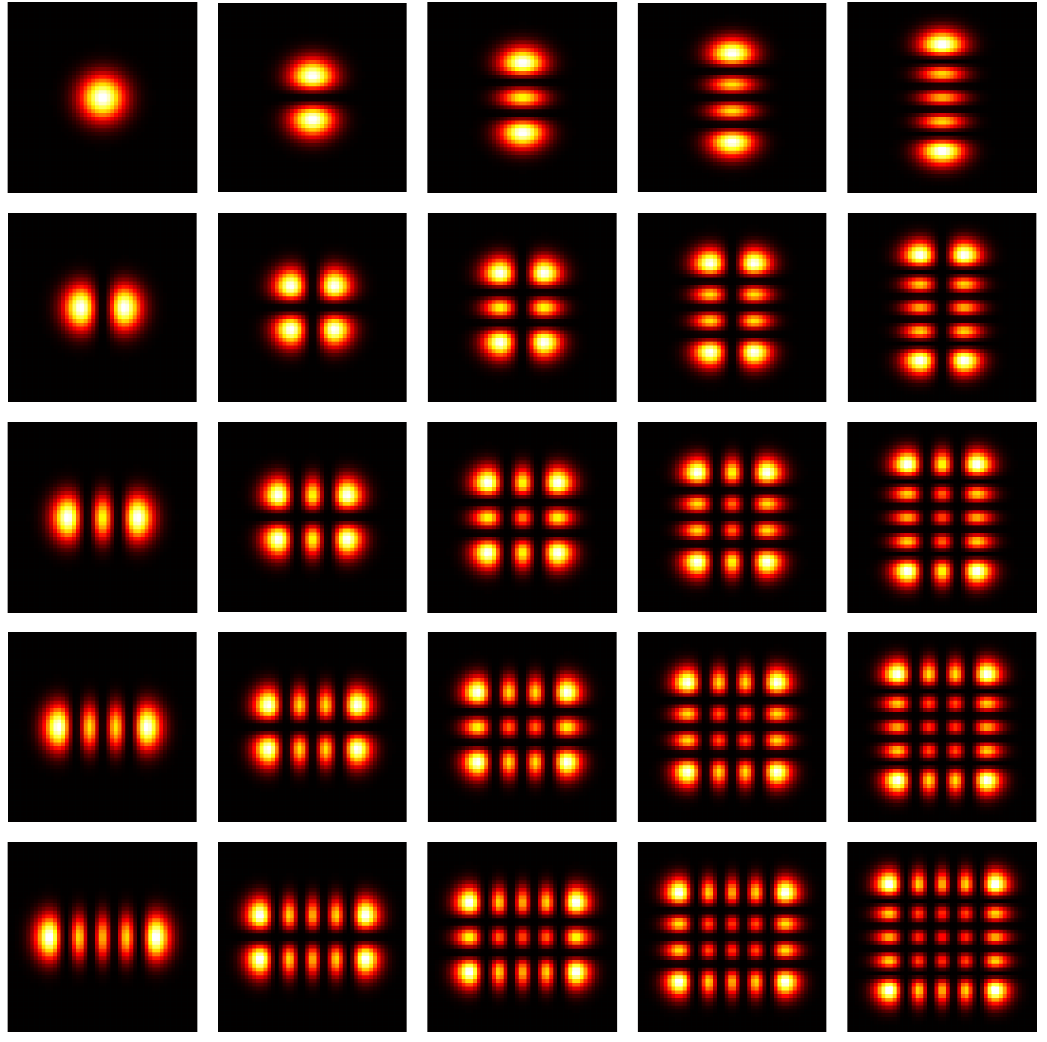} 
    \caption{Transverse profile of the Hermite-Gaussian modes for $m,l=0,1,2,3,4$.}
    \label{fig:w5_HG_modes}
\end{wrapfigure}

which can be solved analytically using Hermite polynomials and get an infinite number of solutions for the electric field according to the equation~\cite{siegman1986lasers}
\begin{equation}\label{eq:l5_gaussian}
\begin{split}
    E_{lm}(x,y,z) &= E_0 \frac{w_0}{w(z)}H_l \left( \frac{\sqrt{2}x}{w(z)} \right) H_m \left( \frac{\sqrt{2}y}{w(z)} \right) \times \\
    &\exp\left( -(x^2+y^2)\left(\frac{1}{w(z)^2} + \frac{ik}{2R(z)} \right) \right) \exp(i \psi(z)),
\end{split}
\end{equation}
where $E_0$ is the field amplitude, $R(z)$ is the radius of curvature of the wavefront at $z$, $w_0$ is the beam size (radius) at its waist position at $z=0$, $w(z)$ is the beam size at position $z$, $H_l$ and $H_m$ are Hermite polynomials of orders $l$ and $m$, and $\psi$ is the Gouy phase of the beam, which is an additional phase shift experienced by a focused beam as it passes through its waist, and increases by $\pi$ radians for a Gaussian beam from far field to far field.

The term $w_0 / w(z)$ shows how the electric field changes in response to the changes in the beam size. Since the total power of the laser beam $P \sim \int EE^*dS$ is conserved, the scaling of the electric field with the beam size as $1/w(z)$ is natural. Hermite polynomials with $l,m=0,1,2,...$ determine the profile of the electric field in the XY-plane. For example, Hermite polynomials $H_l$ and $H_m$ cross zero $l$ and $m$ times, as shown in Fig.~\ref{fig:w5_HG_modes}. The next exponential term in Eq.~(\ref{eq:l5_gaussian}) corresponds to the exponential decay of the electric field in the XY-plane and to the additional phase of the beam due to its curvature.

Three key parameters of the Gaussian beam given by Eq.~(\ref{eq:l5_gaussian}) are its beam radius, wavefront radius of curvature, and Gouy phase, defined as
\begin{equation}\label{eq:l5_w_R}
    w(z) = w_0 \sqrt{1 + \left( \frac{z}{z_R} \right)^2} \hspace{1cm}
    R(z) = z + \frac{z_R^2}{z} \hspace{1cm}
    \psi(z) = \arctan\left(\frac{z}{z_R}\right).
\end{equation}
Here, $z_R = \pi n w_0^2 / \lambda$ is the Rayleigh range of the beam, where $\lambda$ is the vacuum wavelength and $n$ is the refractive index of the medium. These expressions describe how the beam expands, how its wavefront curvature evolves, and how it accumulates the Gouy phase as it propagates along the Z-axis.

\subsection{Laser beam propagation}

Our next goal is to find how beam parameters change when we propagate laser beams through optical systems. We can use the ray-trace matrices as we considered in Lecture~\ref{Lecture1} for Gaussian beam propagation if we introduce the q-parameter given by the equation
\begin{equation}
    \frac{1}{q} = \frac{1}{R} - \frac{i \lambda}{\pi n w^2},
\end{equation}
where $n$ is the index of refraction of the medium.

If the $q$-parameter is known at a particular point $z$ then we can find the beam radius of curvature and size by taking the real and imaginary parts of $1/q$. If we know a $q$-parameter at one point and need to compute it at another point, then we can utilise the ray trace matrices according to the equation
\begin{equation}
    \begin{pmatrix}
    q_2 \\
    1
    \end{pmatrix}
    =
    k M
    \begin{pmatrix}
    q_1 \\
    1
    \end{pmatrix},
\end{equation}
where $k$ is an unknown parameter that we need to find from the equation that corresponds to the second row, $M$ is the total ABCD matrix of the path between the propagation points. For example, for free space propagation of distance $L$, the equations read
\begin{equation}
    \begin{pmatrix}
    q_2 \\
    1
    \end{pmatrix}
    =
    k
    \begin{pmatrix}
    1 & L \\
    0 & 1
    \end{pmatrix}
    \begin{pmatrix}
    q_1 \\
    1
    \end{pmatrix}.
\end{equation}
Therefore, for the system of equations $k=1$ and $q_2 = q_1 + L$ that corresponds to Eq.~(\ref{eq:l5_w_R}) for the beam size and radius of curvature.

We can design optical systems to control the size and radius of curvature of a laser beam at a desired location using the q-parameter formalism. Tightly focused beams achieve high intensities and can heat and cut materials. The scattering of laser light by the atmosphere, known as Rayleigh scattering, has been exploited in visual displays such as laser shows. Focused laser beams can also serve as optical tweezers for trapping and manipulating microscopic objects, including bacteria and atoms.

\subsection{Cutting and welding with laser beams}

When a tightly focused laser beam is absorbed by a material, such as a metal sheet, the absorbed energy raises the temperature of the target. For sufficiently powerful beams, the temperature can exceed the material’s melting and boiling points, causing localised melting and evaporation, which produces holes or cuts. By guiding the laser along carefully designed trajectories, precise shapes can be fabricated~\cite{steen2010laser}. The main advantages of laser cutting are its accuracy and small material waste. The energy required to evaporate a given portion of the material of mass $m$ can be estimated using the following equation:
\begin{equation}
    E = m (C_s(T_m - T_0) +  L_f +  C_l(T_v - T_m) + L_v), 
\end{equation}
where $T_0 < T_m$ is the initial temperature of the heated object, $T_m$ and $T_v$ are the melting and boiling temperatures of the material, $C_s$ and $C_l$ are the specific heat capacities of the material in its solid and liquid states, and $L_s$ and $L_v$ are the latent heats of melting and evaporation.

The heat $E$ required to make a hole in the object is provided by the optical beam according to the equation
\begin{equation}
    E = P \tau,
\end{equation}
where $P$ is the optical power and $\tau$ is the exposure time. In practice, optical power $P$ can be large to achieve a small $\tau$ and fast cutting. Short exposure times are advantageous because the laser energy is deposited before significant heat can diffuse away from the illuminated spot via thermal conduction or radiation. The evaporated mass $m$ can be estimated from the equation
\begin{equation}
    m = \rho \pi w^2 h,
\end{equation}
where $\rho$ is the density of the material, $w$ is the laser beam size, and $h$ is the thickness of the material. It is important to keep the laser beam size small to reduce the amount of lost material and reduce the heat loss from the thermal radiation that is proportional to the area of the heated area.

In the calculations above, we assumed that all the laser light is absorbed by the material. In practice, however, part of the incident beam can be reflected, transmitted, or scattered. Therefore, selecting an appropriate wavelength is crucial to maximise absorption for cutting applications. Absorption can vary by orders of magnitude depending on the material and wavelength. For example, the absorption of glass at 1550\,nm is very low, which makes it ideal for optical communications, as we will discuss in Lecture~\ref{Lecture10}, but this wavelength is unsuitable for cutting or welding. In contrast, glass absorbs strongly at wavelengths $\lambda \sim 10$\,um, making these wavelengths effective for material processing.

The physical basis of absorption is the excitation of electrons, atoms, and molecules in a material by the electric field of the laser beam. The photon energy is transferred to the material’s vibrational modes and heats it locally. For instance, a molecule composed of three atoms can be modeled as three masses connected by springs and has multiple longitudinal and rotational vibrational modes. The absorption spectrum typically appears continuous over a certain wavelength range because the number of modes is large. Just as the atmosphere absorbs light at specific wavelengths, each material exhibits characteristic absorption properties determined by its chemical structure.

For example, laser eye surgeries are typically performed with ultraviolet light, which is strongly absorbed by the cornea. The cornea contributes about two-thirds of the optical power of the combined cornea–lens system, and reshaping it can effectively alter the eye’s focal length and improve vision. Another example is tattoo removal. Tattoo ink particles are too large for the immune system to eliminate from the dermis (the lower layer of the skin). To remove a tattoo, a laser is used to break the ink into smaller particles without damaging the surrounding skin. This requires selecting a wavelength that is strongly absorbed by the ink but minimally absorbed by the skin. A third example is welding. In Advanced LIGO, fused silica (high-purity SiO$_2$) suspensions are monolithic~\cite{Aasi_2015}: the mirrors are welded to silica fibers using a CO$_2$ laser operating at $\approx 10$\,um, which melts the glass locally to create strong, precise joints.

\subsection{Scattering of light}

Laser beams interacting with matter can undergo several types of scattering~\cite{bohren2008absorption}, which can be broadly classified as elastic or inelastic. In elastic scattering, the photon energy remains unchanged: for example, Rayleigh scattering occurs when light is scattered by particles much smaller than the wavelength, and Mie scattering involves particles comparable to the wavelength and often leads to directional effects. In inelastic scattering, the photon exchanges energy with the medium, shifting its frequency. Brillouin scattering involves energy transfer to acoustic phonons, slightly shifting the light frequency, whereas Raman scattering transfers energy to molecular vibrations or rotations, producing larger frequency shifts characteristic of the material. At much higher energies, Compton scattering occurs when photons collide with free or loosely bound electrons, for example, in tungsten sheets in gamma-ray detectors as we considered in Lecture~\ref{Lecture4}, and lose energy.

In this section, we focus on Rayleigh scattering of laser beams because it determines how light propagates through the clear atmosphere and how much signal is lost due to interactions with molecules. Although Rayleigh scattering is weak for visible and infrared laser light, it becomes significant over long distances and at shorter wavelengths, affecting applications such as free-space optical communication, LIDAR, and atmospheric sensing. It also enables laser shows and useful diagnostics: by analysing the scattered light, we can extract information about air density, temperature, and composition. The underlying reason for scattering light on molecules comes from an induced dipole moment $\VF{p}$ given by the equation
\begin{equation}\label{eq:l6_dipole}
    \VF{p} = \alpha \VF{E},
\end{equation}
where $\alpha$ is polarisability of air molecules and $\VF{E}$ is the electric field of the laser beam. Although the polarisability varies slightly between different molecular species in air (N$_2$, O$_2$, etc), these values are of the same order of magnitude, and it is reasonable to adopt an effective average polarisability representative of air as a whole.

The induced dipole moment from Eq.~(\ref{eq:l6_dipole}) is time-dependent because it follows the oscillating electric field of the laser at optical frequencies. A time-varying dipole radiates electromagnetic waves, meaning that air molecules act as secondary sources of radiation at the same frequency as the incident laser beam. However, unlike the highly directional laser beam, dipole radiation is emitted over a wide range of angles. As a result, the scattered light is distributed according to the characteristic angular pattern of dipole radiation, described by the equation
\begin{equation}\label{eq:w5_dipole_rad}
    \langle S \rangle = \frac{\mu_0 p_0^2 \omega^4}{32 \pi^2 c} \frac{\sin^2 \theta}{r^2} \sim \frac{1}{\lambda^4},
\end{equation}
where $\langle S \rangle$ is the time-averaged Poynting vector, $\mu_0$ is the magnetic permeability of free space, $p_0$ is the amplitude of the induced dipole moment oscillating at angular frequency $\omega$, $\theta$ is the angle between the dipole moment and the direction of observation, and $r$ is the distance from the dipole to the observer.

The $1/\lambda^4$ scaling from Eq.~\ref{eq:w5_dipole_rad} enhances the scattering of shorter wavelengths compared to longer ones. This explains why the sky appears blue, for example. The same principle is important in remote sensing and atmospheric diagnostics, where shorter wavelengths provide higher sensitivity to small particles and molecular composition. In laser propagation and free-space communication, this scaling implies that shorter-wavelength beams suffer greater attenuation due to scattering, influencing the choice of operating wavelength.

We can derive Eq.~\ref{eq:w5_dipole_rad} using the retarded potentials~\cite{jackson1999classical}. For the vector potential $\VF{A}$ we have
\begin{equation}
\VF{A} = \frac{\mu_0}{4\pi} \int_V \frac{\VF{j}\!\left(\VF{r'},\, t-\frac{|\VF{r}-\VF{r'}|}{c}\right)}{|\VF{r}-\VF{r'}|}\, d^3\VF{r'},
\end{equation}
where the integration is over the volume $V$ in which the current density $\VF{j}$ is non-zero. The vectors $\VF{r}$ and $\VF{r'}$ denote the observation and source position vectors, respectively. Since we are in the radiation zone (e.g.\ distances of order 100\,m to kilometres in laser scattering experiments) and the dipole is spatially small, we may assume $r \gg r'$. This allows the approximation $|\VF{r}-\VF{r'}| \approx r$ and $t - |\VF{r}-\VF{r'}|/c \approx t - r/c$, so that
\begin{equation}
\VF{A} \approx \frac{\mu_0}{4\pi r} \int_V \VF{j}\!\left(\VF{r'},\, t-\frac{r}{c}\right)\, d^3\VF{r'} 
= \frac{\mu_0}{4\pi r}\, \dot{\VF{p}}\!\left(t-\frac{r}{c}\right),
\end{equation}
where we used $\int \VF{j}\, d^3\VF{r'} = \dot{\VF{p}}$ for an oscillating electric dipole $\VF{p}(t)$.

The magnetic field follows from $\VF{B} = \nabla \times \VF{A}$. Retaining only the leading $1/r$ radiation terms (and neglecting terms of order $1/r^2$ arising from differentiating $1/r$), we obtain
\begin{equation}
\VF{B} = \frac{\mu_0}{4\pi r}\, \nabla \times \dot{\VF{p}}\!\left(t-\frac{r}{c}\right).
\end{equation}

For simplicity, we choose a coordinate system in which $\VF{p} = p\,\VF{e}_z$ and the observation point lies in the $x$--$z$ plane. In this case, $\dot{\VF{p}}(t-r/c)$ has only a $z$-component and depends on $r = \sqrt{x^2+z^2}$, so that the curl is purely along $\VF{e}_y$. Evaluating the derivative gives
\begin{equation}
\VF{B} = -\frac{\mu_0}{4\pi r}\frac{\partial \dot{p}(t-r/c)}{\partial x}\,\VF{e}_y
= \frac{\mu_0}{4\pi r c}\, \ddot{p}(t-r/c)\,\frac{x}{r}\,\VF{e}_y
= \frac{\mu_0}{4\pi r c}\, \ddot{p}(t-r/c)\,\sin\theta\,\VF{e}_y.
\end{equation}

The electric field is orthogonal to $\VF{B}$ and, in the radiation zone, satisfies $E = cB$. The Poynting vector $\VF{S} = \VF{E} \times \VF{B}/\mu_0$ is therefore directed radially outward along $\VF{r}$, with magnitude
\begin{equation}
S = \frac{EB}{\mu_0} = \frac{cB^2}{\mu_0}
= \frac{\mu_0}{16\pi^2 c}\,\frac{\left[\ddot{p}(t-r/c)\right]^2}{r^2}\,\sin^2\theta.
\end{equation}

For a harmonically oscillating dipole $p(t) = p_0 \sin(\omega t)$, we have $\ddot{p}(t) = -\omega^2 p_0 \sin(\omega t)$, and time averaging over one optical cycle yields $\langle \ddot{p}^{\,2} \rangle = \tfrac{1}{2}\omega^4 p_0^2$. This removes the fast $2\omega$ oscillations and produces a steady radiated intensity, leading to Eq.~\ref{eq:w5_dipole_rad}.

The induction of dipoles in response to an electric field occurs in gases and solids. In solids, where the structure is uniform on the scale of the wavelength, the induced dipoles oscillate in a coherent, phase-aligned manner, which leads primarily to a collective response described by the refractive index, with little scattering. In contrast, in gases, like air, molecular positions fluctuate due to thermal motion, and the induced dipoles are randomly phased, resulting in Rayleigh scattering. The key difference lies in the spatial order of the medium, which determines whether light is predominantly transmitted with a phase shift (coherent response) or scattered in different directions (incoherent response).
 
\subsection{Optical tweezers}

\begin{wrapfigure}{r}{0.35\textwidth}
    \centering
    \hspace{-1cm}
    \includegraphics[width=0.33\textwidth]{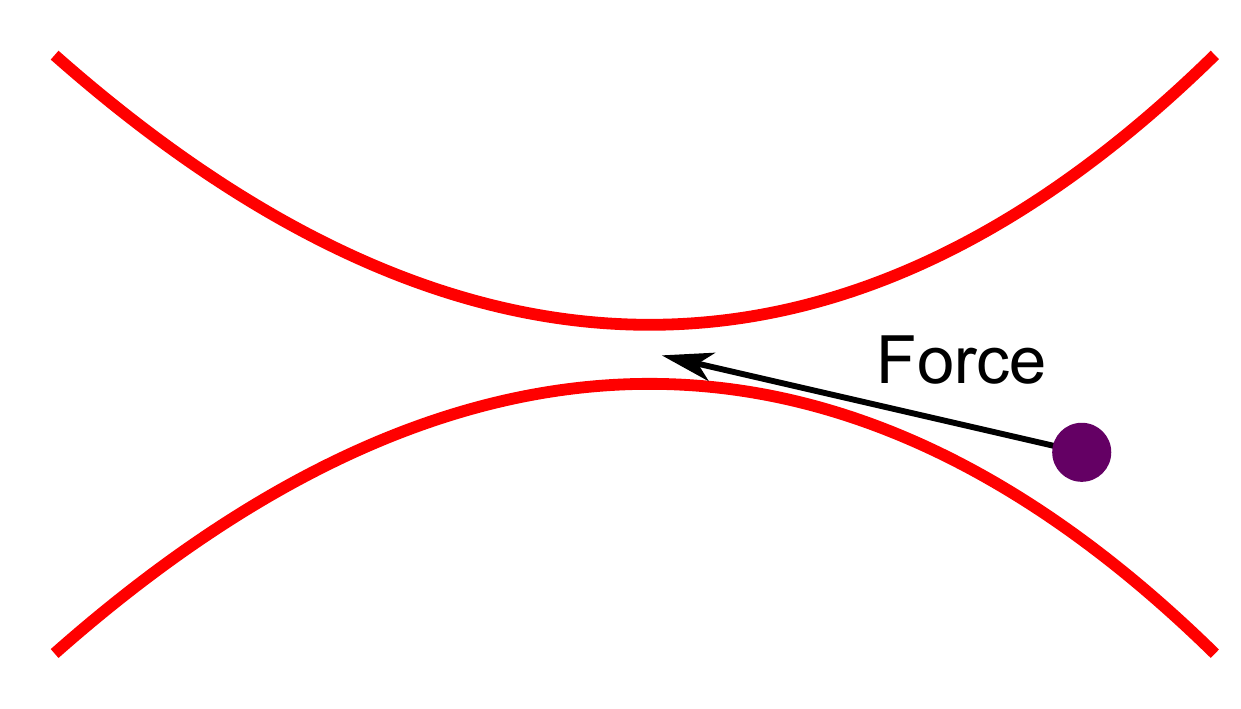} 
    \caption{Sketch of optical tweezers and a dielectric microparticle.}
    \label{fig:w5_tweezers}
\end{wrapfigure}

Next, we consider how laser beams can utilise their optical power to trap matter, acting as optical tweezers~\cite{ashkin1986optical}. When a particle is placed in a focused laser beam, the oscillating electric field induces a dipole moment. Similar to Rayleigh scattering, this induced dipole interacts with the electromagnetic field, and the resulting Lorentz force gives rise to a net force that can confine the particle near the region of highest intensity, as shown in Fig.~\ref{fig:w5_tweezers}. This mechanism enables a wide range of applications, including the levitation of microparticles, the manipulation of biological cells, and the cooling and trapping of atoms in advanced atomic physics experiments. These techniques were recognised by the 2018 Nobel Prize in Physics.

Optical levitation with tweezers is also a powerful platform in quantum optomechanics and enables precise control of isolated particles with minimal environmental coupling~\cite{millen2020optomechanics}. By trapping nanoparticles in a vacuum, we can study the quantum motion of the particle, cooling of the center-of-mass mode toward the quantum ground state, and perform tests of fundamental physics such as wavefunction collapse models and decoherence mechanisms. Levitated systems also serve as highly sensitive probes for force sensing, acceleration, and gravitational effects, and provide a platform for exploring light–matter interactions at the quantum level, including non-classical state preparation and macroscopic quantum superpositions.

The Lorentz force acting on the induced dipole from the laser electric and magnetic fields ($\VF{E}$ and $\VF{B}$) is given by the equation
\begin{equation}\label{eq:w5_tweezer_F}
\begin{split}
    &\VF{F} = q(\VF{E} + \frac{d\VF{x}}{dt} \times \VF{B}) \\
    &\VF{F} = (\VF{p} \cdot \nabla)\VF{E} + \frac{d\VF{p}}{dt} \times \VF{B} \\
   &\VF{F} = \alpha (\VF{E} \cdot \nabla) \VF{E} + \alpha \frac{d \VF{E}}{dt} \times \VF{B},
\end{split}
\end{equation}
where $\VF{p}=\alpha \VF{E}$ is the induced dipole moment. We can simplify Eq.~(\ref{eq:w5_tweezer_F}) by utilising an equation from vector analysis
\begin{equation}
    (\VF{E} \cdot \nabla) \VF{E} = \frac{1}{2} \nabla E^2 - \VF{E} \times (\nabla \times \VF{E})
\end{equation}
and Faraday's law of induction, we find that 
\begin{equation}\label{eq:w5_force}
\begin{split}
    & \VF{F} = \frac{1}{2} \alpha \nabla E^2 + \alpha \VF{E} \times \frac{d\VF{B}}{dt} + \alpha \frac{d \VF{E}}{dt} \times \VF{B} = \frac{1}{2} \alpha \nabla E^2 + \alpha \frac{d\VF{S}}{dt} \\
    & \VF{F} = \frac{1}{2} \alpha \nabla E^2 \sim \nabla I,
\end{split}
\end{equation}
where $I$ is the intensity of light. If we compare the above equation with an equation for force in a scalar field $\phi$: $\VF{F} = -\nabla \phi$, we find that the shapes of the equations are identical and the laser intensity plays the role of a scalar field potential.

The force given by Eq.~(\ref{eq:w5_force}) is trapping near the beam waist for the fundamental Gaussian beam ($l=m=0$) where the intensity is highest. If the beam waist is at position $x=y=z=0$ then the intensity near the beam waist is given by the equation
\begin{equation}
    I(x,y,z) = I_0 \frac{w_0^2}{w(z)^2} \exp \left( -2\frac{x^2 + y^2}{w(z)^2} \right) \approx I_0 \left(1-\left(\frac{z}{z_R}\right)^2 - 2\frac{x^2 + y^2}{w_0^2} \right),
\end{equation}
where $I_0$ is the laser beam intensity at the centre of the beam waist, and $w_0$ is the beam waist.

%% file: week6.tex
\section{Applications of lasers: Laser technologies}
\label{Lecture6}

Instead of relying on filtering thermal light, as discussed in Lecture~\ref{Lecture5}, we can generate coherent light via induced emission. All lasers share three key components: an active medium that provides gain, a pumping mechanism to create a population inversion, and an optical cavity to provide feedback and select the lasing mode. In this lecture, we will explore different types of lasers, starting with solid-state lasers such as ruby, Nd:YAG, and Ti:sapphire, which offer high power and low noise. We will also discuss semiconductor lasers, which are compact, consume less energy, and allow fast modulation for communication applications. Finally, we will cover gas lasers, which are often used as frequency standards due to their stable emission. 

In this lecture, we discuss
\begin{itemize}
\item induced emission,
\item ruby lasers,
\item Nd:YAG lasers,
\item titanium sapphire lasers,
\item semiconductor lasers,
\item gas lasers.
\end{itemize}

\subsection{Induced emission}

An electron in an atom can interact with photons in three ways: (i) absorb a photon and become excited (induced absorption), (ii) spontaneously emit a photon and return to a lower energy state (spontaneous emission), or (iii) emit a photon in the presence of another photon and return to a lower state (induced emission). Spontaneous and induced emission are similar processes because process (ii) is triggered by vacuum fluctuations. The key difference between processes (ii) and (iii) is that spontaneous emission occurs in all directions, and induced emission produces a photon identical to the incident one. This property of induced emission is the foundation of laser light.

To determine the rate of induced emission, we consider a collection of atoms in a black box in thermal equilibrium. In this model, the population of atoms in the ground and excited states is given by the Boltzmann distribution:
\begin{equation}\label{eq:w6_thermal}
    \frac{N_2}{N_1} = \exp \left(-\frac{h \nu}{kT} \right),
\end{equation}
where $N_1$ and $N_2$ are the number of atoms in the ground and excited states, $h$ is the Planck constant, $h \nu$ is the difference between the two atomic levels, which is also equal to the energy of photons with frequency $\nu$. Since the thermal radiation density inside the box is given by the equation~\cite{svelto2010principles}
\begin{equation}
    \rho(\nu) = \frac{8 \pi h \nu^3}{c^3}\frac{1}{\exp \left(\frac{h \nu}{kT} \right)-1}.
\end{equation}
The number of atoms moving from the ground to the excited state is $N_1 B_{12} \rho$ due to induced absorption. The number of atoms moving from the excited to the ground state due to the induced emission is $N_2 B_{21} \rho$ and due to the spontaneous emission is $N_2 A_{21}$. Since the thermal statistics given by the equation Eq.~(\ref{eq:w6_thermal}) stays the same over time, the number of atoms transitioning from the ground state to excited and back must be the same as given by the equation
\begin{equation}
N_1 B_{12} \rho = N_2 B_{21} \rho + N_2 A_{21}.
\end{equation}
Substituting the Boltzmann factor and the expression for the photon density, we find the equations
\begin{equation}
\begin{split}
& B_{12}\exp\left(\frac{h\nu}{kT}\right)  \rho = B_{21} \rho + A_{21} \\
& \rho \left(B_{12}\exp\left(\frac{h\nu}{kT}\right) - B_{21} \right) = A_{21} \\
& \frac{8 \pi h \nu^3}{c^3}\frac{B_{12}\exp \left(\frac{h\nu}{kT}\right) - B_{21}} {\exp \left(\frac{h \nu}{kT} \right)-1} = A_{21}.
\end{split}
\end{equation}
Since $A_{21}$ is temperature independent, the expression yields the Einstein equations
\begin{equation}\label{eq:w6_Einstein}
B_{12} = B_{21}, \hspace{1cm} A_{21} = \frac{8 \pi h \nu^3}{c^3}B_{12},
\end{equation}
which show that the probability for an atom to be excited by photon absorption is equal to the probability of relaxation due to induced emission.

Since the probabilities of induced absorption and induced emission are equal, a light beam propagating through an atomic medium can either be attenuated or amplified, depending on the population of the energy levels. The net power generated or absorbed per unit volume is given by the equation
\begin{equation}
    \frac{P}{V} = (N_2 - N_1) W_i h \nu,
\end{equation}
where $W_i \sim B_{12}, B_{21}$~\cite{svelto2010principles} is the induced rate of atom-photon interaction. The equation shows that if $N_2 < N_1$ the medium absorbs energy from the light field, which is the typical situation in thermal equilibrium given by Eq.~(\ref{eq:w6_thermal}). However, if the system is externally pumped to achieve a population inversion, corresponding to a negative temperature, such that $N_2 > N_1$, the medium amplifies the light.

However, a two-level system (one ground state and one excited state) cannot operate as a laser, even with external pumping, because the pump induces both absorption and stimulated emission at equal rates, as given by Eq.~\ref{eq:w6_Einstein}, and prevents a population inversion. To overcome this limitation, we can use a three-level system: one ground state and two excited states. An external pump excites atoms to a higher excited state, from which they rapidly relax to a lower, metastable excited state that does not interact strongly with the pump. This allows a population inversion to build up between the metastable state and the ground state. The lasing process is initiated by the spontaneous emission of a photon, which is then amplified through stimulated emission as it propagates through the medium.

\subsection{Ruby (Cr:Al$_2$O$_3$) laser}

The ruby laser was the first working laser, demonstrated in 1960 by Theodore Maiman~\cite{maiman1960stimulated}. It operates as a three-level system in which chromium ions (Cr$^{3+}$) are embedded in a solid host and provide the active medium. In ruby, Cr$^{3+}$ ions substitute for about 0.05\% of the Al$^{3+}$ ions in the sapphire lattice~\cite{svelto2010principles}. When optically pumped, typically with a flash lamp, electrons in the Cr$^{3+}$ ions are excited to higher energy levels and then rapidly relax to a long-lived metastable state. From this state, they can undergo stimulated emission to the ground state, producing coherent red light at $694.3$\,nm or at 692.9\,nm. The latter wavelength has a lower gain and is less popular in ruby lasers.

\begin{wrapfigure}{r}{0.37\textwidth}
    \centering
    \includegraphics[width=0.35\textwidth]{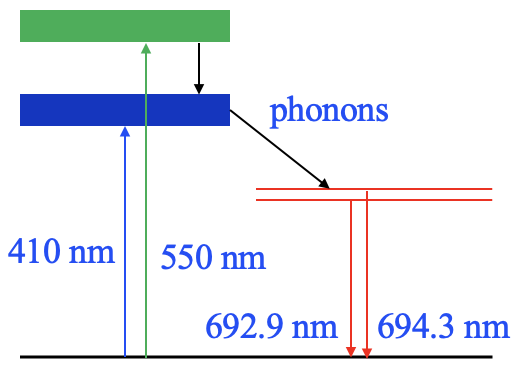} 
    \caption{Simplified diagram of the ruby laser transitions.}
    \label{fig:w6_ruby}
\end{wrapfigure}

The sapphire host lattice plays a key role in enabling laser operation. Its purpose is to broaden the pumping transitions and allow absorption over a wide range of wavelengths. The substrate also keeps the lasing transition narrow, which ensures a well-defined output frequency. In addition, the lattice provides good mechanical and thermal stability and allows the crystal to withstand intense optical pumping without damage.

The ruby laser transitions are shown in Fig.~\ref{fig:w6_ruby}. The broad absorption bands around 410\,nm and 550\,nm in ruby arise because these transitions involve higher excited states of the Cr$^{3+}$ ions that are strongly coupled to lattice vibrations (phonons), leading to significant broadening and a wide range of allowed transition energies. In contrast, the red lines correspond to transitions between the metastable excited state and the ground state, where the electronic states are more weakly coupled to the lattice. As a result, these transitions are less affected by phonons than the pumping transitions and work for laser emission.

\subsection{Nd:Yag laser}

In a three-level system, such as a ruby laser, the lower laser level coincides with the ground state, and more than half of the atoms must be excited to achieve population inversion. The lasing process can therefore be made more efficient by introducing a lower laser level that is separate from the ground state. In such a four-level laser system, population inversion can be achieved and maintained even when only a small fraction of atoms are excited because the lower laser level lies above the ground state and rapidly depopulates.

\begin{wrapfigure}{r}{0.41\textwidth}
    \centering
    \vspace{-4mm}
    \includegraphics[width=0.40\textwidth]{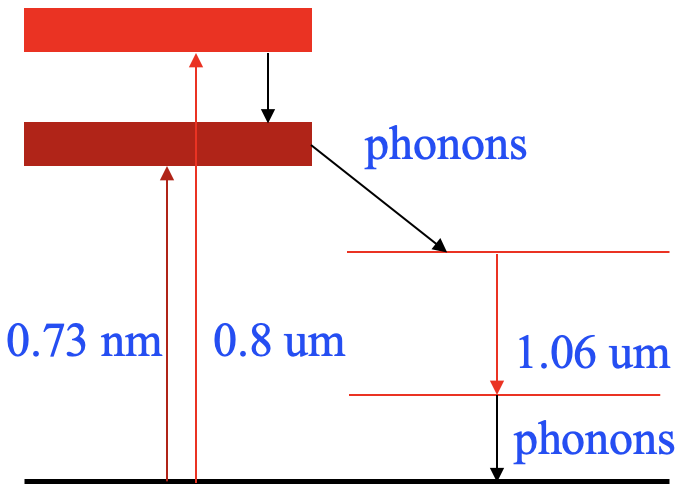} 
    \caption{Simplified diagram of the Nd:YAG pumping and lasing transitions.}
    \label{fig:w6_NdYag}
\end{wrapfigure}

The Nd:YAG laser is a widely used four-level laser system in which neodymium ions (Nd$^{3+}$) are doped into a yttrium aluminum garnet (YAG) crystal host. Optical pumping excites the Nd$^{3+}$ ions to higher energy levels, from which they rapidly relax to a long-lived metastable state. A population inversion is then established between this state and a lower-lying level, enabling stimulated emission, most commonly at a wavelength of $1064$\,nm in the infrared, as shown in Fig~\ref{fig:w6_NdYag}.

The choice of Nd$^{3+}$ ions provides well-defined energy levels suitable for efficient lasing transitions. The YAG substrate offers excellent mechanical strength, thermal conductivity, and optical quality, allowing the crystal to withstand high pump powers. These properties make Nd:YAG lasers practical, with applications ranging from industrial cutting and welding to medical procedures and scientific research, including gravitational-wave detection~\cite{savagesavage2011psl}.

\subsection{Ti:sapphire (Ti:Al$_2$O$_3$) laser}

The titanium–sapphire (Ti:sapphire) laser is a widely tunable laser in which Ti$^{3+}$ ions are doped into a sapphire (Al$_2$O$_3$) crystal lattice~\cite{moulton1986tisa}. Typically, about 0.2\% of the Al$^{3+}$ ions are replaced by Ti$^{3+}$ ions, which act as the active lasing centres. Similar to the ruby laser, the sapphire host lattice provides excellent mechanical and thermal stability, and enables a broad gain bandwidth of approximately $650–1100$\,nm. Ti:sapphire lasers are usually pumped with green light, for example, from a frequency-doubled Nd:YAG laser, which efficiently excites the Ti$^{3+}$ ions to higher energy levels. The ions then relax non-radiatively to a metastable state, from which stimulated emission leads to laser action.

One of the key advantages of Ti:sapphire lasers is their broad tunability, typically around $400$\,nm, which makes them versatile for spectroscopy, ultrafast optics, and nonlinear optical applications. The combination of Ti$^{3+}$ ions and the sapphire lattice allows both high gain and a wide tunable range. Ti:sapphire lasers are commonly used in research laboratories for generating femtosecond pulses, precision spectroscopy~\cite{demtroder2014laser}, and in applications requiring adjustable wavelengths across the near-infrared and visible spectrum.

\subsection{Semiconductor lasers}

Semiconductor lasers are typically compact, with sizes on the order of $100$\,um, and can be directly integrated into optoelectronic circuits, such as Ethernet-to-fiber converters. Unlike lasers based on atomic media, semiconductor diodes are electrically rather than optically pumped. This allows their output power to be modulated at very high frequencies, up to $\approx 20$\,GHz, and makes them suited for high-speed communication applications.

In Lecture~\ref{Lecture3}, we discussed silicon diodes for light detection. In this section, we focus on a different class of diodes, such as GaN, GaAs, and InP, that implement a reverse process and emit light when an electric current passes through them. The key distinction between semiconductors used for light absorption and those used for light emission lies in their electronic band structure, which can feature either an indirect or direct band gap. The minimal-energy state in the conduction band and the maximal-energy state in the valence band each have a specific crystal momentum. If these momenta coincide, the semiconductor has a direct band gap and can efficiently emit photons because photons carry a small momentum. If the momenta differ, the semiconductor has an indirect gap, and photon emission is inefficient because the electron must transfer momentum to the crystal lattice.

\begin{figure}[t!]
\centering
\includegraphics[width=5.5cm]{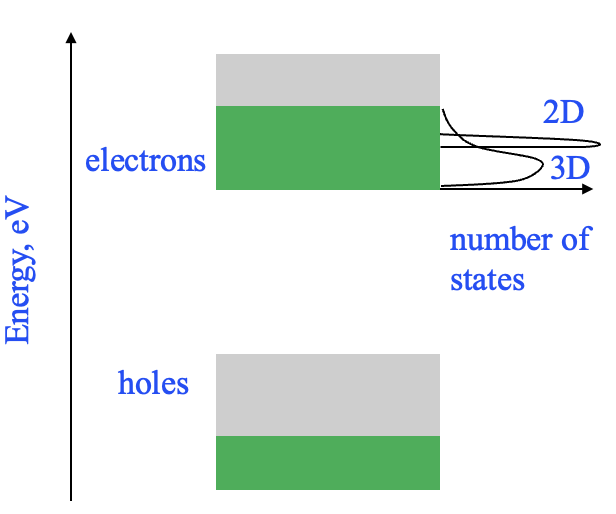} 
\hspace{1mm}
\includegraphics[height=2cm]{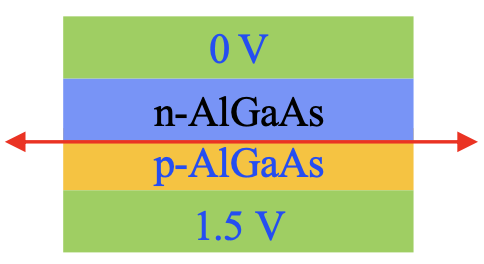}
\hspace{1mm}
\includegraphics[height=2cm]{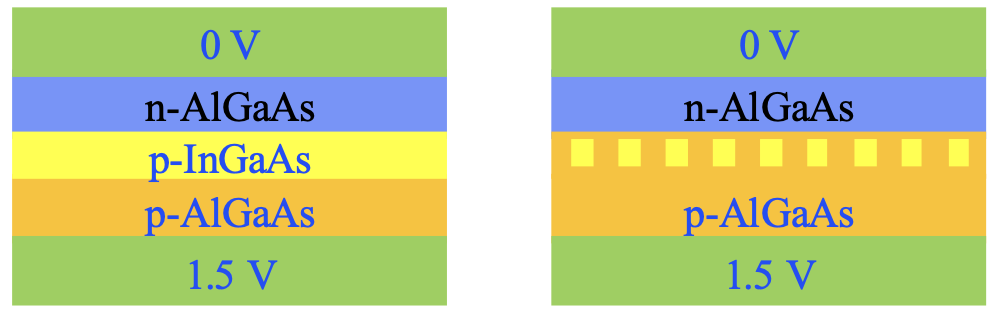} 
\caption{(i) Simplified diagram of a conduction and valence bands in a semiconductor laser with a direct bandgap. Quantum films, wires, and dots constrain the density of states, reduce the threshold current, and sharpen the gain of a laser. (ii) Diagram of a forward-biased AlGaAs PN-junction with a laser beam. (iii-iv) InGaAs quantum wells inside the PN-junction, which are required to sharpen the population of the electrons in the conducting band.}
\label{fig:w6_AlGaAs}
\end{figure}

In semiconductor lasers, the conduction and valence bands form an effective “two-level” system, as shown in Fig.~\ref{fig:w6_AlGaAs}. When a forward bias is applied across the PN-junction, electrons are injected into the conduction band while holes occupy the valence band. If the carrier injection is sufficient, a population inversion is achieved near the band edges, and the probability of electrons occupying the conduction states exceeds that of holes occupying the corresponding valence states involved in the optical transition. Under these conditions, spontaneous emission can trigger induced emission, where photons emitted by one electron stimulate additional emissions from others, leading to coherent light amplification. Lasing occurs when this amplification exceeds the losses in the optical cavity.

A common challenge in conventional PN-junction lasers is the presence of multiple modes, which broaden the emission spectrum and reduce the laser coherence. This arises from the spread of electron energies in the conduction band, allowing transitions over a range of photon energies. One way to reduce this spread is to confine carriers more tightly, which can be achieved using materials with different bandgaps or low-dimensional structures such as quantum wells (films), quantum wires, or quantum dots. These approaches create discrete energy levels, reduce the number of available optical transitions, and favour single-mode operation, as shown in Fig.~\ref{fig:w6_AlGaAs}.

\subsection{Gas lasers}

Unlike in solid-state lasers, the atomic or molecular energy levels in gas lasers are not coupled to a lattice. This property makes gas lasers ideal for applications that require stable frequency standards. Historically, gas lasers were among the earliest types of lasers developed, beginning with the helium-neon (He–Ne) laser demonstrated in 1960~\cite{javan1961gas}, shortly after the first ruby laser.

\begin{wrapfigure}{r}{0.44\textwidth}
    \centering
    \vspace{-1cm}
    \includegraphics[width=0.42\textwidth]{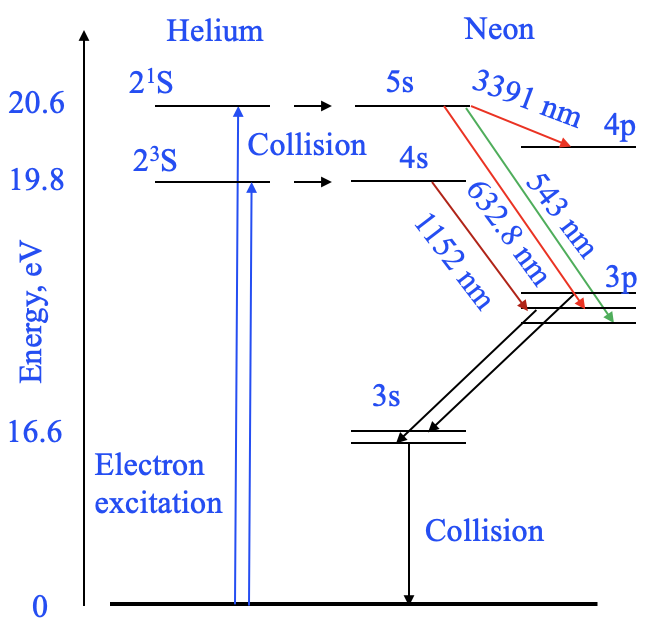} 
    \caption{Simplified diagram of the HeNe laser transitions. The neon s and p states are split into four and ten sublevels because of the interaction with the remaining five electrons in the 2p orbitals. A small number of levels is kept for simplicity.}
    \label{fig:w6_HeNe}
\end{wrapfigure}

Gas lasers are typically electrically pumped, in which an electrical discharge excites the gas atoms or molecules to higher energy levels.
The discharge is initiated when a sufficiently high voltage is applied across the gas, causing a small number of free electrons to accelerate in the electric field. These electrons gain kinetic energy and ionise gas atoms through collisions, creating additional electrons in an avalanche process that sustains the plasma. The energetic electrons excite helium atoms into long-lived metastable states, which store energy due to their slow radiative decay. These metastable helium atoms then transfer their energy via collisions to neon atoms, as shown in Fig.~\ref{fig:w6_HeNe}, and given by the equations
\begin{equation}
\begin{split}
&{\rm He}(2^1 S) + {\rm Ne} = {\rm He} + {\rm Ne}(5s) + \Delta E_1 \\
&{\rm He}(2^3 S) + {\rm Ne} = {\rm He} + {\rm Ne}(4s) + \Delta E_2,
\end{split}
\end{equation}
where $\Delta E_{1,2}$ is the energy difference between the excited states of He and Ne. The upper script in the notation in the helium states is $2S+1$, where $S=0$ or $S=1$ is the total spin. The energy transfer is efficient in a HeNe mixture because $\Delta E = 50$\,meV, and is comparable to the thermal energy of the atoms and enables effective collisional excitation of neon.

He–Ne lasers are more efficient than ruby lasers, despite both being three-level systems, because the lower lasing state is not the ground state of neon, as shown in Fig.~\ref{fig:w6_HeNe}. In He–Ne lasers, helium atoms transfer their energy to neon atoms via near-resonant collisions and populate the upper laser levels directly. The lower laser levels in neon decay rapidly through collisions or spontaneous emission, allowing a stable population inversion to be maintained at relatively low pumping thresholds. In contrast, ruby lasers rely on optical pumping of chromium ions to a short-lived excited state, and the lower laser level is a part of the ground-state manifold.

Another gas laser is the argon-ion laser that emits intense, coherent light at 488\,nm in the blue region of the visible spectrum and is used in scientific instrumentation, medical applications, and laser light shows due to its high beam quality and stability. Its operation is based on electron-impact excitation of argon ions in a gas discharge, producing narrow spectral lines. In contrast, the CO$_2$ laser operates in the infrared at 10.6\,um, and relies on vibrational–rotational transitions of CO$_2$ molecules in a gas mixture. This allows the laser to deliver high continuous-wave power and makes it suitable for industrial applications such as cutting, welding, and materials processing, where high energy density is required.

%% file: week7.tex
\section{Precision measurements: Michelson interferometer}
\label{Lecture7}

In Lecture~\ref{Lecture5}, we examined applications that rely on the intensity of laser beams. But lasers are also extraordinary for their phase stability: the phase of light oscillates rapidly, by $2\pi$ every $500$\,nm for green light and remains predictable over long distances. This makes laser beams some of the most precise rulers in existence, far surpassing the precision of mechanical and electronic measuring devices. In this lecture, and again in Lecture~\ref{Lecture8}, we explore how the coherent phase of laser light can be used to measure distances with precision, reaching scales as small as $10^{-20}$\,m, enabling experiments in gravitational wave detection, interferometry, and fundamental physics.

In this lecture, we discuss
\begin{itemize}
\item interference of light,
\item beam splitters,
\item Michelson interferometers,
\item quantum-limited resolution,
\item temporal coherence in laser interferometers.
\end{itemize}

\subsection{Interference of light}

When two waves overlap in space and time, they form a new wave whose amplitude can be larger or smaller than that of the individual waves. Mathematically, interference arises from the addition of the electric fields of the two waves, not their intensities. This is a direct consequence of the linearity of Maxwell’s equations: if $\VF{E_1}$ and $\VF{E_2}$ are solutions to the wave equation, then their sum $\VF{E_1} + \VF{E_2}$ is also a valid solution. The intensity of the resulting wave is proportional to the square of the total electric field:
\begin{equation}
    I = I_1 + I_2 + 2\epsilon_0 c \,\langle \VF{E}_1 \cdot \VF{E}_2 \rangle,
\end{equation}
where $I_1$ and $I_2$ are the intensity of the first and second waves, $\epsilon_0$ is the vacuum permittivity, $c$ is the speed of light, and $\langle . \rangle$ denotes the time-averaged value. The third term in this expression is called the interference term, which can be positive or negative depending on the relative phase of the waves. Only if the interference term is zero can the total intensity be obtained simply by adding $I_1$ and $I_2$.

One example with a vanishing interference term is when two copropagating light beams have the same frequency but different polarisation states. If the first wave is horizontally polarised and the second wave is vertically polarised, then $\VF{E_1} \cdot \VF{E_2} = 0$ and the waves do not interfere. Another common case occurs with light sources that emit a broad spectrum of frequencies, such as thermal sources. Although each individual frequency component has a nonzero interference term, these terms vary randomly across the spectrum and average out when summed over all frequencies. This is why light from ordinary sources, like torches, does not produce observable interference patterns, and the total intensity is the sum of the intensities of the individual beams.

The next example illustrates a case where the interference term is nonzero and plays a key role in the resulting intensity distribution. Consider two plane waves of the same frequency, $\omega$, intersecting at an angle $\theta$ and forming an interference pattern along the $y$-axis on a screen in the XY-plane. The electric fields of the two waves can be written as
\begin{equation}
    \VF{E}_1 = \VF{E}_{01} \sin(\omega t - \VF{k}_1 \cdot \VF{r} + \phi_1), \hspace{1cm}
    \VF{E}_2 = \VF{E}_{02} \sin(\omega t - \VF{k}_2 \cdot \VF{r} + \phi_2),
\end{equation}
and the intensity at a point on the screen is
\begin{equation}
    I = I_1 + I_2 + 2 \sqrt{I_1 I_2} \cos(k y - \phi),
\end{equation}
where 
\begin{equation}
    k = |\VF{k}_1 - \mathbf{k}_2| = 2 k_1 \sin(\theta/2), \qquad \phi = \phi_1 - \phi_2.
\end{equation}
For small intersection angles, $\theta \ll 1$, the spatial period of the interference fringes is approximately
\begin{equation}
    l = \frac{2 \pi}{k} = \frac{\lambda}{2 \sin(\theta/2)} \approx \frac{\lambda}{\theta}.
\end{equation}
If the two waves have equal intensities, $I_1 = I_2$, the total intensity varies between 0 and $4I_1$ due to the interference term, producing the characteristic bright and dark fringes.

\subsection{Beam splitter}

Since the phase of an electromagnetic wave changes by \(2\pi\) every wavelength, we can use it to measure small distance fluctuations between two objects. But how can we extract the phase if we can only measure the power of optical waves (\(\sim 10^{14}\,\mathrm{Hz}\)) without their phase information? The solution lies in interference: by combining two co-propagating beams, we can make the measured optical power depend on their relative phase.

If the two electric fields have the same amplitude and frequency and propagate along the \(z\)-axis, they can be written as
\begin{equation}
\mathbf{E}_1 = \mathbf{E}_0 \sin(\omega t - kz + \phi_1), \qquad
\mathbf{E}_2 = \mathbf{E}_0 \sin(\omega t - kz + \phi_2),
\end{equation}
and the total power (intensity integrated over the beam area) is
\begin{equation}
P = 2 P_0 \left(1 + \cos(\phi)\right),
\end{equation}
where \(\phi = \phi_1 - \phi_2\) and \(P_0\) is the power of each wave. This shows that the measured power depends directly on the optical phase difference. In computing the optical power, we neglect the rapidly oscillating terms at \(2\omega\) because photodetectors cannot resolve such high frequencies. In practice, the observed power is a time-averaged value over the detector response time \(\tau_r \sim 1~\mathrm{ns}\), which is much longer than the optical period. During this averaging interval, the \(2\omega\) oscillations cancel out.

To make two waves co-propagate and measure their relative phase shift, we can use a 50/50 beam splitter, which divides each incoming wave into two beams of equal amplitude. If two waves are incident on the beam splitter from different directions, each wave is split and co-propagates with a portion of the other wave. Mathematically, the output fields are
\begin{equation}
\begin{split}
E_1 &= \frac{E_0}{\sqrt{2}} \big(\sin(\omega t - kz + \phi) - \sin(\omega t - kz)\big) = \sqrt{2} E_0 \, \sin\frac{\phi}{2} \, \cos\big(\omega t - kz + \frac{\phi}{2}\big),\\
E_2 &= \frac{E_0}{\sqrt{2}} \big(\sin(\omega t - kz + \phi) + \sin(\omega t - kz)\big) = \sqrt{2} E_0 \, \cos\frac{\phi}{2} \, \sin\big(\omega t - kz + \frac{\phi}{2}\big),
\end{split}
\end{equation}
where the factor \(1/\sqrt{2}\) accounts for the 50/50 splitting. The different signs correspond to the relative phase shift introduced by reflection from opposite sides of the beam splitter, as we will discuss in Lecture~\ref{Lecture8}. In practice, which side carries the minus sign is arbitrary, and it does not affect the final measured optical power in interferometric measurements.

\subsection{Michelson interferometer}

\begin{wrapfigure}{r}{0.4\textwidth}
    \centering
    \vspace{-3mm}
    \includegraphics[width=0.38\textwidth]{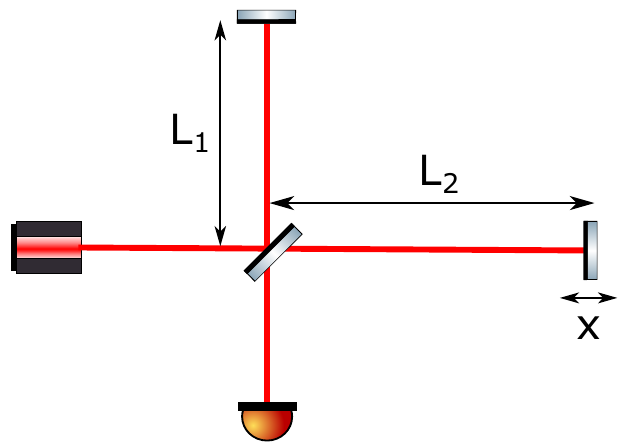} 
    \caption{Michelson interferometer.}
    \label{fig:w7_Mi}
\end{wrapfigure}

We can convert the mechanical displacement of a mirror to an optical signal utilising a Michelson interferometer, which consists of a beam splitter and two mirrors. The incident laser field $E_{\rm in} = E_0 \sin(\omega_0 t)$ is split by the 50/50 beam splitter in two beams of equal power that travel to the end mirrors and back, as shown in Fig.~\ref{fig:w7_Mi}. Upon recombination, the beams interfere, and the resulting light is divided into two output ports. One beam propagates back toward the laser (the symmetric port), and the other exits toward the antisymmetric port, where a photodetector is placed. The electric field at the photodetector depends on the relative phase accumulated in the two arms, and is therefore sensitive to any difference in the optical path lengths caused by mirror displacements, as given by the equation
\begin{equation}\label{eq:l7_E_Mi}
    E = \frac{1}{2}E_0 \left(\sin(\omega_0 t - 2 k L_1) - \sin(\omega_0 t - 2 k L_2) \right) = E_0 \sin\left(\frac{2\pi \Delta L}{\lambda}\right) \cos\left(\omega_0 t - \frac{2\pi (L_1+L_2)}{\lambda}\right),
\end{equation}
where $\omega_0$ is the angular frequency of the laser beam, $L_1$ and $L_2$ are lengths of each of the interferometer arms, $\Delta L = L_1 - L_2$. The measured optical power is then given by the equation
\begin{equation}\label{eq:l7_P_Mi}
    P = P_0 \sin^2 \left(\frac{2 \pi \Delta L}{\lambda}\right) = \frac{P_0}{2} \left(1 - \cos{\frac{4 \pi \Delta L}{\lambda}} \right).
\end{equation}

When computing the optical power from an electric field, we need to evaluate the surface integral of the field intensity given by the equation
\begin{equation}\label{eq:l7:power_as}
\begin{split}
    P = \epsilon_0 c \iint \limits_{-\infty}^{+\infty} E^2 dS & = C E_0^2 \sin^2\left(\frac{2\pi \Delta L}{\lambda}\right) \cos^2\left(\omega_0 t - \frac{2\pi (L_1+L_2)}{\lambda}\right) \\
    &= \frac{C}{2} E_0^2 \sin^2\left(\frac{2\pi \Delta L}{\lambda}\right) \left( 1 - \cos\left(2\omega_0 t - \frac{4\pi (L_1+L_2)}{\lambda}\right) \right)
\end{split}
\end{equation}
where $C$ includes physical constants and a constant proportional to the beam area from the integration. The $2\omega$ term in the equation above vanishes because the measured power is time-averaged over $\sim 1$\,ns due to the limited mobility of electrons in photodetectors, as discussed above. Therefore, the total power measured at the antisymmetric port is given by the equation
\begin{equation}
    P = \frac{C}{2} E_0^2 \sin^2\left(\frac{2\pi \Delta L}{\lambda}\right),
\end{equation}
and the input power is given by the equation
\begin{equation}
    P_0 = \epsilon_0 c \iint \limits_{-\infty}^{+\infty} E_{\rm in}^2 dS = C E_0^2 \sin^2(\omega_0 t) = \frac{C}{2} E_0^2.
\end{equation}
These relations explain why we can move from Eq.~(\ref{eq:l7_E_Mi}) to Eq.~(\ref{eq:l7_P_Mi}) by squaring the electric field, replacing $C E_0^2/2$ with $P_0$, and discarding the rapidly oscillating terms at $2\omega_0$.

Equation~(\ref{eq:l7_P_Mi}) shows that the observed optical power varies between $0$ and $P_0$ as the path length difference $\Delta L$ changes by $\lambda/4$, corresponding to a physical displacement of only a few hundred nanometres for typical optical wavelengths. This illustrates the extreme sensitivity of interferometric measurements to minute displacements.

A prominent application example is the Michelson interferometer configuration used in gravitational-wave detectors such as LIGO~\cite{Aasi_2015}. The design builds on the pioneering interferometric techniques developed by Albert A. Michelson~\cite{michelson1881interference}, for which he was awarded the 1907 Nobel Prize in Physics. More recently, the LIGO Scientific Collaboration, including Rainer Weiss, Barry C. Barish, and Kip S. Thorne, received the 2017 Nobel Prize in Physics for the direct detection of gravitational waves.
In these detectors, a passing gravitational wave induces a differential strain in the interferometer arms, causing one arm to lengthen while the other shortens, and thereby producing a time-dependent path length difference $\Delta L$ proportional to the gravitational-wave amplitude.

Another important application is local displacement sensing, where one arm of the interferometer is fixed ($L_1 = {\rm const}$), and the other mirror is attached to a moving object. Motion of the object changes $L_2$ and modulates the optical power at the antisymmetric port. This principle is widely used in precision positioning systems, for example, in optical lithography, where nanometer- and even sub-nanometer-scale control of wafer and mask positions is required~\cite{mack2007fundamental}. Interferometric sensors monitor the position of the wafer stage in real time, allowing feedback systems to correct for vibrations, and thermal drift. As a result, interferometry plays a crucial role in achieving the extreme alignment accuracy needed for modern semiconductor fabrication.

\subsection{Quantum-limited resolution}

Michelson interferometers provide a strong optical response to even tiny variations in the path length difference $\Delta L$. However, the measurement precision is fundamentally limited by quantum shot noise: fluctuations in the detected optical power can arise either from actual changes in $\Delta L$ or from the quantum nature of light itself. To quantify this limit, our goal is to express the effect of shot noise in units of displacement and estimate the achievable resolution of a Michelson interferometer. In the following, we consider small displacements around a fixed operating point of the interferometer, $x_0$:
\begin{equation}
    \Delta L(t) = x_0 + x(t),
\end{equation}
where $x(t) \ll \lambda$ is the measured signal. We then linearise measured optical power given by Eq.~(\ref{eq:l7_P_Mi}) around $x_0$ and find the derivative of the optical power at $x_0$. We define this quantity as an optical gain $G$ of the interferometer. It shows how much power the photodetector sees in response to the longitudinal displacement of one of the Michelson arms. Calculating the derivative of $P$ over $\Delta L$, we find the optical gain
\begin{equation}
    G \equiv \frac{dP}{d\Delta L} = \frac{2\pi}{\lambda}P_0 \sin \left( \frac{4 \pi \Delta L}{\lambda} \right) \approx \frac{2\pi}{\lambda}P_0 \sin \left( \frac{4 \pi x_0}{\lambda} \right).
\end{equation}

We now compute power fluctuations $\Delta P_{\rm shot}$ due to photon shot noise. For a measurement time $\tau$, the average number of photons on the photodetector is given by the equation
\begin{equation}
    \langle N \rangle = \frac{P \tau}{h\nu},
\end{equation}
where $\nu = \omega_0 / (2\pi)$ is the frequency of light and deviation from $\langle N \rangle$ every measurement is $\sqrt{\langle N \rangle}$. Therefore, the power fluctuations on the photodetector are given by the equation
\begin{equation}
    \Delta P_{\rm shot} = \sqrt{\langle N \rangle} \frac{h \nu}{\tau}= \sqrt{\frac{P(x_0)  h \nu}{\tau}},
\end{equation}
and the quantum-limited resolution of the interferometer is given by the equation
\begin{equation}
    \Delta x_{\rm shot} = \frac{\Delta P_{\rm shot}}{G}.
\end{equation}
Note that $\Delta x_{\rm shot}$ may depend on $x_0$. For example, if $x_0 = \lambda/4$ then $G(\lambda/4) = 0$\,W/m and no signal can be measured. However, if $x_0 = \lambda/8$ then the optical gain is maximised. If $x(t) \ll x_0 \ll \lambda$ then the shot noise limited resolution does not depend on $x_0$ and is given by the equation
\begin{equation}
    \Delta x_{\rm shot} = \frac{\lambda}{4\pi} \sqrt{\frac{h\nu}{P_0 \tau}}
\end{equation}
and is on the order of $10^{-15}$\,m for $P_0 = 1$\,W and $\tau = 1$\,ms, $\lambda=1064$\,nm.

\subsection{Temporal coherence}

Up to this point, we have assumed perfectly monochromatic light throughout this lecture. In practice, however, the optical frequency fluctuates over time, since the temporal coherence of light (introduced in Lecture~\ref{Lecture5}) is finite. To quantify this, we define the coherence time 
$\tau_c$ according to the equation
\begin{equation}
    \tau_c = \frac{2}{\Delta \omega},
\end{equation}
where $\Delta \omega$ is the spectral linewidth (full width at half maximum) of the laser in angular frequency. This definition becomes particularly intuitive when we analyse the measured power in a Michelson interferometer, as it directly determines how long a well-defined phase relationship can be maintained.

When the input electric field has a finite coherence time, its phase becomes time-dependent, and the field can be written as
\begin{equation}
    E_{\rm in} = E_0 \sin(\omega_0 t + \phi(t)),
\end{equation}
where the phase $\phi(t)$ represents the temporal fluctuations of the phase. To analyse the resulting power fluctuations at the photodetector, we note that the two beams traveling along the interferometer arms and recombining at the detector are time-delayed versions of the input field, as described by the equation
\begin{equation}\label{eq:l7_coh}
    E = \frac{1}{2}E_0 (\sin(\omega_0 (t - \tau_1)+\phi(t-\tau_1)) - \sin(\omega_0 (t - \tau_2) + \phi(t-\tau_2)),
\end{equation}
where $\tau_1 = 2L_1 / c$ and $\tau_2 = 2L_2 / c$ are time delays due to the beam travelling in the arms. The above equation is similar to Eq.~(\ref{eq:l7_E_Mi}) because $\omega_0 \tau_1 = 2 k L_1$ and $\omega_0 \tau_2 = 2k L_2$ but also includes phase fluctuations of the input laser beam. The measured optical power is given by the equation
\begin{equation}\label{eq:l7_P_Mi_2}
    P = \frac{P_0}{2} \left(1 - \cos(\omega_0 \Delta \tau + \Delta \phi) \right),
\end{equation}
where $\Delta \tau = \tau_2 - \tau_1$ and $\Delta \phi = \phi(t-\tau_1)-\phi(t-\tau_2) \approx \dot{\phi} \Delta \tau$ accounts for the time-dependent phase of the laser. Since the phase is the time integral of the instantaneous frequency, we can write $\Delta \phi \approx \Delta \omega \Delta \tau$. For $\Delta \tau = \tau_c$, the coherence time, the standard deviation of the phase difference satisfies $\sigma_\phi = 2$\,rad.

Eq.~\ref{eq:l7_P_Mi_2} shows that when the difference in time delays between the two arms approaches the laser’s coherence time, the measured power fluctuates randomly because the phase difference fluctuates by roughly 1\,rad. In this regime, the observed power becomes a random quantity, and no stable interference is observed at the antisymmetric port because the beams no longer maintain a well-defined phase relationship.

Similarly to shot noise, the time-dependent phase term $\Delta \phi$ introduces power fluctuations that are indistinguishable from the motion of the mirror. To achieve high-precision measurements without being limited by the finite coherence time of the input beam, it is necessary to satisfy $\Delta \tau \ll \tau_c$. In practice, this means carefully matching the interferometer arms ($L_1 \approx L_2$) to minimise the coupling of laser frequency noise into the signal measured at the antisymmetric port. The corresponding coherence length, defined as $l_c = c \tau_c$, can vary significantly between lasers: for stabilised Nd:YAG lasers, the coherence length can reach 100\,km, while standard He–Ne lasers have coherence lengths around 20\,cm.

%% file: week8.tex
\section{Precision measurements: Fabry-Perot interferometers}
\label{Lecture8}

Following Lecture~\ref{Lecture7}, we continue our discussion of interferometric measurements and explore how laser light can measure microscopic displacements with high precision. In this lecture, we focus on Fabry–Perot interferometers, which consist of two or more partially transmissive mirrors. If the interferometer is on resonance, the phase of the transmitted or reflected light is far more sensitive to mirror motion than in a standard Michelson interferometer. The quantum-limited displacement sensitivity of Fabry–Perot interferometers (also known as optical cavities) can reach the scale of $10^{-20}$\,m. We also examine their applications, including optical atomic clocks, laser gyroscopes, and optical coatings.

In this lecture, we discuss
\begin{itemize}
\item optical cavities,
\item shot noise limited resolution,
\item review of applications,
\item optical coatings.
\end{itemize}

\subsection{Fields in optical cavities}

We consider a Fabry-Perot interferometer with two mirrors that have identical power transmissivity, $T \ll 1$. The corresponding field transmission and reflection coefficient are $t = \sqrt{T}$ and $r = \sqrt{1-T}$ and satisfy the energy conservation law
\begin{equation}
    r^2 + t^2 = 1.
\end{equation}
This relation ensures that the incident power is either reflected or transmitted by the mirror. In practice, as discussed in Lecture~\ref{Lecture5}, a small fraction of the power can be absorbed or scattered by the mirror surface. For high-quality cavity mirrors, however, these losses are negligible: the surfaces are polished to a roughness below 0.1\,nm and typically scatter only a few parts per million ($\sim 10^{-6}$) of the incident power.

\begin{wrapfigure}{r}{0.4\textwidth}
    \centering
    \vspace{-3mm}
    \includegraphics[width=0.38\textwidth]{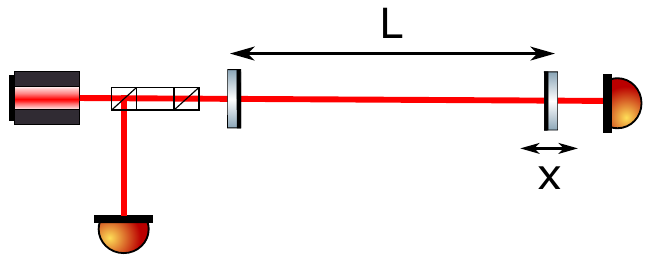} 
    \caption{Linear Fabry-Perot cavity.}
    \label{fig:w8_Cav}
\end{wrapfigure}

Our goal is to derive how laser fields propagate in optical cavities and how mirror motion can be measured using these fields. In this lecture, we represent electric fields as complex numbers, which provides a mathematically convenient formalism. This approach is equivalent to the method discussed in Lecture~\ref{Lecture7}, but it is simpler to apply to optical cavities and optical fibers.

Consider a laser beam propagating along the Z-axis, with the electric field at $z=0$ given by the equation
\begin{equation}
    E(z=0) = E_0 \sin(\omega_0 t + \phi_0) = \frac{E_0 e^{i\phi_0}}{2i} e^{i\omega_0 t} + c.c. = E_1 e^{i\omega_0 t} + c.c.,
\end{equation}
where “c.c.” denotes the complex conjugate. Similar to Lectures~\ref{Lecture2} and~\ref{Lecture4}, the field propagation over a distance 
$z$ is expressed as
\begin{equation}
    E_2 = E_1 e^{-i\phi},
\end{equation}
where $\phi = kz$ is the phase accumulated by the laser beam while travelling a distance $z$ between points 1 and 2 and the total electric field in point 2 is $E(z) = E_2 e^{i\omega_0 t} + c.c.$.

We now determine the electric field inside the optical cavity, $E_{\rm cav}$, near the input coupler. We consider the interference of the laser field with the field reflected from the input mirror on the cavity side of the mirror. In the steady-state regime, this leads to the equation\begin{equation}
    E_{\rm cav} = E_{\rm in} t + r^2 E_{\rm cav} e^{-i\phi_{rt}},
\end{equation}
where $\phi_{rt} = 2kL$ is the round-trip phase accumulated by the laser beam inside the cavity of length $L$. The first term on the right-hand side represents the input laser field transmitted through the mirror, while the second term corresponds to the cavity field circulating inside, which is attenuated by a factor $r^2$ due to reflections from both mirrors. The solution to this equation is then
\begin{equation}
    E_{\rm cav} = \frac{t}{1 - r^2 e^{-i\phi_{rt}}} E_{\rm in}
\end{equation}
and, therefore, the optical power in the cavity, $P_{\rm cav}$, is given by the equation
\begin{equation}
    P_{\rm cav} = \frac{T}{|1 - r^2 e^{-i\phi_{rt}}|^2} P_{\rm in}.
\end{equation}
On resonance, $\phi_{rt} = 2\pi N$, where $N$ is an integer number, and we get a strong power amplification inside the optical cavity, which is given by the equation
\begin{equation}
    P_{\rm cav} = \frac{1}{T} P_{\rm in} \gg P_{\rm in}.
\end{equation}

The power amplification in the cavity does not violate energy conservation, since we are considering the steady-state regime of the electric fields and power. Immediately after the laser is turned on, however, the power inside the cavity builds up from zero to $P_{\rm in}/T$ over a finite buildup time. Once established, the cavity stores and circulates the power accumulated during previous round trips, maintaining the amplified field within the cavity.

Similarly, we can find the transmitted power $P_{\rm tr} = T P_{\rm cav}$ and reflected power $P_{\rm refl}$ from the equations
\begin{equation}\label{eq:l8_measured_P}
\begin{split}
    & P_{\rm tr} = \frac{T^2}{|1 - r^2 e^{-i\phi_{rt}}|^2} P_{\rm in} \\
    & E_{\rm refl} = -r E_{\rm in} + tr E_{\rm cav} e^{-i\phi_{rt}} \\
    & P_{\rm refl} = \left| \frac{-r + re^{-i\phi_{rt}}}{1-r^2e^{-i\phi_{rt}}} \right|^2 P_{\rm in}.
\end{split}
\end{equation}

\begin{figure}[t!]
\centering
\includegraphics[height=4.1cm]{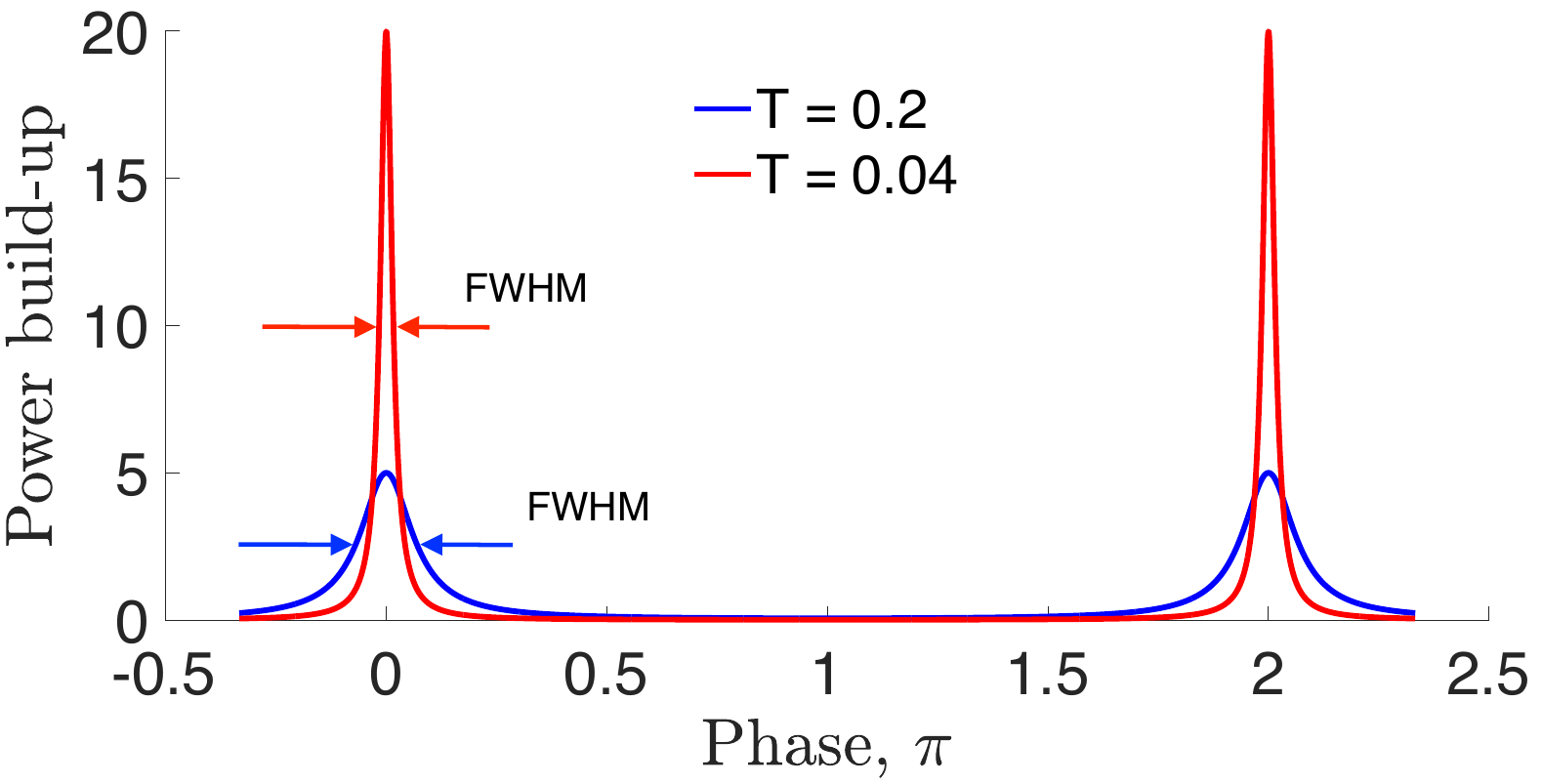} 
\hspace{1mm}
\includegraphics[height=4.1cm]{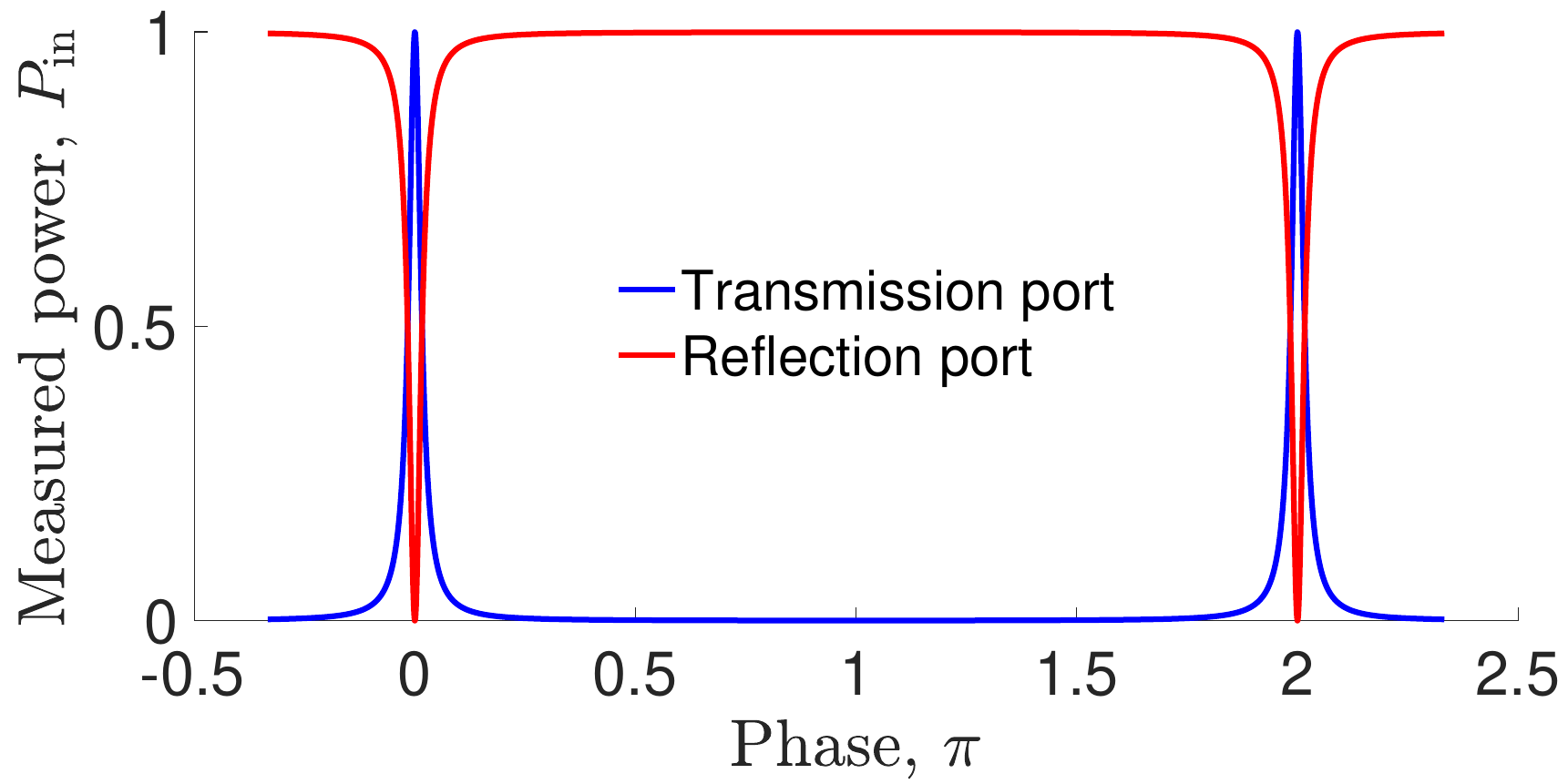} 
\caption{Cavity power build-up factors for two different cavities as a function of the round-trip phase (left). Reducing the transmission increases the intracavity power build-up and narrows the full width at half maximum (FWHM). Measured signals at the transmission and reflection ports, normalised to the input power (right).}
\label{fig:w8_FP_sig}
\end{figure}

In contrast to the Michelson interferometer considered in Lecture~\ref{Lecture7}, the measured photodetector signals $P_{\rm tr}$ and $P_{\rm refl}$ show resonant behaviour, as shown in Fig.~\ref{fig:w8_FP_sig}. This implies that optical cavities tuned near resonance are more sensitive to small displacements than the Michelson interferometer.

The bandwidth of the resonance $\Delta \phi_{rt}$ is defined as the full width at half maximum of the resonant peak. An important property of optical cavities is their finesse, which is given by the equation
\begin{equation}
    \mathcal{F} = \frac{2\pi}{\Delta \phi_{rt}}
\end{equation}
and is proportional to the power build up factor ($B = P_{\rm cav} / P_{\rm in}$) in the cavity. Higher finesse leads to narrower resonant peaks and more sensitive interferometric measurements. However, the gain-bandwidth product of the resonators is $B \Delta \phi_{rt} = {\rm const}$ and is independent of $T$, as shown in Fig.~\ref{fig:w8_FP_sig}.

\subsection{Shot noise limited resolution}

Similar to Michelson interferometers, the displacement measurement in a Fabry–Perot cavity is fundamentally limited by quantum shot noise: fluctuations in the measured power can arise either from changes in the round-trip phase $\phi_{rt}$ or from the quantum nature of light itself. In the following analysis, we assume that the interferometer is detecting small displacements around its operating point, and linearise the response. The round-trip phase is given by the equation
\begin{equation}
    \phi_{rt} = 4\pi\frac{x_0 + x(t)}{\lambda},
\end{equation}
where $x_0$ is the static imbalance of the two arms and $x(t) \ll \lambda$ is the measured signal. We then linearise the measured electric field at the reflection port given by Eq.~(\ref{eq:l8_measured_P}) for $\phi_{rt} \ll 1$ (up to $2\pi N$) and get the equation for the electric field and power
\begin{equation}
    E_{\rm refl} = \frac{-i\phi_{\rm rt}}{T}E_{\rm in} \hspace{1cm} P_{\rm refl} = \frac{\phi_{\rm rt}^2}{T^2}P_{\rm in}.
\end{equation}
Therefore, the optical gain is given by the equation
\begin{equation}
    G(x_0) = \frac{32 \pi^2}{\lambda^2 T^2} P_{\rm in} x_0.
\end{equation}

We then compute the shot noise similar to the case with the Michelson interferometer discussed in Lecture~\ref{Lecture7} and get
\begin{equation}
    \Delta P_{\rm shot} = \frac{4\pi x_0}{\lambda T} \sqrt{\frac{P_{\rm in} h \nu}{\tau}},
\end{equation}
where $\tau$ is the measurement time. We can find the shot noise-limited resolution at $x_0$ from the equation
\begin{equation}
\Delta x_{\rm shot} = \frac{P_{\rm shot}(x_0)}{G(x_0)} = \frac{\lambda T}{8\pi} \sqrt{\frac{h \nu}{P_{\rm in} \tau}}.
\end{equation}
For $T = 10^{-5}$, the shot noise-limited resolution of the optical cavity is $\sim 10^5$ times better than the one of the Michelson interferometer considered in Lecture~\ref{Lecture7} and equals $5.6 \times 10^{-21}$\,m for the same input power and measurement time ($P_0 = 1$\,W, $\tau = 1$\,ms, $\lambda=1064$\,nm).

\subsection{Review of applications}

Optical cavities are utilised in a variety of applications. In this section, we discuss applications in lasers, gravitational-wave detection, optical clocks, and laser gyroscopes.

\subsubsection{Lasers}

As discussed in Lecture~\ref{Lecture6}, an actively pumped atomic medium can generate a laser beam starting from the spontaneous emission of photons. However, the direction of spontaneous emission is random because vacuum fields couple to atoms from all directions. To produce a coherent, directional beam, an optical cavity is employed. The cavity surrounds the active medium, so that only photons resonating between the two mirrors are preferentially amplified. In addition, the bandwidth of the optical cavity is typically much narrower than that of the atomic transition used for lasing, which results in a significant reduction of the laser linewidth.

\subsubsection{Gravitational-wave detectors}

The second application we consider is gravitational wave detection. As discussed in Lecture~\ref{Lecture7}, a Michelson interferometer forms the core of the LIGO detectors. However, the actual detectors are more sophisticated than a simple Michelson interferometer. Their sensitivity is greatly enhanced through the use of four auxiliary optical cavities, as shown in Fig.~\ref{fig:w8_app}\,(left). Two of these cavities are embedded into the interferometer arms and amplify the phase shift caused by passing gravitational waves. The other two cavities, placed around the beam splitter, further increase the circulating laser power and optimise the interferometer’s response to specific gravitational wave frequencies. Together, these optical cavities allow LIGO to detect displacements on the order of $10^{-20}$\,m, and enable the observation of gravitational waves from distant astrophysical events such as black hole~\cite{abbott2016observation} and neutron star~\cite{abbott2017gw170817} mergers.

\subsubsection{Optical atomic clocks}

Next, we consider optical atomic clocks, which stabilise a laser to a narrow atomic transition and achieve a fractional frequency uncertainty at the level of $\Delta f / f \sim 10^{-19} - 10^{-18}$~\cite{huntemann2016single}. The stabilised optical frequency is then converted to the radio-frequency domain using optical frequency combs.

In many atomic clock implementations, however, the atoms must be trapped and prepared before each measurement cycle, which typically takes $\sim 0.1 - 1$\,s. During this dead time, the atomic reference is unavailable, and the laser frequency would drift if left uncontrolled. To overcome this limitation, high-finesse optical cavities act as stable frequency references. By locking the laser to a cavity resonance, its frequency is tied to the cavity length, which can be made stable against environmental perturbations. In this way, optical cavities suppress short-term frequency noise, and the atomic transition provides long-term accuracy.

\begin{figure}[t!]
\centering
\includegraphics[height=3.85cm]{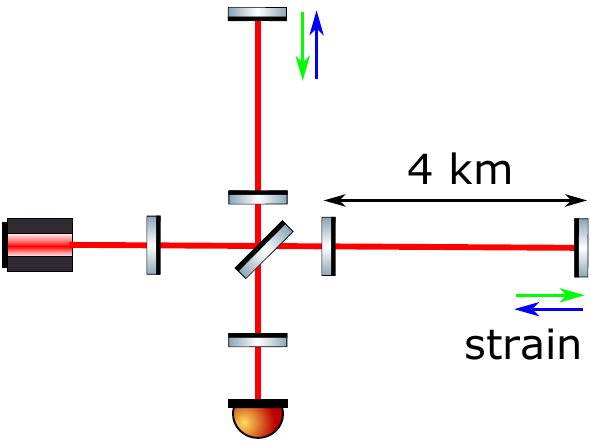} 
\hspace{4mm}
\includegraphics[height=3.85cm]{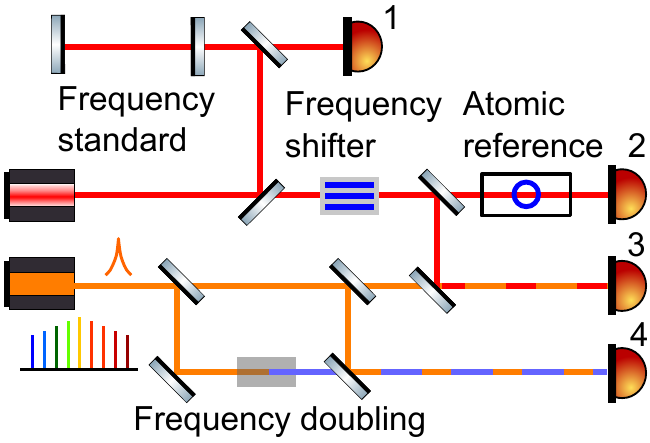} 
\hspace{4mm}
\includegraphics[height=3.85cm]{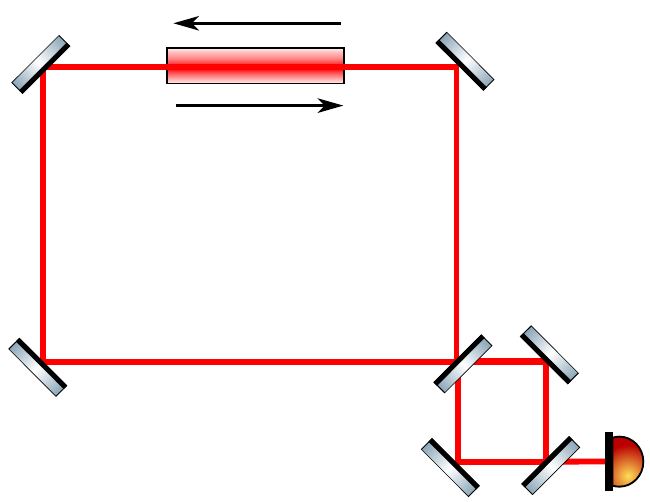} 
\caption{Simplified diagrams of the LIGO interferometer (left), optical atomic clock (centre), and Sagnac interferometer (right).}
\label{fig:w8_app}
\end{figure}

An example of a layout of an optical atomic clock is shown in Fig.~\ref{fig:w8_app}\,(centre). A continuous-wave laser with an eigen mode, $\omega_L$ is set to measure the atomic transition, $\omega_A$. Our goal is to transfer the stability of an optical frequency $\omega_A$ to a signal that can be directly measured and processed by electronic systems in the MHz-GHz domain. Since fluctuations in the laser’s cavity length, refractive index, and gain medium, driven by thermal, mechanical, and quantum effects, continuously perturb its frequency $\omega_L$, we stabilise it to a frequency standard: $\omega_L = \omega_0$, where $\omega_0$ is the resonant frequency of the Fabry-Perot resonator. The stabilisation is achieved by measuring the frequency difference between the laser and the cavity with a photodiode 1 and correcting the laser frequency.

Laser frequency noise and resonator mirror motion are equivalent in interferometric measurements because both manifest as fluctuations in optical phase. In previous discussions, we treated phase variations as arising from the changes of the cavity length. However, laser frequency noise can be reinterpreted as an effective displacement noise of the mirrors, and vice versa. If the cavity mirrors do not move but the laser frequency changes than the total round-trip phase is given by the equation
\begin{equation}\label{eq:w8_freq_phase}
\phi_{rt} = 4\pi \frac{L}{\lambda} = \frac{2L\omega_L}{c} = \frac{\omega_L}{\rm FSR} = \frac{\omega_0}{\rm FSR} + \frac{\Delta \omega_L}{\rm FSR} = 2\pi N + \frac{\Delta \omega_L}{\rm FSR},
\end{equation}
where $L$ is the cavity length, $N$ is an integer, and ${\rm FSR} = c/(2L)$ is the free spectral range of the resonator, which is the frequency spacing between adjacent longitudinal modes. Eq.~\ref{eq:w8_freq_phase} shows that $\phi_{rt}$ is a linear function of $\omega_L$ and any deviations from the resonator eigen mode, $\Delta \omega_L$, lead to an observable signal on photodetector 1. The signal is then conditioned and fed back to the laser control system to stabilise its frequency to $\omega_0$.

At this step, the laser inherits the stability of the frequency standard, but the laser frequency should still be stabilised to the atomic transition, $\omega_A$, with a low frequency bandwidth, typically below $0.1$\,Hz. Since $\omega_A \neq \omega_0$, a frequency shifter is installed to change the laser frequency by $\omega_A - \omega_0$. The frequency shifting can be done with an acousto-optic modulator~\cite{yariv1989quantum}, which consists of a crystal, such as tellurium dioxide or quartz, and a piezoelectric transducer. An RF signal drives the transducer, generating an acoustic wave that modulates the refractive index of the crystal via the photoelastic effect. The process creates a diffraction grating. When the laser beam is diffracted from this moving grating, it experiences a Doppler shift, and the frequency of the first-order diffracted beam is shifted by the RF drive frequency, which is typically tens to hundreds of MHz. By adjusting the RF frequency, we can control the amount of frequency shift applied to the laser beam.

An error signal from an atomic transition for the stabilisation of the laser frequency can be obtained using direct absorption spectroscopy, where the laser frequency is tuned across the resonance and the transmitted optical power is monitored by photodetector 2. Near the atomic resonance, the transmission exhibits a sharp frequency-dependent change, and small deviations of the laser frequency produce measurable variations in the detected signal. This process is similar to observing the Fabry-Perot cavity resonance, considered above and shown in Fig.~\ref{fig:w8_FP_sig}. By operating on the side of the absorption line, the photodetector 2 measures how changes in laser frequency lead to a change in transmitted power. This provides an error signal that can be fed back to the acousto-optic modulator to stabilise the laser frequency to $\omega_A$.

Our laser frequency in transmission of the acousto-optic modulator inherits the stability of the atomic transition, $\omega_A$, and we need to reduce it for electronics systems while maintaining its stability. This can be achieved with a frequency comb produced by a fs-laser. In the frequency domain, the laser output shows many comb lines centered around the frequency $\omega_C$ and spaced by the frequency $\Omega_T$. As we will discuss in Lecture~\ref{Lecture10}, the spacing frequency is determined by the free spectral range of the fs-laser cavity and is typically in the MHz-GHz range. In the time domain, the fs-laser output looks like fs-pulses separated by a time step $2\pi/\omega_T$. The frequency comb can be produced with non-linear materials, as we will consider in Lecture~\ref{Lecture10}. The comb has two degrees of freedom, $\omega_C$ and $\omega_T$, and the optical frequency of the $n$-th line is given by the equation
\begin{equation}
\omega_n = \omega_C + n\omega_T,
\end{equation}
where $n$ can be positive or negative, and the stability of $\omega_C$ and $\omega_T$ is significantly worse than the stability of $\omega_A$ without active control.

The first control loop required to stabilise the frequency comb can be achieved via frequency doubling of the fs pulses. This process is implemented using a nonlinear crystal, as will be discussed in Lecture~\ref{Lecture11}. The nonlinear optical response of the crystal leads to a polarisation that depends nonlinearly on the electric field, giving rise to radiation at twice the input optical frequency. If the frequency comb is sufficiently broad such that its highest-frequency components overlap with the lowest-frequency components of the frequency-doubled spectrum, then the carrier-envelope offset frequency can be stabilised by detecting a beat signal on photodiode 4, according to the equation
\begin{equation}
\omega_C + q \omega_T = 2\omega_C - 2p\omega_T \hspace{5mm} => \hspace{5mm} \omega_T = \frac{\omega_C}{q + 2p},
\end{equation}
where $q$ and $p$ are the comb line numbers, whose frequencies are compared on photodetector 4.

The second control loop for the frequency comb is implemented by comparing $\omega_C$ and $\omega_A$ on photodetector 3. The resulting beat signal provides a measure of their frequency difference, which can be processed and fed back to the femtosecond laser to enforce the condition $\omega_C = \omega_A$. Together, these control loops enable the transfer of stability from the atomic reference frequency $\omega_A$ to the comb and ultimately to the output frequency $\omega_T$ according to the equation
\begin{equation}
\frac{\Delta \omega_A}{\omega_A} = \frac{\Delta \omega_C}{\omega_C} = \frac{\Delta \omega_T}{\omega_T},
\end{equation}
which can then be readily accessed and processed by electronic systems.

\subsubsection{Laser gyroscopes}

The next application we consider in this section is rotation sensing. Folded optical cavities, consisting of three or more mirrors, as shown in Fig.~\ref{fig:w8_app}, can function as laser gyroscopes when two counter-propagating beams circulate within the same resonator. In this configuration, rotation of the cavity induces a phase shift between the two beams via the Sagnac effect~\cite{saleh2019fundamentals}, given by the equation
\begin{equation}
    \Delta \phi = \frac{8 \pi A}{c \lambda} \Omega,
\end{equation}
where $A$ is the area enclosed by the resonator, $\Omega$ is the angular frequency of the gyroscope. 

State-of-the-art devices, such as the Ring G gyroscope with a 4\,m scale, are capable of measuring small variations of Earth rotation such as polar motion and fluctuations in the rotation axis. Recent advances~\cite{schreiber2011earth} have demonstrated that large-scale ring laser gyroscopes can resolve geophysical signals, opening the possibility of detecting subtle rotational effects and contributing to fields such as geodesy and fundamental physics.

\subsection{Optical coatings}

We now consider another important application of optical resonators: optical coatings, which are thin films deposited on substrates to tune their optical properties. For example, anti-reflective coatings are applied to glasses and camera lenses to maximise the transmission of light to the imaging system. Conversely, high-reflectivity coatings are used on mirrors to form high-finesse optical resonators. As discussed above, the performance of an optical cavity is determined by the transmission of its mirrors, which in turn is governed by the structure of the coating layers.

Optical coatings can be viewed as Fabry--Perot resonators at the microscopic scale, where multiple reflections occur at the interfaces between dielectric layers. When two dielectric materials, such as air and fused silica or fused silica and tantala, form a boundary, an incident electromagnetic wave is partially reflected and partially transmitted. This behaviour follows directly from Maxwell’s boundary conditions at the interface. If the refractive indices of the two media are $n_1$ and $n_2$, and the wave is incident from medium 1 onto medium 2, the electric and magnetic fields satisfy
\begin{equation}
    E_{\rm in} + E_r = E_t, \hspace{1cm}
    B_{\rm in} - B_r = B_t,
\end{equation}
\begin{equation}
    B_{\rm in} = \frac{n_1 E_{\rm in}}{c}, \hspace{1cm}
    B_r = \frac{n_1 E_r}{c}, \hspace{1cm}
    B_t = \frac{n_2 E_t}{c},
\end{equation}
from which it follows that
\begin{equation}
    \frac{E_t}{E_{\rm in}} = \frac{2n_1}{n_1 + n_2}, \hspace{1cm}
    \frac{E_r}{E_{\rm in}} = \frac{n_1 - n_2}{n_1 + n_2}.
\end{equation}
These expressions also show that the reflected field acquires a $\pi$ phase shift when light is reflected from a medium of higher refractive index, and no phase flip occurs for reflection from a lower-index medium. The results above are valid for both S- and P-polarised light at normal incidence. In general, however, the reflectivity is polarisation-dependent and is described by the Fresnel equations~\cite{hecht2017optics}
\begin{align}
r_s &= \frac{n_1 \cos\theta_i - n_2 \cos\theta_t}{n_1 \cos\theta_i + n_2 \cos\theta_t}, \hspace{1cm}
t_s = \frac{2 n_1 \cos\theta_i}{n_1 \cos\theta_i + n_2 \cos\theta_t}, \\[10pt]
r_p &= \frac{n_2 \cos\theta_i - n_1 \cos\theta_t}{n_2 \cos\theta_i + n_1 \cos\theta_t}, \hspace{1cm}
t_p = \frac{2 n_1 \cos\theta_i}{n_2 \cos\theta_i + n_1 \cos\theta_t},
\end{align}
where $\theta_i$ is the angle of incidence, $\theta_t$ is the refraction angle, and $r_{s,p}$ and $t_{s,p}$ are the field reflection and transmission coefficients for S- and P-polarised light, respectively.

\subsubsection{Anti-reflective coating}

\begin{figure}[t!]
\centering
\includegraphics[height=3.3cm]{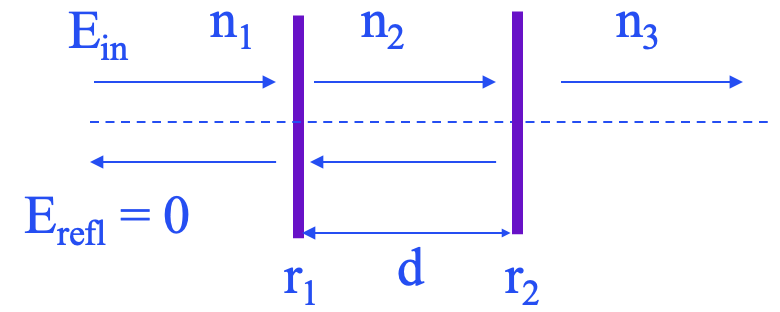} 
\hspace{3mm}
\includegraphics[height=3.3cm]{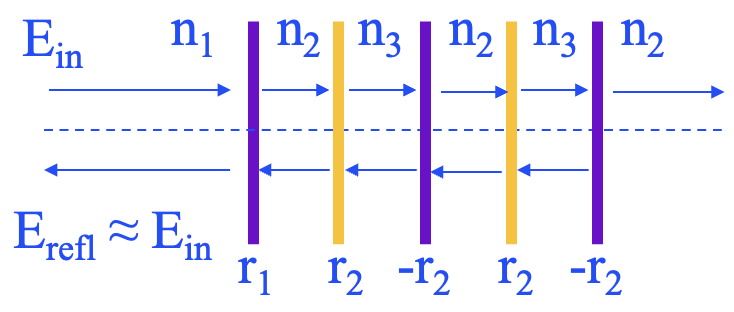} 
\caption{Examples of an anti-reflective (left) and high-reflective coatings (right).}
\label{fig:w8_coatings}
\end{figure}

As an example of a coating that transmits all light through an air-glass interface at a particular wavelength $\lambda$, we consider a dielectric layer of thickness $d$, index of refraction $n_2$, as shown in Fig.~\ref{fig:w8_coatings}. Our goal is to determine $d$ and $n_3$. A Fabry-Perot interferometer is formed by two boundaries between three dielectrics. The field reflectivities are given by the equations
\begin{equation}
    r_1 = \frac{n_2 - n_1}{n_1 + n_2} \hspace{1cm} r_2 = \frac{n_2 - n_3}{n_2 + n_3}.
\end{equation}
As we discussed above, the cavity transmits all incident power if $r_1^2 = r_2^2$. We can check that the case $r_1 = r_2$ does not have any solutions for $n_2$ because $n_1 \neq n_3$. The case $r_1 = -r_2$ leads to the solution $n_2 = \sqrt{n_1 n_3}$. We then set the cavity on resonance and compensate the minus sign in reflectivity by setting $\phi_{rt} = \pi$. This condition is achieved by tuning the thickness of the layer to $\lambda/(4 n_2)$. Therefore, we have an anti-reflective coating for a specific wavelength $\lambda$.

While a single-layer anti-reflective coating can eliminate reflection at a specific wavelength, achieving low reflection over a broad range of wavelengths requires multiple layers~\cite{macleod2010thin}. In such designs, dielectric layers with chosen refractive indices and thicknesses are stacked on the substrate. Each interface produces partial reflections, and by selecting the optical thickness, these reflections can interfere destructively over a wide spectral range rather than at a single wavelength. 

\subsubsection{High-reflective coatings}

To implement a high-reflectivity coating, one can deposit multiple alternating layers of materials with high and low refractive indices. For example, fused silica with $n_2 \approx 1.45$ can serve as the low-index material, while tantala (Ta$_2$O$_5$) with $n_3 \approx 2.1$ provides a high refractive index contrast. Each layer is typically designed to have an optical thickness of $\lambda/4$, such that the physical thicknesses are $\lambda/(4 n_2)$ for fused silica and $\lambda/(4 n_3)$ for tantala. This quarter-wave structure ensures that reflections from successive interfaces interfere constructively, leading to high overall reflectivity. To achieve a transmission as low as $T \sim 10^{-5}$, approximately $20$ bi-layers are required. Despite this large number of layers, the total coating remains thin, on the order of $\approx 6$\,um. This illustrates how thin-film coatings can modify optical properties while occupying only a microscopic thickness.

In addition to fused silica and tantala, other materials are used in optical coatings depending on the application and wavelength range. Common low-index materials include magnesium fluoride (MgF$_2$) and calcium fluoride (CaF$_2$), while high-index materials include titanium dioxide (TiO$_2$), hafnium dioxide (HfO$_2$), and niobium pentoxide (Nb$_2$O$_5$). The choice of coating materials is determined by refractive index contrast, optical absorption, mechanical stability, and thermal noise, which are particularly important in high-precision systems such as gravitational-wave detectors~\cite{harry2007thermal}.

%% file: week9.tex
\section{Communication: Copper cables, phone-to-satellite communication}
\label{Lecture9}

Communication refers to the transmission of information over distance by encoding it onto electromagnetic signals and recovering it at a receiver. This process involves technologies ranging from internet data transfer to mobile networks and satellite links. In classical systems, information is transmitted through copper cables as electrical voltages, which are robust but suffer from resistive losses and limited bandwidth over long distances, as discussed in this lecture. We also consider space-based communication, which enables information transfer without physical connections and does not rely on ground infrastructure, but is subject to significant free-space losses. By contrast, long-distance terrestrial communication typically relies on optical fibres, which will be discussed in Lecture~\ref{Lecture10}.

In this lecture, we discuss
\begin{itemize}
\item copper cables and the telegrapher's equations,
\item encoding of "0" and "1",
\item phone-to-satellite communication.
\end{itemize}

\subsection{Copper cables and the telegrapher's equations}

Telegraphs revolutionised communication in the 19th century by transmitting electrical signals over long distances~\cite{standage1998victorian}. A telegraph system consists of a transmitter, a receiver, and conducting wires forming a series circuit powered by a battery. The telegraph key acts as a manually operated switch: when pressed, it completes the circuit and allows an electrical current to flow; when released, it breaks the circuit and stops the current. At the receiving end, the electrical pulses activate an electromagnet, which drives a mechanical armature connected to a marking mechanism, typically an inked stylus or roller. This mechanism produces a sequence of marks on a moving paper strip, encoding the transmitted information in a time series of short and long pulses, which can then be decoded as text, for example, in Morse code.

Copper-based communication systems are still widely used, particularly in the form of coaxial cables~\cite{collin2001foundations}, which consist of a central conducting wire surrounded by a dielectric insulator and an outer shielding conductor, as shown in Fig.~\ref{fig:w9_coax}. This geometry protects against electromagnetic interference and allows transmission of high-frequency signals. Coaxial cables are commonly used for local network connections, cable television distribution, and antenna feed lines, where they connect antennas to receivers or transmit signals within a local area. Practical transmission lengths are limited to the range from about 100\,m up to 1\,km without amplification because of the resistive losses. 

\begin{figure}[t!]
\centering
\includegraphics[height=4cm]{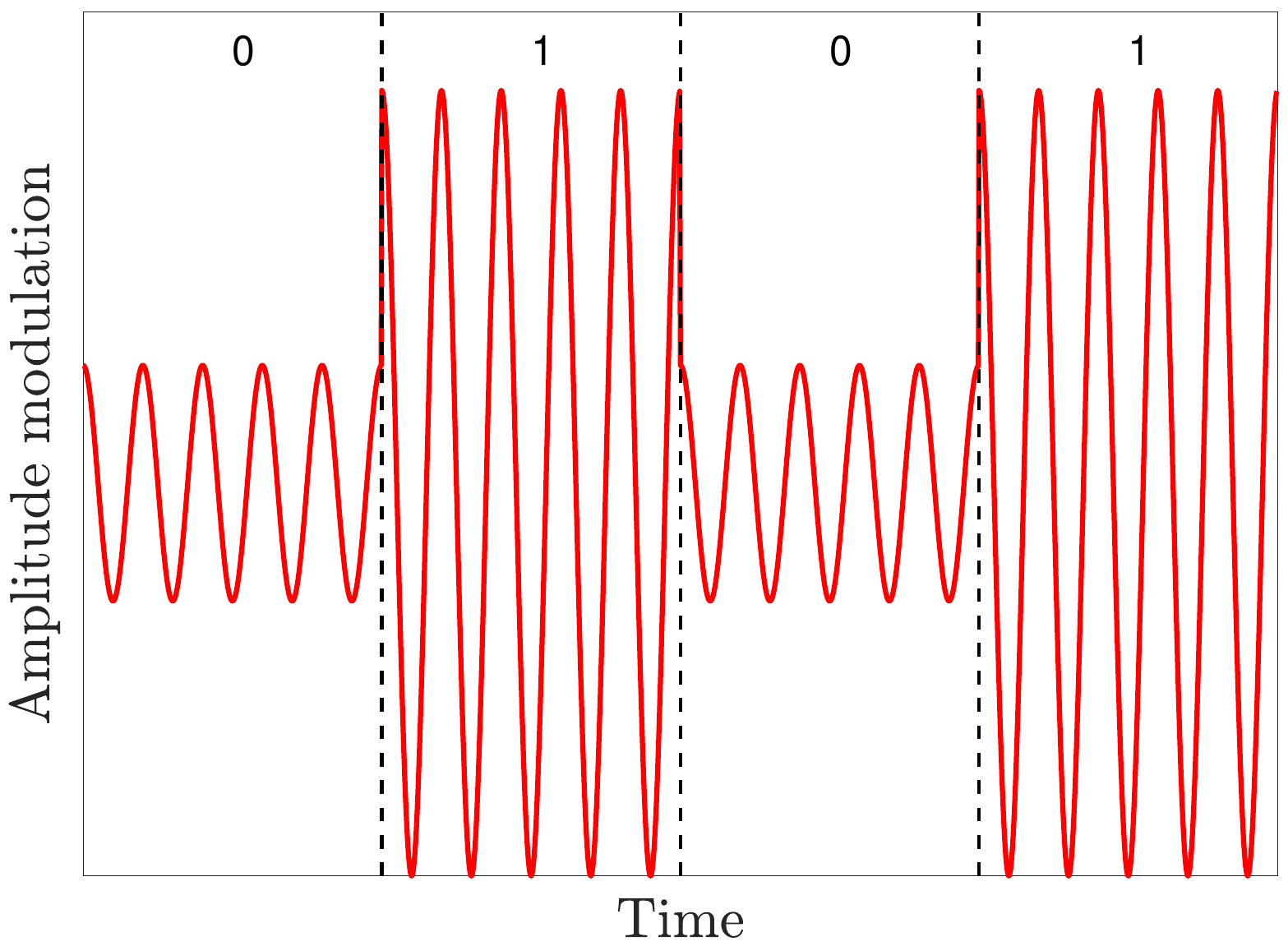} 
\hspace{1mm}
\includegraphics[height=4cm]{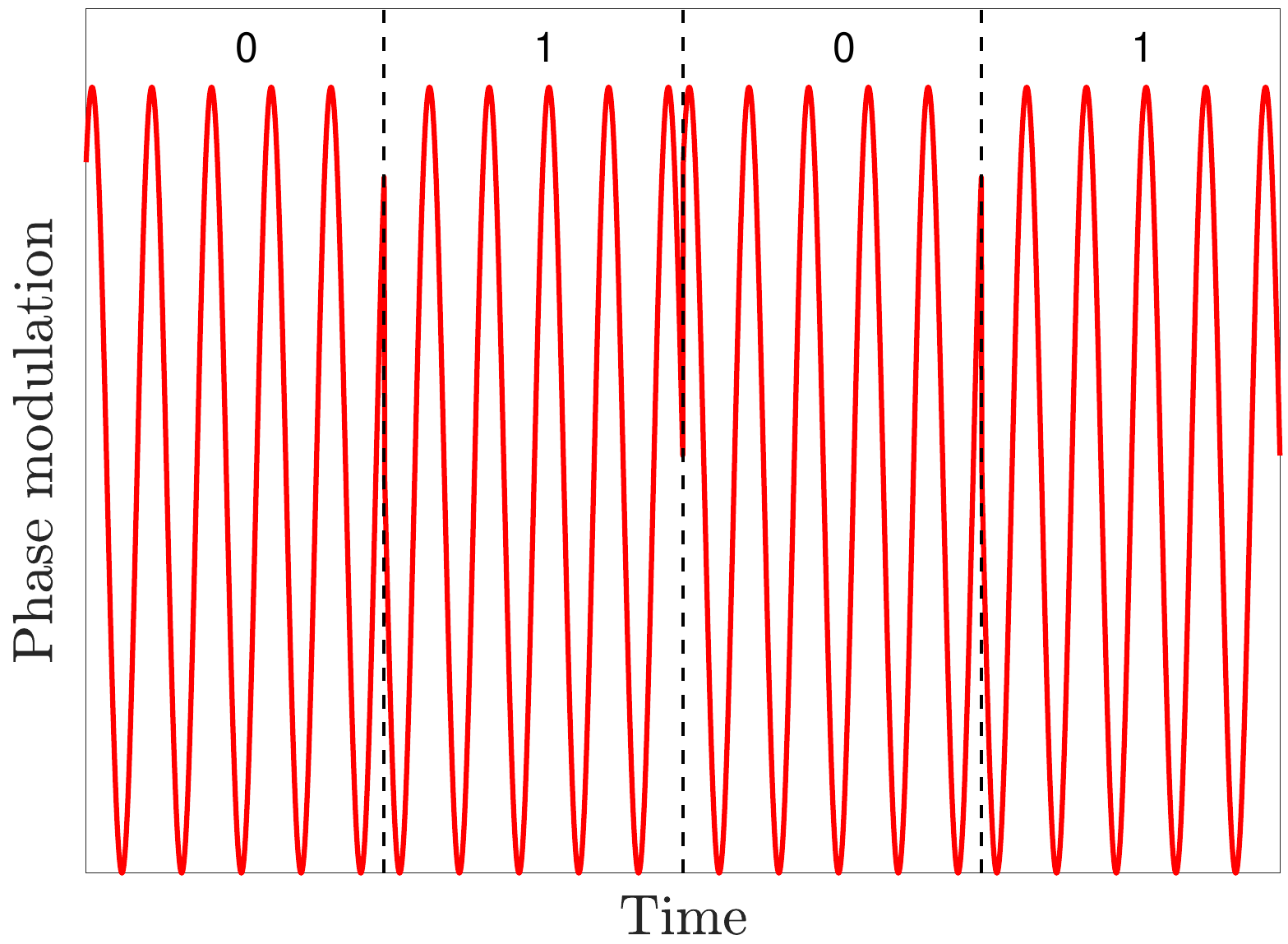} 
\hspace{1mm}
\includegraphics[height=4cm]{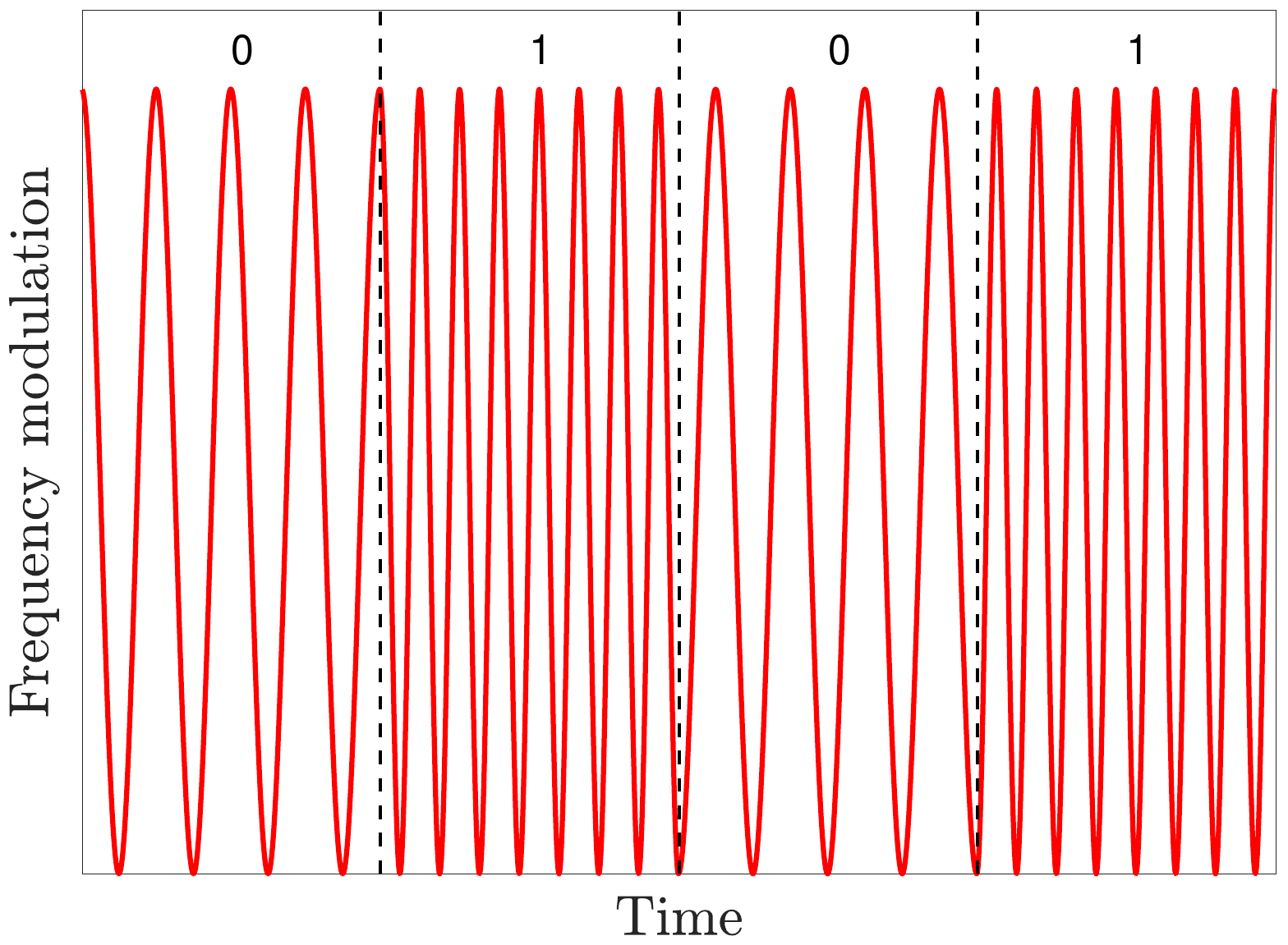} 
\caption{Examples of encoding bits with an analog signal: amplitude modulation (left), phase modulation (centre), and frequency modulation (right).}
\label{fig:w9_bits}
\end{figure}

Digital communication is based on encoding information into discrete binary values, typically “0” and “1”~\cite{proakis2007digital}. A “0” can be represented, for example, by a low voltage level (e.g., below $\sim 0.8$\,V DC), low-frequency, carrier phase shift, or absent oscillations, or in optical systems by negligible optical power. A “1” corresponds to a higher voltage level (e.g., above $\sim 2$\,V DC), higher-frequency electrical oscillation, an opposite phase shift to "0", or optical power above a defined detection threshold. Examples of the amplitude, phase, and frequency encoding of bits are shown in Fig.~\ref{fig:w9_bits}. The key advantage of discrete states over continuous amplitudes is their resilience to noise: moderate distortion cannot swap “1” and “0”. The trade-off is that digital encoding typically requires higher bandwidth compared to analog signals.

\subsubsection{Telegrapher's equation}

\begin{figure}[t!]
\centering
\includegraphics[height=2.5cm]{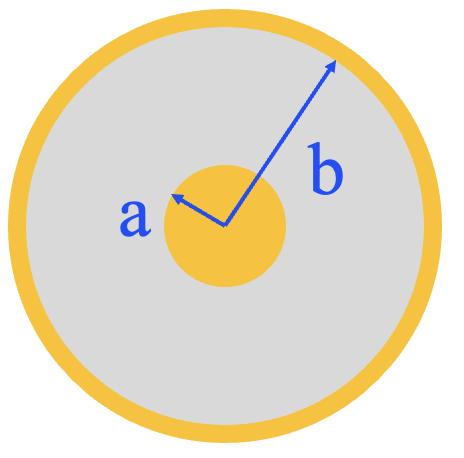} 
\hspace{5mm}
\includegraphics[height=3.2cm]{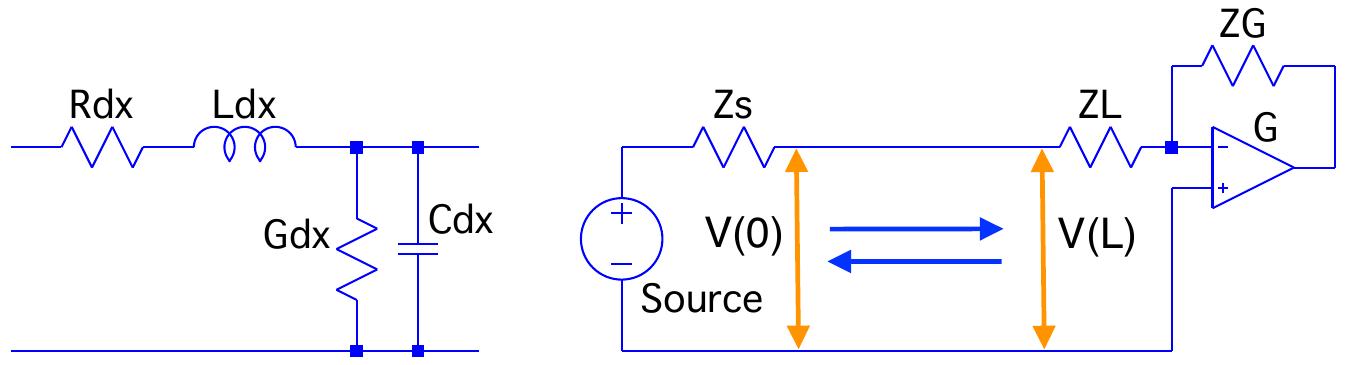}
\caption{Cross-section of a coaxial cable (left). An infinitesimal segment of the transmission line characterised by resistance, inductance, capacitance, and conductance per unit length (centre). Back-reflections in the transmission line arising from impedance mismatch at the load and source (right). The transmitted signal is amplified with a gain given by the ratio $-Z_G / Z_L$.}
\label{fig:w9_coax}
\end{figure}

The telegrapher’s equations describe how voltage and current propagate along a transmission line, such as a coaxial cable, taking into account the distributed resistance ($R$), inductance ($L$), capacitance ($C$), and conductance ($G$), as shown in Fig.~\ref{fig:w9_coax}. The equations show that electrical signals propagate as damped waves with finite speed and attenuation. Considering an infinitesimal segment of the transmission line and applying Kirchhoff’s laws yields
\begin{equation}
\frac{dI}{dx} = -V(i\omega C + G) \hspace{1cm} \frac{dV}{dx} = -I(R + i\omega L),
\end{equation}
where $\omega$ is the angular frequency of the transmitted signal, $I(x)$ and $V(x)$ denote the position-dependent current and voltage along the line. The circuit parameters of a cylindrical coaxial cable are given by the following equations
\begin{equation}\label{eq:w9_RLCG}
R = \frac{\rho}{A}, \hspace{1cm} L = \frac{\mu \mu_0}{2 \pi} \ln{\frac{b}{a}}, \hspace{1cm} C = \frac{2\pi \epsilon \epsilon_0}{\ln{\frac{b}{a}}}, \hspace{1cm} G = \omega C\tan\delta,
\end{equation}
where $\rho$ is the resistivity of the conductor, $A$ is its cross-sectional area, $b$ and $a$ denote the radii of the outer and inner conductors. The quantities $\mu$ and $\epsilon$ are the relative permeability and permittivity of the dielectric medium, and $\delta$ is the dielectric loss angle.

In coaxial cables, the dielectric material between the inner and outer conductors is chosen to combine low loss, mechanical stability, and desired electrical properties. Common materials include polyethylene, cross-linked polyethylene, polytetrafluoroethylene (Teflon), and foamed polyethylene. In most practical coaxial cables, the dielectric is non-magnetic, $\mu \approx 1$, and the relative permittivity is typically $\epsilon \approx 2$. Real dielectrics exhibit losses due to microscopic lag in molecular polarisation and small amounts of ionic conduction, which are captured by the loss tangent. This parameter quantifies the phase lag between the electric field and polarisation response. The loss is typically small, $\delta \sim 10^{-3}$ or below, and allows efficient transmission over 100\,m with limited signal degradation.

\subsubsection{Signal reflections}

For simplicity, we first consider wave propagation via the transmission line in the lossless case, $\rho = 0$, $\delta = 0$. The solutions to the telegrapher's equations are
\begin{equation}\label{eq:w9_lossless_sol}
\begin{split}
& V(x,t) = V_1 e^{i\omega t} e^{-i k x} + V_2 e^{i\omega t} e^{i k x} \\
& I(x, t) = \frac{V_1}{Z_0} e^{i\omega t} e^{-i k x} - \frac{V_2}{Z_0} e^{i\omega t} e^{i k x},
\end{split}
\end{equation}
where $k=\omega\sqrt{LC}$ is the wave number, and $Z_0 = \sqrt{\frac{L}{C}}$ is the characteristic impedance of the line. In coaxial cables, the most common standard values are $50$\,Ohms and $75$\,Ohms~\cite{pozar2011microwave}. The former cables are widely used in RF and microwave systems, laboratory instrumentation, and radio communications because they offer a good compromise between low loss and high power-handling capability. In contrast, the latter cables are optimised for lower attenuation and are commonly used in video transmission, cable television, and broadband internet distribution. The group velocity of the wave is given by the equation
\begin{equation}
v_{\rm gr} = \frac{d\omega}{dk} = \frac{1}{\sqrt{LC}} = \frac{1}{\sqrt{\epsilon \epsilon_0 \mu \mu_0}} = \frac{c}{\sqrt{\epsilon}} \approx \frac{c}{1.5}.
\end{equation}

Eq.~\ref{eq:w9_lossless_sol} shows two travelling waves: in the positive and negative directions of the X-axis. If we consider a transmission line of length $L$ loaded by an impedance $Z_L$ and powered by the source with impedance $Z_s$ then the boundary conditions are given by the equations
\begin{equation}
V(L) = I(L) Z_L \hspace{1cm} I(0) Z_s = V_{\rm in} - V(0),
\end{equation}
where $V_{\rm in}$ is the source voltage. The amplitudes of the forward and backward propagating waves are then given by the equations
\begin{equation}\label{eq:w9_V12}
\begin{split}
&V_1= V_{\rm in} \frac{\frac{Z_0}{Z_0 + Z_s}}{1 - \frac{Z_s - Z_0}{Z_s + Z_0} \frac{Z_L - Z_0}{Z_L + Z_0}e^{-2i k l}} = V_{\rm in} \frac{t_1}{1-r_1 r_2 e^{-2i k l}} \\
&V_2 = V_1 e^{-2i k x} \frac{Z_L - Z_0}{Z_L + Z_0} = r_2 V_1 e^{-2i k l},
\end{split}
\end{equation}
where we introduced field reflectivity and transmissivity coefficients $r_1 = \frac{Z_s - Z_0}{Z_s + Z_0}$,  $r_2 = \frac{Z_L - Z_0}{Z_L + Z_0}$, and $t_1 = \frac{Z_0}{Z_0 + Z_s}$. Eq.~\ref{eq:w9_V12} is formally identical to the field propagation in a Fabry–Perot interferometer, as discussed in Lecture~\ref{Lecture8}. In this analogy, the source and load impedances play the role of partially reflecting mirrors, and may lead to multiple reflections and interference of the propagating wave along the transmission line. We can eliminate back-reflections by impedance matching, choosing the source and load impedances such that $R_L = Z_0 = R_s$. In this case, the reflection coefficients vanish, $r_1 = r_2 = 0$, and the transmission line is perfectly matched, enabling maximum power transfer from the source into the line.

\subsubsection{Lossy transmission line}

In the case $\phi \neq 0$ and $\delta \neq 0$, the wave equation is 
\begin{equation}
V'' + V(\omega^2 CL - GR -i\omega(LG + GR)) = 0
\end{equation}
and the signal is attenuated in the transmission line according to the equation
\begin{equation}
V(x,t) = V_1 e^{i\omega t} e^{-i k x} \exp(-\alpha x),
\end{equation}
where the attenuation coefficient $\alpha$ is given by the equation
\begin{equation}
\alpha = \frac{1}{2}\left(G Z_0 + \frac{R}{Z_0} \right).
\end{equation}

The attenuation of signals in transmission lines increases with frequency due to two main effects: the skin effect and dielectric losses. As frequency rises, alternating currents in conductors are confined to an increasingly thin surface layer known as the skin depth~\cite{jackson1999classical}, given by the equation
\begin{equation}
\sigma = \sqrt{\frac{2 \rho}{\omega \mu_0}}.
\end{equation}
The skin effect reduces the effective cross-sectional area available for current flow, and increases the effective resistance per unit length, $R$, of the conductor. In addition, the dielectric contribution to losses also grows with frequency: the effective conductance is given by Eq.~\ref{eq:w9_RLCG}. Higher frequency increases energy dissipation in the insulating material due to lagging polarisation response. Together, the increase in resistive losses from skin effect and the frequency-dependent dielectric heating limit the frequencies of the propagating signals to $\sim 1$\,GHz.

\subsection{Frequency multiplexing and Shannon theorem}

Frequency multiplexing in coaxial cables allows a single physical transmission line to be shared by many independent users by assigning each user a separate frequency band within the available spectrum. In practical cable systems, the total available bandwidth is divided into channels of typical width $B\approx 8$\,MHz, which are allocated sequentially to different users. For example, User 1 may be assigned $800–808$\,MHz, User 2 $808–816$\,MHz, and so on, up to User 25 occupying $992–1000$\,MHz.

For a particular allocated bandwidth $B$, no matter how cleverly information is encoded or compressed, it is impossible to exceed the theoretical upper bound on the communication rate~\cite{shannon1948mathematical} given by the Shannon theorem:
\begin{equation}
C = B \log_2\left(1 + \frac{P_L}{P_n} \right),
\end{equation}
where $P_L$ is the signal power on the load, $P_n$ is the noise power, and $C$ in bits/s is the maximum communication rate achievable in the presence of noise. Shannon capacity represents a fundamental ceiling on communication performance. In practice, real communication systems operate below this limit and typically require a signal-to-noise ratio greater than $\approx 10$ to ensure reliable decoding with manageable error rates.
For example, the maximum communication rate over a 100-m-long coaxial cable is $\sim 300$\,Mb/s if the source power is 10\,W, the channel bandwidth is 8\,MHz, and the detector noise power is 0.1\,pW.

\begin{figure}[t!]
\centering
\includegraphics[height=6cm]{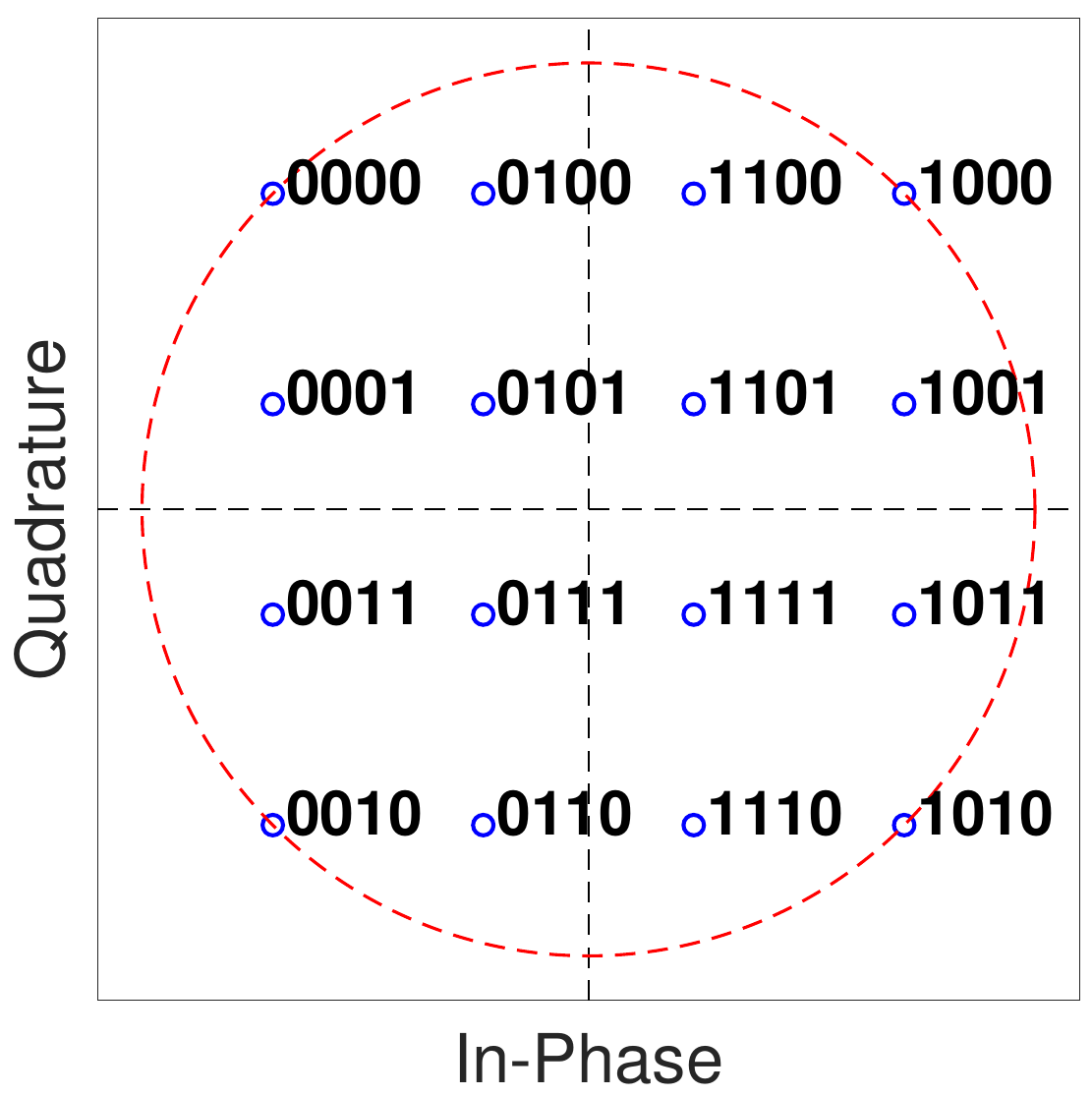} 
\hspace{5mm}
\includegraphics[height=6cm]{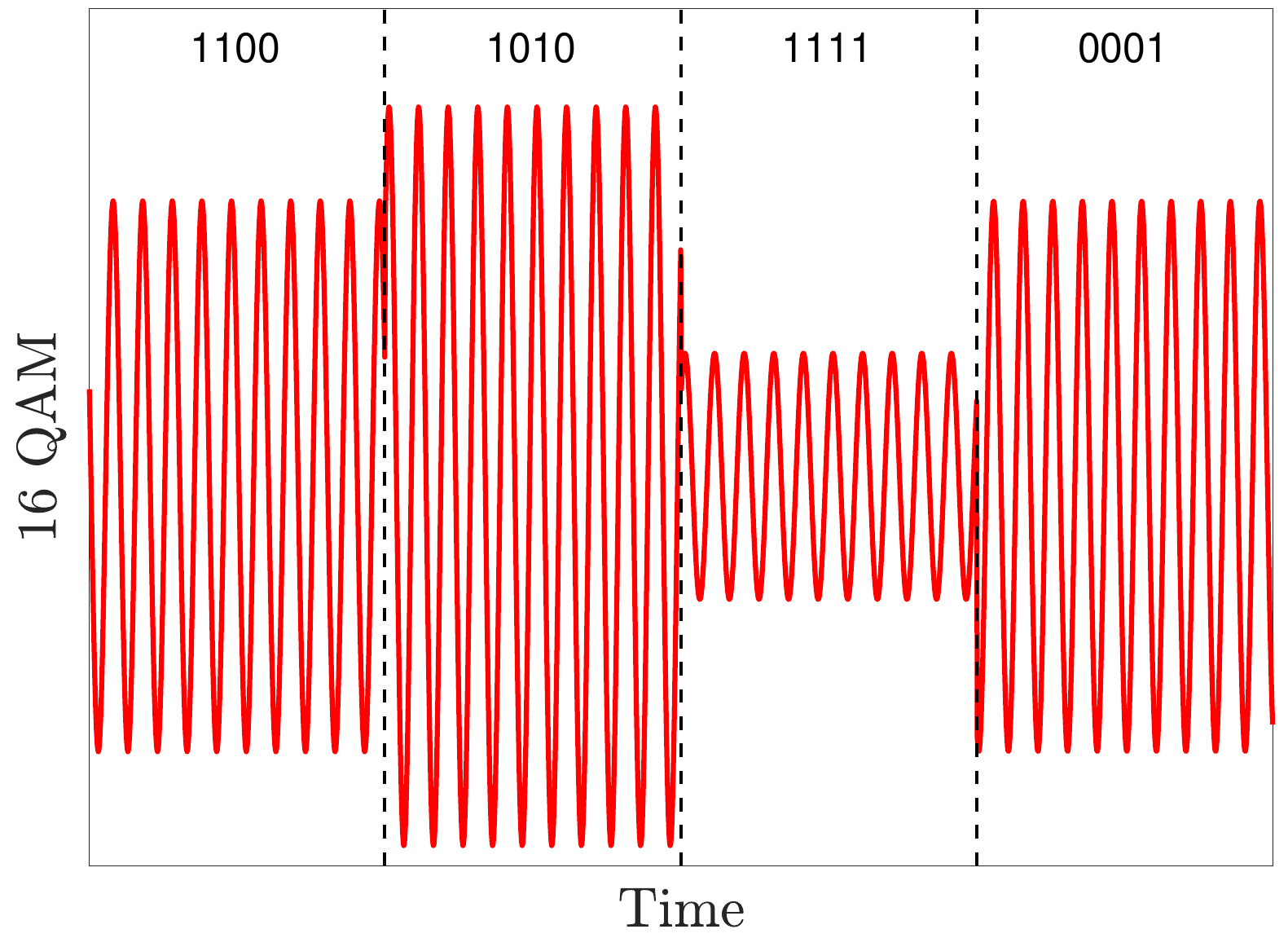} 
\caption{Quadrature Amplitude Modulation (16-QAM): constellation diagram showing bit mapping, where the circle radius corresponds to the maximum signal amplitude, the amplitude of each symbol is given by the magnitude of the vector from the origin, and the phase is given by the angle of that vector relative to the $x$-axis (left), together with an example of a signal encoded using 16-QAM (right).}
\label{fig:w9_qam}
\end{figure}

Modern encoding schemes approach the Shannon limit. For example, Quadrature Amplitude Modulation (QAM) is more efficient than using only amplitude, phase, or frequency modulation, shown in Fig.~\ref{fig:w9_bits}, because QAM combines both amplitude and phase variations to encode information in a two-dimensional signal space. The scheme packs more bits per symbol than single-parameter schemes, as shown in Fig.~\ref{fig:w9_qam}. This allows QAM to achieve high spectral efficiency and makes it the dominant scheme in high-performance communication systems such as Wi-Fi, LTE, and 5G.

\subsection{Cellphone-to-satellite communication}

An emerging technology is the cellphone-to-satellite communication, which enables mobile devices to connect directly to satellites without the need for additional ground-based infrastructure, and offers several important advantages. A key benefit is global coverage, allowing connectivity in remote regions. It also requires no extra hardware, such as external terminals, and provides improved resilience in cases when ground networks may be absent. However, this technology also faces significant challenges. The received signals typically have a low signal-to-noise ratio due to large transmission distances and limited antenna sizes in mobile devices. As a result, the system has limited communication capacity and relatively low uplink data rates compared to terrestrial networks. In addition, maintaining a stable satellite link from a phone requires higher energy consumption, as the phone must transmit at higher power to overcome path losses.

The power transmission in free space is governed by the Friis equation~\cite{friis1946note}
\begin{equation}
P_L = P_{\rm in} \frac{G_1 A_{\rm eff}}{4\pi d^2} = P_{\rm in} G_1 G_2 \left( \frac{\lambda}{4\pi d} \right)^2,
\end{equation}
where $G_1$ and $G_2$ are the gains of the transmission and receiving antennas, $A_{\rm eff}$ is the effective area of the receiving antenna, $d$ is the separation between the antennas. The Friis transmission equation reflects how electromagnetic waves spread in free space: as a transmitter radiates power, the energy is distributed over the surface of an expanding sphere whose area grows as $4\pi d^2$.

Antenna gain quantifies how effectively an antenna radiates or receives power in a given direction compared to an ideal isotropic radiator. The gain is closely related to directivity, which describes how strongly the radiation is concentrated in a particular direction, as we discussed in Lecture~\ref{Lecture2}. Typical cellphone antenna gain is $G \approx 1$ because mobile phone antennas are omnidirectional. Since a phone can be held in any position and must communicate reliably with base stations from different directions, its antenna cannot focus power into a narrow beam. In addition, size constraints and proximity to the user’s hand and body introduce losses, further limiting achievable gain.

Satellite antenna gains are typically very high~\cite{maral2011satellite}, $G \sim 10^2-10^5$, because satellites use directional antennas, such as parabolic reflectors or phased arrays, to focus electromagnetic energy into narrow beams. The large physical size of the antenna relative to the wavelength allows it to achieve high directivity. Such a gain is necessary to compensate for the large free-space path loss over distances of hundreds of kilometres, and ensure sufficient signal strength at the receiver.

Similar to coaxial cables, Shannon’s theorem provides a guideline for the achievable data rate in satellite–cellphone links by relating the channel capacity. For typical systems operating around 2\,GHz with a per-user bandwidth of about 1\,MHz, the achievable rate is primarily limited by the signal-to-noise ratio. Although the Friis equation is symmetric with respect to antenna gains, meaning that propagation losses are the same in both directions, the transmit powers are asymmetric: a satellite can transmit with much higher power than a cellphone. As a result, the downlink (satellite to phone) achieves a significantly higher power at the receiver than the uplink (phone to satellite). Consequently, even though the channel bandwidth is similar, the uplink data rate is an order of magnitude or more lower than the downlink rate.

%% file: week10.tex
\section{Communication: Optical fibres}
\label{Lecture10}

Following Lecture~\ref{Lecture9}, we continue our discussion of communication technologies. High-performance long-distance communication relies on optical fibres, where information is encoded onto laser light. It enables a low-loss transmission over thousands of kilometres with high data rates. At the source, lasers provide the coherent light required for dense wavelength multiplexing, where many signals are transmitted simultaneously at different wavelengths. Additional technologies, such as optical amplifiers, modulators, and photodetectors, enable encoding, boosting, and decoding of signals, making optical communication systems a key technology of the internet.

In this lecture, we discuss
\begin{itemize}
\item optical modes in 2D waveguides,
\item modal dispersion,
\item chromatic dispersion,
\item polarisation dispersion, 
\item signal attenuation in fibres,
\item frequency multiplexing,
\item frequency comb generation.
\end{itemize}

\subsection{Optical modes in fibres}

\begin{wrapfigure}{r}{0.4\textwidth}
    \centering
    \vspace{-3mm}
    \includegraphics[width=0.38\textwidth]{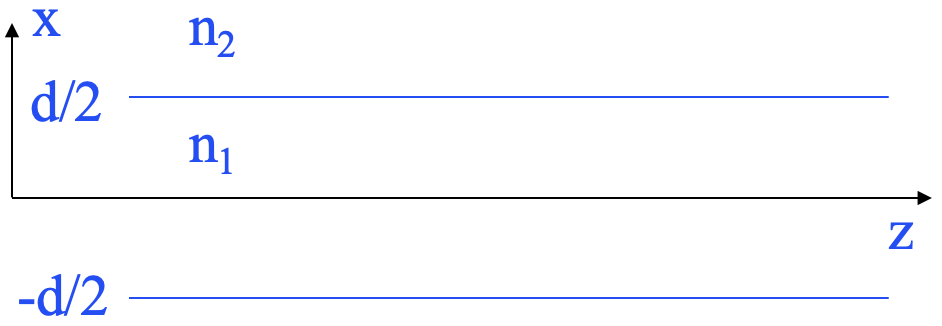} 
    \caption{Diagram of a 2D waveguide of width $d$. The beam propagates along the Z-axis.}
    \label{fig:w10_WG}
\end{wrapfigure}

In the geometrical optics model, a light ray can undergo total internal reflection at the interface between two dielectric media with refractive indices $n_1$ and $n_2$, provided $n_1 > n_2$. In this lecture, we consider a planar dielectric waveguide formed by two interfaces~\cite{yariv2007photonics}, where the refractive index is $n_2$ in the cladding regions $x > d/2$ and $x<-d/2$ and $n_1$ in the core region $-d/2 \leq x \leq d/2$. The structure is translationally invariant along the Y- and 
Z-axes. Within the geometrical optics picture, if the angle of incidence exceeds the critical angle for total internal reflection, the ray in the XZ-plane is confined to the core and undergoes repeated reflections at the interfaces, resulting in net propagation along the Z-direction.

We now consider a wave description of beam propagation in dielectric waveguides. Starting from Maxwell’s equations, we derive the wave equation and apply it to 2D optical waveguides. This reduces the problem by eliminating any dependence on the Y-coordinate. Further simplification is obtained by considering transverse electric (TE) modes, in which only the Y-component of the electric field is non-zero, while the magnetic field has both 
X- and Z-components. Under these assumptions, we seek solutions to the wave equation in the form
\begin{equation}
\begin{split}
    \VF{E} &= E_y(x) \VF{e_y} \exp(i(\omega t - \beta z))  \\
    \VF{H} &= (H_x(x) \VF{e_x} + H_z(x) \VF{e_z}) \exp(i(\omega t - \beta z)),
\end{split}
\end{equation}
where $\beta$ is the Z-component of the wave vector. From the wave equation, we get equations for the electric field in the core and cladding according to
\begin{equation}\label{eq:l11_wave_E}
    \frac{\partial^2 E_y}{\partial x^2} + (k_0^2 n_{1,2}^2 - \beta^2)E_y = 0,
\end{equation}
where $k_0 = \omega/c$ is the magnitude of the wave vector in a vacuum and index 1 corresponds to the wave equation in the core, $-d/2 \leq x \leq d/2$, where the index of refraction is $n_1$, and index 2 corresponds to the wave equation in the cladding, where the index of refraction is $n_2$.

By examining Eq.~(\ref{eq:l11_wave_E}), we find that the equation has the form $E_y'' + w_{1,2} E_y = 0$ which are second order equations. If $w_{1,2} = k_0^2 n_{1,2}^2 - \beta^2 > 0$ then the solution are trigonometric functions. If $w_{1,2} < 0$, then the solutions are exponential functions. Quantitatively, $w_1$ must be larger than zero to satisfy the boundary and symmetry conditions, as discussed below. At the same time, $w_2<0$ in the lower-index cladding ensures that the field decays exponentially away from the core. This exponentially decaying tail is known as the evanescent field, and it is responsible for the confinement of the mode within the waveguide.

We search for the solutions to Eq.~(\ref{eq:l11_wave_E}) in the form
\begin{equation}\label{eq:l11_sol}
    \begin{split}
        E_y(x) &= A \sin(hx) + B \cos(hx), \hspace{1cm} -d/2 \leq x \leq d/2 \\
        E_y(x) &= C e^{-qx}, \hspace{3.7cm} d/2 \leq x \\
        E_y(x) &= D e^{qx}, \hspace{3.9cm} -d/2 \geq x,
    \end{split}
\end{equation}
and find $h$ and $q$ by substituting the above expressions to Eq.~(\ref{eq:l11_wave_E}):
\begin{equation}
    h = \sqrt{k_0^2 n_{1}^2 - \beta^2} \hspace{1cm}  q = \sqrt{\beta^2 - k_0^2 n_{2}^2}.
\end{equation}

Both $h$ and $q$ depend on the wave vector component along the waveguide, $\beta$. We can exclude $\beta$ from the $V$-parameter given by the equation
\begin{equation}
    V^2 = u^2 + v^2 = \frac{d^2}{4}(h^2 + q^2) = \left( \frac{\pi d}{\lambda} \right)^2 (n_1^2-n_2^2),
\end{equation}
where $V$ depends on the waveguide properties, such as indices of refraction and size of the core, and the wavelength, but does not depend on the Z-component of the wave vector. Dimensionless quantities $u \equiv hd/2$ and $v \equiv qd/2$ are introduced for simplicity.

We now apply the Maxwell boundary conditions at the core-cladding interfaces. Since the electric and magnetic fields must be continuous at the boundaries, we get a set of equations
\begin{equation}
\begin{split}
x=\frac{d}{2}:\hspace{1cm} & A\sin u + B\cos u = C e^{-v} \\
& uA\cos u - u B\sin u = -v C e^{-v} \\
x=-\frac{d}{2}:\hspace{1cm} & -A\sin u + B\cos u = D e^{-v} \\
& uA\cos u + u B\sin u = v D e^{-v}.
\end{split}
\end{equation}

The waveguide structure is symmetric about $x=0$, meaning the refractive index satisfies $n(x)=n(-x)$. As a result, the wave equation is invariant under the transformation $x \to -x$, and its solutions can be chosen to have definite parity: either symmetric or antisymmetric. In the core, the general solution $E_y(x) = A\sin(hx) + B\cos(hx)$ contains both odd and even components. Enforcing symmetry requires that the field be either purely even or purely odd, which is only possible if one of the coefficients vanishes. Thus, symmetric modes correspond to $A=0$ (cosine solutions), and antisymmetric modes correspond to $B=0$ (sine solutions). From the boundary conditions, we find the solutions for the symmetric and antisymmetric modes:
\begin{equation}\label{eq:l11_uv}
\begin{split}
    &A = 0, \hspace{1cm} C=D, \hspace{1cm} u \tan(u) = v \\
    &B = 0, \hspace{1cm} C=-D, \hspace{1cm} u \cot(u) = -v.
\end{split}
\end{equation}

\begin{figure}[t!]
\centering
\includegraphics[height=9cm]{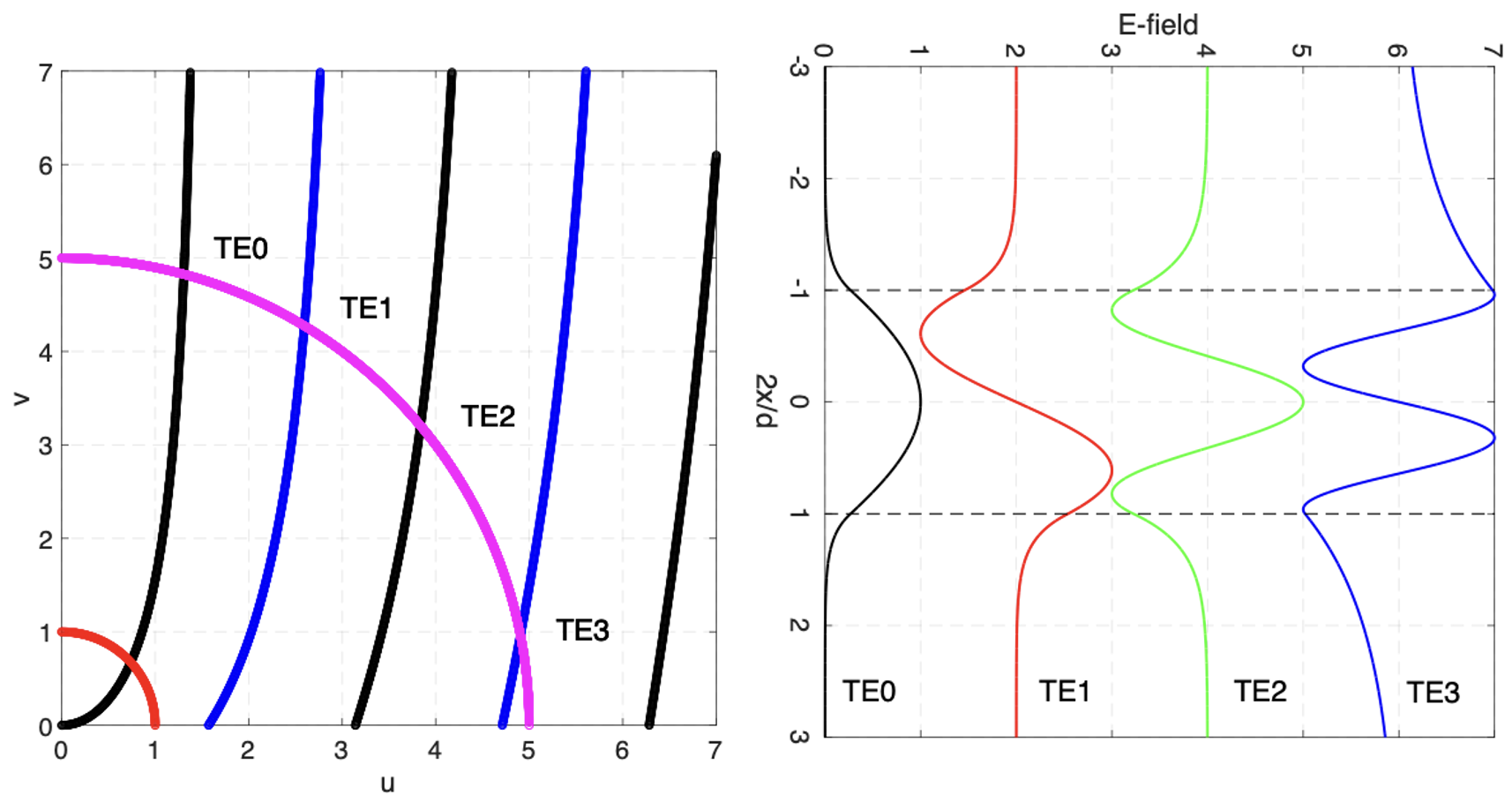} 
\caption{Solutions to the Maxwell's wave equation in a 2D waveguide for $V=1$ and $V=5$ (left). Profiles of the modes in the waveguide (right). The fields are plotted with an offset along the X-axis for clarity.}
\label{fig:w10_2D_sol}
\end{figure}

We can then find solutions to equations $u \tan(u) = v$ and $u \cot(u) = -v$ with a constraint that $u^2 + v^2 = V^2$. It is not feasible to solve these equations analytically, but we can approximately determine the solutions graphically, as shown in Fig.~\ref{fig:w10_2D_sol}. The number of solutions depends on $V$: the larger $V$, the more solutions (or modes) we have in the waveguide. In particular, the waveguide supports only one transverse electric mode if the V-parameter satisfies the inequality
\begin{equation}
    V < \frac{\pi}{2}
\end{equation}
and constrains the diameter of the core to $d<4.3$\,um for waveguide parameters $n_1=1.6$, $n_2=1.59$, and wavelength $\lambda=1550$\,nm. The waveguide supports $N$ modes if the V-parameter satisfies the inequality
\begin{equation}
(N-1)\frac{\pi}{2} \leq V < N\frac{\pi}{2}
\end{equation}
and since $V \sim \lambda^{-1}$, the same waveguide may be single-mode for one wavelength and support several modes for a shorter wavelength.

The communication signal propagates in the waveguide with a group velocity, which we can find from the dispersion relationship
\begin{equation}
    \frac{\omega^2 n_1^2}{c^2} = \beta_i^2 + \left( \frac{2 u_i}{d} \right)^2.
\end{equation}
The group velocity is given by the equation
\begin{equation}
    v_{\rm gr,i} = \frac{d\omega}{d\beta_i} = \frac{c}{n_1^2}n_{\rm eff,i},
\end{equation}
where we introduced an effective index of refraction for a particular mode $i$ according to the equation
\begin{equation}
    n_{\rm eff, i} = \frac{\beta}{k_0} = \sqrt{n_1^2 - \left( \frac{2 u_i}{k_0 d} \right)^2 },
\end{equation}
where $u_i$ are solutions to Eq.~(\ref{eq:l11_uv}). All effective indices of refraction satisfy the inequality $n_2 < n_{\rm eff,i} < n_1$ and show that different modes propagate with different group velocities in the waveguide. This property is the key to modal dispersion and limits the communication bandwidth in waveguides.

\subsection{Modal dispersion}

Modal dispersion is undesirable in fibre communication because different modes propagate with different group velocities in the waveguide~\cite{yariv2007photonics}. If a transmitted pulse excites multiple modes, each mode arrives at the fibre output at a different time, leading to temporal broadening of the signal. As a result, even a very short input pulse spreads as it propagates. The resulting pulse width at the output can be estimated as
\begin{equation}
    \Delta \tau = t_{\rm n} - t_0 =  \frac{L}{v_{\rm gr,q}} -  \frac{L}{v_{\rm gr,0}} = \frac{Ln_1^2}{c}\left( \frac{1}{n_{\rm eff, q}} - \frac{1}{n_{\rm eff, 0}}\right),
\end{equation}
where $n_{\rm eff,q}$ is the effective refractive index of the highest-order mode. The fundamental (lowest-order) transverse electric mode propagates fastest, and higher-order modes travel more slowly, leading to the overall pulse dispersion.

As an example, consider an optical fibre with a core index of refraction $n_1 = 1.6$, a length $L=100$\,km and a width $d=10$\,um can support three transverse electric modes $i=0,1,2$ with solutions $u_0 = 1.1$, $u_1 = 2.4$, and $u_2 = 3.2$ at a wavelength of $\lambda = 0.5$\,um. The delay in arrival times between the fastest and slowest modes is given by the equation
\begin{equation}
    \tau = t_2 - t_0 = \frac{L}{v_{\rm gr,2}} - \frac{L}{v_{\rm gr,1}} \approx \frac{L}{c}\frac{2(u_2^2 - u_0^2)}{n_1 (d k_0)^2} =  238\,{\rm nsec},
\end{equation}
which limits the communication bandwidth to $1/\tau = 4.2$\,Mb/s.

\subsection{Chromatic dispersion}

Modal dispersion limits the communication bandwidth to $\sim 10$\,Mb/s in multi-mode fibres. This limitation can be overcome by using single-mode fibres, which support only the fundamental mode and therefore eliminate modal dispersion. However, single-mode fibres are typically more expensive to manufacture, as their core diameter is only a few micrometres, requiring high precision, whereas multimode fibres have much larger cores ($\sim 100$\,um).

In single-mode fibres, pulse broadening still occurs due to chromatic dispersion, which arises from the wavelength dependence of the refractive indices $n_1$ and $n_2$. As discussed in Lecture~\ref{Lecture5}, the optical properties of materials vary with wavelength, so different frequency components of a pulse propagate with different group velocities, leading to temporal spreading of the signal.

Chromatic dispersion persists even when using nominally monochromatic light because any finite-duration pulse must contain a range of frequencies, regardless of how narrow the laser linewidth is. This is a direct consequence of the time–frequency uncertainty principle: shaping a signal in time (for example, to encode information in pulses) necessarily introduces a finite spectral bandwidth. These different frequency components propagate with slightly different group velocities due to the wavelength dependence of the refractive index, leading to temporal broadening of the pulse. In practice, this modulation-induced bandwidth is typically much larger than the intrinsic linewidth of the laser, and therefore dominates the effect of chromatic dispersion in communication systems. This phenomenon can be understood using the time–frequency uncertainty principle,
\begin{equation}
    \Delta \omega \tau \geq \frac{1}{2},
\end{equation}
where $\tau$ denotes the measurement time or pulse duration and $\Delta \omega$ is the uncertainty of the pulse angular frequency.

The pulse dispersion can be evaluated by taking the Fourier transform of the pulse and determining $\Delta \omega$ as the full width at half maximum (FWHM) of its spectral peak. This can then be converted to a wavelength bandwidth using
\begin{equation}
    \Delta \lambda = \frac{\lambda}{\omega}\Delta \omega,
\end{equation}
allowing one to incorporate the dispersive properties of the medium. The resulting temporal broadening of the pulse after propagation through a fibre of length $L$ is given by
\begin{equation}
    b = a + D\,\Delta \lambda\, L,
\end{equation}
where $b$ is the final pulse width, $a$ is the initial width, and $D$ is the group-velocity dispersion coefficient, typically expressed in units of $\mathrm{ps/(nm\cdot km)}$. The chromatic dispersion is significantly smaller than the model dispersion, and single-mode fibres can achieve a communication bandwidth of $\sim 30$\,Gb/s over a 1\,km fibre link.

\begin{figure}[t!]
\centering
\includegraphics[height=4.2cm]{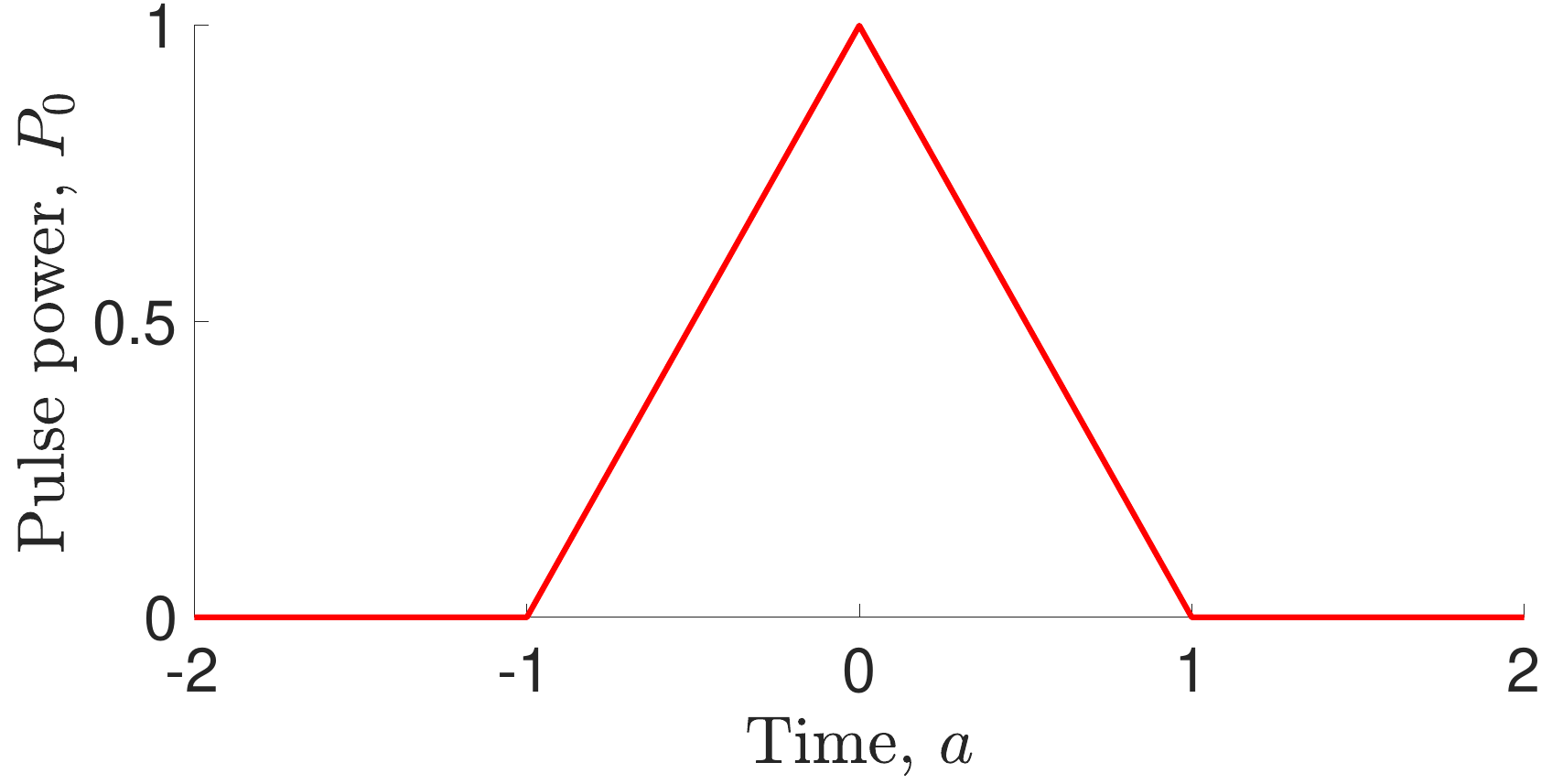} 
\hspace{1mm}
\includegraphics[height=4.2cm]{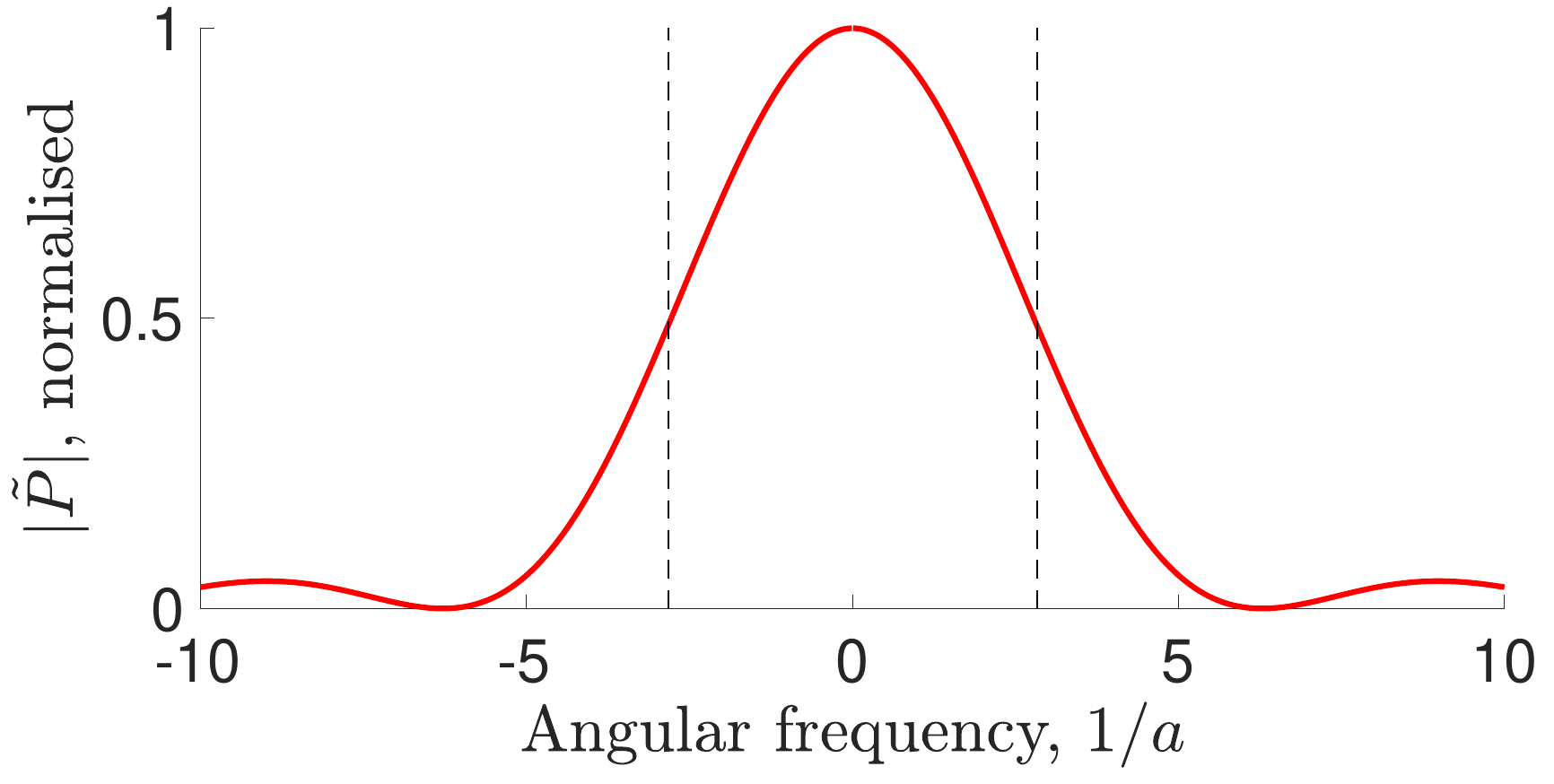} 
\caption{Pulse power in the time (left) and frequency (right) domain.}
\label{fig:w10_chromatic}
\end{figure}

For a 100-km link with $D=17$\,ps/(nm km), as an example, consider a pulse whose power temporal profile is given by the equation
\begin{equation}
P(t) = P_0 (1 - \frac{|t|}{a})
\end{equation}
for $-a<t<a$ and $P(t) = 0$ for all other times, where $a = 100$\,ps, as shown in Fig.~\ref{fig:w10_chromatic}. The laser wavelength is $\lambda=1550$\,nm. The Fourier transform of the pulse power is given by the equation
\begin{equation}
\begin{split}
    \hat{P}(\omega) &= \int_{-\infty}^{\infty} P(t) e^{-i \omega t} dt = P_0 \int_{-a}^{a} (1 - \frac{|t|}{a}) e^{-i \omega t} dt = P_0 \int_{-a}^{0} (1 + \frac{t}{a})e^{-i \omega t} dt + P_0 \int_{0}^{a} (1 - \frac{t}{a})e^{-i \omega t} dt = \\
    &= P_0 \int_{-a}^{a} e^{-i \omega t} dt + P_0 \int_{-a}^{0} \frac{t}{a}e^{-i \omega t} dt - P_0 \int_{0}^{a} \frac{t}{a} e^{-i \omega t} dt = \\
    &= P_0\frac{e^{i\omega a} - e^{i\omega a}}{i\omega} - P_0\frac{e^{i\omega a}}{i\omega} + P_0\frac{1-e^{i\omega a}}{\omega^2 a} + P_0\frac{e^{-i\omega a}}{i\omega} - P_0\frac{e^{-i\omega a}-1}{\omega^2 a} = \\
    & = \frac{P_0}{\omega^2 a} (2 - e^{i\omega a} - e^{-i\omega a}) = \frac{P_0}{\omega^2 a}4 \sin^2{\frac{\omega a}{2}} =a P_0 {\rm sinc}^2\left(\frac{\omega a}{2}\right).
\end{split}
\end{equation}

The half maximum is achieved for ${\rm sinc}\left(\frac{\omega a}{2}\right) = \frac{1}{\sqrt{2}}$. The full-width at half-maximum is $\Delta \omega_{\rm fw} \approx 5.6 / a = 56 \times 10^9$\,rad/s. We find the wavelength uncertainty from the equation 
\begin{equation}
    \frac{\Delta \lambda}{\lambda} = \frac{\Delta \omega_{\rm fw}}{\omega_0},
\end{equation}
and get $\Delta \lambda = 0.07$\,nm. Therefore, the pulse width at the end of the fibre is $b = 2a + DL\Delta \lambda = 312$\,ps and the maximum bit-rate is $1/b = 3.2$\,Gb/s.

\subsection{Polarisation dispersion}

\begin{wrapfigure}{r}{0.45\textwidth}
    \centering
    \vspace{-3mm}
    \includegraphics[width=0.43\textwidth]{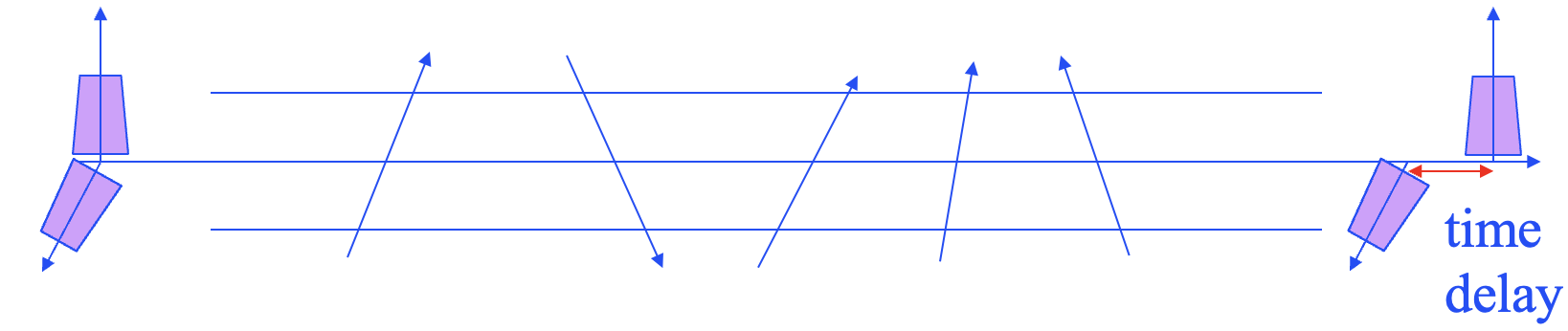}
    \caption{Pulse broadening due to the polarisation dispersion. Arrows inside the waveguide show the direction of the fast axis.}
    \label{fig:w10_pol_dispersion}
\end{wrapfigure}

In free space, light can be decomposed into vertical (S) and horizontal (P) polarisations, or any linear combination of the two. Similarly, dielectric waveguides support two fundamental polarisation states: transverse electric (TE) modes, which we have considered above, and transverse magnetic (TM) modes, in which the magnetic field lies entirely in the transverse plane. In birefringent media, these two polarisation states propagate with different group velocities due to a difference in effective refractive index. Although fused silica is only weakly birefringent, mechanical stress can induce a small anisotropy in the refractive index, typically of order $\Delta n \sim 10^{-5}$ between the fast and slow axes~\cite{yariv2007photonics}. 

In this section, we estimate the effect of polarisation-mode dispersion on pulse broadening. Since mechanical stress is applied randomly along the fibre, the fast and slow axes of birefringence are also randomly oriented, as shown in Fig.~\ref{fig:w10_pol_dispersion}. We therefore divide the fibre into segments of length $l$, within which the orientation of the principal axes can be assumed constant. The differential delay between the two polarisation states in each segment is then given by the equation
\begin{equation}
 \tau_i = l \left(\frac{1}{v_{\mathrm{gr},1}} - \frac{1}{v_{\mathrm{gr},2}} \right) \approx \frac{l}{v_{\mathrm{gr}}^2} \frac{d v_{\mathrm{gr}}}{d n}\,\Delta n_i,
\end{equation}
where $\Delta n_i$ denotes the birefringence experienced in the $i$-th segment. Because the orientation of the fast and slow axes varies randomly from segment to segment, the individual delays add statistically rather than coherently. As a result, the total timing spread is obtained by adding contributions in quadrature:
\begin{equation}
 \tau^2 = \sum_{i=1}^N \tau_i^2 = N l^2 \left( \frac{1}{v_{\mathrm{gr}}^2} \frac{d v_{\mathrm{gr}}}{d n} \right)^2 \Delta n^2 = L l \left( \frac{1}{v_{\mathrm{gr}}^2} \frac{d v_{\mathrm{gr}}}{d n} \right)^2 \Delta n^2,
\end{equation}
where $\Delta n^2$ denotes the mean-squared birefringence averaged over random orientations, and $L = Nl$ is the total fibre length.

The proportionality coefficient, $D_{\rm PMD}$, between the time delay $\tau$ and $\sqrt{L}$ is the coefficient of the polarisation mode dispersion, and it is typically measured in units of ${\rm ps} / \sqrt{\rm km}$. The resulting root-mean-square time delay is given by the equation
\begin{equation}
    \sqrt{\tau^2} = D_{\rm PMD} \sqrt{L},
\end{equation}
and is typically smaller than chromatic dispersion over kilometre-scale distances.

\subsection{Signal attenuation}

Dielectric waveguides enable signal transmission over long distances, with performance primarily limited by the absorption and scattering properties of the guiding medium. As discussed in Lecture~\ref{Lecture5}, optical materials exhibit strong wavelength-dependent absorption, so the choice of operating wavelength is crucial for low-loss communication. In optical fibres, one aims to minimise attenuation in both the core and cladding materials. A key material is fused silica, which exhibits exceptionally low loss around the telecommunications wavelength of $1550$\,nm. At this wavelength, Rayleigh scattering is already significantly reduced due to the $1/\lambda^4$ scaling, and absorption from the fused silica impurities dominates the absorption spectrum starting from $1600$\,nm.

In practical dielectric waveguides, such as optical fibres, the attenuation at 1550\,nm can be as low as a few percent per kilometre. The propagation of optical power along the fibre of length $L$ is well described by an exponential decay law,
\begin{equation}
P(L) = P_0 \exp(-\alpha L),
\end{equation}
where $\alpha$ is the attenuation coefficient (optical depth per unit length). This model captures the cumulative effect of scattering and absorption processes and sets the limit on how far optical signals can be transmitted before amplification is required. In quantum communications, which we will consider in Lecture~\ref{Lecture11}, the exponential loss in the communication channel is a central challenge, since quantum signals cannot be amplified without destroying their quantum state.

\subsection{Wavelength multiplexing}

Similar to the frequency multiplexing in coaxial cables, discussed in Lecture~\ref{Lecture9}, wavelength multiplexing in optical fibres allows a single fibre to carry many independent data streams simultaneously by using different wavelengths of light~\cite{agrawal2012fiber}. In modern systems, typically 100–200 distinct wavelengths can be transmitted through the same fibre. At the transmitter, multiple laser sources generate the different wavelengths, which are then combined into a single beam and injected into the fibre. At the receiver end, the combined signal is separated back into individual channels using optical devices such as diffraction gratings, arrayed waveguide gratings, or thin-film filters.

An emerging approach in optical communications is the use of a single optical frequency comb generated by a microresonator~\cite{pfeifle2014coherent}, replacing the need for many individual lasers. In these systems, a continuous-wave pump laser is coupled into a high-quality microresonator, where nonlinear effects (such as Kerr nonlinearity) generate a broad, equally spaced set of optical lines known as a Kerr frequency comb. Each comb line acts as a carrier that can be independently modulated, enabling wavelength-division multiplexing from a single compact device. 

\subsection{Frequency comb generation}

Optical frequency combs enable a direct link between optical and radio frequencies, as discussed in Lecture~\ref{Lecture8} in the context of optical atomic clocks. Their development was recognised with the Nobel Prize in Physics 2005, awarded to Theodor W. Hänsch and John L. Hall. A frequency comb is typically produced by a mode-locked laser, in which many longitudinal resonator modes are excited with fixed phase relationships, resulting in a coherent set of equally spaced frequencies~\cite{hall2000optical}. 

Mode-locking can be understood from the laser rate equations,
\begin{equation}
\begin{split}
& \frac{dN_a}{dt} = R - B N_a N_\nu - \frac{N_a}{\tau_a}, \\
& \frac{dN_\nu}{dt} = B N_a N_\nu - \frac{N_\nu}{\tau_c},
\end{split}
\end{equation}
where \(N_a\) is the population of the upper lasing level, \(R\) is the pumping rate, \(B\) is the stimulated emission coefficient, \(\tau_a\) is the spontaneous emission lifetime of the atoms, \(N_\nu\) is the number of photons in the cavity, and \(\tau_c\) is the photon lifetime in the resonator. The cavity supports modes separated by the free spectral range \(c/L_{\rm cav}\), where \(L_{\rm cav}\) is the optical round-trip length of the cavity. If the laser parameters are stationary, the longitudinal modes compete for the available gain. In this regime, small fluctuations are amplified, and typically one mode depletes the gain more efficiently than the others, eventually dominating the cavity and suppressing all remaining modes. As a result, the laser operates in a single-frequency (or few-mode) regime, as we discussed in Lecture~\ref{Lecture6}.

In contrast, when the system parameters, such as the pump rate \(R\) or cavity loss \(\tau_c\), are modulated at the cavity free spectral range, the mode competition is suppressed, and the modes become phase-locked. Under these conditions, many longitudinal modes can coexist and remain stable, leading to the formation of a frequency comb and a train of ultrashort pulses as given by the equation
\begin{equation}
E(t) = \sum_{-n}^{+n} E_0 \exp(i(\omega_0 + m\Delta \omega)t + m\phi) = A(t)\exp(i\omega_0 t),
\end{equation}
where $\omega_0$ is the central frequency, and the phases of all waves are related to each other (locked) via the term $m\phi$. The amplitude $A(t)$ of the field is given by the equation
\begin{equation}
A(t) = E_0 \frac{\sin((2n+1)(\Delta \omega t+\phi)/2))}{\sin((\Delta \omega t+\phi)/2))}.
\end{equation}

Mode locking in lasers can be achieved through either active or passive modulation techniques~\cite{svelto2010principles}. In active modulation, an external signal is used to periodically control the laser parameters at the cavity free spectral range, for example, through amplitude modulation, phase modulation, or direct modulation of the laser gain, thereby enforcing phase coherence between longitudinal modes. In contrast, passive modulation relies on intrinsic nonlinear effects within the cavity to achieve mode-locking. Common mechanisms include the use of a saturable absorber, which preferentially transmits high-intensity light and suppresses low-intensity fluctuations, and Kerr nonlinearity, which induces intensity-dependent phase shifts as given by the equation
\begin{equation}
n_0 = n_1 + n_2 I,
\end{equation}
where $n_1$ is the linear index of refraction, $I$ is the beam intensity, and $n_2$ is the Kerr coefficient.

\begin{figure}[t!]
\centering
\includegraphics[height=3.57cm]{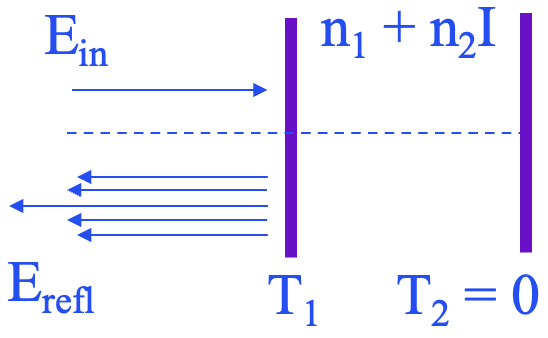} 
\hspace{1mm}
\includegraphics[height=3.57cm]{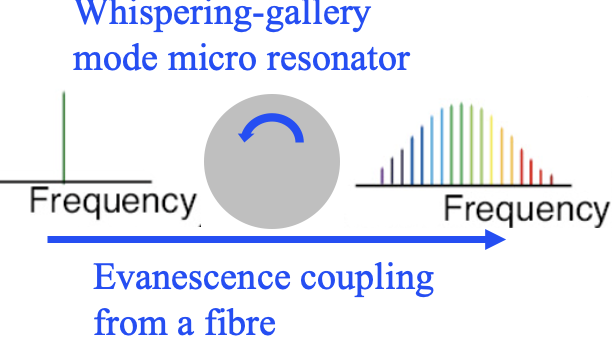} 
\hspace{1mm}
\includegraphics[height=3.57cm]{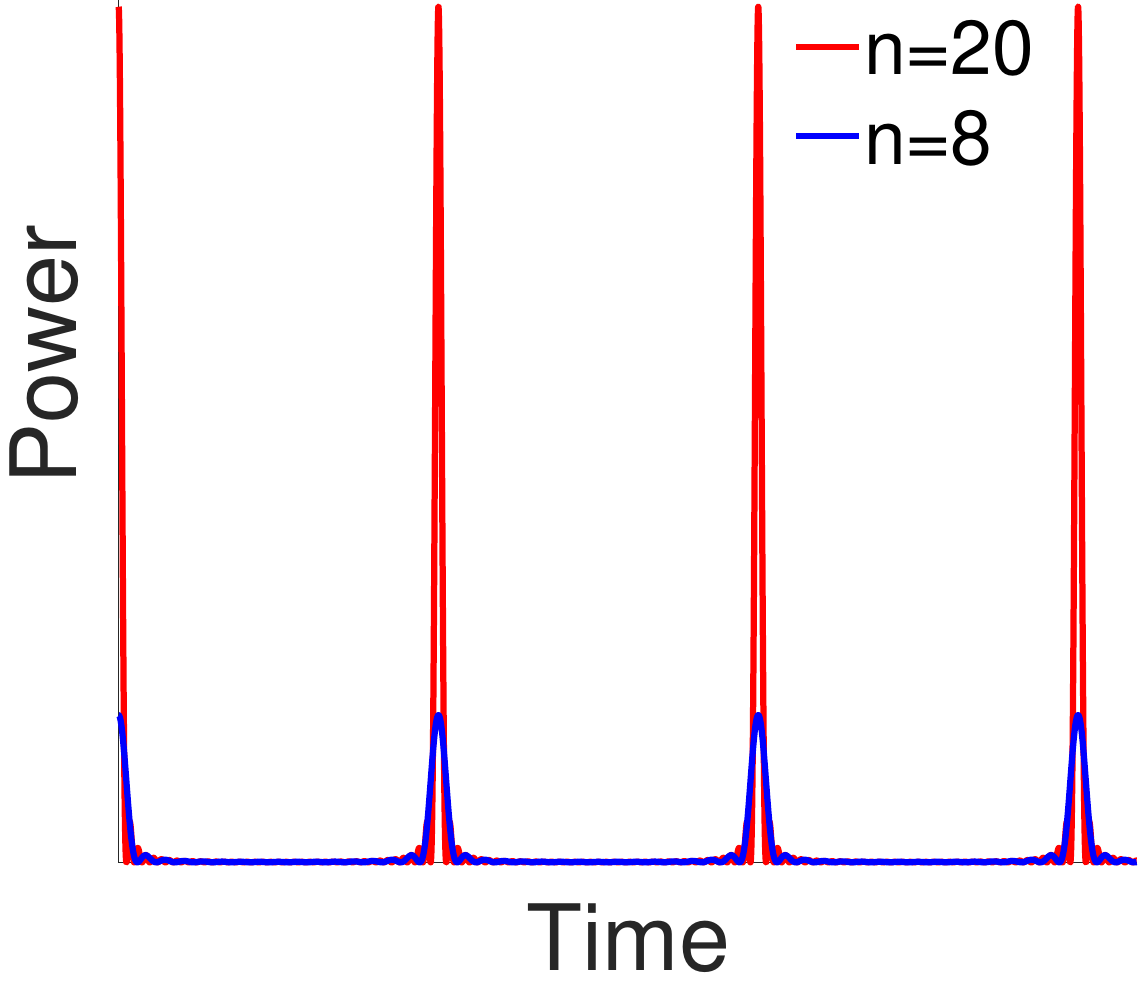} 
\caption{Passive linear resonator with a Kerr nonlinearity for the frequency comb generation (left). Practical implementation with a whispering-gallery-mode resonator (centre). Example of pulses in the time domain for a different number of the comb teeth, $n=8$ and $n=20$ (right).}
\label{fig:w10_comb}
\end{figure}

In this section, we consider frequency comb generation with a passive microresonator, as shown in Fig.~\ref{fig:w10_comb}. Passive microresonators used for frequency comb generation are typically high-Q integrated photonic structures such as whispering-gallery-mode resonators, made from materials like silicon nitride, silica, magnesium fluoride, or crystalline fluorides (e.g.\ CaF$_2$). These platforms combine low optical loss with strong optical confinement, enhancing nonlinear interactions~\cite{boyd2020nonlinear} and enabling Kerr comb generation. The Kerr nonlinearity coefficient $n_2$ depends on the material and is typically in the range $10^{-20}$--$10^{-19}\,\mathrm{m}^2/\mathrm{W}$.

Similar to Lecture~\ref{Lecture8}, we consider a linear cavity tuned on resonance and seek solutions for the electric field in the form
\begin{equation}\label{eq:w10_comb_cav}
\sum_{-n}^{n} E_m (t+\tau) e^{i m \Delta \omega t} = \sqrt{T_1} E_{\rm in}(t+\tau) + r_1 \sum_{-n}^{n} E_m(t) e^{i m \Delta \omega t} e^{-i\phi_m},
\end{equation}
where $T_1 \ll 1$ is the power transmissivity of the input coupler, $r_1 = \sqrt{1-T_1}$ is the corresponding field reflectivity, $n$ is the number of modes, $\tau$ is the cavity round trip time, $\Delta \omega = 2\pi/\tau$ is the free spectral range of the resonator (in rad/s), and $\phi_m$ is the round-trip phase accumulated by the $m$-th mode. 

Since the resonator is pumped by a single-frequency field, Eq.~\ref{eq:w10_comb_cav} admits only a single non-zero solution for $m=0$ in the linear case ($n_2 = 0$). However, in the presence of Kerr nonlinearity, the modes become coupled through the intensity-dependent phase shift,
\begin{equation}
\phi_m = \frac{4 \pi L}{\lambda_m} \left( n_1 + n_2 I \right)
= 2 \pi N + \frac{2 \pi \omega_m n_2}{\Delta \omega} \left( \sum_{-n}^{n} E_m (t) e^{i \omega_m t} + {\rm c.c.} \right)^2,
\end{equation}
where $N$ is an integer and $\omega_m = \omega_0 + m\Delta \omega$ is the angular frequency of the $m$-th mode, and $\omega_0$ is the pump frequency.

In general, Eq.~\ref{eq:w10_comb_cav} requires numerical methods to solve. However, the start of comb generation can be understood analytically in the weakly nonlinear regime, $n_2 I \ll 1$, by considering only the modes $m = 0, \pm 1$ with a strong pump field ($m=0$) and weak sideband fields ($m =\pm 1$).

The growth of the $m \pm 1$ fields is driven by the modulation of the resonant pump field, $E_0$ and $E_0^*$, by the time-dependent component of the phase, as given by the equations
\begin{equation}\label{eq:w11_E_m_1}
\begin{split}
& E_{+1} (t+\tau) e^{i \omega_{+1} t} = r_1 E_{+1}e^{i \omega_{+1} t}  -i E_0 e^{i \omega_0 t} \phi_{0+} + i E_0^* e^{-i \omega_0 t} \phi_{0+}, \\
& E_{-1} (t+\tau) e^{i \omega_{-1} t} = r_1 E_{-1}e^{i \omega_{-1} t}  - i E_0 e^{i \omega_0 t} \phi_{0-} + i E_0^* e^{-i \omega_0 t} \phi_{0-},
\end{split}
\end{equation}
where $\phi_{0+}$ and $\phi_{0-}$ are the time-dependent components of the nonlinear phase shift acquired by the pump field inside the resonator that drive the generation of sidebands at frequencies $\omega_{+1}$ and $\omega_{-1}$, respectively. Both phases satisfy $\phi_{0+}, \phi_{0-} \ll 1$, and we use the approximation $\exp(i\phi_{0\pm}) \approx 1 + i\phi_{0\pm}$. The static term, $1$, contributes to the steady-state pump field, yielding $E_0 \approx 2E_{\rm in}/\sqrt{T_1}$ on resonance.

Up to the first order in $E_{+1}$ and $E_{-1}$, the relevant oscillating terms of the round-trip phase that convert the pump fields to the sidebands are given by the equation
\begin{equation}
\begin{split}
& \phi_{0,+} \approx \frac{2 \pi \omega_0 n_2}{\Delta \omega} \left(E_{+1} E_0^* e^{i \Delta \omega t} + E_{-1}^* E_0 e^{i \Delta \omega t} + E_{+1} E_0 e^{i \Delta \omega t} e^{2i \omega_0 t}  \right), \\
& \phi_{0,-} \approx \frac{2 \pi \omega_0 n_2}{\Delta \omega} \left(E_{-1} E_0^* e^{-i \Delta \omega t} + E_{+1}^* E_0 e^{-i \Delta \omega t} + E_{-1} E_0 e^{-i \Delta \omega t} e^{2i \omega_0 t}  \right),
\end{split}
\end{equation}
where the first two terms in each expression correspond to the nonlinear mixing processes that couple the pump field $E_0$ to the sideband fields at frequencies $\omega_{+1}$ and $\omega_{-1}$, while the third terms arise from interactions involving the complex conjugate field $E_0^*$. Substituting these phase terms into Eq.~\ref{eq:w11_E_m_1} and retaining only the contributions oscillating at the sideband frequencies, we obtain the following equations:
\begin{equation}\label{eq:w11_E_m_ph}
\begin{split}
& E_{+1} (t+\tau) = r_1 E_{+1}(t)  -i E_0 \frac{2 \pi \omega_0 n_2}{\Delta \omega} \left(E_{+1}(t) E_0^* + E_{-1}^*(t) E_0 \right) + i E_0^* \frac{2 \pi \omega_0 n_2}{\Delta \omega} E_{+1}(t) E_0 , \\
& E_{-1} (t+\tau) = r_1 E_{-1}(t) -i E_0 \frac{2 \pi \omega_0 n_2}{\Delta \omega} \left(E_{-1}(t) E_0^* + E_{-1}^*(t) E_0 \right) + i E_0^* \frac{2 \pi \omega_0 n_2}{\Delta \omega} E_{-1}(t) E_0.
\end{split}
\end{equation}

Since the cavity has high finesse ($T_1 \ll 1$), the cavity fields $E_m(t)$ cannot change significantly on time scales shorter than the round-trip time of the cavity $\tau$, and we can approximate $E_m(t+\tau) = E_m(t) + \tau dE_m/dt$. The equations for the sidebands ($m=\pm 1$) reduce to
\begin{equation}\label{eq:w10_comb_1}
\begin{split}
& \tau \frac{dE_{+1}}{dt} = -\frac{T_1}{2} E_{+1} - i \gamma E_{-1}^*, \\
& \tau \frac{dE_{-1}}{dt} = -\frac{T_1}{2} E_{-1} - i \gamma E_{+1}^*,
\end{split}
\end{equation}
where the effective pumping rate $\gamma$ is given by
\begin{equation}
\gamma \approx \frac{2\pi \omega_0 n_2 I}{\Delta \omega}.
\end{equation}
These equations describe a parametric interaction in which two photons from the pump field combine with one sideband photon to generate the opposite sideband, as given by the equation
\begin{equation}
2\omega_0 \rightarrow \omega_{+1} + \omega_{-1}.
\end{equation}

The first term on the right-hand side of Eq.~\ref{eq:w10_comb_1} represents cavity losses, while the second term describes nonlinear parametric gain. The solutions take the form
\begin{equation}
E_{+1} = A_{+1} e^{\alpha t} + c.c., \hspace{1cm}
E_{-1} = A_{-1} e^{\alpha t} + c.c.,
\end{equation}
with growth rate
\begin{equation}
\alpha = \frac{\gamma - \frac{T_1}{2}}{\tau}.
\end{equation}
Thus, the $m =\pm 1$ sidebands grow exponentially when the nonlinear gain exceeds the cavity losses, i.e.\ when $\gamma > \frac{T_1}{2}$.

In practice, this exponential growth occurs only in the weak-field regime. As the sidebands increase in amplitude, energy is transferred to higher-order modes ($m=\pm 2, \pm 3, \dots$), leading to saturation and the formation of a full frequency comb. The resulting comb lines can then be separated and individually modulated, enabling applications such as wavelength-division multiplexing in optical communication systems.

The number of modes generated in a microresonator frequency comb is primarily limited by dispersion and phase matching~\cite{kippenberg2011microresonator}. For efficient Kerr comb generation, the resonator modes must remain equally spaced so that nonlinear mixing can cascade coherently across many longitudinal modes. However, similar to chromatic dispersion in optical fibres, microresonators exhibit group-velocity dispersion~\cite{herr2012temporal}, meaning the free spectral range is not perfectly constant with frequency. As the comb broadens, this detuning accumulates and eventually destroys the phase-matching condition required for continued sideband growth. Note that Eq.~\ref{eq:w10_comb_cav} assumes the same $\Delta \omega$ independent of $m$. In typical silicon nitride or crystalline microresonators, this results in combs spanning from tens to a few hundred modes in narrowband regimes, while broadband “octave-spanning” combs can reach several hundred to over a thousand modes in optimised devices~\cite{delhaye2007optical}.

%% file: week11.tex
\section{Quantum communication}
\label{Lecture11}

In Lectures~\ref{Lecture5} and~\ref{Lecture10}, we considered applications of light related to its optical power (or amplitude quadrature of the electric field). In Lectures~\ref{Lecture7} and~\ref{Lecture8}, we considered applications of laser beams to precision measurements related to the phase quadrature of the electric field. In this lecture, we explore applications related to correlations (or entanglement) of two spatially separated laser beams. Entangled photon pairs can be distributed between two users, who measure random but correlated outcomes. This property is required for protocols such as quantum key distribution, where entanglement guarantees that any eavesdropping attempt disturbs the correlations and can therefore be detected, providing secure communication based on quantum mechanics rather than computational complexity.

In this Lecture, we discuss
\begin{itemize}
\item quantisation of light,
\item secure communication,
\item quantum key distribution,
\item nonlinear crystals,
\item quantum internet.
\end{itemize}

\subsection{Quantisation of electromagnetic fields}

The quantum nature of light was first convincingly demonstrated through the photoelectric effect~\cite{einstein1905photoelectric}. In this experiment, monochromatic light of frequency $\nu$ is directed onto a metal surface inside a vacuum tube, causing the emission of electrons from the cathode. A collector electrode is biased with a variable voltage, which can either accelerate the emitted electrons (positive bias) or oppose their motion (negative bias). As the retarding voltage is increased, only the most energetic photoelectrons reach the collector. When the photocurrent drops to zero, the applied stopping potential $V_0$ is sufficient to prevent even the fastest electrons from arriving, implying a maximum kinetic energy $K_{\max} = eV_0$. This provides a direct electrical measurement of the energy carried by the emitted electrons.

A key observation is that, for a given material, there exists a threshold optical frequency below which no photoelectrons are emitted, regardless of the intensity of the incident light. Above this threshold, the maximum kinetic energy of the electrons increases linearly with the light frequency, and the intensity primarily affects the number of emitted electrons. This behaviour cannot be explained by classical wave theory and led Einstein in 1905 to propose that light is quantised into discrete energy packets~\cite{einstein1905photoelectric}, each carrying energy $E = h\nu$, building on Planck’s earlier work on blackbody radiation. In this picture, a single photon transfers its energy to a single electron, with part of the energy used to overcome the material work function and the remainder appearing as kinetic energy of the electron.

We have already explored quantum properties of light in Lectures~\ref{Lecture3},~\ref{Lecture7}, and~\ref{Lecture8}, where we considered quantum shot noise in imaging and laser position measurements. In this Lecture, we discuss useful properties of the quantisation of light related to the photon's entanglement. We treat the entanglement between two photons as a quantum correlation in which the joint state of the pair cannot be written as a product of independent states for each photon. Instead, the photons share a single inseparable quantum state, so that measurements on one photon are correlated with measurements on the other, even when they are spatially separated. We first discuss classical communication algorithms and then introduce quantum key distribution.

\subsection{Secure communication}

Unencrypted classical communication is vulnerable because information is ultimately encoded as binary data, typically using standards such as UTF-8 or ASCII, and transmitted as a sequence of bits over physical channels. These signals propagate through communication links where they can be intercepted at many points, often without disrupting the transmission in any noticeable way. In practice, this is particularly problematic because signal losses in the channel naturally occur due to attenuation, as we discussed in Lectures~\ref{Lecture9} and~\ref{Lecture10}, making it difficult to distinguish legitimate transmission degradation from deliberate eavesdropping. Furthermore, communication infrastructure such as coaxial cables, optical fibre links, routers, amplifiers, and servers provides multiple access points where signals may be copied.

\begin{figure}[t!]
\centering
\includegraphics[height=7cm]{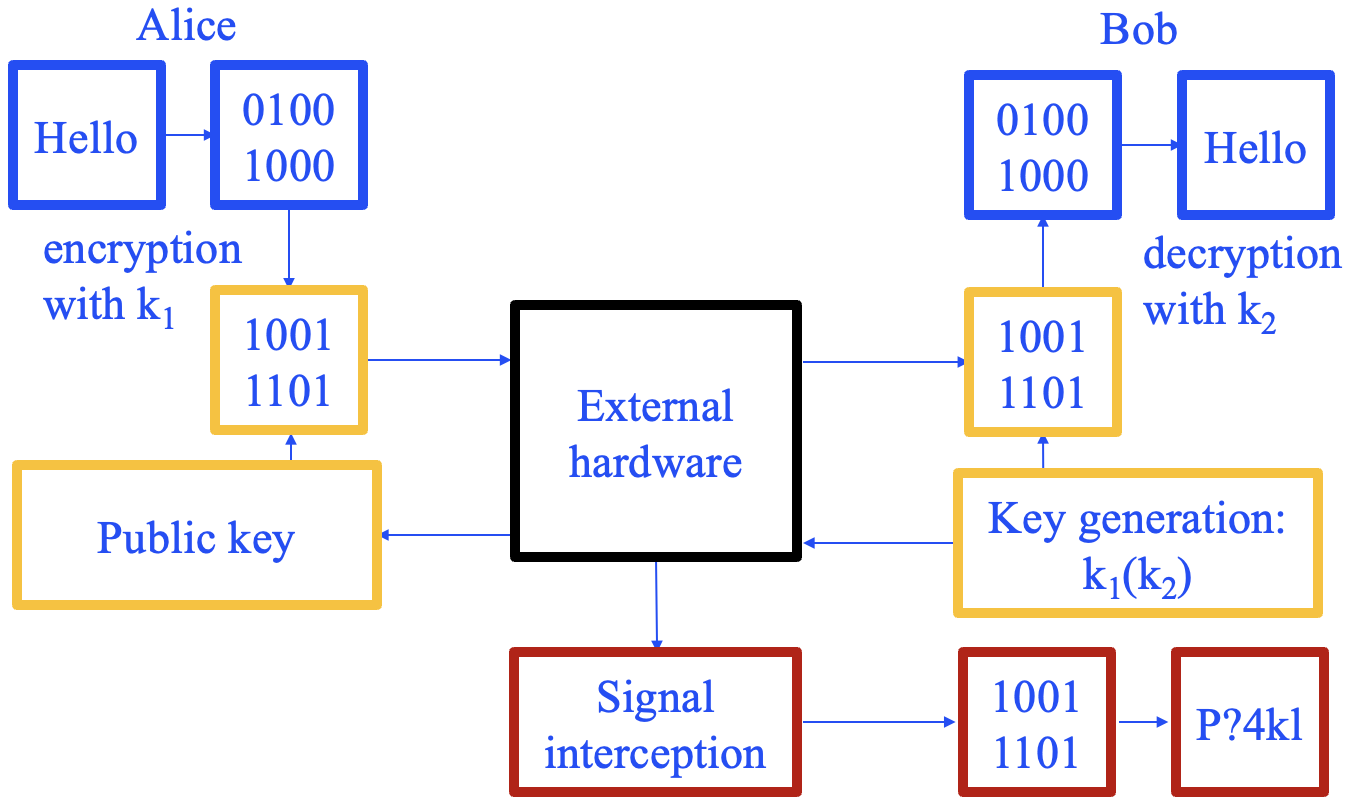} 
\caption{Example of encrypted communication between Alica and Bob. Even in the presence of the signal interception by an eavesdropper, the message cannot be decrypted without the private key, $k_2$, because the public key, $k_1$, is used only to encrypt the message but cannot decrypt it.}
\label{fig:w11_encrypted}
\end{figure}

In encrypted communication, the original message is first transformed into a scrambled form before transmission using an encryption function. A message $m$ is encoded into a ciphertext $c$ according to the equation
\begin{equation}
c = f(m, k_1),
\end{equation}
where $k_1$ is an encryption key. After transmission, the receiver applies a decryption function to recover the original message via
\begin{equation}
m = g(c, k_2),
\end{equation}
where $k_2$ is the corresponding decryption key. In cryptographic protocols, $k_1$ and $k_2$ are typically related as a public--private key pair, where the encryption key may be publicly known while the decryption key remains private, as shown in Fig~\ref{fig:w11_encrypted}. The functions $f$ and $g$, together with the keys, are constructed such that they satisfy the consistency condition
\begin{equation}
g(f(m, k_1), k_2) = m
\end{equation}
for any message $m$ and ensure that only the intended receiver can reliably recover the message.

\subsubsection{RSA algorithm}

Classical public-key cryptography relies on the computational difficulty of deriving a private key from a publicly shared key, a task believed to be infeasible for sufficiently large parameters. A central example is the RSA algorithm, which is based on number-theoretic properties of large prime numbers~\cite{rivest1978method}. In key generation, two large primes $p$ and $q$ are chosen, and their product $n = pq$ is computed. The following steps then generate the private key:
\begin{itemize}
\item evaluate $x = \mathrm{lcm}(p-1, q-1)$, the least common multiplier of $p-1$ and $q-1$,
\item select an integer $e$ such that $1 < e < x$ and $e$ is coprime to $x$,
\item determine $d$ as the modular multiplicative inverse of $e$ modulo $x$, satisfying $ed \equiv 1 \ (\mathrm{mod}\ x)$.
\end{itemize}
The public key is then $(n, e)$, and the private key is $d$. Encryption of a message $m$ is performed via $c(m) = m^e \ (\mathrm{mod}\ n)$, and decryption is achieved using $m(c) = c^d \ (\mathrm{mod}\ n)$. The security of the scheme relies on the fact that recovering $d$ from the public information requires factoring $n$ into $p$ and $q$, which is computationally hard for classical algorithms when $n$ is large.

As an example, consider the key generation with parameters: $p = 7$ and $q = 11$. It is straightforward that these numbers are prime. For large numbers, we may check that the numbers are prime with several iterations of Fermat's little theorem: $a^{p-1} = 1\,(\mathrm{mod}\ p)$ and $a^{q-1} = 1\,(\mathrm{mod}\ q)$ for any $a$ that is not a multiple of $p$ and $q$~\cite{hardy2008introduction}.

The least common multiple of $p-1=6$ and $q-1=10$ is equal to the product of these numbers divided by the greatest common divisor, which we can find using the Euclidean algorithm:
\begin{itemize}
\item 10:6 = 1, remainder 4
\item 6:4 = 1, remainder 2
\item 4:2 = 2, remainder 0
\item ${\rm gcd}(6,10) = 2$ and ${\rm lcm}(10,6) = 10 \times 6 / 2 = 30$.
\end{itemize}

Then we need to select $e$ such that ${\rm gcd}(e, x) = 1$. We first need to decompose $x$ into the prime numbers and get $x = 2 \times 3 \times 5$. Therefore, we may select $e = 7, 11, 17, 19, ...$. In this problem, we select $e=7$.

We can find $d$ by the Extended Euclidean Algorithm that finds $d$ and $f$ that satisfy the Bézout’s identity $ed + fx = {\rm gcd(e,x)} = 1$ as was enforced by the selection of $e$. We find that $d = 13$ and check that $ed = 3x + 1$.

If we like to send a message $m=8$ then the encrypted message is $c(m) = 8^7\,(\mathrm{mod}\ 77) = 57$.
We apply a similar approach to decrypt the message and find that $57^{13}\,(\mathrm{mod}\ 77) = 8$. It is important to note that the powers make the number very large quickly, and we need to keep track only of the remainders. We can prove this by presenting $m^k = c \times n + z$, where $c$ and $z$ are some integers. Therefore, $m^{k+1} = cm \times n + zm$. Since we are interested in the residual after our division by $n$, there is no need to keep track of the first term in the equation.

\subsubsection{Post quantum algorithms}

On classical computers, certain problems are easy to verify but hard to solve: for example, checking whether a number is prime can be done efficiently with polynomial complexity in the number of bits, whereas factoring a large composite number into its prime components is believed to be computationally intractable. This asymmetry supports the security of many public-key schemes, including RSA. However, this assumption is challenged by quantum computing. In particular, Shor’s algorithm~\cite{shor1997polynomial} demonstrates that a sufficiently powerful quantum computer could factor large integers and break widely used cryptosystems.

To address this threat, researchers have developed alternative schemes collectively known as post-quantum cryptography, which are based on mathematical problems believed to be hard even for quantum computers~\cite{bernstein2009post}, such as lattice-based, code-based, and hash-based constructions. These systems aim to provide long-term security in the presence of quantum adversaries while remaining implementable on classical hardware. However, there is a possibility that new quantum algorithms will be discovered in the future, potentially breaking currently proposed schemes, making this an active and evolving area of research.

\subsection{Quantum key distribution}

\begin{figure}[t!]
\centering
\includegraphics[height=7cm]{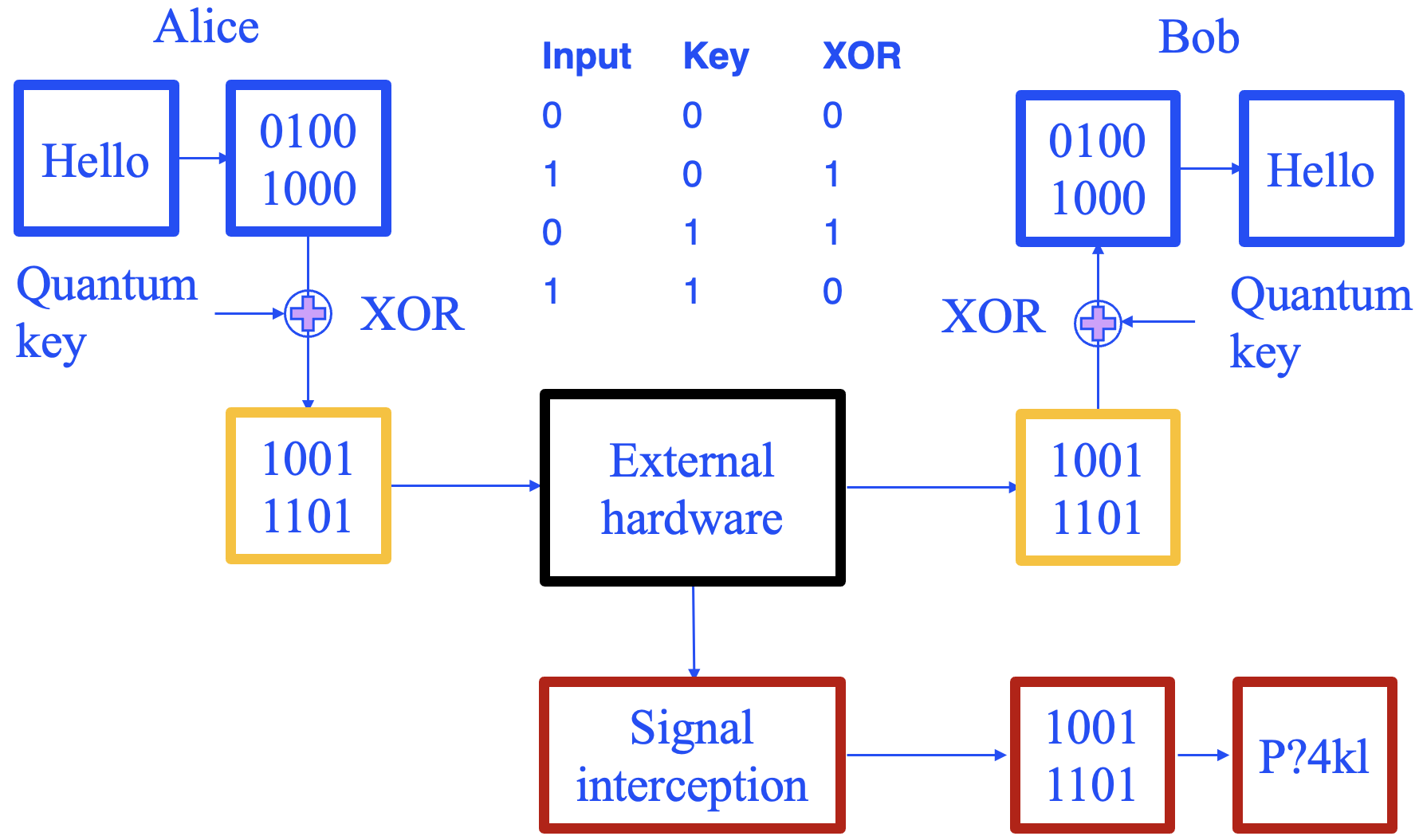} 
\caption{Example of encrypted communication between Alica and Bob with a quantum key and XOR operation. Since the photons are entangled, the quantum key is fundamentally random but is identical in the encryption and decryption processes done by Alice and Bob.}
\label{fig:w11_xor}
\end{figure}

Quantum key distribution provides a fundamentally secure method for sharing cryptographic keys by exploiting the principles of quantum mechanics. In this approach, only the key is transmitted over a quantum channel, while the encrypted message itself is sent over a classical channel. The security arises from the fact that unknown quantum states cannot be perfectly copied (the no-cloning theorem), and any eavesdropping attempt inevitably disturbs the transmission and can be detected. Once a shared secret key is established, it can be used to encrypt the message using a simple bitwise XOR operation, shown in Fig.~\ref{fig:w11_xor}. The bits are combined with the key bits to produce the encoding,
\begin{equation}
c = m \oplus k,
\end{equation}
and the original message is recovered by applying the same operation again,
\begin{equation}
m = c \oplus k.
\end{equation}

Quantum key distribution requires a range of photonic technologies to generate and detect single photons. On the source side, both continuous-wave and pulsed lasers are used to produce stable optical fields, which are then converted into quantum states using nonlinear optical media~\cite{boyd2020nonlinear}. Entangled photon pairs are commonly generated via spontaneous parametric down-conversion in nonlinear crystals such as BBO (beta-barium borate) and KTP (potassium titanyl phosphate), or alternatively through four-wave mixing in integrated platforms such as silicon photonics. 

\begin{figure}[t!]
\centering
\includegraphics[height=6cm]{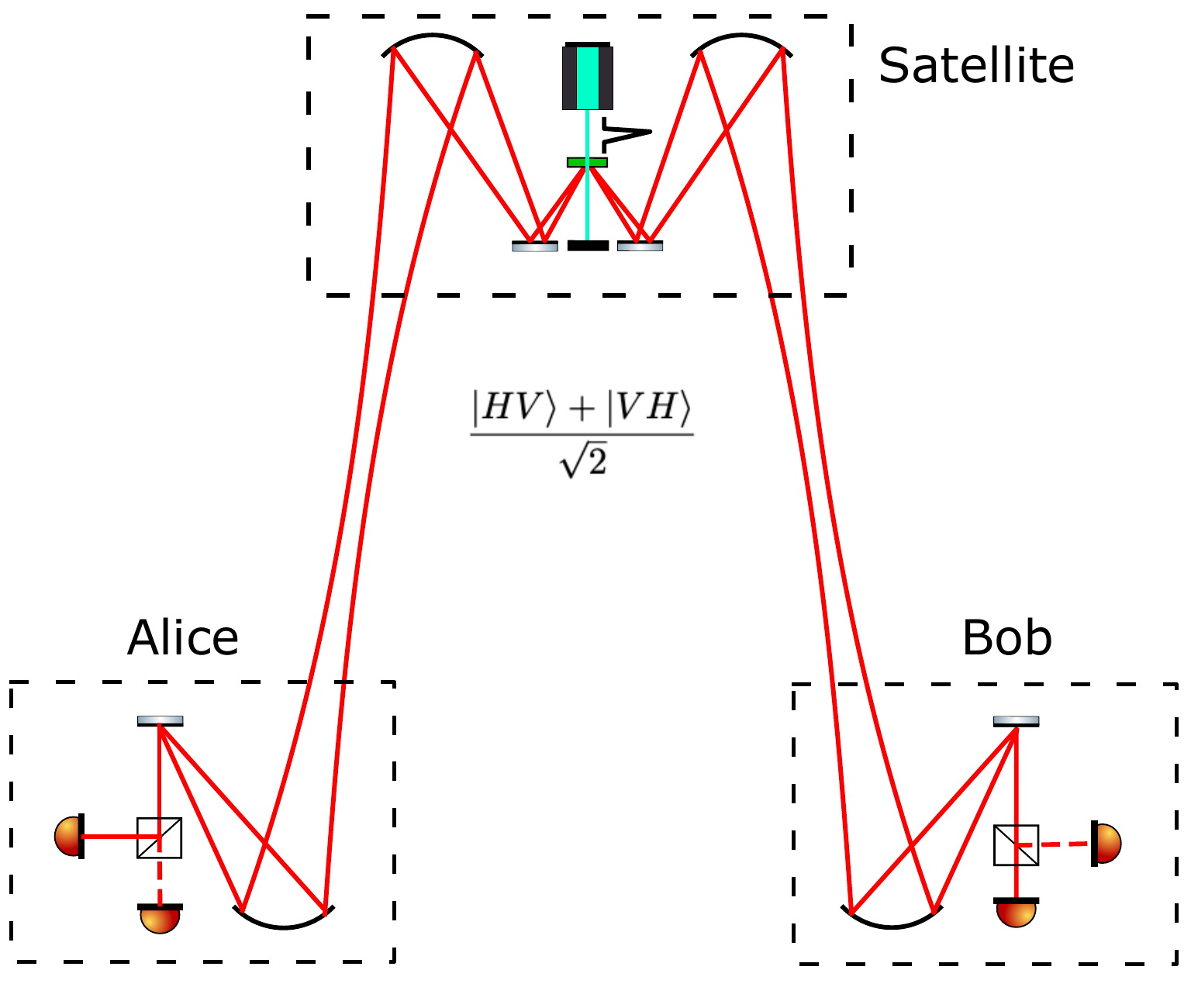} 
\hspace{2mm}
\includegraphics[height=6cm]{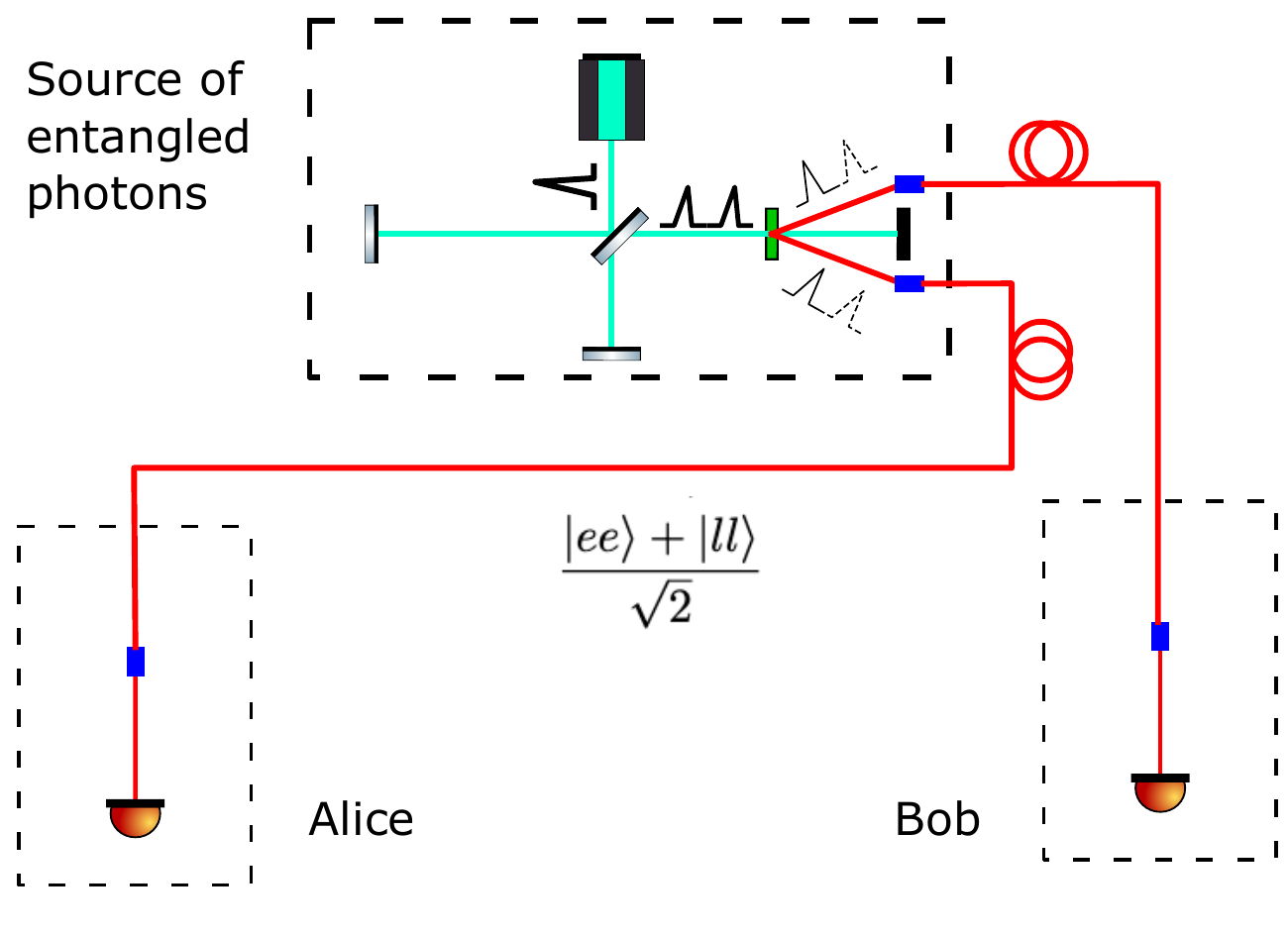} 
\caption{Quantum key distribution from a satellite (left): entangled photon pairs are generated using a nonlinear crystal and transmitted to Alice and Bob via optical telescopes. Located on Earth, Alice and Bob analyse the state of their photons using polarising beam splitters and pairs of photodetectors. Quantum key distribution in optical fibres (right): time-bin entanglement is generated using a Michelson interferometer in combination with a nonlinear crystal. Alice and Bob then measure the arrival times of their photons within each time window and determine whether they correspond to early or late time bins.}
\label{fig:w11_qkd}
\end{figure}

On the detection side, quantum key distribution relies on single-photon detectors with high quantum efficiency capable of resolving individual quantum events~\cite{gisin2002quantum}. The most widely used technologies include avalanche photodiodes, which operate in Geiger mode for near-single-photon sensitivity, and superconducting nanowire single-photon detectors, which offer higher efficiency, lower noise, and faster timing resolution.

In free-space quantum key distribution, particularly in satellite-based links, as shown in Fig.~\ref{fig:w11_qkd}\,(left), entanglement can be encoded in several degrees of freedom, including polarisation, spatial (momentum) modes, and orbital angular momentum. Polarisation entanglement is typically generated using a nonlinear crystal, where a pump photon is converted into a pair of photons with correlated, orthogonal polarisation states. As a result, if Alice measures one photon to be vertically polarised, Bob will find the other to be horizontally polarised, and vice versa. However, free-space propagation introduces significant challenges, including atmospheric turbulence, absorption, and beam wandering, all of which can degrade the entanglement and reduce the transmission fidelity over long distances.

In fibre-based quantum key distribution, different encoding schemes are typically used to ensure robustness against environmental perturbations. Common approaches include time-bin encoding, where information is carried in photon arrival times, and frequency encoding, which utilises different spectral components. Unlike polarisation entanglement, these methods are less sensitive to fibre random polarisation drifts due to stress, as we discussed in Lecture~\ref{Lecture10}. Time-bin encoding~\cite{brendel1999pulsed} is typically implemented using an interferometric setup, such as a Michelson interferometer, which splits an initial laser pulse into two pulses separated by a short time delay, as shown in Fig.~\ref{fig:w11_qkd}\,(right). Each pulse is then directed into a nonlinear crystal, where it can probabilistically generate an entangled photon pair. Since the pair-generation probability is kept low, typically only one of the two pulses produces a pair. As a result, each photon is prepared in a coherent superposition of two temporal modes, corresponding to early and late arrival times, denoted by the states $\ket{e}$ and $\ket{l}$, respectively.

\subsection{Nonlinear crystals}

In a nonlinear crystal, the polarisation, $\VF{P}$, responds nonlinearly to the applied electric field, $\VF{E}$, and the displacement vector, $\VF{D}$, is given by the equation
\begin{equation}
    \VF{D} = \epsilon_0 \VF{E} + \VF{P} = \epsilon_0 (1 + \chi) \VF{E} + \VF{P}^{\rm NL},
\end{equation}
where $\chi = \epsilon - 1$ is electric susceptibility of the material and $\VF{P}^{\rm NL}$ is the nonlinear component of the polarisation vector. In this Lecture, we focus on the second-order nonlinearity, $\xi$. The nonlinear component of the polarisation vector is related to the electric field by the tensor equation
\begin{equation}
    P^{\rm NL}_i = \epsilon_0 \xi_{ikm}E^k E^m,
\end{equation}
which states that different components of the electric field contribute to $P^{\rm NL}_i$, and $i=1,2,3$. Components of the tensor $\xi$ are determined by the crystal properties. In a KTP crystal, for example, the largest component is $\xi_{333} \approx 17$\,pm/V, which implies that the electric field along the vertical axis creates the largest contribution to the nonlinear polarisability, $\VF{P}^{\rm NL}$.

Nonlinear crystals lead to nonlinear effects when the frequency of light changes, including frequency doubling, as we discussed in Lecture~\ref{Lecture8}. We show these nonlinear effects by starting with Maxwell's equations in a dielectric

\begin{equation}\label{eq:Maxwell_diff}
\nabla \cdot \VF{D} = 0 \hspace{1cm} \nabla \cdot \VF{B} = 0 \hspace{1cm} \nabla \times \VF{E} = -\frac{\partial \VF{B}}{\partial t} \hspace{1cm} \nabla \times \VF{H} = \frac{\partial \VF{D}}{\partial t},
\end{equation}
where $\VF{H}$ is the magnetisation vector, and $\VF{B}$ is the magnetic field. We can relate the magnetisation vector $\VF{H}$ to the magnetic field $\VF{B}$. Since we only consider non-magnetic dielectric materials, such as BBO and KTP crystals, we get $\VF{H} = \VF{B} / \mu_0$, where $\mu_0$ is the vacuum magnetic permeability. We then solve Maxwell's equations by taking the curl of the Faraday law and get
\begin{equation}
\nabla \times \nabla \times \VF{E} = -\frac{\partial \VF{\nabla \times B}}{\partial t} = -\mu_0\frac{\partial^2 \VF{D}}{\partial t^2} = -{\mu_0\epsilon_0(1+\chi)}\frac{\partial^2 \VF{E}}{\partial t^2} -\mu_0\frac{\partial^2 \VF{P^{\rm NL}}}{\partial t^2},
\end{equation}
which we simplify by noting that $\mu_0\epsilon_0 = 1/c^2$, $1+\chi = \epsilon = n^2$, where $n$ is the refractive index of the crystal. Similar to Lecture~\ref{Lecture2}, we further simplify the equation by applying the curl of curl rule, and get the equation
\begin{equation}
\nabla \times \nabla \times \VF{E} = \nabla(\nabla \cdot \VF{E}) - \nabla^2 \VF{E} = \frac{1}{\epsilon_0 n^2}\nabla(\nabla \cdot \VF{D}) - \frac{1}{\epsilon_0 n^2}\nabla(\nabla \cdot \VF{P}^{\rm NL}) - \nabla^2 \VF{E}.
\end{equation}
The first term in the last expression vanishes because of Gauss' law, and the second term vanishes because $\frac{\partial P^{\rm NL}_y}{\partial y} = 0$ for a plane wave. In practice, even if we consider a Gaussian beam instead of a plane wave, the term is still negligible because it equals to $\sim \frac{1}{w^2 \epsilon_0 n^2} P^{\rm NL}$, where $w$ is the beam size. This term is much smaller than $\mu_0 \partial^2 P^{\rm NL} / \partial t^2$ because $w \gg \lambda$, where $\lambda$ is the wavelength of light. Therefore, we get the wave equation
\begin{equation}\label{eq:w11_wave_eq}
\frac{n^2}{c^2}\frac{\partial^2 \VF{E}}{\partial t^2} - \nabla^2 \VF{E} = -\mu_0\frac{\partial^2 \VF{P^{\rm NL}}}{\partial t^2},
\end{equation}
which has a standard wave equation on the left and a driving term on the right that leads to the generation of nonlinear effects in the crystals:
\begin{itemize}
\item Frequency doubling, or second-harmonic generation, is a nonlinear optical process in which two photons of the same frequency interact in a nonlinear medium to produce a single photon with twice the frequency (half the wavelength):
\begin{equation}
2\omega_1 = \omega_2,
\end{equation}
where $\omega_1$ is the frequency of the input wave and $\omega_2$ is the frequency of the generated second-harmonic wave. The nonlinear polarisation $\mathbf{P}^{\mathrm{NL}}$ oscillates at $\omega_2$ and acts as the source term that drives the frequency-doubled field.
\item Spontaneous parametric down conversion is a nonlinear process in which an input photon spontaneously splits into a pair of lower-energy photons while conserving energy and crystal momentum, as given by the equation
\begin{equation}
\omega_1 = \omega_2 + \omega_3, \hspace{1cm} \VF{k_1} = \VF{k_2} + \VF{k_3},
\end{equation}
where $\omega_2$ and $\omega_3$ are the angular frequencies of the produced photons, and $\VF{k_{1,2,3}}$ are wave vectors of the photons, which are not necessarily coaligned.
\item Sum-frequency generation is a nonlinear process when two input photons interact in a nonlinear crystal and produce a higher energy photon according to the equation
\begin{equation}
\omega_1 + \omega_2 = \omega_3,
\end{equation}
where $\omega_1$ and $\omega_2$ are angular frequencies of the interacting photons which make the nonlinear polarisation $\VF{P^{\rm NL}}$ oscillate at frequency $\omega_1 + \omega_2$ and generate a new wave of frequency $\omega_3$.
\item The electro-optic effect is defined as the change in the refractive index (more generally, the refractive index tensor) of a material in response to an applied external electric field. In the linear electro-optic (Pockels) regime, this can be expressed as
\begin{equation}
\Delta n \sim \xi E_{\rm ext},
\end{equation}
where $E_{\rm ext}$ is the externally applied electric field, typically applied via electrodes. The effect enables external control of the phase, polarisation, or amplitude of light propagating through the medium.
\end{itemize}

\subsubsection{Second harmonic generation}

\begin{wrapfigure}{r}{0.33\textwidth}
    \centering
    \vspace{-4mm}
    \includegraphics[width=0.31\textwidth]{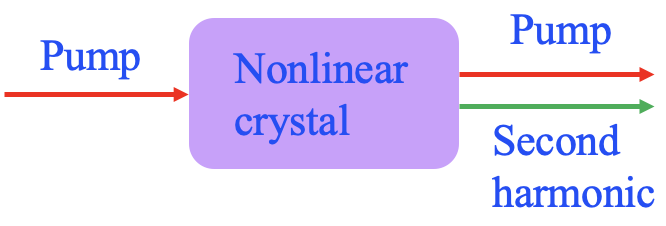} 
    \caption{Diagram of the second harmonic generation.}
    \label{fig:w11_SHG}
\end{wrapfigure}

As an example, we consider the frequency-doubling process, which was utilised in Lecture~\ref{Lecture8} in the context of optical atomic clocks and is shown in Fig.~\ref{fig:w11_SHG}. We begin with a strong pump field at frequency $\omega_0$ and generate radiation at the second-harmonic frequency $2\omega_0$. Both fields are assumed to be S-polarised and to propagate along the Z-axis. The total electric field can be written as
\begin{equation}
E(t,z) = E_1(z) e^{i (\omega_0 t - k_1 z)} + E_2(z) e^{i (2\omega_0 t - k_2 z)} + {\rm c.c.},
\end{equation}
where $k_1 = 2 \pi n_1 / \lambda_1$ and $k_2 = 2 \pi n_2 / \lambda_2$ are the wave numbers of the pump and second-harmonic fields, respectively, and $n_1$ and $n_2$ are the refractive indices of the nonlinear crystal at frequencies $\omega_0$ and $2\omega_0$.

The field amplitudes $E_1(z)$ and $E_2(z)$ are assumed to vary slowly with propagation distance (i.e.\ $|dE_{1,2}/dz| \ll |E_{1,2}|/\lambda_{1,2}$). The initial conditions are $E_1(0) = E_{10}$ for the pump field and $E_2(0) = 0$ for the second-harmonic field. The goal is to determine the fields after propagation through a crystal of length $L$, namely $E_1(L)$ and $E_2(L)$.

The components of the $P^{\rm NL}$ that oscillate at $\omega_0$ and $2\omega_0$ are given by the equation
\begin{equation}
    P^{\rm NL} = \epsilon_0 \xi E^2 = \epsilon_0 \xi \left(E_1^2 e^{i (2\omega_0 t - 2k_1 z)} + E_1^* E_2 e^{i (\omega_0 t - (k_2 - k_1) z)} \right) + {\rm c.c.}
\end{equation}

By substituting the electric field $E(t,z)$ and the nonlinear polarisation $P^{\rm NL}$ into Eq.~\ref{eq:w11_wave_eq}, and neglecting second-order derivatives $E''_{1,2}(z)$, the coupled wave equations for the pump and second-harmonic fields can be written as
\begin{equation}\label{eq:shg_main_fields}
\begin{split}
& \frac{dE_1}{dz} = -\frac{i \pi \xi}{\lambda_1 n_1} E_1^* E_2 \, e^{-i \Delta k z}, \\
& \frac{dE_2}{dz} = -\frac{i \pi \xi}{\lambda_2 n_2} E_1^2 \, e^{i \Delta k z},
\end{split}
\end{equation}
where $\Delta k = k_2 - 2 k_1 \neq 0$ because $\lambda_1 = 2\lambda_2$ but $n_1 \neq n_2$. The first equation describes the evolution (depletion) of the pump field due to the nonlinear interaction, while the second equation governs the generation and growth of the second-harmonic field. We need numerical tools to solve Eqs.~\ref{eq:shg_main_fields}. However, we solve them analytically in the approximation of a strong pump field and a weak second harmonic field. In this case, we approximate $E_1(z) \approx E_{10}$ and solve the second equation for the second harmonic field $E_2(z)$.

The phase factor $e^{i \Delta k z}$ oscillates with a period $2\pi/\Delta k$, causing the growth rate of $E_2(z)$ to alternate between positive and negative values. As a result, no significant net growth of the second-harmonic field occurs over long propagation distances. This phase-mismatch problem can be overcome using quasi-phase matching~\cite{fejer1992quasi}, in which the sign of the nonlinear coefficient $\xi$ is periodically reversed every segment of length $l_0 = \pi/\Delta k$. The length is determined by $\Delta n = n_2 - n_1$ in a particular crystal and is typically $\sim 10$\,um. This sign flip compensates for the phase slippage between the interacting waves and supports the growth of the generated second-harmonic field. In practice, quasi-phase matching is implemented using periodically poled nonlinear crystals, such as periodically poled lithium niobate (PPLN) or potassium titanyl phosphate (PPKTP), where the crystal domains are fabricated with alternating orientation to achieve the required modulation of the nonlinear response.

We first integrate Eq.~\ref{eq:shg_main_fields} over a single domain of length $l_0$ and obtain
\begin{equation}
E_2(l_0) = -\frac{\pi \xi}{\lambda_2 n_2} E_1^2 \frac{e^{i \Delta k l_0} - 1}{\Delta k}
= \frac{2\pi \xi}{\lambda_2 n_2 \Delta k} E_1^2
= \frac{2 \xi}{\lambda_2 n_2} E_1^2 l_0,
\end{equation}
where we have used the quasi-phase-matching condition $\Delta k l_0 = \pi$. By summing the contributions from all domains, taking into account the periodic sign reversal of $\xi$, the second-harmonic field grows constructively along the crystal, yielding
\begin{equation}
E_2(L) = \frac{2 \xi}{\lambda_2 n_2} E_1^2 L.
\end{equation}
This result is smaller by a factor of $\pi/2$ compared to the ideal case of perfect phase matching ($\Delta k = 0$), where the growth is fully coherent over the entire crystal length.
We can estimate the power of the second-harmonic field generated in the crystal as
\begin{equation}
P_2 \approx 2\,\epsilon_0 c n_2 S |E_2(L)|^2 \;\approx\; \frac{2}{\epsilon_0 S c n_2} \left( \frac{\xi P_1 L}{\lambda_2 n_1} \right)^2 
\;\approx\; 25\,{\rm mW} \left( \frac{40\,\mu{\rm m}}{w} \right)^2 \left( \frac{P_1}{1\,{\rm W}} \right)^2 \left( \frac{L}{3\,{\rm cm}} \right)^2,
\end{equation}
where $P_1$ is the pump power and $S = \pi w^2$ is the beam area. This expression highlights the quadratic dependence of the second-harmonic power on both the pump power and the interaction length, as well as its inverse dependence on the beam area.

\subsubsection{Spontaneous parametric down conversion}

\begin{wrapfigure}{r}{0.33\textwidth}
    \centering
    \vspace{-4mm}
    \includegraphics[width=0.31\textwidth]{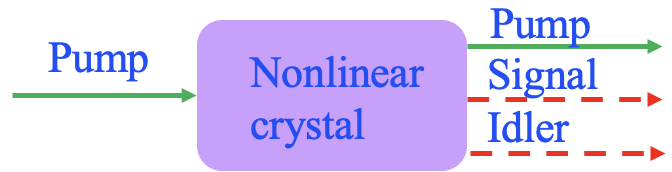} 
    \caption{Generation of entangled photon pairs.}
    \label{fig:w11_SPDC}
\end{wrapfigure}

In this section, we consider a pump field at frequency $\omega_p$ which generates pairs of entangled photons via parametric down-conversion, as shown in Fig.~\ref{fig:w11_SPDC}. The generated photons have frequencies $\omega_s$ and $\omega_i$, and are commonly referred to as the signal and idler fields, respectively. The total electric field inside the nonlinear crystal can be written as
\begin{equation}
\VF{E}(t,\VF{r}) =
\VF{E}_s\, e^{i \left(\omega_s t - \VF{k}_s\cdot \VF{r}\right)}
+
\VF{E}_i\, e^{i \left(\omega_i t - \VF{k}_i\cdot \VF{r}\right)}
+
\VF{E}_p\, e^{i (\omega_p t - k_p z)}
+ \text{c.c.},
\end{equation}
where $\VF{E}_s$ and $\VF{E}_i$ denote the complex amplitudes of the signal and idler fields, while $\VF{E}_p$ is the strong pump field at frequency $\omega_p$. The pump propagates along the Z-axis, and the signal and idler waves may propagate at non-collinear angles with respect to Z, as required by momentum conservation (phase matching) in the nonlinear medium.

In addition to quasi-phase matching via periodic poling, efficient nonlinear interaction can be achieved through intrinsic phase-matching geometries in birefringent crystals. In birefringent nonlinear crystals, light can propagate in two distinct polarisation eigenmodes: the ordinary (o) and extraordinary (e) waves. The ordinary wave experiences a refractive index that is independent of propagation direction, and the extraordinary wave experiences a direction-dependent refractive index due to the crystal anisotropy. This difference allows one to engineer phase-matching conditions by choosing appropriate propagation angles and polarisation combinations such that the interacting waves satisfy momentum conservation in nonlinear processes.

These processes are commonly classified into three types:
\begin{itemize}
\item \textbf{Type 0 (co-polarised):} all interacting waves share the same polarisation state (e.g.\ $e \rightarrow e + e$), which maximises the effective nonlinear coefficient but typically requires periodic poling to satisfy the quasi-phase-matching condition.
\item \textbf{Type I:} the pump field has one polarisation, while the signal and idler share the same orthogonal polarisation (e.g.\ $e \rightarrow o + o$ or $o \rightarrow e + e$). This configuration is commonly used due to its relatively high efficiency and more straightforward phase-matching conditions compared with Type 0.
\item \textbf{Type II:} the signal and idler have orthogonal polarisations (e.g.\ $e \rightarrow o + e$), enabling direct generation of polarisation-entangled photon pairs and making it widely used in quantum optics experiments~\cite{kwiat1995new}.
\end{itemize}

In the previous section, we considered second-harmonic generation under a Type~0 quasi-phase-matching condition. In this section, as an example, we consider Type~I phase matching, but the equations governing pair production are analogous. For simplicity, we choose the pump propagation direction such that the phase-matching condition
\begin{equation}
k_p = k_s + k_i
\end{equation}
is satisfied, where the pump is extraordinarily polarised, and the signal and idler are ordinarily polarised, i.e.\ $n_p = n_e(\omega_p)$, $n_s = n_o(\omega_s)$, and $n_i = n_o(\omega_i)$. We also assume the undepleted pump approximation, such that the pump field remains approximately constant along the crystal, $E_p(z) \approx E_p(0)$.

The components of the nonlinear polarisation $P^{\rm NL}$ that oscillate at the signal and idler frequencies are given by the equation
\begin{equation}
P^{\rm NL} = \epsilon_0 \xi \left(
E_i^* E_p\, e^{i (\omega_s t - (k_p - k_i) z)}
+
E_s^* E_p\, e^{i (\omega_i t - (k_p - k_s) z)}
\right) + \text{c.c.}
\end{equation}

Substituting into the wave equation and applying the slowly varying envelope approximation, the coupled equations for the signal and idler fields become
\begin{equation}\label{eq:opo_main_fields}
\begin{split}
\frac{dE_s}{dz} &= -\frac{i \xi \omega_s}{2 c n_s}\, E_i^* E_p \, e^{-i \Delta k z}, \\
\frac{dE_i}{dz} &= -\frac{i \xi \omega_i}{2 c n_i}\, E_s^* E_p \, e^{-i \Delta k z},
\end{split}
\end{equation}
where $\Delta k = k_p - k_s - k_i$. Under perfect phase-matching conditions ($\Delta k = 0$), the exponential factors reduce to unity, and the signal and idler fields grow coherently along the propagation direction.

By differentiating each equation over $z$ and substituting the derivative from the other equation, we find that 
\begin{equation}\label{eq:spdc_diff}
E''_s - \alpha^2 E_s =0, \hspace{1cm} E''_i - \alpha^2 E_i =0,
\end{equation}
where the coefficient, $\alpha$, is given by the equation
\begin{equation}
\alpha = \sqrt{\frac{\xi^2 \omega_s \omega_i E_p E_p^*}{4c^2n_sn_i}}.
\end{equation}

The solutions to the signal and idler field equations are given by the squeezing transformation
\begin{equation}\label{eq:opo_simple}
\begin{split}
& E_s(z) = C_1\cosh{(\alpha z)} + C_2\sinh{(\alpha z)} \\
& E_i(z) = C_3\cosh{(\alpha z)} + C_4\sinh{(\alpha z)},
\end{split}
\end{equation}
where $C_{1,2,3,4}$ are constants of integration, which we find from the initial conditions
\begin{equation}
C_1 = E_s(0), \hspace{5mm} C_2 = \frac{-iE_p}{\sqrt{E_pE_p^*}}\sqrt{\frac{\omega_s n_i}{\omega_i n_s}}E_i^*(0), \hspace{5mm} C_3 = E_i(0), \hspace{5mm} C_4 = \frac{-iE_p}{\sqrt{E_pE_p^*}}\sqrt{\frac{\omega_i n_s}{\omega_s n_i}}E_s^*(0).
\end{equation}
We choose the pump phase such that $E_p = i |E_p|$, and normalise the signal and idler fields as
\begin{equation}\label{eq:w11_E_norm}
E_{s,i} = \sqrt{\frac{\hbar \omega_{s,i}}{2 \epsilon_0 n_{s,i} V}}\, a_{s,i},
\end{equation}
where $V$ is the volume of the interaction. This volume cancels out in the final transformation equations but is required to ensure that the fields $a_{s,i}$ are dimensionless and correspond to the ladder operators in quantum mechanics.

We acknowledge that our normalisation of the electric fields given by Eq.~\ref{eq:w11_E_norm} is not a standard one in quantum optics and is introduced here for simplicity. Instead of considering a continuum of the signal and idler modes at a broad range of frequencies, we adopt a single-mode normalisation. We assume that the nonlinear interaction is effectively restricted to well-defined modes for the signal and idler fields that are selected by the pump frequency and crystal phase matching. Under this approximation, the field can be projected onto a single effective mode with volume $V$, allowing the continuous frequency to be replaced by a discrete harmonic oscillator description.

The squeezing transformation for the normalised fields is then given by the equations
\begin{equation}\label{eq:opo_a}
\begin{split}
& a_s(L) = a_s(0)\cosh{r} + a_i^*(0)\sinh{r} \\
& a_i(L) = a_i(0)\cosh{r} + a_s^*(0)\sinh{r},
\end{split}
\end{equation}
where the squeezing parameter $r$ is given by the equation
\begin{equation}
r = \alpha L  = \sqrt{\frac{\xi^2 \omega_s \omega_i E_p E_p^*}{4c^2n_sn_i}} L.
\end{equation}

So far, we obtained the squeezing transformation given by Eq.~\ref{eq:opo_a} for the classical field amplitudes. To obtain the corresponding quantum description~\cite{walls2008quantum}, these amplitudes are promoted to operators via the correspondence
\begin{equation}
a_s \rightarrow \hat{a}_s, \hspace{1cm} a_i \rightarrow \hat{a}_i, \hspace{1cm}
a_s^* \rightarrow \hat{a}_s^\dagger, \hspace{1cm} a_i^* \rightarrow \hat{a}_i^\dagger,
\end{equation}
where $\hat{a}_s, \hat{a}_i$ and their Hermitian conjugates are the annihilation and creation operators for the signal and idler modes, respectively. These operators obey the bosonic commutation relations
\begin{equation}
[\hat{a}_s, \hat{a}_s^\dagger] = 1, \hspace{1cm} [\hat{a}_i, \hat{a}_i^\dagger] = 1, \hspace{1cm}
[\hat{a}_s, \hat{a}_i] = [\hat{a}_s, \hat{a}_i^\dagger] = 0,
\end{equation}
which ensure the correct quantum statistics of the electromagnetic field. The introduction of these operators is required to describe intrinsically quantum phenomena such as photon-number quantisation, spontaneous emission, and the generation of nonclassical states of light, including entangled photon pairs and squeezed states, which have no classical analogue.

The operators act as ladder operators of the quantum harmonic oscillator on the Fock states according to the equations
\begin{equation}
\hat{a}_{s,i}\ket{n}_{s,i} = \sqrt{n_{s,i}}\,\ket{n-1}_{s,i}, \hspace{1cm}
\hat{a}_{s,i}^\dagger\ket{n}_{s,i} = \sqrt{n_{s,i}+1}\,\ket{n+1}_{s,i},
\end{equation}
where $\ket{n}_{s,i}$ denotes the state with $n_{s,i}$ photons in the signal or idler mode.

With this quantisation, the classical squeezing transformation becomes a Bogoliubov transformation~\cite{bogoliubov1958new} acting on the operators,
\begin{equation}
\begin{split}
\hat{a}_s(L) &= \hat{a}_s(0)\cosh r + \hat{a}_i^\dagger(0)\sinh r, \\
\hat{a}_i(L) &= \hat{a}_i(0)\cosh r + \hat{a}_s^\dagger(0)\sinh r,
\end{split}
\end{equation}
which is generated by the unitary two-mode squeezing operator
\begin{equation}
\hat{S}(r) = \exp \left[ r \left( \hat{a}_s^\dagger \hat{a}_i^\dagger - \hat{a}_s \hat{a}_i \right) \right].
\end{equation}
This can be verified by direct substitution into the Heisenberg evolution,
\begin{equation}
\hat{a}_{s,i}(L) = \hat{S}^\dagger(r)\, \hat{a}_{s,i}(0)\, \hat{S}(r),
\end{equation}
which reproduces the Bogoliubov transformation above. The transformation describes the coherent creation of photon pairs in the signal and idler modes while preserving the bosonic commutation relations.

We send a strong pump field into the crystal while the signal and idler modes are initially in the vacuum state, $\ket{0,0}$. In the weak-squeezing regime, \(r \ll 1\), the action of the squeezing operator on the vacuum produces the output state according to
\begin{equation}
\hat{S}(r)\ket{0,0} \approx \left(1 + r \left( \hat{a}_s^\dagger \hat{a}_i^\dagger - \hat{a}_s \hat{a}_i \right)\right)\ket{0,0} \approx \ket{0,0} + r\, \ket{1,1},
\end{equation}
where \(\ket{n,n}\) denotes a state with \(n\) photons in both the signal and idler modes. This expression shows explicitly that, to leading order, the nonlinear interaction generates correlated photon pairs.

The squeezing transformation generates photon pairs whose quantum state depends on the degrees of freedom selected by the phase-matching configuration of the nonlinear interaction. While the operator \(S(r)\) produces correlated pairs of excitations, the physical encoding of these pairs, such as polarisation, spatial mode, or time-bin, is determined by how phase matching constrains the allowed modes. For example, in Type II phase matching, the signal and idler photons are generated in orthogonal polarisation states, \((\ket{HV} + \ket{VH})/\sqrt{2}\), as we considered in the case of quantum key distribution in free-space. In time-bin schemes, the same squeezing process can produce superpositions, such as \((\ket{ee} + \ket{ll})/\sqrt{2}\), corresponding to early and late emission times, as we considered in the case of quantum key distribution in fibres.

Since the amplitude for generating a single photon pair is \(r\), the corresponding probability per effective interaction window (or per pump pulse in the pulsed case) is given by the squared modulus of this amplitude, \(P_{1\,\mathrm{pair}} = |r|^2\). As an illustrative example, for a typical spontaneous parametric down-conversion process in a $\mathrm{BBO}$ crystal pumped at $775\,\mathrm{nm}$ with pulse energy of $1\,\mathrm{nJ}$ and pulse duration of $200\,\mathrm{fs}$, one obtains a squeezing parameter of order \(r \sim 10^{-1}\), corresponding to a single-pair generation probability \(r^2 \sim 10^{-2}\).

\subsection{Quantum internet}

The quantum internet~\cite{kimble2008quantum} aims to enable the distribution of entangled states over long distances and form the foundation for distributed quantum computing. A central challenge in achieving this goal is the exponential loss of photons in optical fibres, where transmission scales as $\exp(-\alpha L)$ with distance $L$, as we discussed in Lecture~\ref{Lecture10}. This exponential attenuation significantly reduces the probability of successfully transmitting quantum information, leading to an exponentially decreasing communication rate, $C$, between Alice and Bob according to the equation
\begin{equation}
C = C_0 \exp(-\alpha L),
\end{equation}
where $C_0$ is the rate of entangled photon production. The exponential loss makes direct long-distance quantum communication impractical.

\begin{figure}[t!]
\centering
\includegraphics[height=4.5cm]{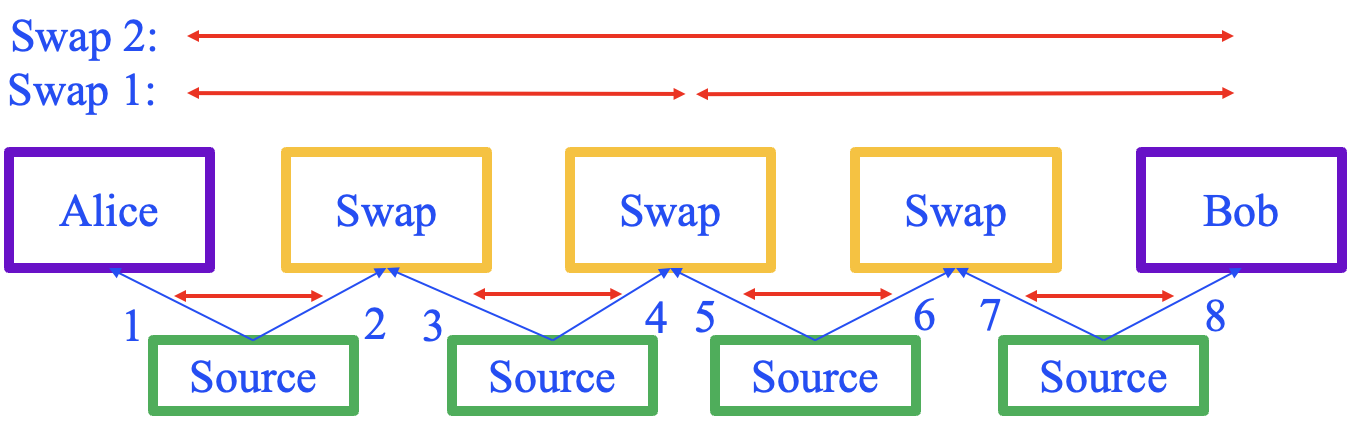} 
\caption{Quantum channel with $N=4$ sources of entangled photons. Each link has a transmission probability of $\frac{1}{2}$.}
\label{fig:w11_inet}
\end{figure}

To overcome this limitation, the key objective is to transform this exponential scaling into a polynomial scaling with distance~\cite{briegel1998quantum}. This requires the development of quantum repeaters, which divide the communication channel into shorter segments, as shown in Fig.~\ref{fig:w11_inet}, and use entanglement swapping to extend quantum correlations over long distances. Essential components of such repeaters include quantum memories, which can store quantum states while waiting for successful entanglement generation in neighbouring links, and quantum non-demolition measurements, which enable the detection of photons without destroying their quantum state.

As an illustrative example of the exponential loss problem, we consider a quantum communication channel shown in Fig.~\ref{fig:w11_inet}, where the total link is divided into $2N = 8$ shorter segments. Each segment has a transmission probability of $P_0 = \exp(-\alpha L/8) = \tfrac{1}{2}$. If only a single source of entangled photons is used across the entire link, the probability of a successful joint detection by Alice and Bob is given by
\begin{equation}\label{eq:w11_prob_0}
P = P_0^{2N} = \left(\tfrac{1}{2}\right)^8 = \frac{1}{256},
\end{equation}
where $P$ represents the probability of simultaneous photon detection at both ends of the channel.

Next, we introduce $N=4$ sources of entangled photon pairs along the channel. In the first swapping step, joint measurements are performed on photon pairs $(2,3)$ and $(6,7)$, which project photons $(1,4)$ and $(5,8)$ into entangled states. In the second stage, a joint measurement on photons $(4,5)$ extends the entanglement further, resulting in photons $(1,8)$ becoming entangled. This establishes the required long-distance entanglement between Alice and Bob. 

In this section, we evaluate the probability of simultaneous photon detection by Alice and Bob in the presence of loss, and discuss enabling technologies for quantum repeaters, including quantum memories and quantum non-demolition measurements.

\subsubsection{Entanglement swapping}

\begin{wrapfigure}{r}{0.33\textwidth}
    \centering
    \vspace{-4mm}
    \includegraphics[width=0.31\textwidth]{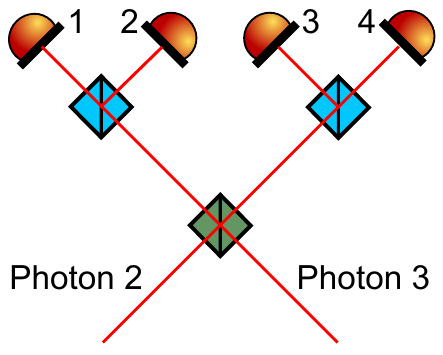} 
    \caption{Entanglement swapping with linear optics and photodetectors.}
    \label{fig:w11_swap}
\end{wrapfigure}

In entanglement swapping, two independent entangled pairs are converted into a longer-distance entangled pair via a joint measurement. One of the simplest entanglement swapping schemes involves a linear-optical implementation, as shown in Fig.~\ref{fig:w11_swap}. This is achieved using beam splitters, phase shifters, and single-photon detectors. However, due to the lack of deterministic photon-photon interactions, such schemes are intrinsically probabilistic, with a maximum success probability of $50\%$ when using only passive linear optics.

As an example, consider polarisation entanglement, which can be utilised in free-space links as we discussed above. We start with the entangled photon pairs $(1,2)$ and $(3,4)$, which are in the quantum states gievn by the equations
\begin{equation}
\ket{\psi}_{1,2} = \frac{\ket{HV} + \ket{VH}}{\sqrt{2}} \hspace{1cm} \ket{\psi}_{3,4} = \frac{\ket{HV} + \ket{VH}}{\sqrt{2}} 
\end{equation}
and the total state is given by the equation
\begin{equation}\label{eq:w11_1234}
\ket{\psi}_{1,2,3,4} = \frac{\ket{HVHV} + \ket{HVVH} + \ket{VHHV} + \ket{VHVH}}{2}.
\end{equation}

We then implement the measurement on photons $(2,3)$ with a nonpolarising 50/50 beam splitter, two polarising beam splitters, and four photodetectors, as shown in Fig.~\ref{fig:w11_swap}.

A 50:50 non-polarising beam splitter acts on the spatial modes of the incoming photons while leaving their polarisation unchanged. For two photons with identical polarisation (e.g.\ both horizontally polarised), the input state can be written as $\hat{a}_H^\dagger \hat{b}_H^\dagger \ket{0, 0}$, where $\hat{a}_H$ and $\hat{b}_H$ are the ladder operators for photons 2 and 3, respectively. Similar to the electric fields, as we discussed in Lecture~\ref{Lecture7}, a 50/50 beam splitter transforms the input mode operators according to
\begin{equation}
\hat{a}_{H,V}^\dagger \rightarrow \frac{1}{\sqrt{2}}(\hat{c}_{H,V}^\dagger + \hat{d}_{H,V}^\dagger), \hspace{3mm}
\hat{b}_{H,V}^\dagger \rightarrow \frac{1}{\sqrt{2}}(\hat{c}_{H,V}^\dagger - \hat{d}_{H,V}^\dagger),
\end{equation}
where $\hat{c}_{H,V}$ and $\hat{d}_{H,V}$ are the ladder operators for the transmission ports of the beam splitter for photons in the H and V polarisation states, respectively.

Applying this transformation, the output state of photons $(2,3)$ becomes
\begin{equation}
\frac{1}{2}(\hat{c}_H^\dagger + \hat{d}_H^\dagger)
(\hat{c}_H^\dagger - \hat{d}_H^\dagger)\ket{0,0} = 
\frac{1}{2}\left(\left(\hat{c}_H^\dagger\right)^2 - \left(\hat{d}_H^\dagger\right)^2\right)\ket{0,0} = 
\frac{1}{\sqrt{2}}\left(\ket{HH,0} - \ket{0,HH}\right),
\end{equation}
where the terms correspond to both photons exiting the same output port. The coincidence term $\ket{H,H}$ vanishes due to destructive interference, resulting in photon bunching. This is the well-known Hong--Ou--Mandel effect~\cite{Hong1987HOM} and arises from the indistinguishability of the photons.

In contrast, for photons $(2,3)$ with orthogonal polarisations, for example $\ket{HV}$, the input state $\hat{a}_H^\dagger \hat{b}_V^\dagger \ket{0}$ evolves into a superposition of four possible output configurations
\begin{equation}
\begin{split}
\frac{1}{2}(\hat{c}_H^\dagger + \hat{d}_H^\dagger)
(\hat{c}_V^\dagger - \hat{d}_V^\dagger)\ket{0,0} =& 
\frac{1}{2}(\hat{c}_H^\dagger \hat{c}_V^\dagger - \hat{c}_H^\dagger \hat{d}_V^\dagger + \hat{d}_H^\dagger \hat{c}_V^\dagger - \hat{d}_H^\dagger \hat{d}_V^\dagger)\ket{0,0} \\
=&\frac{1}{2}\left(
\ket{HV, 0}
- \ket{H, V}
+ \ket{V, H}
- \ket{0,HV}
\right).
\end{split}
\end{equation}
In this case, no destructive interference occurs because the photons have different polarisation states, as we discussed in Lecture~\ref{Lecture7}.

The polarising beam splitters separate the photons according to their polarisation and may entangle photons $(1,4)$. This leads to the following table, which summarises all possible input states of photons $(2,3)$, the corresponding detection outcomes at the photodetectors, and the resulting post-measurement state of photons $(1,4)$ after the measurement-induced projection.
\begin{table}[h]
\centering
\begin{tabular}{c c c c}
 Photon 2 & Photon 3 & Photodetector measurements & Photons 1,4 \\ 
 $\ket{H}$ & $\ket{H}$ & $(2,0,0,0)$ or $(0,0,0,2)$ &  $\ket{VV}$\\ 
  
 $\ket{H}$ & $\ket{V}$ & $(1,1,0,0)$ or $(1,0,1,0)$ or $(0,1,0,1)$ or $(0,0,1,1)$ & $\frac{\ket{HV} + \ket{VH}}{\sqrt{2}}$ \\
  
 $\ket{V}$ & $\ket{H}$ & $(1,1,0,0)$ or $(1,0,1,0)$ or $(0,1,0,1)$ or $(0,0,1,1)$ & $\frac{\ket{HV} + \ket{VH}}{\sqrt{2}}$ \\
  
 $\ket{V}$ & $\ket{V}$ & $(0,2,0,0)$ or $(0,0,2,0)$ & $\ket{HH}$
\end{tabular}
\caption{Input states of photons $(2,3)$ and possible measurements with photodetectors $(1-4)$.}
\label{table:w11_swap}
\end{table}

During the experiment, the state of photons $(2,3)$ is not directly accessible; instead, only the photodetector click patterns are observed. If the measurement outcome corresponds to $(2,0,0,0)$ or $(0,0,0,2)$ photons, then the input state of photons $(2,3)$ must be $\ket{HH}$, and consequently photons $(1,4)$ collapse to $\ket{VV}$ according to Eq.~\ref{eq:w11_1234}. Similarly, if the detected pattern is $(0,2,0,0)$ or $(0,0,2,0)$, then photons $(2,3)$ were in state $\ket{VV}$, resulting in photons $(1,4)$ being projected onto $\ket{HH}$. In both cases, the resulting state of photons $(1,4)$ is separable, and no entanglement swapping is achieved. However, if the detection outcomes correspond to $(1,1,0,0)$, $(1,0,1,0)$, $(0,1,0,1)$, or $(0,0,1,1)$, then photons $(2,3)$ are projected onto the subspace spanned by $\ket{HV}$ and $\ket{VH}$, leading to an entangled state of photons $(1,4)$ as shown in Table~\ref{table:w11_swap}. Since 4 out of the 8 possible detection outcomes result in successful entanglement swapping, the overall success probability is $\tfrac{1}{2}$.

To overcome the limitation of linear optics in entanglement swapping, alternative approaches have been proposed and demonstrated~\cite{sangouard2011quantum}, including the use of nonlinear optical interactions, matter–light interfaces (such as atoms, ions, or solid-state emitters), and measurement-based schemes assisted by quantum memories. These platforms can, in principle, enable deterministic or near-deterministic entanglement swapping, but often at the cost of increased experimental complexity.

The entanglement swapping alone is not sufficient to improve the quantum key distribution rate. Even assuming a deterministic production of entangled photons and entanglement swapping with a 100\% success rate, the probability that each operation in step 1 of Fig.~\ref{fig:w11_inet} succeeds is $\frac{1}{4}$ per swap due to the transmission losses. Since the photon transmission through the channel is independent and random, the overall coincidence probability factorises over all links. In addition, the probability that photons $(1,4,5,8)$ are successfully transmitted through their respective channel segments is $\frac{1}{2^4}$. Therefore, the total success probability is given by
\begin{equation}\label{eq:w11_P_swap}
P_{1} = \frac{1}{4^2}\,\frac{1}{2^4} = \frac{1}{256},
\end{equation}
which is identical to the result obtained without entanglement swapping in Eq.~\ref{eq:w11_prob_0}.

\subsubsection{Quantum memory}

The primary purpose of the quantum memory is to store quantum states to synchronise probabilistic events across a distributed network. Since entanglement generation and swapping processes are inherently probabilistic, quantum memories allow successful events in different segments of a link to be “held” until neighbouring segments also succeed. The process removes the need for simultaneous transmission success, required in Eq.~\ref{eq:w11_P_swap}. Promising physical platforms for quantum memory~\cite{Lvovsky2009quantummemory} include atomic ensembles, rare-earth-doped crystals, cold atoms, and solid-state spin systems, each offering different trade-offs between storage time, efficiency, bandwidth, and operational complexity.

In the presence of quantum memory, the production of an entangled pair does not have to be deterministic, and the entanglement swapping operations associated with measurements on photons $(2,3)$ and $(6,7)$ do not need to occur simultaneously, since successfully generated entangled pairs can be stored until neighbouring links are ready. As a result, the success probability of step 1 reduces to a single swapping probability of $\frac{1}{4}$. As in the previous case, the probability that photons $(1,4,5,8)$ are successfully transmitted through their respective channel segments is $\frac{1}{2^4}$. Therefore, the total success probability is given by
\begin{equation}
P_{2} = \frac{1}{4}\,\frac{1}{2^4} = \frac{1}{64},
\end{equation}
which represents an improvement in the success rate by a factor of 4 compared to Eq.~\ref{eq:w11_prob_0}.

More generally, for a quantum repeater architecture consisting of $N$ entangled photon-pair sources assisted by quantum memories, the protocol requires $\log_2(N)$ entanglement-swapping operations to extend entanglement across the full link. The corresponding total success probability can be expressed as
\begin{equation}
P_{2}(N) = \exp \left(-\alpha \frac{L}{N} \big(\log_2(N) + 1\big) \right),
\end{equation}
which is significantly larger than the direct transmission probability $\exp(-\alpha L)$, demonstrating the exponential suppression of loss achieved by segmenting the channel and using quantum repeaters.

\subsubsection{Quantum non-demolition measurements}

Quantum non-demolition (QND) measurements allow one to measure the presence of a photon without destroying its quantum state. For example, we may measure the presence of a photon without observing its polarisation state or arrival time and preserve coherence for subsequent quantum operations. In quantum communication, QND measurements enable the verification of successful photon transmission and entanglement distribution without absorbing the photon. The process is needed for synchronising probabilistic processes in quantum repeaters. One prominent example is a Kerr-nonlinearity-based QND interaction, where a signal photon induces a phase shift on a strong probe beam because the index of refraction depends on the total light intensity according to the equation $n_1 + n_2 I$, as we considered in Lecture~\ref{Lecture10}. The presence of a single photon thus produces a measurable phase shift in the probe field and leaves the signal photon intact. 

Promising implementations of QND measurements~\cite{Grangier1998QNDreview} include cavity systems, where single atoms coupled to a high-finesse Fabry–Perot resonator induce a state-dependent phase shift on a probe field. Another important platform is atomic ensembles, where collectively enhanced light–matter interactions enable nondestructive readout of collective spin or stored photonic states via probe beams.

In the presence of entanglement swapping, quantum memories, and quantum non-demolition measurements, the success probability of establishing entanglement between Alice and Bob can be further improved. Since QND measurements allow us to witness the successful transmission of photons at each station, entanglement swapping in step 1 is performed only when all four photons $(1,2,3,4)$ and, independently, $(5,6,7,8)$ have been successfully transmitted through their respective lossy channel segments. Because quantum memories remove the requirement for simultaneous successful events, the swapping operations in step 1 do not need to be coincident. The resulting total success probability is therefore
\begin{equation}
P_{3} = \frac{1}{2^4} = \frac{1}{16},
\end{equation}
and, for a general number of segments $N \gg 1$, this scaling becomes
\begin{equation}
P_{3}(N) = \exp\left(-2 \alpha \frac{L}{N} \right) \approx 1 - 2 \alpha \frac{L}{N}.
\end{equation}
This represents a significant improvement compared to Eq.~\ref{eq:w11_prob_0}, reducing the exponential loss behaviour to a polynomial scaling with distance.

The above scaling assumes idealised conditions, namely perfect entanglement swapping, perfect quantum memories, and ideal quantum non-demolition measurements with unit efficiency. In this limit, all successfully heralded events are assumed to be stored and processed without loss or error, and the only remaining source of inefficiency arises from the channel losses. In practical implementations, however, finite memory coherence times, non-unit retrieval efficiencies, detector inefficiencies, and imperfect entanglement swapping reduce the overall success probability and introduce additional loss channels~\cite{sangouard2011quantum}. These imperfections typically degrade the ideal polynomial scaling, leading to a lower effective communication rate, although the repeater architecture can still provide a significant advantage over direct transmission in realistic parameter regimes. This is an active area of research.

%% file: exercises.tex
\section{Non-assessed problems}
\label{Problems}

\section*{Week 1}

\textbf{Problem 1}

An imaging system consists of a lens and a recording surface. We need to take an image of an object located $d_1 = 5$\,m away from the lens. What should be the distance from the lens to the recording surface and the focal length of the lens to image the object with a magnification $m = -0.2$?

\textbf{Problem 2}

Find the magnification factor of a microscope that consists of two lenses if $d_1 = 6$\,cm, $L = 30.9$\,cm, $f_1 = 5$\,cm, $f_2 = 1$\,cm.

\begin{figure}[h]
\centering
\includegraphics[width=0.6\columnwidth]{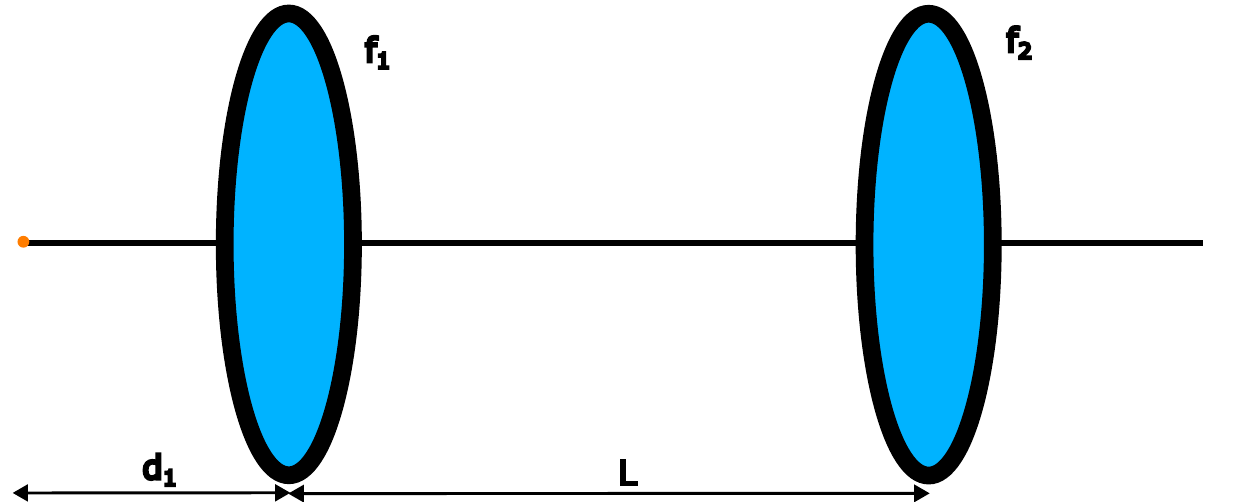} 
\end{figure}

\textbf{Problem 3}

A person can distinguish two dots on a wall if these dots are separated by no more than 2 cm and the wall is 10\,m away. The person’s iris has a diameter of 4\,mm during the test. Estimate the glass prescription in diopters required for the person.

\textbf{Problem 4}

An optical telescope observes the Earth in the visible band from a low orbit (height of 500 km) and consists of a lens and a sensing element that are separated by 30 cm.

a) Estimate the diameter of the lens required to distinguish cars on Birmingham roads.

b) The telescope is mounted on an aluminium frame with an index of thermal expansion of $2 \times 10^{-5}$ / K. To what precision should the temperature of the frame be controlled to achieve the appropriate depth of focus?

\textbf{Problem 5}

The James Webb telescope was launched on Dec 25, 2021, to observe the most distant events in the universe at infrared wavelengths. Calculate the angular resolution of the telescope at 10 um given its diameter of 6.5 m.

\newpage

\section*{Week 2}

\textbf{Problem 1}

A Starlink dish consists of a few thousand emitters of radio waves at frequencies around 10 GHz. Estimate the timing precision of emitters required to broadcast data to Starlink satellites that move along the sky.

\textbf{Problem 2}

Show that if the incident wave on the circular aperture is planar, then the expression for the electric field in polar coordinates is given by the equation
\begin{equation}
    E(\theta) \sim \frac{J_1(\xi)}{\xi}, \hspace{1cm} \xi=\frac{\pi D \sin(\theta)}{\lambda} \approx \frac{\pi D \theta}{\lambda}
\end{equation}
where $\theta = \sqrt{x^2 + y^2}/z$ is the angle at which point $x,y$ is seen from the aperture relative to the Z-axis and $J_1$ is the Bessel function of the first kind.

\textbf{Problem 3}

Find the minimal beam spot size in focus of a lens with a diameter of 1 cm and a focal length of 5 cm. Assume a wavelength of 500 nm.

\textbf{Problem 4}

Prove the following property of the Fourier Transform: inversion of the coordinate,
\begin{equation}
   {\rm FT}(f(-x)) = \tilde{f}(-k).
\end{equation}

\textbf{Problem 5}

Compute the integral
\begin{equation}
I(k) = \int\limits_{-A}^{A} \sin(k_1 x + \phi) \sin(kx) dx
\end{equation}
and explore the final answer for different k for $A \rightarrow \infty$.

\newpage

\section*{Week 3}

\textbf{Problem 1}

An optical system consists of a lens and a sensing surface. The diameter of the lens is 3\,cm, and its focal length is 5\,cm. Find the maximum pixel size required to achieve the diffraction-limited angular resolution for visible light (with wavelength in the range from 380\,nm up to 780\,nm).

\textbf{Problem 2}

Consider the human visual system as a simplified model of colour perception:

a) If an equal number of S- and L-cones are activated, what colour is perceived?

b) If an equal number of L- and M-cones are activated, what colour is perceived?

\textbf{Problem 3}

a) The bandgap of AgBr is 2.5 eV. Estimate the maximum wavelength of light that can excite an electron from the valence to the conducting band.

b)Estimate the number of electrons in the conducting band due to the thermal energy per grain.

c) A photon excited an electron to the conducting band. Estimate the spread of the electron’s wavefunction in the conducting band.

\textbf{Problem 4}

Estimate the responsivity of a silicon pn-junction (photodiode or camera pixel) in units of A/W for different wavelengths. Assume that one photon with energy larger than silicon’s bandgap of 1.1 eV promotes one electron to the conductive band.

\textbf{Problem 5}

If you sample the signal
\begin{equation}
V(t) = \cos(20\pi \times t)
\end{equation}
with a sampling time of 1\,sec starting at $t=0$, what values will you get?

\textbf{Problem 6}

Derive the computational acceleration of the compression process of a 256 by 256 image due to 8 by 8 block splitting.

\textbf{Problem 7}

A telescope with a primary mirror diameter of 5 m and a focal length of 10 m takes a picture of stars in the Andromeda galaxy (2.5 million l.y. away from Earth). Estimate the signal-to-noise ratio for an image of a Sun-like star from the galaxy if an exposure time is 1 sec. The telescope camera uses a silicon CMOS detector with a pixel size of 1\,um, carrier concentration of $10^7$\,cm$^{-3}$, carrier generation time of 10\,us, and thickness of the depletion layer of 1\,um. Assume the Sun’s luminosity near Earth is 1350\,W/m$^2$, the separation between Earth and Sun is 150 million km, the diameter of the Sun is 1.4 million km, average separation between stars in the galaxy is 5 l.y.

\newpage

\section*{Week 4}

\textbf{Problem 1}

An X-ray telescope with a primary mirror diameter $D = 30$\,cm, a focal length of 3\,m, and pixel size of 1\,um observes two sources in the sky separated by 10 nrad:

Source A: energy = 5\,keV, flux = 100\,photons m$^{-2}$ s$^{-1}$.

Source B: energy = 30\,keV, flux = 50\,photons m$^{-2}$ s$^{-1}$.

How can the telescope resolve these sources? 

\textbf{Problem 2}

Optical and infrared cameras have the same apertures. Compare the power of light hitting these two cameras when the optical camera takes a picture of a 10\,W light bulb and the infrared camera takes a picture of a person. The total area of the person is 1.9\,m$^2$. Assume that both objects are 10\,m away from the cameras.

\textbf{Problem 3}

Compare the size of the M87 black hole with the resolution of the Einstein Horizon Telescope. The mass of the M87 black hole is $6.5 \times 10^9 M_\odot$ and is 55 million light-years away from us. The telescope operates at a wavelength of 1.3\,mm and has telescopes all over Earth.

\textbf{Problem 4}

The synthetic aperture radar is installed on a satellite with an orbit height of 500\,km and operates with a wavelength of 3 cm. Estimate the required exposure time of a particular Earth surface element to achieve a spatial resolution of 1\,m. 

\newpage

\section*{Week 5}

\textbf{Problem 1}

Estimate the power of the Sun's light that comes out after spatial and spectral filtering with a pinhole and brazing grating if we need to achieve the relative spectral width
of $10^{-3}$. Assume that the solar irradiance is 1350\,W/m$^2$
and peaks at 500\,nm. The separation between Earth and Sun is 150 million km, and the diameter of the Sun is 1.4 million km.

\textbf{Problem 2}

Consider a monochromatic beam with an unknown
transverse profile in the z=0 plane. In the Fraunhofer
(far-field) limit, the diffraction pattern is observed to
have the same functional form as the beam at the
aperture, apart from scaling and a phase factor. What is
the transverse profile of the beam?

\textbf{Problem 3}

Find $d_2$ and $w_f$ assuming $z_0 = \frac{\pi w_0^2}{\lambda} \gg f$.

\begin{figure}[h]
\centering
\includegraphics[width=0.7\columnwidth]{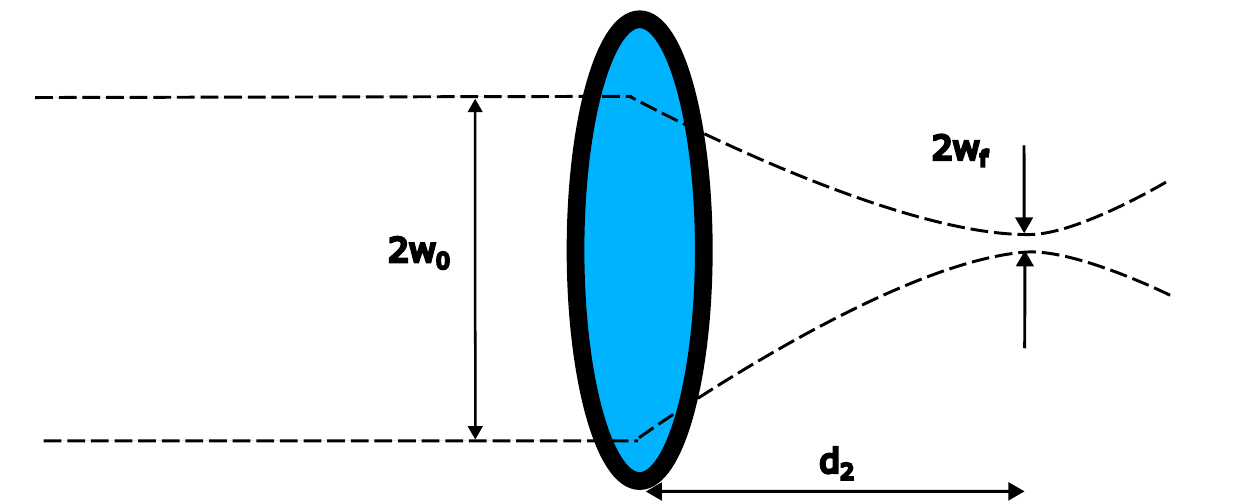} 
\end{figure}

\textbf{Problem 4}

Find the beam size and its curvature, assuming the laser wavelength of 1064\,nm and $w_0 = 200$\,um.

\begin{figure}[h]
\centering
\includegraphics[width=0.8\columnwidth]{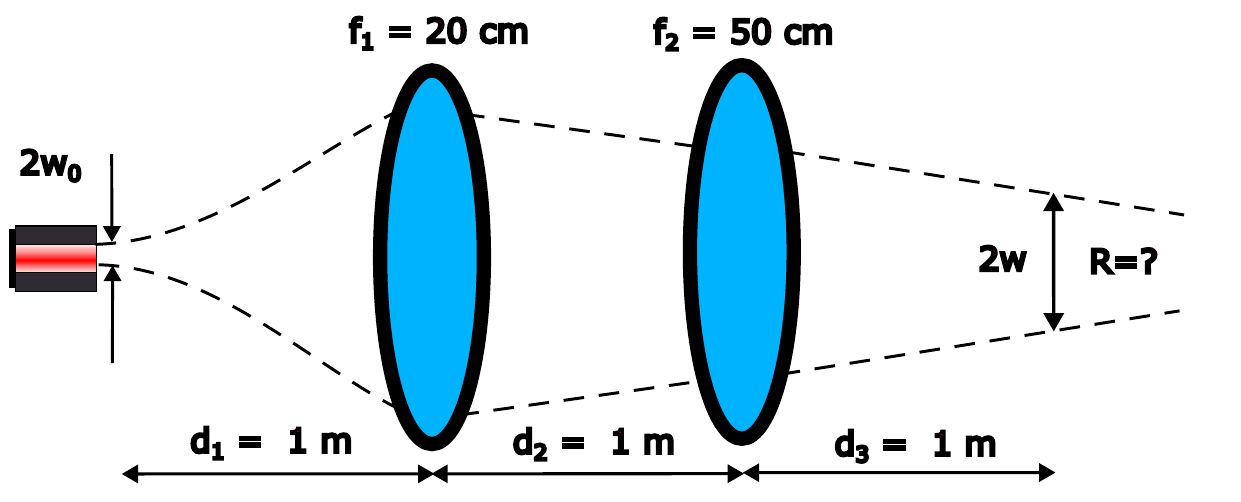} 
\end{figure}

\textbf{Problem 5}

a) An iron sheet of thickness $h = 1$\,mm is cut with a CO$_2$ laser beam ($\lambda = 10.6$\,um). The beam power is 10\,W. If the lens focal length $f=2.5$\,cm and the separation between the laser and the sheet is $d_1 + d_2 = 10$\,cm then find $d_1$ that minimises the beam size $w$. 

b) Estimate the cutting speed if the iron melting and evaporation temperatures are $T_m = 1810$\,K and $T_v = 3160$\,K, its specific heat capacities in the solid and liquid forms are $C_s = 449$\,J/kg/K and $C_l = 900$\,J/kg/K and latent heat of vaporization is $L_v = 6.8$\,MJ/kg and is much larger than the latent heat of melting. The iron density $\rho = 7870$\,kg/m$^3$.

\begin{figure}[h]
\centering
\includegraphics[width=0.8\columnwidth]{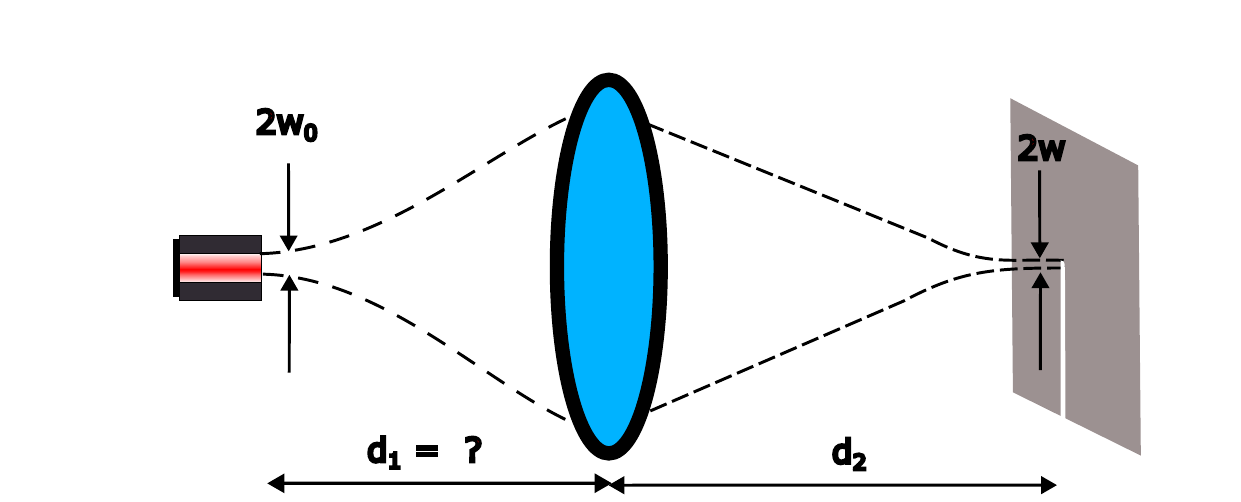} 
\end{figure}

\textbf{Problem 6}

A green laser beam (lambda = 532\,nm) is fully enclosed on an optical table apart from a length of 1\,cm where the beam radius is 1\,mm. The beam is observed from a distance of 3\,m. Estimate the degree of coherence of the laser beam at the observational point, both temporal and spatial.

\textbf{Problem 7}

Calculate the ratio of axial and radial stiffnesses of a laser trap with a beam waist $w_0$.

\newpage

\section*{Week 6}

\textbf{Problem 1}

Atoms are in a thermal equilibrium with thermal radiation inside a box. Assuming atoms have two energy levels, prove that the coefficients of spontaneous absorption and emission are identical, $B_{12} = B_{21}$, and find the rate of spontaneous emission $A_{21}$.

\textbf{Problem 2}

Estimate the output power of an Nd:Yag laser if it is powered by sunlight collected with a lens. The diameter of the lens is 30\,cm and the sunlight intensity is 1350\,W/m$^2$. Assume that 0.1\% of sunlight photons excite Nd ions to the lasing transition. Assume that the laser operates in the saturated regime: the pumping rate equals the induced emission rate.

\textbf{Problem 3}

Estimate the energy states of an electron in the quantum well formed by AlGaAs (bandgap of 1.8\,eV) and InGaAs (bandgap of 1\,eV) if the thickness of the InGaAs layer is 10\,nm.

\newpage

\section*{Week 7}

\textbf{Problem 1}

Two monochromatic waves propagate at an angle 1 degree relative to each other and interfere on a screen. A camera tries to resolve the fringes from a distance of 5\,m away from the screen. What should be the diameter of the camera lens to resolve the fringes?

\textbf{Problem 2}

Calculate the quantum-limited resolution of a Michelson interferometer with the following parameters: operating point $x_0 = 0.03 \lambda$, where $\lambda = 1064$\,nm, measurement time $\tau = 1$\,msec, and the input laser power is $P_{\rm in} = 1$\,W.

\textbf{Problem 3}

Find the coherence length of a laser with a linewidth of 1\,MHz. What condition should the Michelson interferometer satisfy to allow precision measurements with such a laser?

\newpage

\section*{Week 8}

\textbf{Problem 1}

Find the electric field at the antisymmetric port of the Michelson interferometer by propagating complex fields.

\textbf{Problem 2}

Find the maximum and minimum power amplification factors in a resonator with two identical mirrors with a power transmissivity of $T=10^{-5}$.

\textbf{Problem 3}

Prove that the product of the cavity power build-up factor and its FWHM is a constant (independent of the mirror transmissivity).

\textbf{Problem 4}

Show that the sum of the reflected and transmitted power equals the input laser power for all round-trip phases. Why does this happen?

\textbf{Problem 5}

Calculate the shot noise-limited resolution of a Fabry-Perot interferometer with identical mirrors of power transmissivity of T=10\,ppm. Input laser power is 1\,W, wavelength is 1064\,nm, measurement time is 1\,msec.

\textbf{Problem 6}

Find the amount of light transmitted through an uncoated glass-air interface at normal incidence. Assume the air index of refraction $n_1 = 1$ and that of glass $n_2 = 1.45$.

\textbf{Problem 7}

Light is incident on a glass plate from the air. Find the incident angle (Brewster’s angle) when all light is transmitted for P- and S-polarisation.

\newpage

\section*{Week 9}

\textbf{Problem 1}

A 4K video has a resolution of 4096 by 2160 pixels, 24 frames per second, and 256 colors per pixel. Calculate the required rate (bits per second) of the communication channel to stream the video with a compression ratio of 75:1.

\textbf{Problem 2}

Find the propagation velocity of a wave in a lossless coaxial cable (R=G=0) with a Teflon insulation (epsilon = 2.1, non-magnetic).

\textbf{Problem 3}

Find the source and load impedances required to avoid back-reflections and maximise the average signal power coupling (V(0) x I(0) / 2) to a lossless coaxial cable with a Teflon insulation and a = 0.75\,mm, b = 2.5\,mm.

\textbf{Problem 4}

Find the power attenuation per 100\,m in a copper coaxial cable for a signal frequency of 1\,GHz. Copper resistivity is $1.7 \times 10^{-8}$\,Ohm\,m.
Inner and outer conductor radii are $a=0.75$\,mm and $b = 2.5$\,mm.
The cable utilises non-magnetic ($\mu = 1$) teflon insulation with $\epsilon = 2.1$ and loss angle $\delta = 10^{-3}$.

\textbf{Problem 5}

Estimate the communication rate over a 100-m-long cable from Ex. 4 if the source power is 10\,W, the detector noise power in the channel bandwidth of 8\,MHz is 0.1\,pW.

\textbf{Problem 6}

Estimate the upload and download rates for a cellphone–satellite channel using the Shannon theorem. A typical emitting power of a cellphone is 0.25\,W, and its antenna is omnidirectional (gain of 1). The emitting power of the satellite is 30\,W, and its antenna is directional with a gain of 200. Both antennas have a receiving noise level of 15\,fW in the communication bandwidth of B = 1\,MHz. The carrier frequency is 2\,GHz, and the satellite is in low-Earth orbit, 550\,km from Earth.

\newpage

\section*{Week 10}

\textbf{Problem 1}

a) Find $h$ and $q$ from the wave equations in the waveguide.

b) Show that $V^2 = u^2 + v^2 = (n_1^2 – n_2^2)\left(\frac{\pi d}{\lambda} \right)^2$, where $u = hd/2$ and $v = qd/2$.

\textbf{Problem 2}

A 1550-nm laser excites four TE modes in a waveguide with $n_1 = 1.6$, thickness $d=10$\,um, and length of 1\,km. Estimate the communication rate that can be achieved using the fibre.

\textbf{Problem 3}

Find the maximum width d of a waveguide with $n_1 = 1.6$, $n_2 = 1.59$ that can support only one TE mode of a 1550-nm laser.

\textbf{Problem 4}

Consider a pulse given by $I(t) = \cos\frac{\pi t}{2b}$ for $–b<t<b$ and 0 elsewhere, b = 10\,ps. Estimate the communication bandwidth of a km-long fibre with D = 17\,ps/nm/km. Assume the laser wavelength of 1550 nm.

\textbf{Problem 5}

Consider a pulse from Ex 4. Estimate the polarisation dispersion-limited communication bandwidth of a 100-km-long fibre with $D_{\rm PMD} = 0.2$\,ps/$\sqrt{\rm km}$.

\textbf{Problem 6}

Find power attenuation over a fibre with a length of 100\,km and an alpha of 0.04/km.

\textbf{Problem 7}

A fibre link works in the C-band (1530$–$1565\,nm). 

a) Estimate the wavelength multiplexing factor in the link if the frequency comb is produced via the Kerr effect in a whispering-gallery-mode silicon nitride ($n_1=2$) resonator of radius 1\,mm.

b) Estimate the minimum input laser power required to produce the frequency comb if the input coupling is $T_1 = 1$\,\%, the effective beam size equals to the internal wavelength, and $n_2 = 2.4 \times 10^{-19}$\,m$^2$/W.

\newpage

\section*{Week 11}

\textbf{Problem 1}

Consider the key generation with parameters:
\begin{itemize}
\item $p = 61$, $q = 53$
\item $n = pq = 3233$
\item $x = {\rm lcm}(p-1, q-1) = 780$
\item $e = 17$
\item $d = 413$
\item Public keys: 3233, 17
\item Private key: 413
\end{itemize}

Show that the encoding and decoding procedures work for $m=65$:
\begin{itemize}
\item Encryption: $c(65) = 65^{17}\,{\rm mod}\,3233 = 2790$
\item Decryption: $m(2790) = 2790^{413}\,{\rm mod}\,3233 = 65$.
\end{itemize}

\textbf{Problem 2}

Alice and Bob communicate with a public channel by encrypting their messages with quantum keys. A key consists of 1024\,bits. Alice has a source of entangled photons that produces entangled pairs at a rate of 10\,MHz. Find how many quantum keys per second Alice and Bob can generate if Bob is located 30\,km away from Alice and the optical loss of the fibre is $4 \times 10^{-2}$\,/km.

\textbf{Problem 3}

A nonlinear crystal converts one photon with a wavelength of 775 nm to two photons. Find the wavelength of the first photon if the wavelength of the second photon is 1560 nm in vacuum.

\textbf{Problem 4}

Find the angle between the signal and idler photons in a type I process if the signal and idler have the same wavelength and the index of refraction of the crystal at the pump wavelength is 1.53 and at the signal wavelength is 1.66.

\textbf{Problem 5}

Prove that the squeezing operator
\begin{equation}
S(r) = \exp \left(r \hat{a}_s^\dagger \hat{a}_i^\dagger - r \hat{a}_s \hat{a}_i \right)
\end{equation}
implements the squeezing transformation on the ladder operators.

\textbf{Problem 6}

Pulses from a 775-nm fs-laser hit a 10-mm-long BBO crystal with the second-order nonlinearity coefficient of 2.5\,pm/V and index of refraction of 1.6. Estimate the rate of production of entangled photon pairs. The pulse width is 200\,fs, the pulse repetition rate is 10\,MHz, the pulse energy is 1\,nJ, and the beam diameter inside the crystal is 1\,mm.

\textbf{Problem 7}

Find the ratio of the quantum key distribution rates if only one source of photons is used and if 32 sources of photons with quantum memories and nondemolition measurements are utilised in a 500-km fibre link with a photon loss rate of 0.04/km.